\documentclass[fleqn,usenatbib]{mnras}

\usepackage{newtxtext,newtxmath}
\usepackage{siunitx}
\usepackage{caption}

\usepackage[T1]{fontenc}
\usepackage{ae,aecompl}
\usepackage{graphicx,epsf}
\usepackage{subfigure}
\usepackage{tablefootnote}
\usepackage{threeparttable}

\usepackage{graphicx}	
\usepackage{amsmath}	
\usepackage{amssymb}	
\usepackage{units}

\newcommand{\Lrad}{L_{\rm rad}}
\newcommand{\Ldep}{L_{\rm dep}}
\newcommand{\LHe}{L_{\rm He}}

\newcommand{\MHe}{M_{\rm He}}
\newcommand{\MNi}{M_{\rm Ni}}
\newcommand{\Msun}{{\rm M}_\odot}
\newcommand{\kms}{\textrm{km}\,\textrm{s}^{-1}}

\title[Helium in Type Iax Progenitor Systems]{Detection of Circumstellar Helium in Type Iax Progenitor Systems}

\author[W. V. Jacobson-Gal\'an]{Wynn V. Jacobson-Gal\'an,$^{1}$\thanks{E-mail: wjacobso@ucsc.edu (UCSC)}
Ryan~J.~Foley$^{1}$,
Josiah~Schwab$^{1,7}$,
Georgios~Dimitriadis$^{1}$,
\newauthor
Shawfeng~Dong$^{2}$,
Saurabh~W.~Jha$^{3}$,
Daniel~Kasen$^{4,5,6}$,
Charles~D.~Kilpatrick$^{1}$,
\newauthor
Rollin~Thomas$^{4}$
\\
\\
$^{1}$Department of Astronomy and Astrophysics, University of California, Santa Cruz, CA 95064, USA\\
$^{2}$Department of Applied Mathematics, University of California, Santa Cruz, CA 95064, USA\\
$^{3}$Department of Physics and Astronomy, Rutgers, the State University of New Jersey, 136 Frelinghuysen Road, Piscataway, NJ 08854 USA\\
$^{4}$Lawrence Berkeley Laboratory, Berkeley, CA, USA\\
$^{5}$Department of Physics, University of California, Berkeley, CA 94720, USA\\
$^{6}$Department of Astronomy and Theoretical Astrophysics Center, University of California, Berkeley, CA 94720, USA\\
$^{7}$Hubble Fellow\\
}

\date{Accepted XXX. Received YYY; in original form ZZZ}

\pubyear{2018}

\begin{document}
\label{firstpage}
\pagerange{\pageref{firstpage}--\pageref{lastpage}}
\maketitle

\begin{abstract}
We present direct spectroscopic modeling of 44 Type Iax supernovae (SNe~Iax) using spectral synthesis code \texttt{SYNAPPS}. We confirm detections of helium emission in the early-time spectra of two SNe~Iax: SNe~2004cs and 2007J. These \ion{He}{i} features are better fit by a pure-emission Gaussian than by a P-Cygni profile, indicating that the helium emission originates from the circumstellar environment rather than the SN ejecta. Based on the modeling of the remaining 42 SNe~Iax, we find no obvious helium features in other SN~Iax spectra. However, $\approx 76\%$ of our sample lack sufficiently deep luminosity limits to detect helium emission with a luminosity of that seen in SNe~2004cs and 2007J. Using the objects with constraining luminosity limits, we calculate that 33\% of SNe~Iax have detectable helium in their spectra. We examine 11 SNe~Iax with late-time spectra and find no hydrogen or helium emission from swept up material. For late-time spectra, we calculate typical upper limits of stripped hydrogen and helium to be $2 \times 10^{-3} \ \Msun$ and $10^{-2} \ \Msun$, respectively. While detections of helium in SNe~Iax support a white dwarf-He star binary progenitor system (i.e., a single-degenerate [SD] channel), non-detections may be explained by variations in the explosion and ejecta material. The lack of helium in the majority of our sample demonstrates the complexity of SN~Iax progenitor systems and the need for further modeling. With strong independent evidence indicating that SNe~Iax arise from a SD channel, we caution the common interpretation that the lack of helium or hydrogen emission at late-time in SN~Ia spectra rules out SD progenitor scenarios for this class. 
\end{abstract}




\section{Introduction}

Type Iax supernovae (SNe~Iax) are a recently defined class of stellar
explosion \citep{foley13} that share similar characteristics to their
common cousins, SNe~Ia. However, SNe~Iax exhibit lower peak
luminosities and ejecta velocities than SNe~Ia. There exist ${\sim}
50$ confirmed SNe~Iax to date \citep{jha17}, all spectroscopically
similar to the prototypical SN~Iax, SN~2002cx \citep{li03}. As is the
case for SNe~Ia, the progenitor system and explosion mechanisms of
SNe~Iax are still unclear.

Due to their spectroscopic agreement with SNe~Ia near maximum light,
SNe~Iax are generally considered to be thermonuclear explosions that
result from a white dwarf (WD) in a binary system. Based on stellar
population estimates \citep{foley09, mccully14}, SNe~Iax are found in
younger stellar populations with ages ${\leq}100$~Myr
(\citealt{lyman13}; \citealt{lyman18}; \citealt{takaro19}). In
terms of a progenitor channel, these short evolutionary timescales
indicate that SNe~Iax may be the result of more massive WDs in a
binary system with a non-degenerate companion such as a He star
\citep{iben94, hachisu99, postnov14, jha17}. This type of system has been
modeled by \citet{liu10} where the final binary stage consists of a $1
\ \Msun$ C/O WD and a $2 \ \Msun$ He star \citep[see also][]{yoon03,
  wang15, brooks16}. The WD then accretes helium from its binary
companion and explodes as it approaches the Chandrasekhar mass
\citep{jha17}. From model predictions, SN~Iax explosions most resemble
pure deflagrations in which the explosion energy is not enough to
completely unbind the star \citep{jordan12, kromer13, fink14}.

The recent pre-explosion detection of a blue point source coincident
with SN~2012Z provides strong support for the C/O WD + He star
progenitor scenario for SNe~Iax \citep{mccully14}. Such a system would
have a significant amount of He on the WD surface, in the outer layers
of the donor, and possibly in the circumstellar environment.
Therefore under certain conditions, one might detect He features in a
SN~Iax spectrum.  In fact, He lines have been detected in the spectra
of two SNe~Iax: SN~2004cs and SN~2007J \citep{filippenko07, foley09,
  foley13, foley16}. Both objects being spectroscopically SNe~Iax,
their spectra display prominent \ion{He}{i} $\lambda\lambda$ 5875,
6678, 7065, and 7281 emission lines and minor \ion{He}{i} features at
bluer wavelengths.

Since the classification of SNe~2004cs and 2007J as SNe~Iax, there
have been arguments put forward that these objects should be
classified as SNe~IIb \citep{white2015}. However, this claim is
addressed in detail by \citet{foley16}, who compare the spectral
epochs of SNe~2007J and 2002cx to that of SNe~IIb, including
SN~1999cb, which had an extremely weak H-$\alpha$ emission at late
times. \citet{foley16} showed noticeable differences between SN~2007J
and SNe~IIb and detect no hydrogen lines in any SN~2007J spectra.
Additionally, \citet{foley16} show that the light curve of SN~2004cs
does not match that of any SN~IIb template, further differentiating
SNe~Iax with helium lines and SNe~IIb.

\begin{figure}

\begin{center}
	\includegraphics[width=0.47\textwidth]{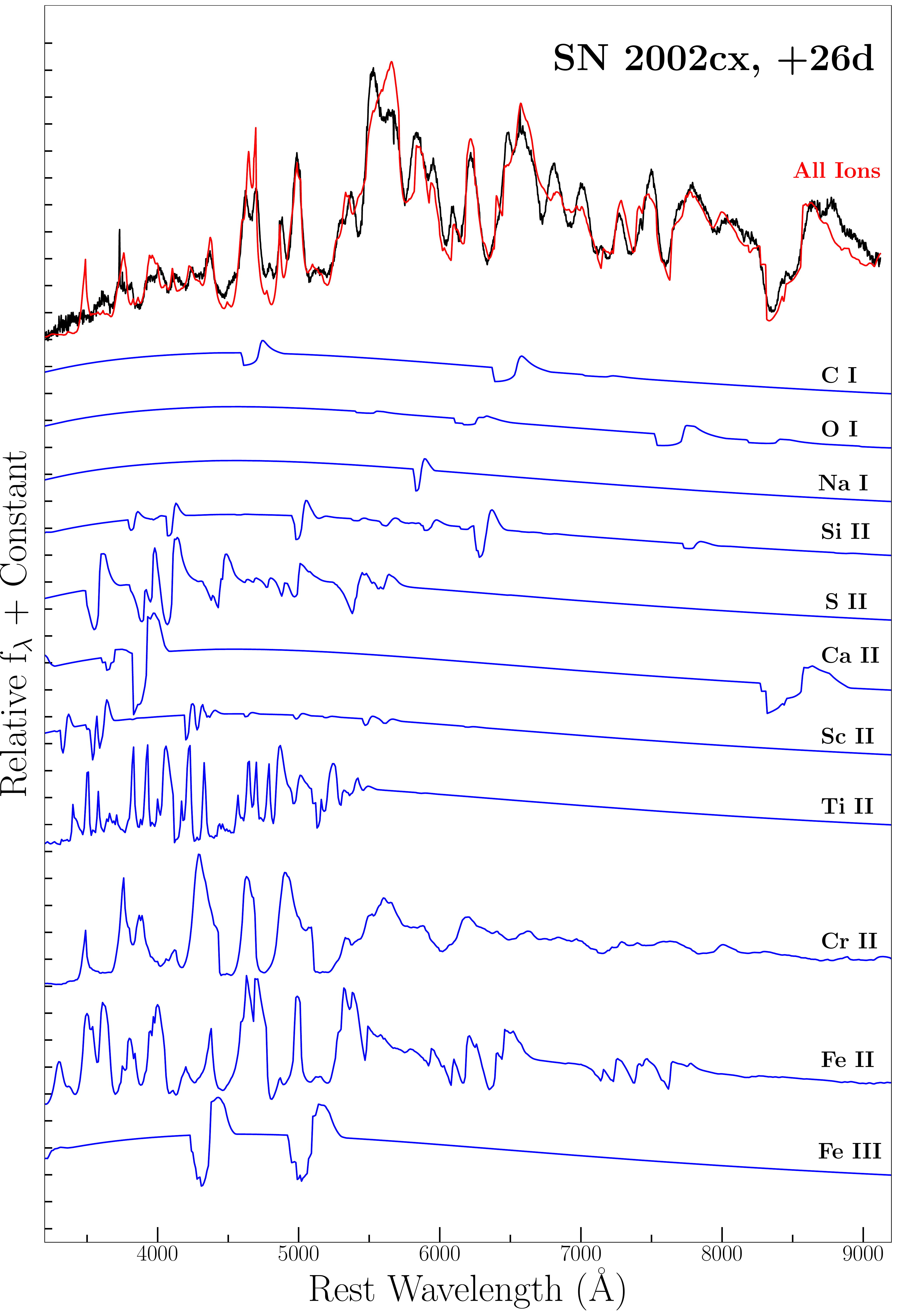}
    
	\caption{Decomposition of all active ions in a \texttt{SYNAPPS} fit shown in blue. In black, spectral data SN~2002cx at +26d after maximum light. \texttt{SYNAPPS} fit to data shown in red.} \label{fig:combo_02cx}
\end{center}
\vspace*{-5mm}
\end{figure}

The detection of He emission in two SNe that, other than their He
emission, are spectroscopically and photometrically similar to other
SNe~Iax suggest that faint, thus-far undetected He features may be
present in spectra of other SNe~Iax. One method for detecting helium
in SNe~Iax is through spectral fitting by radiative transfer
simulations such as \texttt{SYNOW} \citep{parrent14},
\texttt{SYN++/SYNAPPS} \citep{synapps} and \texttt{TARDIS}
\citep{kerzendorf14}. Such an analysis has proven effective in
modeling SN~Iax spectra and is a viable option in determining the
relative strengths of ions in these systems \citep{branch04, jha06, stritzinger14,
  szalai15, barna18}. Furthermore, \cite{magee19} used \texttt{TARDIS} to model the spectra of six SNe Iax, including one spectrum of SN~2007J, to search for signatures of photospheric helium in these objects. \cite{magee19} was able to reproduce the basic \ion{He}{i} features in the first spectral observation of SN~2007J by including significant amounts of helium in the photosphere. We discuss the relevance and implications of these models in Section~\ref{subsec:magee19}. By applying a similar analysis (i.e., parameterized spectral synthesis codes) here, we can comprehensively
search for helium emission for most of the SN~Iax sample.


In Section~\ref{sec:sample}, we introduce our early- and late-time
sample of SNe~Iax. In Section~\ref{sec:method}, we outline our method
for searching for \ion{He}{I} emission in SNe~Iax at early-times. In
Section~\ref{sec:analysis}, we examine the late-time spectra for
hydrogen and helium emission as it relates to the SN~Iax progenitor
system.  We discuss our results in Section~\ref{sec:discussion} and
conclude in Section~\ref{sec:conclusions}.

\section{SN\lowercase{e}~I\lowercase{ax} Sample}
\label{sec:sample}

Through \citet{jha17} and the Open Supernova Catalog
\citep{guillochon17}, we identified 54 spectroscopically confirmed
SNe~Iax, and we use this list to build our sample.  We were able to
obtain 110 spectra of 44 SNe~Iax.  We performed \texttt{SYNAPPS}
fitting for all spectra with a phase of $\lesssim$78~days relative to peak
brightness. We present all spectra with phases and references in Tables \ref{tab:spectra_table1} \& \ref{tab:spectra_table2}. The majority of spectral phases shown are relative to B band maximum. If B band photometry is unavailable for a given object, we use the relations presented in Table 6 of \cite{foley13} to convert all spectral phases to be relative to maximum light in B band. For objects without adequate photometry, we perform a spectral comparison to other SNe Iax using the Supernova Identification package (SNID; \citealt{blondin07}) to estimate each spectral phase. We caution the direct use of spectral phases calculated using SNID due to the small sample size of SNe Iax able to be used in determining these phase range estimates in addition to the overall uncertainty associated with the identification package.

SN~2007J is one such object that does not have as constraining of a light curve as its spectral epochs. For this object, we attempt to match its clear band photometry to the light curve of SN~2004cs in order to calculate a range of possible dates for maximum light. We then use SNID to further constrain the time of maximum when estimating the phases of each spectrum presented.

From this subsample, there are 25~SNe~Iax (and 69 spectra) with sufficient
photometric data to properly flux calibrate the spectra.  We also
examined 24 late-time spectra of 11 SNe~Iax with sufficient photometry
to flux calibrate the spectra.  The late-time spectra range in phase
from 103 to 461~days after peak brightness.

\section{Early-time Helium Emission}
\label{sec:method}
\subsection{Spectroscopic Analysis}
\label{subsec:fitting}

To model the spectral features in early-time, ``photospheric'' SN~Iax
spectra, we apply radiative transfer codes \texttt{SYNAPPS} and
\texttt{SYN++} \citep{synapps}. Both being parameterized spectral
synthesis codes, \texttt{SYNAPPS} performs an automated, $\chi^{2}$
minimization fit to the data while \texttt{SYN++} requires manual
adjustment of available parameters. For each active ion in a given
fit, the parameters used include: optical depth ($\tau$), specific
line velocity limits ($v_{\textrm{min}}/v_{\textrm{max}}$), e-folding
length for the opacity profile, and Boltzmann excitation temperature
($T_{\textrm{exc}}$). General input parameters for
\texttt{SYN++}/\texttt{SYNAPPS} are photospheric velocity
($v_{\textrm{phot}}$), outermost velocity of the line forming region
($v_{\textrm{outer}}$) and blackbody temperature of the photosphere
($T_{\textrm{BB}}$). While \texttt{SYN++}/\texttt{SYNAPPS} are built
on multiple assumptions about the SN explosion such as spherical
symmetry, local thermal equilibrium, and homologous expansion of
ejecta \citep{synapps,parrent14}, they are both excellent tools for
accurate line identification. 

\texttt{SYNAPPS} models photospheric species in LTE and because of its assumption that the lines are formed above an expanding photosphere, the code produces P-Cygni profiles. Because of its general assumptions about the modeled explosion, \texttt{SYNAPPS} may at times be unable to reproduce the line strengths of ions within the photosphere. However, as discussed above, its accuracy in modeling specific P-Cygni line profiles allows for direct identification of all photospheric species present within a given SN. Furthermore, because of its LTE condition, \texttt{SYNAPPS} cannot accurately reproduce the relative fluxes in NLTE species such as \ion{He}{i}, but should be able to reproduce the profile shape. A complete treatment of the NLTE \ion{He}{i} species requires specific ionization conditions in order for accurate line strength reproduction through radiative transfer \citep{hachinger12, dessart15, boyle17}.

For our spectral modeling, we primarily use \texttt{SYNAPPS} in order
to find the best, unbiased fit to the data. Each free parameter discussed above has a range of values that \texttt{SYNAPPS} fits freely without prior constraint. The ranges for each parameter are as follows: $v_{\textrm{min}} = 0.20-5 \times 10^3 \ \kms$, $v_{\textrm{max}} = 5-15 \times 10^3 \ \kms$, $v_{\textrm{phot}} = 1-15 \times 10^3 \ \kms$, $v_{\textrm{outer}} = 1-20 \times 10^3 \ \kms$, $T_{\textrm{BB}} = 1-15 \times 10^3$ K, \ $\tau = 10^{-3} - 10^{5} $, \ aux $= 0.10 - 15$, and $T_{\textrm{exc}} = 1-25 \times10^3$ K. Multiple iterations are performed for a given spectrum if a local minimum appears to be reached rather than a global minimum. In this case, we test a range of starting values for each parameter to confirm that their convergence to a given value is authentic. In each fit, we include
the following ions: \ion{C}{ii}, \ion{O}{i}, \ion{Na}{i},
\ion{Si}{ii}, \ion{S}{ii}, \ion{Ca}{ii}, \ion{Sc}{ii}, \ion{Ti}{ii},
\ion{Cr}{ii}, \ion{Fe}{ii} and \ion{Fe}{iii}. All ions are weighted equally and are given identical upper and lower boundaries for each free parameter when fitting. An example of spectral
decomposition of fitted ions is shown in Figure~\ref{fig:combo_02cx}.

\subsection{Helium Line Profile Shape}
\label{subsec:emission_absorption}

\begin{figure}
\begin{center}
	\includegraphics[width=0.49\textwidth]{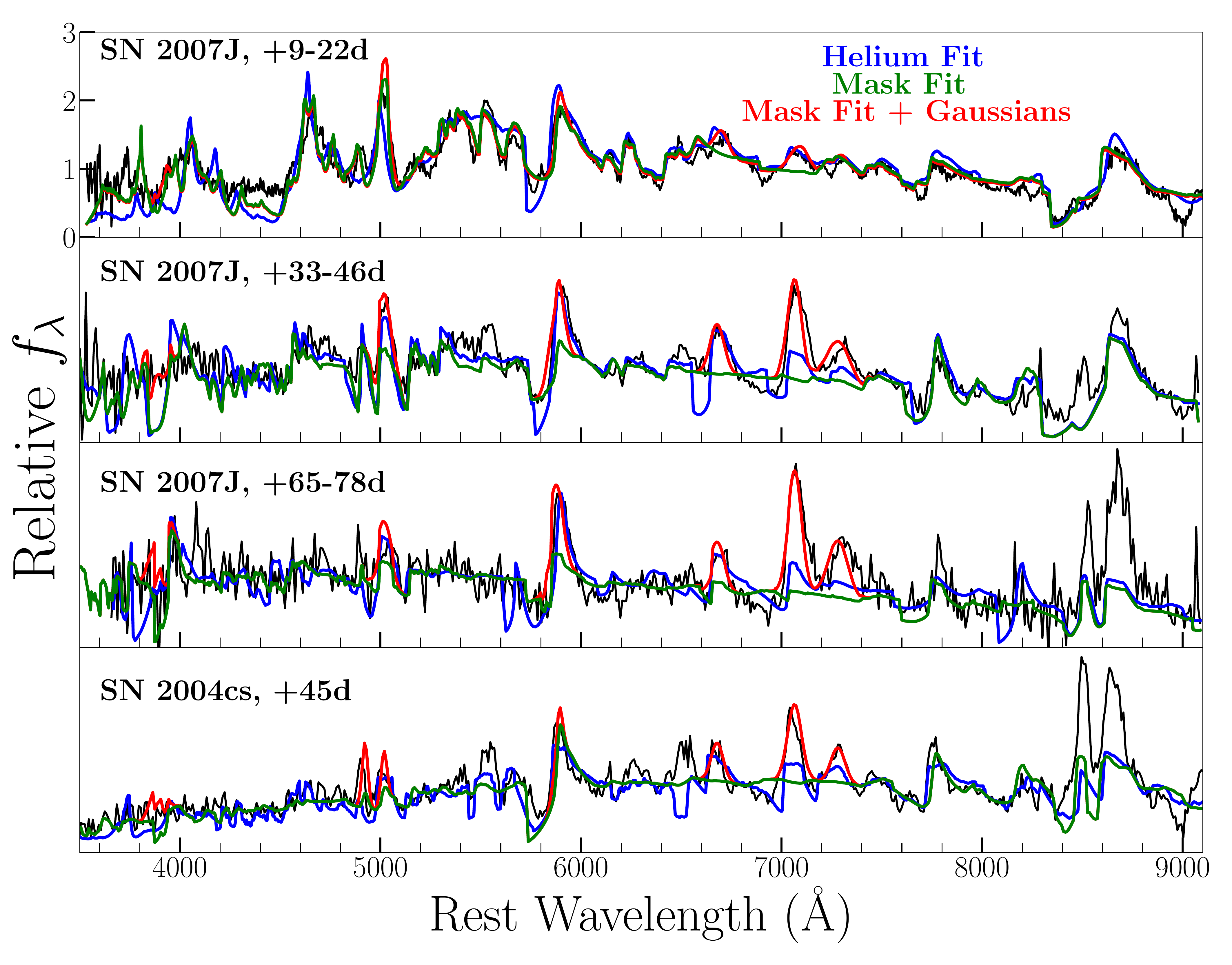}
	\caption{In blue, \texttt{SYNAPPS} fit with active \ion{He}{i} ion. In red, \texttt{SYNAPPS} fit with masked out \ion{He}{i} regions plus Gaussian emission profiles.} \label{fig:abs_vs_emission}
\end{center}
\end{figure}

\begin{figure*}

\subfigure[]{\includegraphics[width=\textwidth]{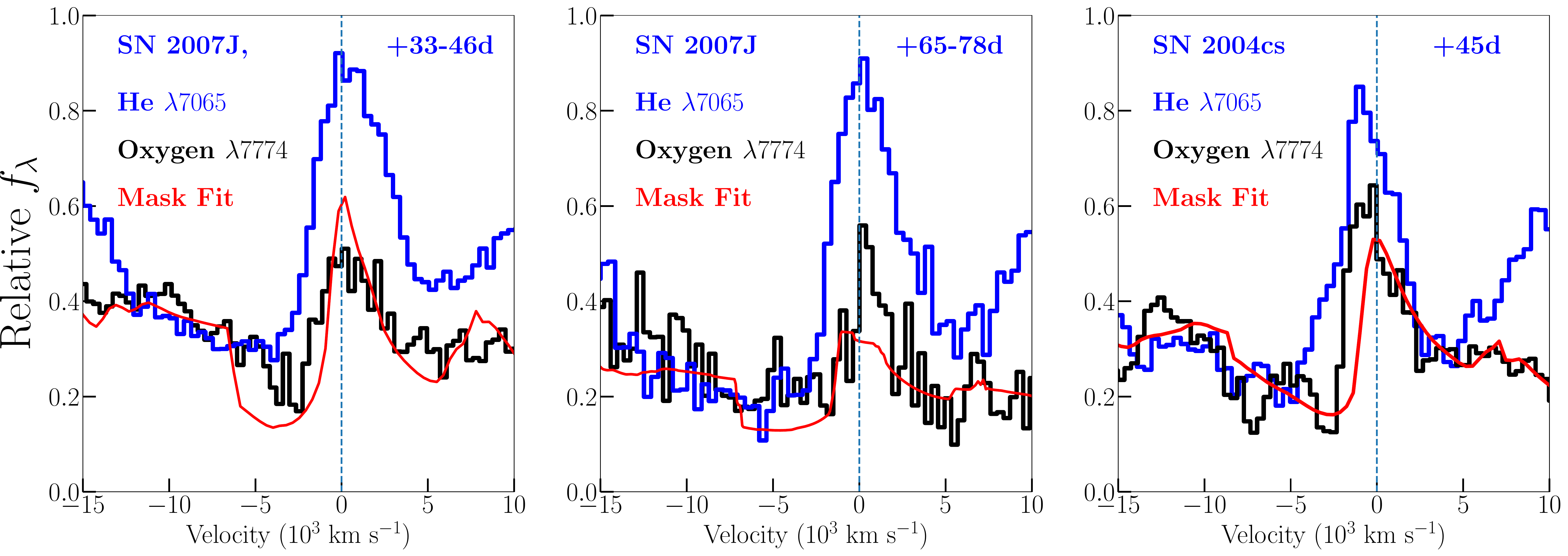}}
\subfigure[]{\includegraphics[width=\textwidth]{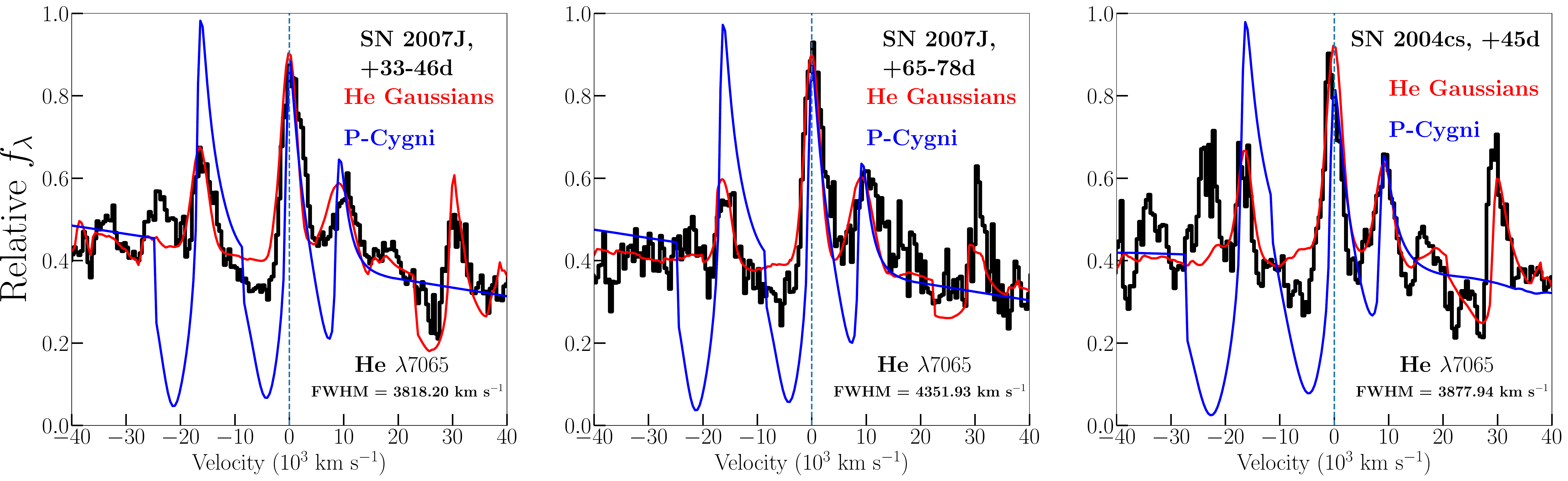}}

\caption{\textbf{(a)} Histogram velocity plot of He $\lambda 7065$ and O  $\lambda 7774$ spectral lines for comparison of absorption profiles. Red line is \texttt{SYNAPPS} mask fit plus helium Gaussians, centered on the O  $\lambda 7774$ line. \textbf{(b)} $\lambda 7065$ helium line velocity for SN~2007J and SN~2004cs. Red line shows \texttt{SYNAPPS} mask fit with all active ions plus Gaussian profiles at prominent \ion{He}{i} lines. Blue line is manually generated \texttt{SYN++} fit to the He $\lambda 7065$ line with no other active ions present. Refer to the blue line in Figure \ref{fig:abs_vs_emission} for the \texttt{SYNAPPS} fit with all active ions and helium. \label{fig:vels_combo} }
\end{figure*}

We first examine the \ion{He}{i} features in the spectra of SNe~2004cs
and 2007J to further understand what might be present in other
spectra.  When fitting these spectra with \texttt{SYNAPPS}, we noticed
that the \ion{He}{i} features were poorly fit. In particular, the
P-Cygni absorption should be much stronger (or the emission should be
much weaker) than observed. Furthermore, the \ion{He}{i} features are
more symmetric, relative to zero velocity, than other features in the
spectrum. While the synthetic \texttt{SYNAPPS} spectra (blue line in
Figure \ref{fig:abs_vs_emission}) are generally reasonable matches to
the data, the helium P-Cygni profiles cannot reproduce the large
emission features for \ion{He}{i} $\lambda\lambda$7065,7281 --- these
helium lines are best fit with Gaussian emission features. Other
species (e.g., \ion{O}{i}; see Figure~\ref{fig:vels_combo}) are,
however, well matched to a P-Cygni profile.

For reference, we plot the velocities of \ion{O}{i} $\lambda$7774 and
\ion{He}{i} $\lambda$7065 in Figure~\ref{fig:vels_combo}(a).  We show
that the absorption profile of oxygen is best represented by a P-Cygni
profile while \ion{He}{i} is best described by a Gaussian for the
$+33-46$-day spectrum of SN~2007J and the +45-day spectrum of SN~2004cs. We
also note that for SN~2007J at a later phase of $+65-78$~days, both
\ion{O}{i} and \ion{He}{i} $\lambda$7065 are best defined as Gaussian
profiles as the SN ejecta becomes more optically thin.

Because a standard \texttt{SYNAPPS} fit cannot reproduce the
helium emission observed, we can fit the spectra with certain regions
masked, where \texttt{SYNAPPS} will predict the spectrum in the masked
regions.  This process allows for a way to examine the complicated
continuum at the position of helium features without being influenced
by any possible helium emission or absorption. This is applied as we
fit the SNe~2004cs and 2007J spectra with \texttt{SYNAPPS}, masking
the \ion{He}{i} features, and then including additional Gaussian line
profiles for each \ion{He}{i} feature. The result is presented as the
red line in Figure~\ref{fig:abs_vs_emission}.  The resulting profile
parameters are presented in Table \ref{tab:line_table}.

As an example, we plot the velocities of the 7065\AA\ \ion{He}{i} line
for both objects at multiple epochs in Figure~\ref{fig:vels_combo}(b).
Gaussian emission and P-Cygni profiles are over-plotted in red and
blue, respectively.  We see that the Gaussian profiles are a better
description for all \ion{He}{i} lines in both objects. In some cases,
the P-Cygni profiles are also too narrow to fit the \ion{He}{i}
emission.

Furthermore, the peaks of the \ion{He}{i} lines in SNe~2004cs and
2007J are both redshifted and blueshifted about rest velocity. This
indicates that the helium emitting region may be kinematically
decoupled from the SN ejecta and that the system's helium is not
constrained to the photosphere as is needed for a P-Cygni profile.
This then strengthens the case for pure \ion{He}{i} emission
manifesting as a Gaussian line profile in both SNe~Iax. All helium
line widths and shifts are presented in Table \ref{tab:line_table}.

\begin{table*}
	\caption{Helium Line Luminosities and Velocities for SNe~2007J and 2004cs}
	\label{tab:line_table}
	\begin{tabular}{cccccccccccc}
		\hline
		Object & Phase & $\lambda 5876$ Lum. & $\lambda 6678$ Lum. & $\lambda 7065$ Lum. & $\lambda 7065$ Lum. & FWHM $\lambda 5876$ & FWHM $\lambda 6678$\\
		\ & (days) & ($10^{38} \rm{ergs \ s}^{-1}$) & ($10^{38} \rm{ergs \ s}^{-1}$) & ($10^{38} \rm{ergs \ s}^{-1}$) & ($10^{38} \rm{ergs \ s}^{-1}$) & ($\kms$) & ($\kms$) \\
		\hline
		SN~2007J & 5-18 & 1.027 & 3.364 & 7.893 & N/A & 458.8 & 887.6 \\
		SN~2007J  & 9-22 & 15.50 & 10.16 & 16.97 & 2.214 & 8367 & 4402 \\
		SN~2007J  & 33-46 &12.37 & 8.453 & 19.25 & 9.979 & 6918 & 4673 \\
		SN~2007J &  65-78 & 2.319 & 1.532 & 7.135 & 4.389 & 4558 & 3879 \\
		SN~2004cs & 45 & 4.389 & 4.066 & 10.21 & 5.523 & 3465 & 2933 \\
		\hline
		\hline
		Object & Phase & Shift $\lambda 5876$ & Shift $\lambda 6678$ & Shift $\lambda 7065$ & Shift $\lambda 7281$ & FWHM $\lambda 7065$ & FWHM $\lambda 7281$\\
		\ & (days) & ($\kms$) & ($\kms$) & ($\kms$) & ($\kms$) & ($\kms$) & ($\kms$)\\
		\hline
		SN~2007J & 5-18 & 27.48 & -204.1 & -270.8 & N/A & 1407 & N/A\\
		SN~2007J  & 9-22 & 2986 & 1337 & 397.4 & 1305 & 4193 & 3735\\
		SN~2007J  & 33-46 & 2385 & 691.2 & 744.6 & 1121 & 4352 & 3596\\
		SN~2007J & 65-78 & 1708 & 839.9 & 588.9 & 1201 & 4725 & 4379\\
		SN~2004cs & 45 &938.6 & -294.4 & -449.8 & -72.11 & 3878 & 4446\\
		\hline
	\end{tabular}
\end{table*}

\begin{figure}
\begin{center}
	\includegraphics[width=0.49\textwidth]{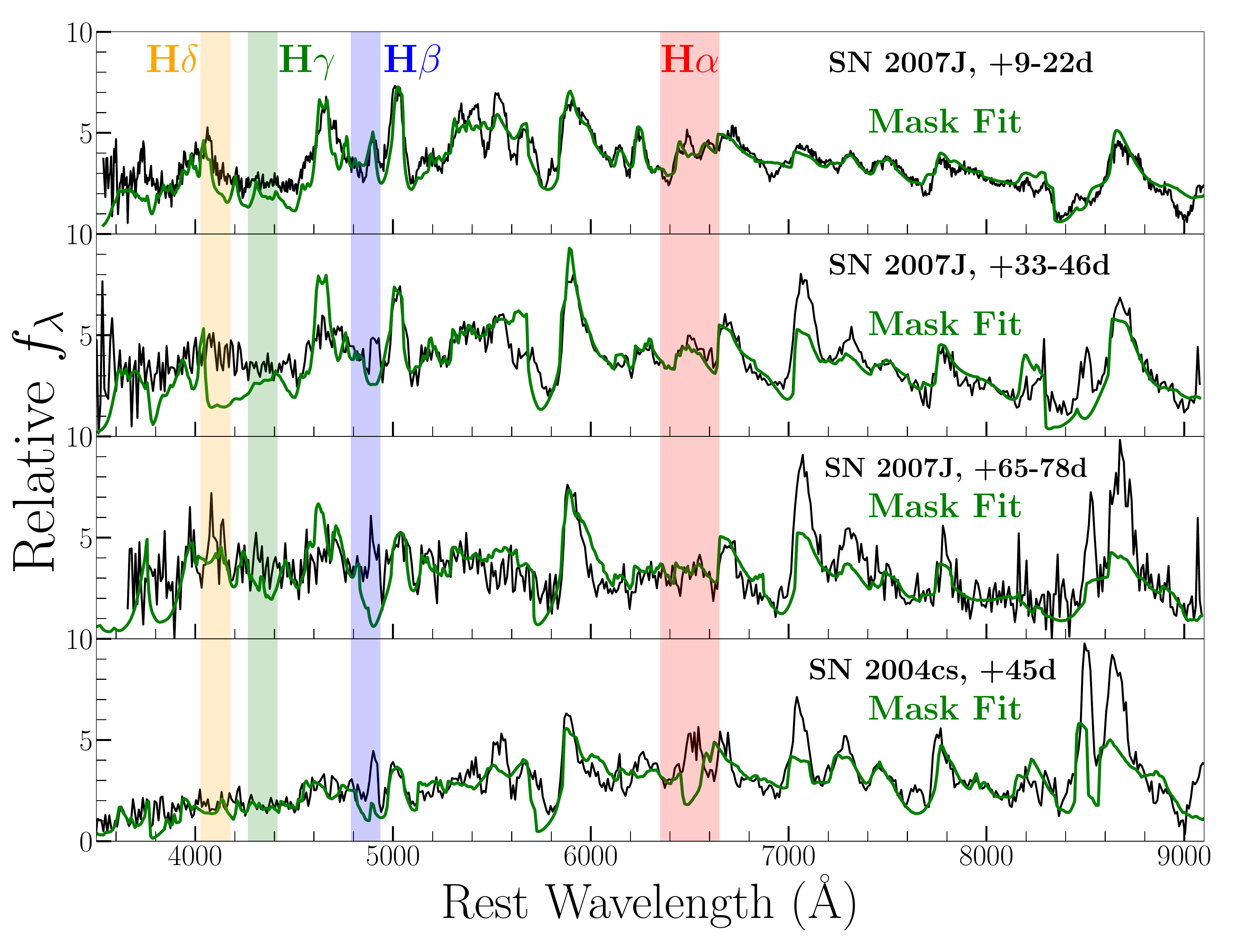}
	\caption{In green, \texttt{SYNAPPS} fits to all epochs of SN~2004cs and SN~2007J with prominent Balmer lines masked out. Region about \ion{H}{$\alpha$} shown in red, \ion{H}{$\beta$} in blue, \ion{H}{$\gamma$} in green, and \ion{H}{$\delta$} in orange. } \label{fig:balmer_lines}
\end{center}
\end{figure}

To test the \citet{white2015} suggestion that SNe~2004cs and 2007J are
SNe~IIb, we perform \texttt{SYNAPPS} fits with prominent Balmer lines
masked out. We fit the $+9-22$, $+33-46$, and $+65-78$-day spectra for SN~2007J and
the +45-day spectrum of SN~2004cs. These fits include all ions listed
in Section~\ref{subsec:fitting} as well as \ion{He}{i}. From
Figure~\ref{fig:balmer_lines}, we see no hydrogen feature in any
spectrum (excluding contaminating galactic H-$\alpha$ emission feature
for SN~2004cs).

\subsection{Fitting Early-time Spectra}
\label{subsec:lum_limits}

\begin{figure}
\begin{center}
	\includegraphics[width=0.49\textwidth]{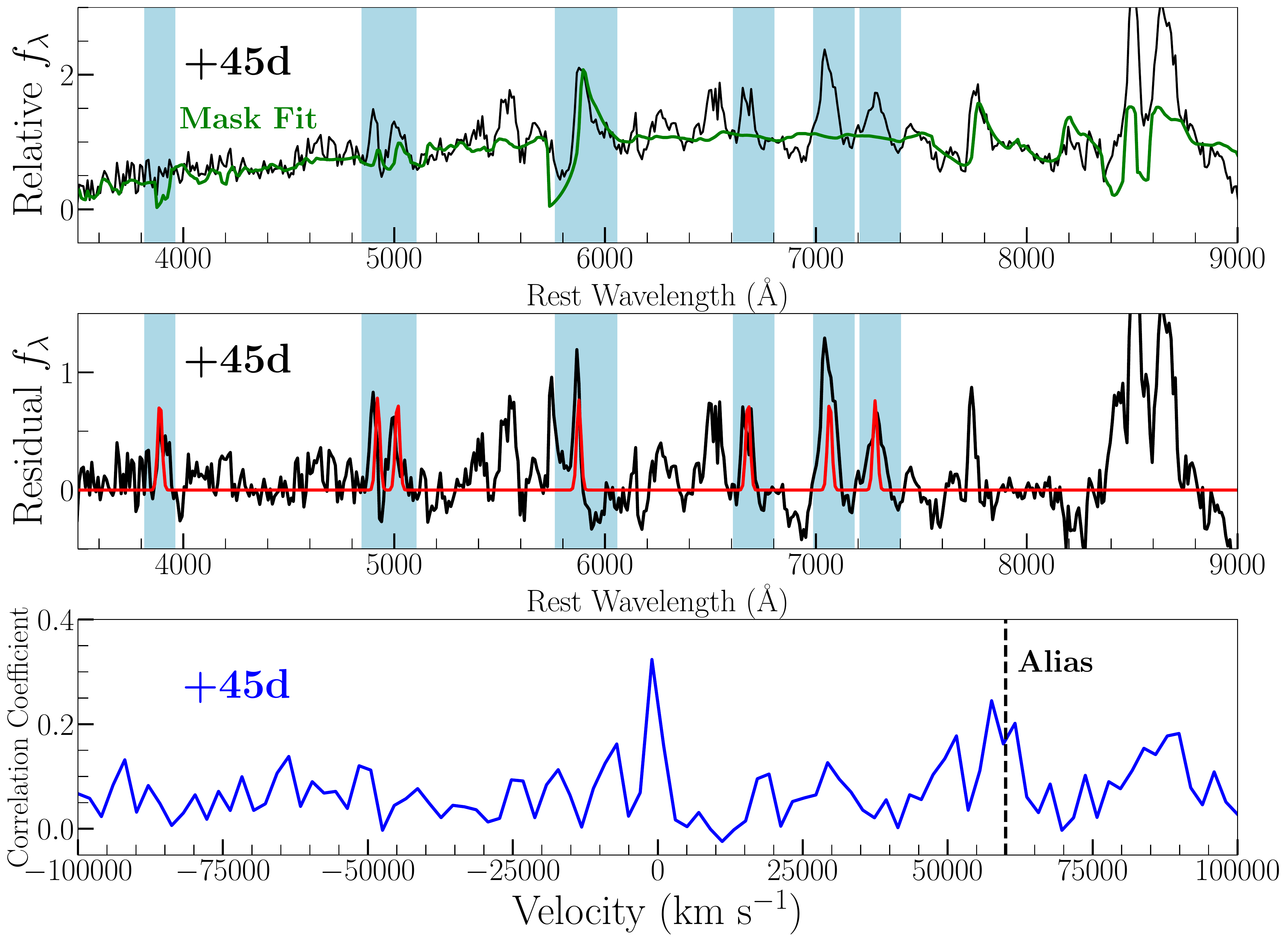}
	\caption{\textit{Top:} \texttt{SYNAPPS} fit to SN~2004cs spectral data at +45d. Phase relative to clear band maximum. Blue regions are masked out in fit shown in green. \textit{Middle:} In black, residuals of data and mask fit. In red, Gaussian profiles centered at prominent \ion{He}{i} emission lines. \textit{Bottom:} Cross correlation of Gaussian profiles and residual flux. Alias correlations marked by dashed black lines at $60$,000~$\kms$. } \label{fig:combo_04cs}
\end{center}
\end{figure}

We now turn to fitting all available early-time SN~Iax spectra to
determine if any have \ion{He}{i} emission.  Since the helium features
in the SNe~2004cs and 2007J spectra have Gaussian line profiles, we
assume that features in other spectra would be similar.  To properly
fit for these features, we mask out regions around each prominent
\ion{He}{i} emission line and only fit the portions of the spectrum
without potential \ion{He}{i} emission. The wavelength ranges that are
applied as masks in each fit are shown in blue in
Figure~\ref{fig:combo_04cs}.  Each given fit to the data with masked
out regions are plotted as green lines in
Figures~\ref{fig:02cx_combo}-\ref{fig:15H_combo}.

Once our spectral models are complete, we subtract the mask fit
spectrum from the data and examine the residuals in each wavelength
region where helium emission is known to occur. We then perform a
visual inspection of each residual spectrum for obvious helium
emission lines. Because pure emission lines can be represented as
Gaussian profiles centered at a given central wavelength, we also
cross-correlate the data residuals with a set of Gaussians (all with
the same height and a superficial FWHM of 20~\AA \ to prevent overlap of \ion{He}{i} lines) centered at the rest wavelength
of prominent optical \ion{He}{i} lines: 3888.65, 4921.93, 5015.68,
5875.63, 6678.15, 7065.19, and 7281.35~\AA.  We cross-correlate by
shifting the Gaussians for range of velocities.  While the height and
width of each Gaussian affects the specific correlation coefficient,
these choices do not impact the relative correlation coefficient for a
range of velocities. Furthermore, we tested the effect of binning on each analyzed spectrum and find that it does not impact the relative correlation coefficient. Examples of prominent residual emission in
SNe~2004cs and 2007J are shown in Figures~\ref{fig:combo_04cs}
and~\ref{fig:07J_combo}, respectively.

\begin{figure}
\begin{center}
	\includegraphics[width=0.49\textwidth]{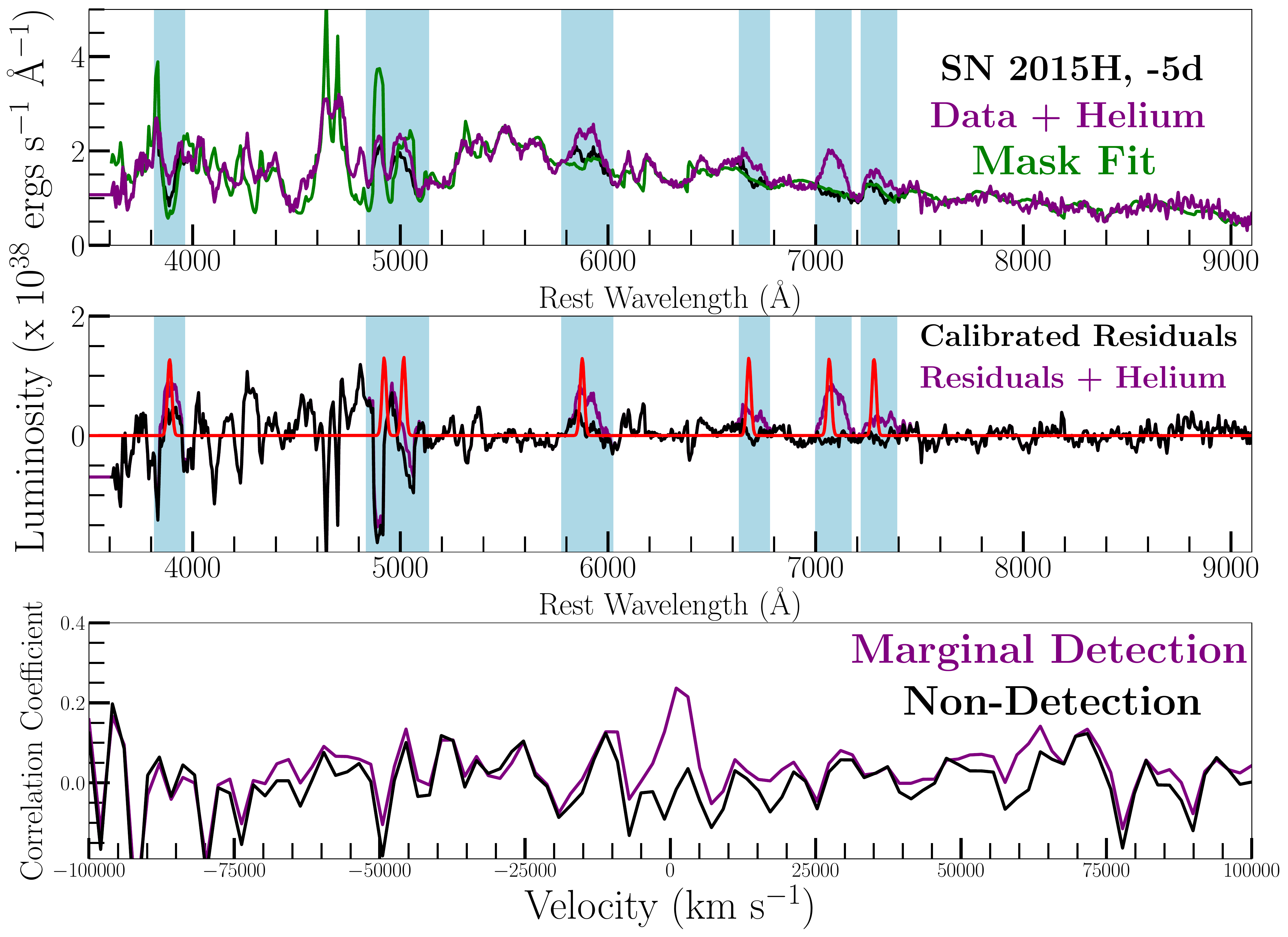}
	\vspace*{-5mm}
	\caption{\textit{Top:} In black, calibrated spectrum of SN 2015H at -5d with \texttt{SYNAPPS} mask fit shown in green. In purple, calibrated spectrum with \ion{He}{i} lines from SN~2007J added in. \textit{Middle:} In black, residuals of SN~2015H and mask fit. In purple, residuals of mask fit and SN~2015H plus \ion{He}{i} lines. In red, Gaussian profiles centered at prominent \ion{He}{i} emission lines. \textit{Bottom:} Cross correlation of Gaussian profiles and residual luminosity for original spectrum (black) and spectrum with helium added (purple). Purple line shows a positive detection of helium. } \label{fig:15H_helium_limit}
\end{center}
\end{figure}

To determine if a spectrum has a significant detection of helium
emission, we cross-correlate all residual spectra with the helium line
Gaussian profiles described above.  We use a window of
${\pm}10^{6}$~$\kms$, which is much larger than physically possible
for helium lines, to estimate the false-detection level.  We define a
significant detection if the correlation coefficient function has its
global maximum within 1000~$\kms$ of zero velocity. The numerical peak in the correlation coefficient function is variable between spectra and thus needs to be analyzed on a case-by-case basis once a visually significant peak is detected with the cross-correlation. At certain velocities that correspond to distinct velocity shifts between the \ion{He}{i} lines, there can be a large correlation coefficient.  We
examine these aliased peaks in detail, and if there is both a peak
near zero velocity and at an aliased velocity, we examine the spectrum
in more detail.

Using this method, we robustly detect \ion{He}{i} features in the
$+9-22$, $+33-46$, and $+65-78$-day spectra of SN~2007J
(Figure~\ref{fig:07J_combo}c) and the single +45-day spectrum of
SN~2004cs (Figure~\ref{fig:combo_04cs}).

For the $+9-22$-day spectrum, there is a positive correlation peak around
$-90$,000~$\kms$ (marked by a dashed line in
Figure~\ref{fig:07J_combo}c) that is similar in strength to the peak
near zero velocity.  This correlation is caused by aliasing between
the different helium lines (as well as other features in the residual
spectrum).  Examining the spectrum in detail, there is clear emission
corresponding to the \ion{He}{i} 6678.15 and 7065.19 lines.  We
therefore consider the main peak at zero velocity to be genuine and
the detection of \ion{He}{i} to be significant.

For the $+5-18$-day SN~2007J spectrum, we do not detect \ion{He}{i}
features by either visually inspecting the residual spectrum or by
examining the correlation coefficients.  \citet{filippenko07} first
saw the increasing helium emission with phase.

Examining all additional early-time spectra, we do not detect any
helium lines for any of the additional 42 SNe~Iax examined.  We
present the all spectra and accompanying fits in
Appendix~\ref{sec:all_spec}.

Since these spectra do not have any detected \ion{He}{i} lines, we
estimate the luminosity limit for these features.  To do this, we
first scale the flux in each spectrum to match the corresponding
broad-band photometry as determined by images, and then add helium
features until the features can be detected in each spectrum.  To do
this, we add the $33-46$-day SN~2007J residual spectrum, which has
particularly high signal-to-noise ratio \ion{He}{i} features, to each
spectrum. This residual spectrum has the flux at all wavelengths other than those coincident with the \ion{He}{i} features set to zero and is a representative helium emission spectrum for SNe Iax. Each spectrum's original continuum and features remain unchanged when this helium emission spectrum is added. We note, however, that this specific helium residual spectrum from SN~2007J is not ideal for such a method due to its difference in phase from other SNe Iax in the sample. However, because these differences would at most make the helium more difficult to detect, the $33-46$-day residual spectrum of SN~2007J is a conservative upper limit for helium detection in other objects.

\begin{figure*}
\begin{center}
	\includegraphics[width=\textwidth]{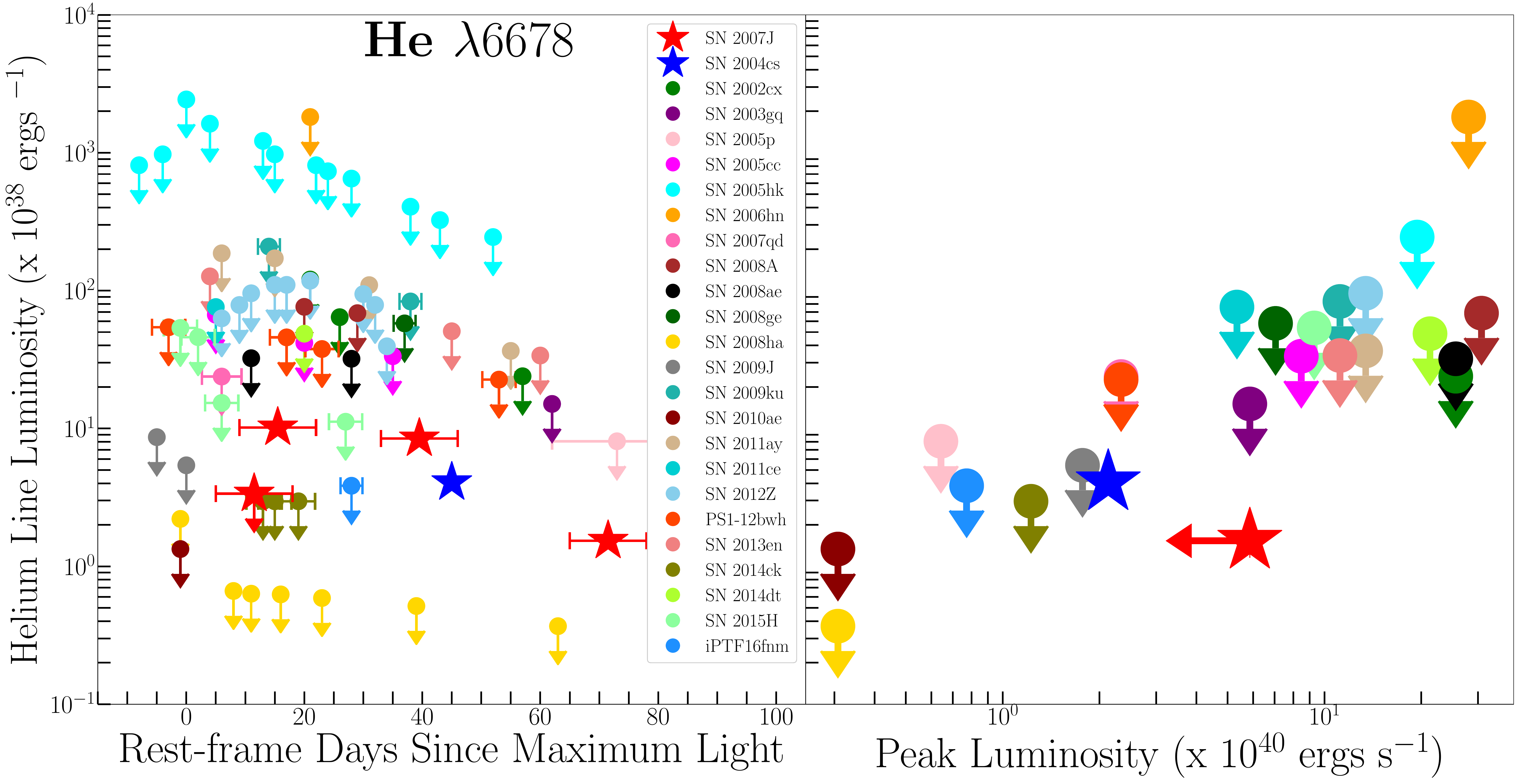}
	\caption{\textit{Left:}  Luminosity limits needed for detection of $6678$\AA\ helium line. All SNe~Iax shown are flux calibrated. True detections in SNe~2004cs and 2007J are shown as blue and red stars, respectively.  \textit{Right:} Most constraining luminosity limits of $6678$\AA\ line with respect to peak luminosity.} \label{fig:lum_limits}
\end{center}
\end{figure*}

\begin{figure*}

\subfigure[]{\includegraphics[width=.48\textwidth]{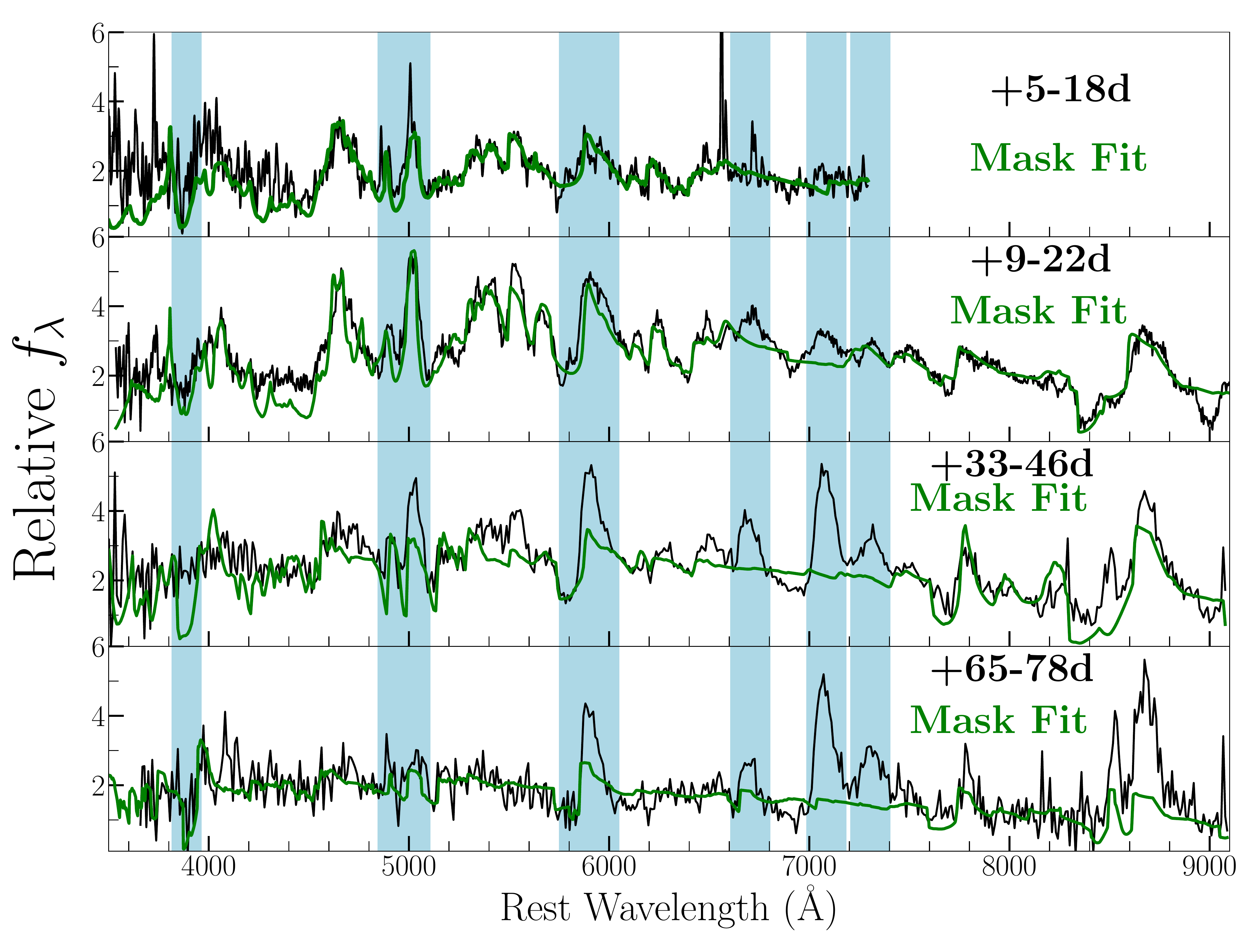}}
\subfigure[]{\includegraphics[width=.48\textwidth]{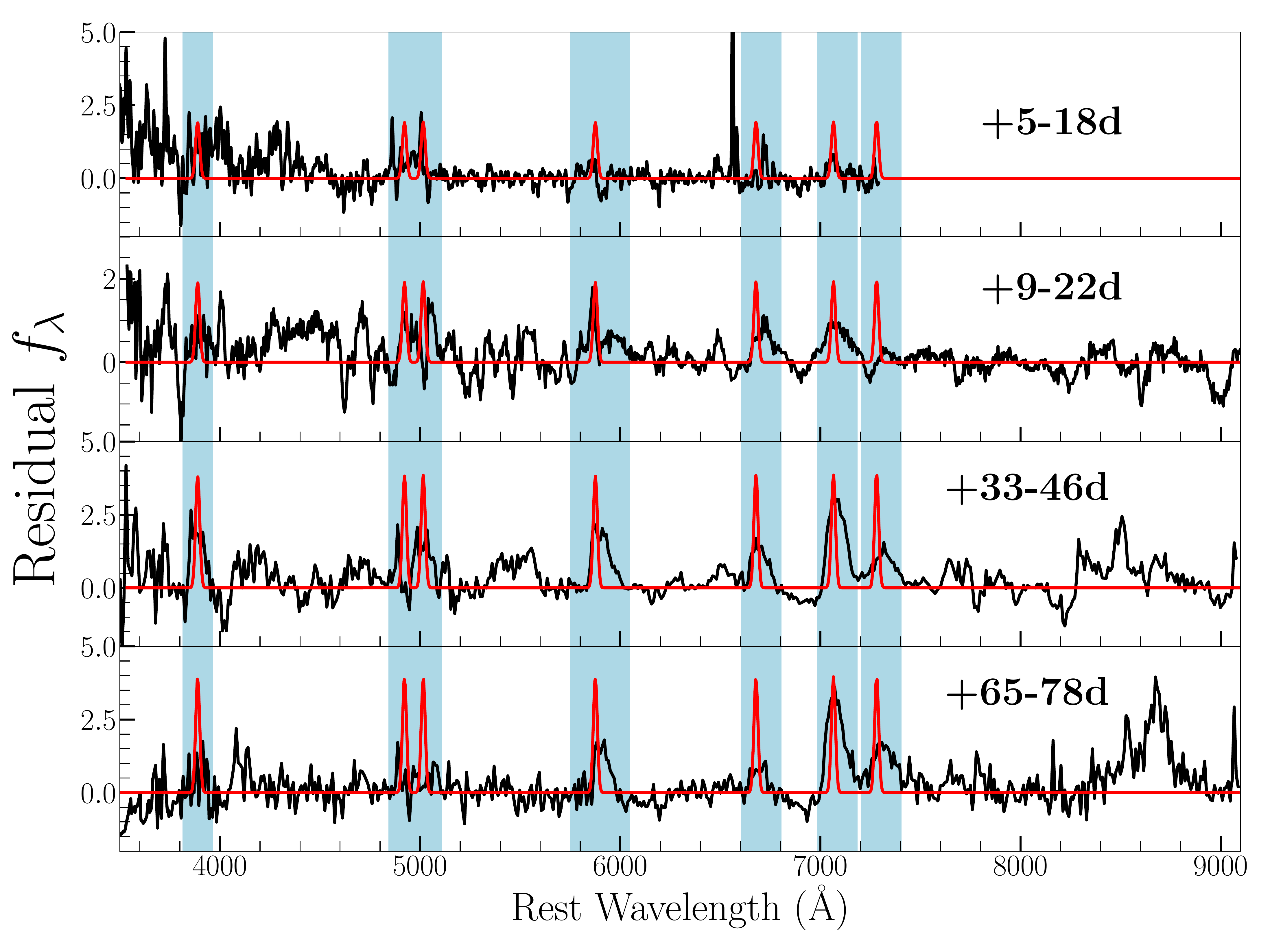}}\\[1ex]
\subfigure[]
{\includegraphics[width=0.6\textwidth]{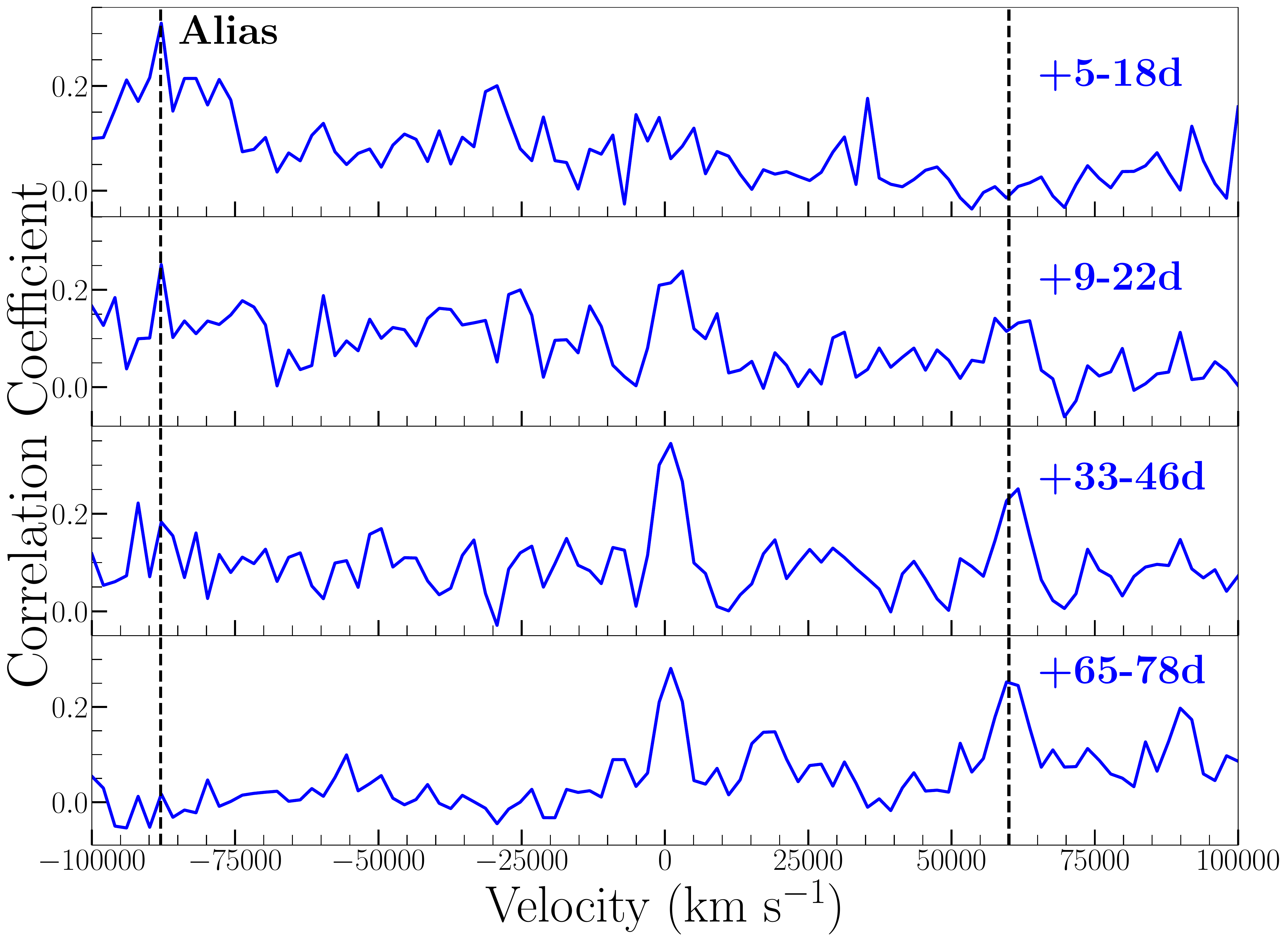}}
\caption{\textbf{(a)} \texttt{SYNAPPS} fits to SN~2007J spectral data at $+5-18$, $+9-22$, $+33-46$ and $+65-78$d. Phase relative to clear band maximum. Blue regions are masked out in fit shown in green. \textbf{(b)} In black, residuals of data and mask fit. In red, Gaussian profiles centered at prominent \ion{He}{i} emission lines. \textbf{(c)} Cross correlation of Gaussian profiles and residual flux. Alias correlations marked by dashed black lines at $-88$,000 $\&$ $60$,000~$\kms$. \label{fig:07J_combo} }
\end{figure*}

Once the helium residuals are added, we then scale the flux of the SN~2007J residual spectrum until
the features are detected, and record the corresponding line flux as
the detection limit. Same as the true detections in SNe 2004cs and 2007J, a detection here is also defined as a peak in the correlation coefficient function within 1000~$\kms$ of zero velocity. An example of this calculation for a marginal detection (i.e. same as $+9-22$d detection in SN~2007J) of helium is shown for SN~2015H in
Figure~\ref{fig:15H_helium_limit}.  The luminosity limits for helium
detection in the 6678.15 $\mbox{\AA}$ line are shown in
Figure~\ref{fig:lum_limits}. Since we use a single spectrum to
determine the luminosity limits, limits for all lines are perfectly
correlated.  The lines have scaling factors of L$_{\lambda5876}$ =
1.47 $\cdot$ L$_{\lambda6678}$ and L$_{\lambda7065}$ = 2.33 $\cdot$
L$_{\lambda6678}$.

As seen in Figure~\ref{fig:lum_limits}, SNe~2004cs and 2007J have
detected helium at a luminosity significantly below the limits for
most SNe~Iax in our sample.  Additionally, as seen for SN~2007J, the
helium luminosity is expected to vary with phase.  Combined, we can
use the detection luminosity as a function of phase to define an
envelope where we would expect to detect helium features similar to
those seen in SNe~2004cs and 2007J.  Of the 23 SNe~Iax examined, 17
do not have sufficiently deep limits to rule out helium features as
luminous as the SN~2004cs and 2007J features.  As such, conclusions
drawn from those SNe are limited.

\begin{figure}
\begin{center}
	\includegraphics[width=0.45\textwidth]{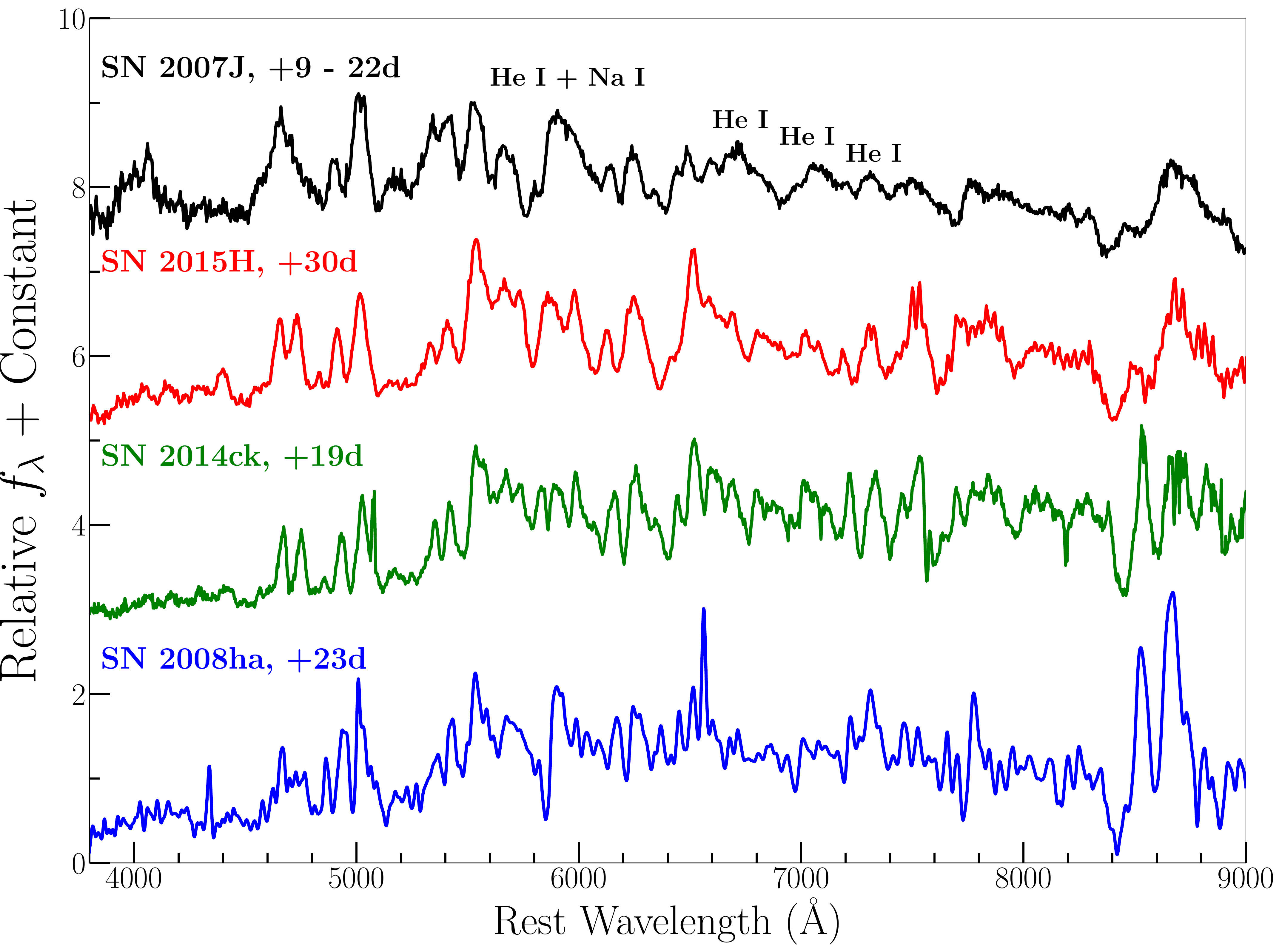}
	\caption{Spectral comparison of SN~2007J, SN~2008ha, SN~2014ck
          and SN~2015H at approximately the same phase. The spectra
          are ordered by decreasing \ion{He}{i} luminosity limits.} \label{fig:compare_07J_08ha_14ck_15H}
\end{center}
\end{figure}

However, 4 SNe~Iax, SNe~2008ha, 2014ck, 2015H and iPTF16fnm,
have spectra over the appropriate phase range and with limits as deep
or deeper than the \ion{He}{i} luminosity detections for SNe~2004cs
and 2007J.  Therefore these 6 SNe~Iax represent the sample of SNe~Iax
where we can directly compare to known detections. The spectra of SNe 2008ha, 2014ck and 2015H at phases near $\sim22$ days are presented in Figure \ref{fig:compare_07J_08ha_14ck_15H} for comparison to the $9-22$-day epoch of SN~2007J.\\ 

In Figure~\ref{fig:lum_limits}, we compare the \ion{He}{i}
$\lambda$6678 luminosity to the SN~Iax peak luminosity\footnote{The
  peak luminosity is occasionally calculated from single photometry
  points.  The SNe have their peak luminosity calculated in a variety
  of filters. Combined, the luminosity presented is not especially
  precise and should be viewed with some caution.}  \citet[and
references therein]{jha17}.  There is a general trend between the peak
luminosity and the \ion{He}{i} luminosity limit.  This is primarily
driven by the signal-to-noise ratio of the spectra (which is usually
similar for all spectra) and the ability for \texttt{SYNAPPS} to
accurately reproduce a spectrum.

\section{Search for Late-time Emission Features}
\label{sec:analysis}

In single-degenerate (SD) progenitor models for SNe~Ia, we expect the
SN to sweep up material ablated from a close companion star and/or in
the circumstellar environment.  This material should emit and be
detectable in late-time spectra as narrow emission lines
\citep{marietta00, mattila05, liu10, liu12, liu13, pan12,
  lundqvist13}. While there have been no detected hydrogen or helium
emission lines in any late-time SN~Ia spectrum
\citep[e.g.,][]{mattila05, leonard07, shappee13, lundqvist13, maguire16, graham17, sand18, shappee18, dimitriadis18},
we attempt to calculate upper limits for the amount of stripped
hydrogen and helium mass present in SNe~Iax late-time spectra.

Upon visual inspection of all late-time SN~Iax spectra, we find no
obvious hydrogen or helium narrow-line emission. As in
\citet{lundqvist15}, we also examine each late-time spectra for
\ion{Ca}{ii} and \ion{O}{i} emission but find no visual detections.

Below, we provide quantitative mass limits on swept-up hydrogen and
helium material.

\subsection{Stripped Hydrogen in Late-time Spectra}
\label{subsec:neb_spectra_h}

\begin{figure}
\subfigure[]{\label{fig:a}\includegraphics[width=.44\textwidth]{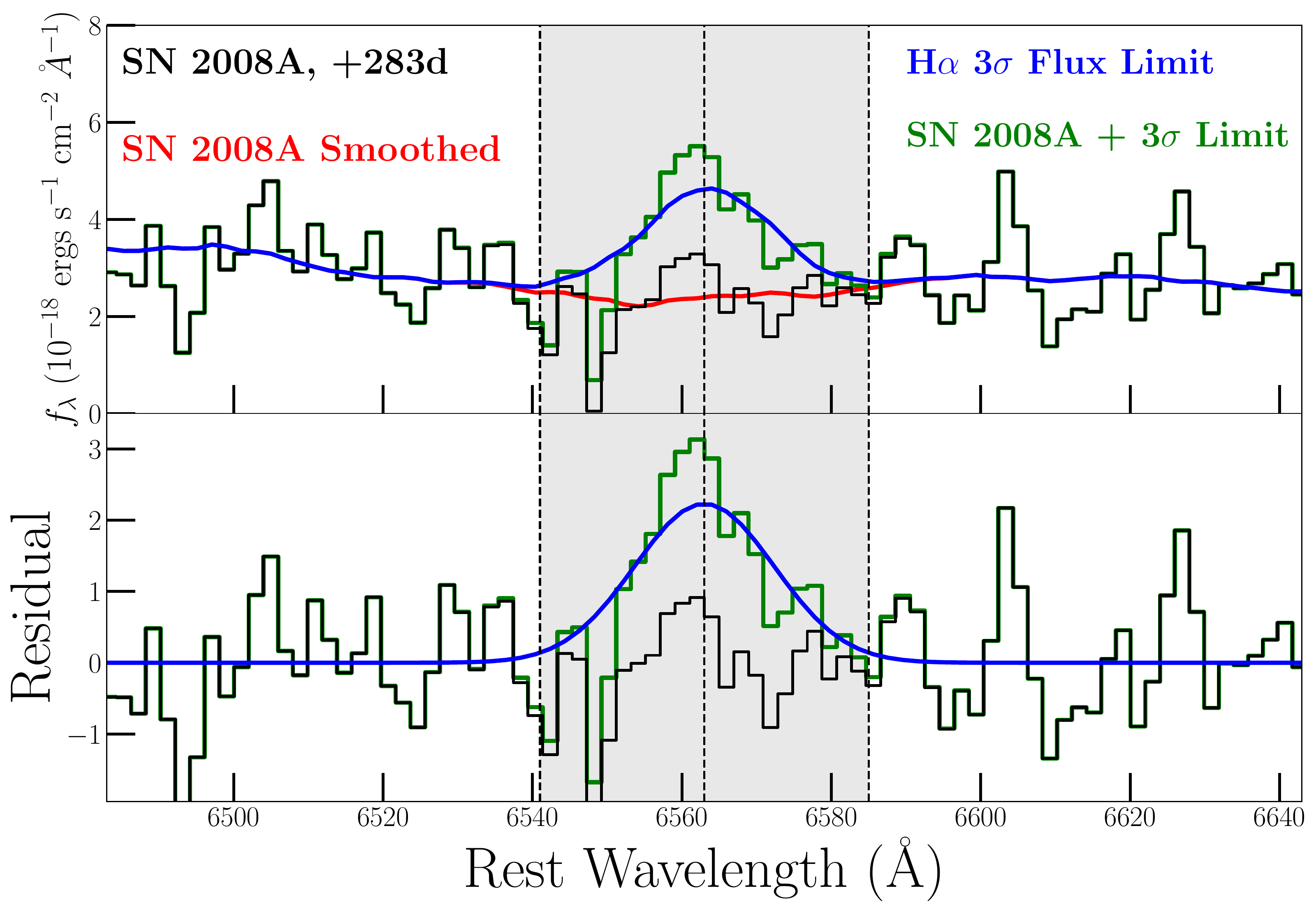}}
\subfigure[]{\label{fig:b}\includegraphics[width=.44\textwidth]{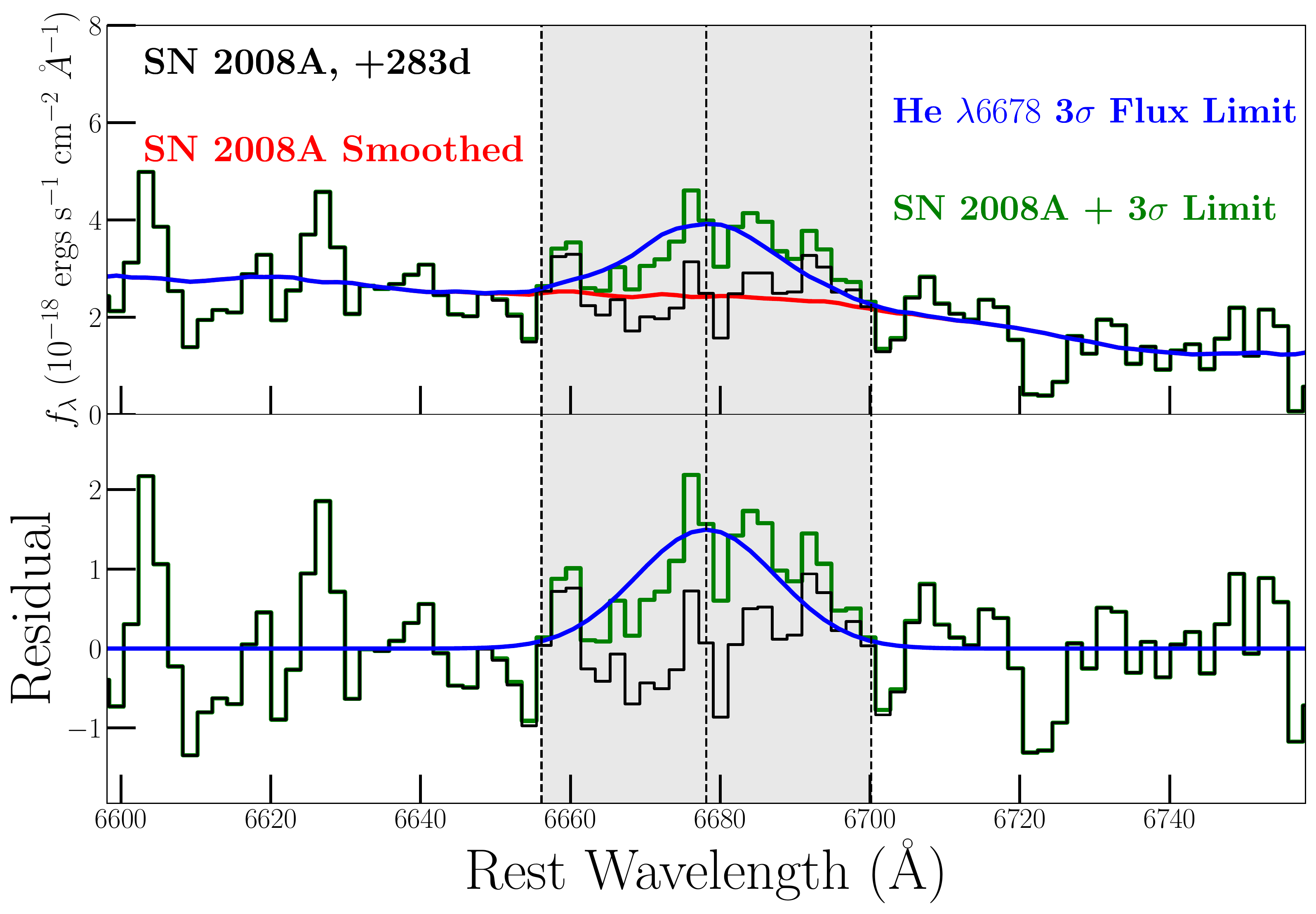}}
\caption{\textbf{(a)} \textit{Top:} In black, flux calibrated late-time data of SN~2008A at +283d with no apparent H-$\alpha$ emission. Shown in red is the smoothed continuum with the $3\sigma$ RMS flux limit for marginal detection of H-$\alpha$ shown in blue. Data plus $3\sigma$ flux limit shown in green. Grey shaded region represents a wavelength range of 22\AA\ ($\sim 1000$ km s$^{-1}$). \textit{Bottom:} In green, residuals of data plus the $3\sigma$ limit and smoothed data. In black, residuals of data and smoothed continuum.  H-$\alpha$ $3\sigma$ flux limit shown in blue. 
\textbf{(b)} Same method as for H$\alpha$, but with marginal detection of $6678$\AA\ \ion{He}{i} emission line. \label{fig:marginal_detection} }
\end{figure}

\begin{figure*}
\subfigure[]{\label{fig:a}\includegraphics[width=.45\textwidth]{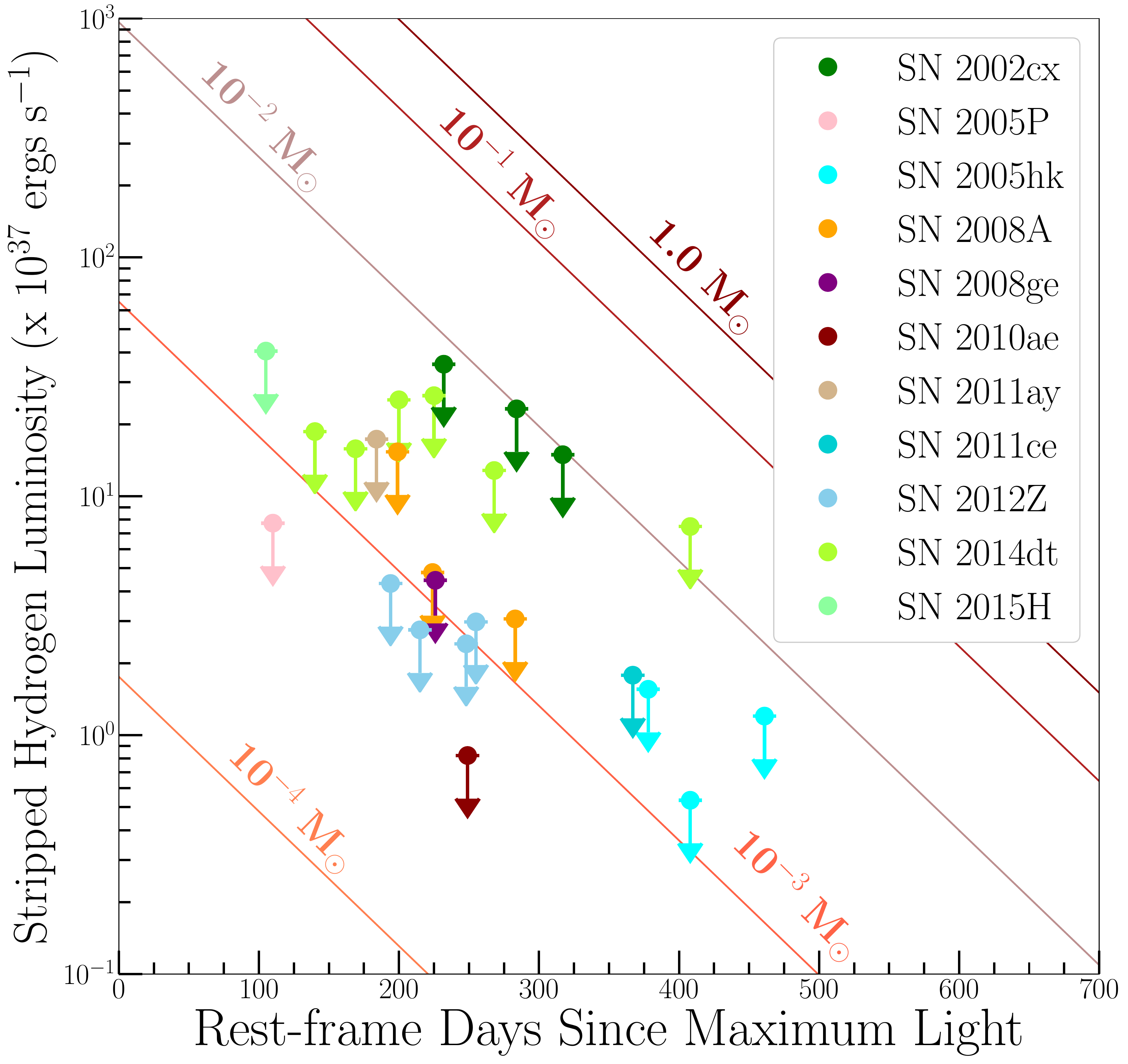}}
\subfigure[]{\label{fig:b}\includegraphics[width=.43\textwidth]{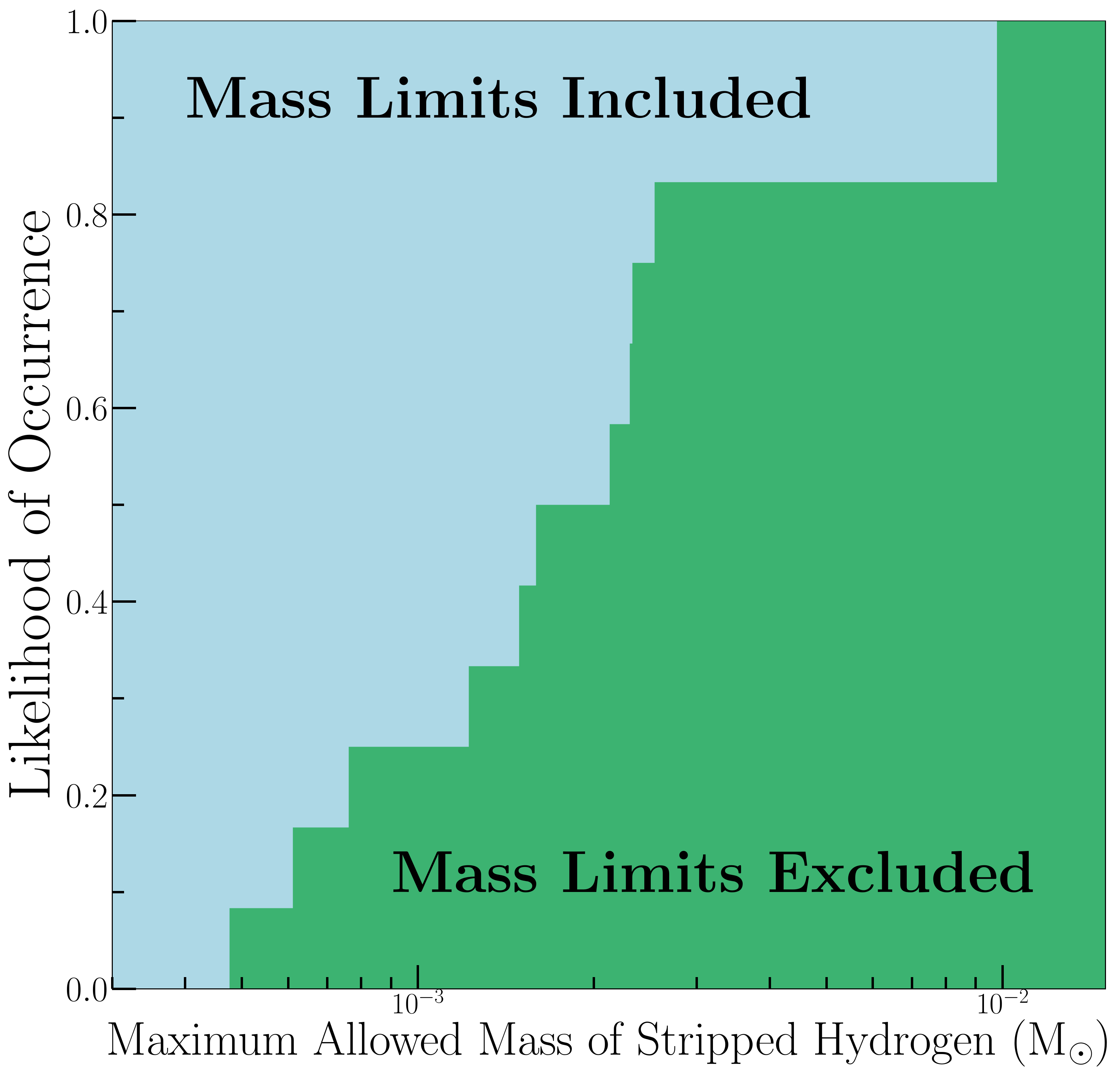}}
\caption{\textbf{(a)} Luminosity limits for marginal detection of hydrogen emission in late-time SN~Iax spectra. Contours are stripped masses calculated using Equation~\ref{eq:ha_lum} with an exponential time-dependence of SNe~Iax at late-times. \textbf{(b)} Cumulative distribution of converted hydrogen luminosity limits into masses using Equation~\ref{eq:ha_lum}. Blue portion represents included stripped mass limits for hydrogen while green portion holds excluded limits. \label{fig:strip_hydro_janos} }
\end{figure*}

To determine the line luminosity limit for each spectrum, we employ
the same method as \citet{sand18}. By representing the H-$\alpha$
emission lines as Gaussian profiles, we calculate the flux limit
needed for marginal detection. As in \citet{sand18}, we estimate the
peak flux of the emission line to be three times the spectrum's
root-mean-square (RMS) flux, with the spectral feature having a FWHM
of 1000~$\kms$. From our 3-$\sigma$ flux limit, we calculate a
luminosity limit for marginal detection of an H-$\alpha$ emission
line. An example of simulated marginal detection of H-$\alpha$ is
shown in Figure~\ref{fig:marginal_detection}(a) for SN~2008A at a
phase of +281~days.  We present the H-$\alpha$ luminosity limits as a
function phase for all SNe Iax with late-time spectra in Figure~\ref{fig:strip_hydro_janos}(a).

We then use the \citet{botyanszki18} relation to convert the
luminosity limits into stripped-mass limits\footnote{We use the
  equation from \citet{sand18}, which is an updated version of that
  presented by \citet{botyanszki18}.},
\begin{equation}\label{eq:ha_lum}
  \log_{10}(L_{\rm{H\alpha}}) = -0.2 M_{1}^2 + 0.17 M_{1} + 40,
\end{equation}
where $L_{\rm{H\alpha}}$ is the H-$\alpha$ luminosity in cgs units at
200~days after peak, $M_{1} = \log_{10}(M_{\rm H}/M_{\odot})$, and
$M_{\rm H}$ is the stripped hydrogen mass. Since this relation is
only valid at 200 days after peak, we scale the luminosity limits to
that epoch, assuming an exponential decline seen for SNe~Iax
\citep{stritzinger15}.  We fit this exponential to the observed late-time luminosities of SNe~2005hnk and 2012Z (see Figure 8 of \citealt{stritzinger15}) and apply this function to other SNe Iax in order to calculate line luminosities at 200 days. For each SN, we determine the
most-constraining mass limit.  We display the cumulative distribution
of mass limits in Figure~\ref{fig:strip_hydro_janos}(b).

It should be noted that the models presented in \citet{botyanszki18} are designed for SNe Ia and not explicitly for SNe Iax. Thus these models are most notably different than SNe Iax in their energetics, explosion asymmetry and total Ni mass produced. As a result, the SN ejecta density and the mixing of H or He may be impacted by the lower explosion energies in SNe Iax, but the differences in Ni mass between the two explosion types will scale directly with the luminosity. Additionally, any potential explosion asymmetry is seemingly negligible given the current consistency in polarization between SNe Iax and SNe Ia \citep{chornock06, maund10}. Nevertheless, the relation derived in \citet{botyanszki18} from radiative transfer simulations is currently the most applicable model for this analysis if we assume that SNe Iax arise from a SD scenario where the impact of the SN ejecta with the companion star results in stripped or swept up H- or He-rich material.

For our late-time SN~Iax sample, the mass limits range from $5 \times
10^{-4}$ to $1 \times 10^{-2} \ \Msun$ with a typical limit of $2
\times 10^{-3} \ \Msun$.  This is about an order of magnitude larger
than that expected for the amount of stripped hydrogen expected to be
swept-up by a SN~Ia in a Roche-lobe filling progenitor system
\citep[$1.4 \times 10^{-2}$ to $0.25 \ \Msun$;][]{pan12,
liu12, boehner17}. \citet{liu13b} estimates that a SN~Iax with a
similar progenitor system would have  $1 \times 10^{-2} - 1.6 \times
10^{-2} \ \Msun$ of stripped hydrogen, also significantly higher than
the limits for our sample.

\subsection{Stripped Helium in Late-time Spectra}
\label{subsec:strip_mass_calc}

\begin{figure*}
\subfigure[]{\label{fig:a}\includegraphics[width=.45\textwidth]{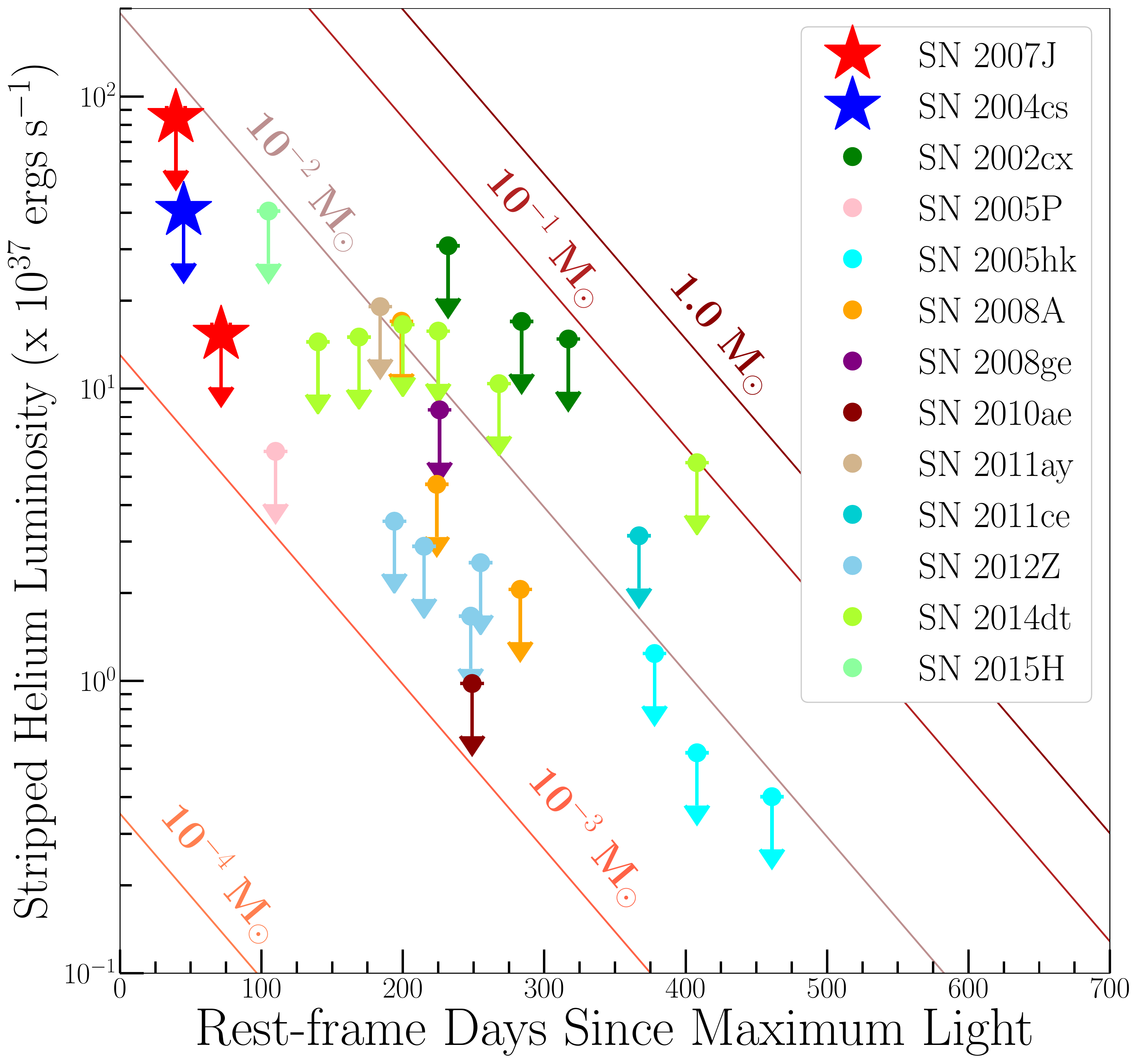}}
\subfigure[]{\label{fig:b}\includegraphics[width=.43\textwidth]{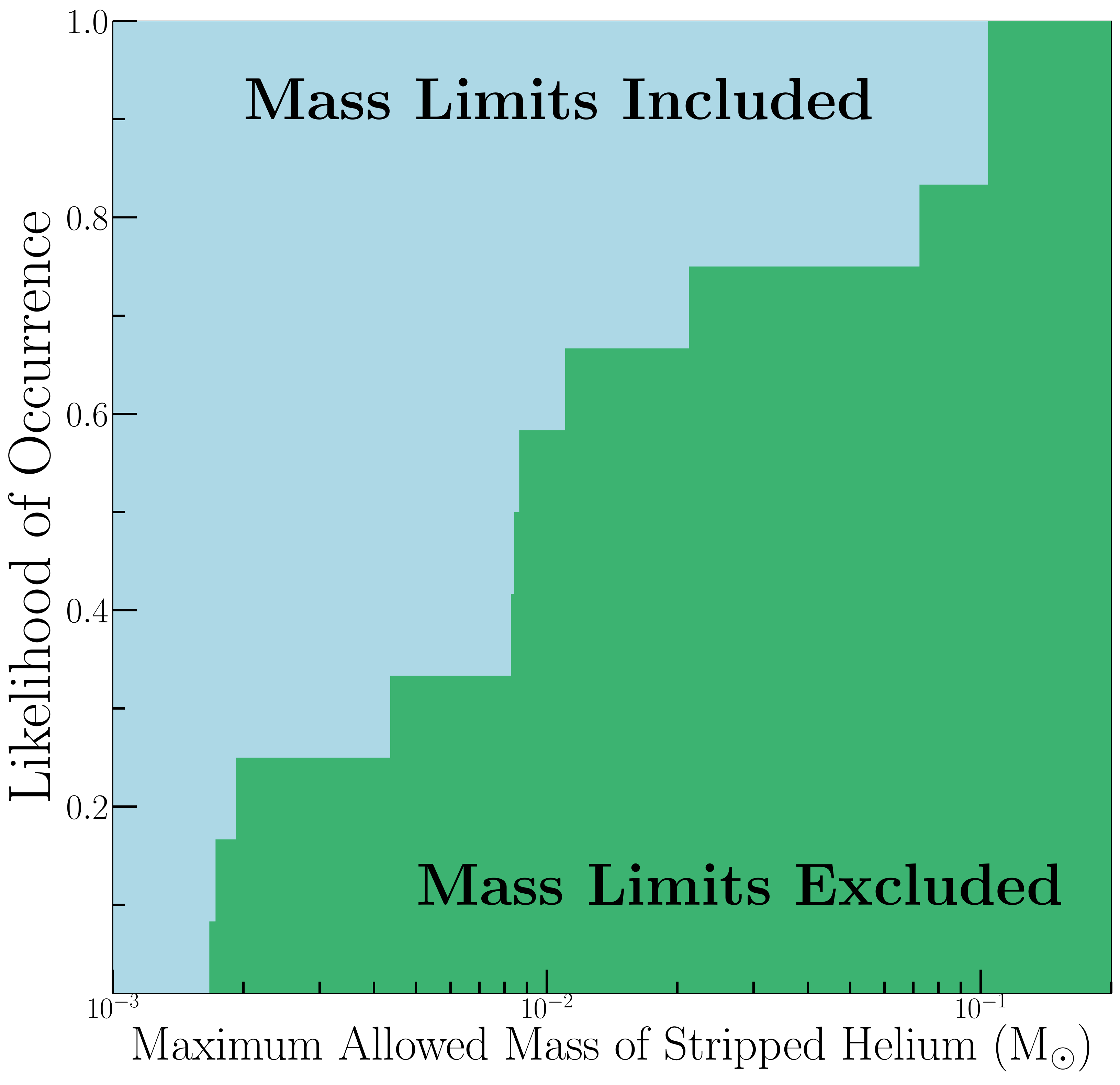}}
\caption{\textbf{(a)} Luminosity limits for marginal detection of helium emission in late-time SN~Iax spectra. Contours are stripped masses calculated using Equation~\ref{eq:he_lum} with an exponential time-dependence of SNe~Iax at late-times. Red and blue stars are the early-time detections of helium and the luminosities of the He $\lambda6678$ line. These are shown as upper limits because the are not true detections of late-time, stripped helium, and are placed for reference to late-time flux limits for marginal detection of stripped helium. \textbf{(b)} Cumulative distribution of converted helium luminosity limits into masses using Equation~\ref{eq:he_lum}. Blue portion represents included stripped mass limits for helium while green portion holds excluded limits.\label{fig:strip_helium_janos} }
\end{figure*}

\begin{figure}
\begin{center}
	\includegraphics[width=0.46\textwidth]{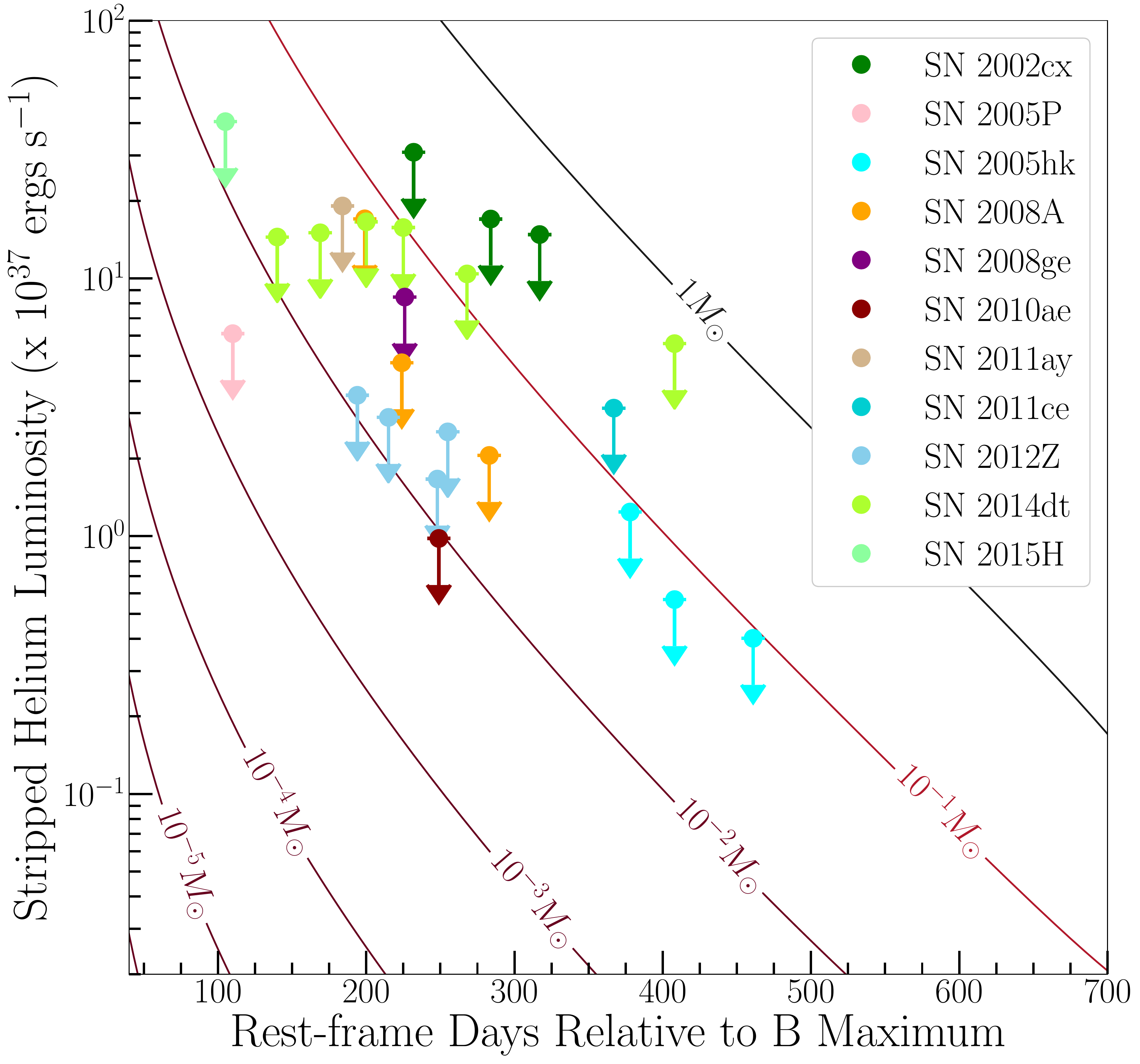}
	\caption{Luminosity upper limits for marginal detection of helium emission in late-time SN~Iax spectra. Contours are stripped masses calculated using Equation~\ref{eq:M_he3}. Masses are calculated with a velocity of 5000 km s$^{-1}$ and $\MNi = 0.1 \Msun$.} \label{fig:strip_mass_analytic}
\end{center}
\end{figure}

We also examine all late-time SN~Iax spectra for helium emission that
could result from companion interaction. Upon visual inspection, we
find no detections of any prominent \ion{He}{i} emission lines.
Following the same method as described in
Section~\ref{subsec:neb_spectra_h}, we calculate luminosity limits for
the \ion{He}{i} $\lambda$6678 line. An example of the calculation of
the 3-$\sigma$ limit for SN~2008A is shown in
Figure~\ref{fig:marginal_detection}(b).

Using Equation~\ref{eq:ha_lum}, \citet{sand18} equates H-$\alpha$
emission to \ion{He}{i} emission to calculate the upper limit of
helium emission in the late-time spectra of SN~2017cbv (i.e., $L_{\rm{H\alpha}} = L_{\rm{He}}$). We use a similar method but scale the input helium luminosity based on the
relative luminosities of H-$\alpha$ and \ion{He}{i} emission found by
\citet{botyanszki18}. To do this, we compare the H-$\alpha$ line
luminosity in Figure~1 of \citet{botyanszki18} to the \ion{He}{i}
$\lambda$6678 line luminosity shown in their Figure~4 for the MS38
model. Based on the relative peaks and FWHM of each respective
emission line, the MS38 model produces an H-$\alpha$ emission line
that is ${\sim}5$ times more luminous than the \ion{He}{i}
$\lambda$6678 emission line for the same mass. As a result, we
use 
\begin{equation}\label{eq:he_lum}
  \log_{10}(L_{\rm{He}}) = -0.2 M_{1}^2 + 0.17 M_{1} + 39.3,
\end{equation}
where $M_{1} = \log_{10}(M_{\rm He}/M_{\odot})$, and
$M_{\rm He}$ is the stripped helium mass, to calculate the amount of
stripped helium in SNe~Iax. This estimate comes with significant caveats.  The efficiency of non-thermal excitation of He is dependent on the degree of Ni mixing in the SN ejecta \citep{lucy91}.  This poses a significant challenge for this type of modeling, as it means that the strength of the He lines is dependent on the distribution of the Ni within the explosion itself and the distribution of the entrained He within the ejecta.  This illustrates the potential limitations of the application of the \cite{botyanszki18} result to SN Iax with He-rich companions: the Ni distribution is from a Ia-like explosion and the He distribution comes from a simulation of interaction with an H-rich companion, but with the entrained H artificially converted to He.  However, 
\cite{botyanszki18} is at present the most applicable model available in the literature.

We display the helium luminosity limits for our SN~Iax sample as a
function of phase and compare to expected helium line luminosity for
various amounts of stripped helium in
Figure~\ref{fig:strip_helium_janos}(a). We also plot a cumulative
distribution of calculated helium masses, using the most constraining
value for each SN, in Figure~\ref{fig:strip_helium_janos}(b). For
reference, we add the early-time He $\lambda$6678 luminosities for
SNe~2004cs and 2007J as upper limits in
Figure~\ref{fig:strip_helium_janos}(a). Without a method for converting early-time helium luminosity to mass, these points represent upper limits on the amount stripped helium that could
be visible at early-times if we assume that the helium detected
in SNe~2004cs and 2007J is the result of companion interaction. Nonetheless, we caution the use of these specific limits in constraining companion interaction models because
the photospheric heating mechanism and specific excitation energy of
helium in SNe~Iax is still unclear at this time.

For our SN~Iax sample, we find a range of stripped helium upper limits
of $2 \times 10^{-3}$ to $1 \times 10^{-1} \ \Msun$. Comparing these
stripped helium mass limits to those predicted by SN~Ia explosion
models, we find that some of our limits are consistent with predicted
stripped masses, but the most constraining limit of $2 \times 10^{-3}
\ \Msun$ is lower than that presented in companion interaction models.
For stripped helium, \citet{pan12} predicts a range of $2.45 \times
10^{-3} - 1.34 \times 10^{-2} \ \Msun$ and \citet{liu13} predicts a
range of $9.5 \times 10^{-3} - 2.8 \times 10^{-2} \ \Msun$.

In addition to using a (modified) relation between line flux and
stripped mass determined by simulations \citep{botyanszki18}, we
present an alternate, analytic method of calculating stripped helium
mass, derived from radioactive-decay powered SN ejecta. We first
assume that the helium is contained in a homogeneous shell about the
SN with mass $\MHe$ and a radius $\Delta r$ that is changing with the expanding ejecta. The shell's
density can then be defined as:
\begin{equation}
  \rho = \frac{\MHe}{4\pi r^2 \Delta r}. \label{eq:rho} 
\end{equation}

The optical depth of $\gamma$-rays from radioactive decay in SNe is 
\begin{equation}
  \tau_{\gamma} = \kappa \rho \Delta r, \label{eq:tau_1}
\end{equation}
where $\kappa = 0.05$~cm$^{2}$~g$^{-1}$ is used for $\gamma$-rays.

Combining Equations~\ref{eq:rho} and \ref{eq:tau_1}, we find a relation for
optical depth of $\gamma$-rays as a function of $\MHe$:
\begin{equation}
  \tau_{\gamma} = \frac{\kappa \MHe}{4\pi r^2}. \label{eq:tau_2} 
\end{equation}
We define the luminosity absorbed by the helium shell to be: 
\begin{equation}
    \Ldep = \Lrad \left (1 - e^{- \tau_{\gamma}} \right ), \label{eq:Lum_dep} 
\end{equation}
where $\Lrad$ is the luminosity of radioactive ${}^{56}$Ni and
${}^{56}$Co. We assume that the ejecta is optically thin to gamma-rays
and thus does not get absorbed in the ${}^{56}$Ni region before
reaching the helium shell. The $\Lrad$ relation is from Equation~19 of
\citet{nadyozhin94},
\begin{equation}
    \Lrad = 1.45\times10^{43} \left [ 4.45\cdot
      e^{-(t/8.8\rm{~d})} + e^{-(t/111.3\rm{~d})} \right ]\frac{\MNi}{\Msun}\label{eq:L_rad}.
\end{equation}

We define the helium line luminosity to some fraction of the deposited helium energy,

\begin{equation}
    L_{\textrm{dep,He}} = \frac{L_{\textrm{He,line}}}{f}, \label{eq:L_he1} 
\end{equation}
where $f$ is the fraction of deposited energy emerging in the helium line. There are $\approx 10$ \ion{He}{i} lines in the optical. With the conservative assumption that the helium luminosity is equally distributed across these lines, we adopt a rough estimate of $f=0.1$. Combining Equations~\ref{eq:tau_2}, \ref{eq:Lum_dep} and \ref{eq:L_he1}, we find
\begin{equation}
    L_{\textrm{dep,He}} = \Lrad \left(1 - e^{-\frac{\kappa M_{\rm He}}{4\pi r^2}}   \right). \label{eq:L_he2} 
\end{equation}

Solving for $\MHe$ and replacing the radius, $r$, with the product of ejecta velocity
and time since explosion, we have

\begin{equation}
  \MHe = -\frac{4\pi (vt)^2}{\kappa} \ln \left [ 1 -
    \left ( \frac{L_{\textrm{He,line}}}{\Lrad} f^{-1}
    \right ) \right ] \Msun  \label{eq:M_he1}
\end{equation}

Picking nominal values for each quantity and assuming that ${}^{56}$Co decay dominates the energy deposition at the epochs of interest, we find

\begin{align}
	\begin{aligned}
	 &\MHe = -\unit[2.35]{\Msun} \left(\frac{v}{\unit[5000]{\kms}}\right)^2  \left(\frac{t}{\unit[100]{d}}\right)^2 \left(\frac{\kappa}{\unit[0.05]{cm^2\,g^{-1}}}\right)^{-1} \times\\ 
	 &\ln \left[1 - 6.9\times 10^{-4} \left(\frac{f}{0.1}\right)^{-1}\left(\frac{\LHe}{\unit[10^{38}]{erg\,s^{-1}}}\right) \left(\frac{\MNi}{\unit[0.1]{\Msun}}\right)^{-1} \exp \left(\frac{t/\unit[100]{d}}{1.113}\right) \right]  \label{eq:M_he2}
	\end{aligned}
\end{align}

Because the quantity in the square brackets is approximately one, we can Taylor expand the natural logarithm to find a final
expression for $\MHe$,

\begin{align}
	\begin{aligned}
	 &\MHe = \unit[1.6\times 10^{-3}]{\Msun} \left(\frac{v}{\unit[5000]{\kms}}\right)^2  \left(\frac{t}{\unit[100]{d}}\right)^2   \left(\frac{\kappa}{\unit[0.05]{cm^2\,g^{-1}}}\right)^{-1}\times \\ 
	 &\left(\frac{f}{0.1}\right)^{-1}\left(\frac{\LHe}{\unit[10^{38}]{erg\,s^{-1}}}\right) \left(\frac{\MNi}{\unit[0.1]{\Msun}}\right)^{-1} \exp \left(\frac{t/\unit[100]{d}}{1.113}\right) \label{eq:M_he3}
	 \end{aligned}
\end{align}

We note that Equation~\ref{eq:M_he3} yields a reasonable mass estimate
for the nominal values.  However, if one requires
estimates far from the nominal values, Equation~\ref{eq:M_he2} may be more appropriate.

In Figure \ref{fig:strip_mass_analytic}, we plot the stripped helium
luminosity limits as a function of phase for our late-time SN~Iax
sample.  For comparison, we also display lines of equal stripped
helium mass as derived with Equation~\ref{eq:M_he3} and assuming an
ejecta velocity of 5000~$\kms$ and a ${}^{56}$Ni mass of $0.1 \
\Msun$. However, these are average values based on the known range of ejecta velocities and ${}^{56}$Ni masses in SNe Iax \citep{foley13, stritzinger15, jha17}. We note that Figure \ref{fig:strip_mass_analytic} is illustrative and makes various assumptions about the physical conditions of the system. Consequently, we suggest calculating $\MHe$ from Equation~\ref{eq:M_he3} with values for ejecta velocity and $\MNi$ specific to a given SN Iax. Under those assumptions, we can place stripped helium mass
limits of $\sim{}3 \times 10^{-3} - 7 \times 10^{-1} \ \Msun$,
comparable to the values derived from the numerical simulations above.

\section{Discussion}
\label{sec:discussion}
\subsection{Rate of Helium Detection in SNe~Iax}
\label{subsec:detection_helium}

We have modeled the early-time spectra of 44 classified SNe~Iax and
find prominent helium emission in the spectra of only two objects,
SNe~2004cs and 2007J. In order to understand this observational
result, we must account for selection effects.  In particular, the
luminosity of the detected helium lines was significantly lower than
the limits for the majority of the sample; 19/25 objects ($76\%$) do
not have deep enough limits to detect lines as faint as in SNe~2004cs
and 2007J.  Furthermore, we could not significantly detect helium in
the first spectrum of SN~2007J, despite there being very strong lines
in its later spectra.  Therefore, temporal evolution is also likely
important for determining the detectability of helium for our sample.

Using the subset of SNe~Ia where we could detect helium lines with a
similar luminosity as for SNe~2004cs and 2007J, we see that 2/6
($33^{+11}_{-7}$\%) SNe~Iax have helium detections.  Any SN~Iax model
must allow for similar fractions of SNe where helium is and is {\it
  not} observable in their spectra.  We examine some of the possible
physical models below.

\subsection{Explanations for Helium in the SN~Iax System}
\label{subsec:helium_novae}
We find that the helium profiles in SNe~2004cs and 2007J are
inconsistent with P-Cygni profiles and are well described by a
Gaussian profile.  This indicates that the helium is {\it not} in the
photosphere, and instead in the circumstellar environment.  We find
that the \ion{He}{i} $\lambda$7065 line in SN~2007J has a FWHM of 4352
and 4725~$\kms$ at $+33-46$ and $+65-78$~days, respectively. The same line has a
FWHM of 3878~$\kms$ for SN~2004cs at a phase of +45~days.
Additionally, there is He-rich material at velocities as high as
8367~$\kms$ (with the lines having a full-width zero intensity of
${\sim}$12,000~$\kms$).  These velocities are higher than the orbital motion and
escape velocity of the system, indicating that an explosive mechanism
is necessary to eject helium into the SN~Iax progenitor system.

In the WD+He star channel for SNe~Iax progenitors, the accretion rate
of helium onto the surface of the C/O WD modulates the amount of
He-rich material in the circumstellar environment.  If the accretion
rate is low ($\dot{M} < 10^{-6} \ \Msun$yr$^{-1}$), it can lead to helium
flashes \citep[e.g.,][]{piersanti14, brooks16} resulting in helium
novae. A possible explanation for the small fraction of SNe~Iax with
helium detections could be the occurrence of such pre-explosion helium
nova eruptions. In the C/O WD + He star progenitor system, the helium
mass transfer rate from the He star onto the WD varies based on the
size of the non-degenerate companion. Through modeling of these
systems for an initial $1 \ \Msun$ WD, \citet{brooks16} show that
systems with lower mass companion stars also experience lower mass
transfer rates prior to explosion. Consequently, systems with
1.3--$1.4 \ \Msun$ He stars undergo helium nova eruptions as the WD
approaches the Chandrasekhar mass (see Figure~3 of
\citealt{brooks16}). Occurring in only a fraction of such scenarios,
these nova eruptions would enrich the progenitor system with helium in
the final years before explosion. However, the ejected He-rich
material may be hidden in cavities of the progenitor system \citep{wood06} and could
lead to an even lower rate of helium detection.

Higher-mass He star companions have higher mass transfer rates,
allowing for steady helium burning on the WD surface and does not
produce helium flashes. For these high accretion rates, a
super-Eddington wind may be formed, which will remove accreted helium
from the system thus making it undetectable \citep{wang15}.
Consequently, the limited fraction of C/O WD + He star systems with
helium novae could explain the detection of helium emission in only
the spectra of SNe~2004cs and 2007J.

Interestingly, the single Galactic helium nova, V445 Puppis, has
helium at velocities as high as 6000~$\kms$ \citep{woudt09},
consistent with the helium velocity seen for SNe~2004cs and 2007J.

\subsection{Photospheric Helium Models}
\label{subsec:magee19}

Using a \texttt{TARDIS} model with significant photospheric helium, \cite{magee19} are able to fit the prominent \ion{He}{i} features in the +9-22d spectrum of SN~2007J. However, as shown in Figure \ref{fig:abs_vs_emission}, we also reproduce all \ion{He}{i} features by using a masked \texttt{SYNAPPS} fit with added helium Gaussian profiles. This indicates that while the interpretation of \cite{magee19} may be correct, our model of circumstellar helium in SN~2007J is not obviously ruled out. However, all later SN~2007J spectra have clear Gaussian \ion{He}{i} line profiles and thus cannot have a photospheric origin. This is not necessarily at odds with the \cite{magee19} result, but it should be noted that they do not attempt to fit these later spectra of SN 2007J nor the SN 2004cs spectrum. Nonetheless, it is possible that there exists both photospheric and circumstellar helium in SNe Iax; this being consistent with both results as well as having a dependence on the quantity of helium at a given location within the SN.

\subsection{Comparison to Single-Degenerate Explosion Models}
\label{subsec:model_compare}

The leading model for the SN~Iax progenitor system is a WD accreting
mass from a non-degenerate He star companion. Based on the calculated
fraction of SNe~Iax with helium, the companion to the WD could also be
another type of non-degenerate star such as a Main Sequence, Red
Giant, or He star. Support for the He-star scenario comes from
the pre-explosion observation of the SN~2012Z progenitor system, which
is consistent with a He-star companion despite the lack of helium
lines in its spectra (\citealt{mccully14}; \citealt{takaro19}; McCully et al., in prep). 

Even for the WD+He star progenitor channel, there are a variety of
explosive mechanisms that could trigger a SN~Iax explosion as the He
star fills its Roche Lobe and transfers matter onto the WD. The
typical Chandrasekhar-mass model involves accretion of He-rich
material and steady burning on the WD surface until the WD reaches
M$_{\textrm{ch}}$ and explodes.  The detection of strong Ni features
in the late-time spectra of SNe~Ia indicate a high central density for
the WD, consistent with a M$_{\textrm{ch}}$ explosion \citep{stritzinger15, foley16}.

Alternatively, a sub-Chandrasekhar mass model with helium detonation on the surface of the WD may
be more appropriate in describing these objects \citep{hillebrandt00}. \citet{wang13} examines this double-detonation model for SNe Iax and finds that such a scenario reproduces the observed $M_{V, {\rm ~peak}}$ for these objects, as well as allows for SNe with helium lines in their
spectra as in SNe~2004cs and 2007J. This type of explosion in the WD+He star progenitor channel could explain the lowest luminosity SNe~Iax such as SNe~2008ha and 2010ae, but ultimately cannot reproduce the low ejecta velocities seen in all SNe~Iax. As stated above, the Nickel abundances found in late-time SN~Iax spectra by \citet{foley16} indicates that the majority of SNe~Iax are
the result of M$_{\textrm{ch}}$ explosions rather than sub-Chandrasekhar mass
explosions.

In each explosion model discussed, the type of helium burning may also affect the
detection of helium within SNe~Iax.  The lack of helium in most
SNe~Iax spectra can be explained by an explosion that completely burns
the surface helium. Complete burning may generate a more luminous explosion that could result in a lack of detected helium in the most-luminous SNe~Iax. Within the limits of detection, the presence of detectable helium in 33\% of SNe~Iax could be explained by incomplete burning where unburned helium remained in a fraction of systems after explosion.

The SD channel, such as a WD+He star progenitor system, has also been examined in the context
of companion interaction. In the pure deflagration explosion, the SN
ejecta forms a shock wave that hits the companion He star and strips
material from the stellar surface. As discussed in
Sections~\ref{subsec:neb_spectra_h} and \ref{subsec:strip_mass_calc},
this material could be hydrogen or helium (depending on the
composition of the companion star) and the amount of stripped mass
depends on the exact explosion model and companion star.
Hydrodynamical simulations done by \citet{liu13} and \citet{pan12}
show that the H- or He-rich material is removed due to ablation (SN
heating) and mass stripping (momentum transfer). These models
demonstrate that the impact of SN ejecta with the non-degenerate
companion causes a hole in the ejecta of $\sim \ang{35}$ for the WD+He
star system. While some of the stripped material will reside within
this ejecta hole, the rest will be swept up by the expanding ejecta
and may become visible as the SN luminosity fades at late-time.

However, we find no evidence of such material in the SN~Iax late-time
spectra despite the stripped mass predictions for H- and He-rich
material from companion interaction. This indicates a discrepancy
between SD model predictions and SN~Iax (and SN~Ia) observations, that
perhaps could be reconciled by having ``hidden'' hydrogen/helium or a
smaller amount of stripped material than predicted.

For SNe~Iax in particular, \citet{liu13b} show that H-rich material
can be hidden in late-time spectra due to the inefficient mass
stripping ($<$0.01$\Msun$). In their pure deflagration model, \citet{liu13b} demonstrate that the relatively low kinetic energy released in the explosion could result in both a low mass of stripped material and potentially no hydrogen emission signatures at late-times. Such a scenario is broadly consistent with the lack of hydrogen emission in late-time SNe Iax spectra, but the mass limit is still higher than those calculated in Section~\ref{subsec:neb_spectra_h}.

Another potential reason for a lack of visible helium emission is that
the high excitation energy of helium may prevent the formation of the
predicted narrow emission lines at late times. As is shown for Type Ic SNe, the degree of Ni mixing in the SN ejecta significantly influences the efficiency of non-thermal helium excitation \citep{lucy91, dessart11, dessart12}.
For SNe Iax, weak Ni mixing in the ejecta could result in helium lines below the limit of detection in late-time spectra.

Furthermore, the specific type of explosion responsible for SNe~Iax
could result in a lack of detectable stripped material from a
companion star. The partial deflagration scenario in particular can
produce a highly asymmetric explosion in which the ejecta either does
not impact the companion at all, or causes much less impact than a
spherical explosion \citep{jordan12, kromer13, fink14}. Such a scenario would strip less H- or He-rich material from the companion and would then produce material in the system below the limit of detection.

Finally, the strength of hydrogen or helium emission features may
depend on the SN viewing angle and it could be the case that observed
SNe~Iax may have undetectable stripped material due to the angle of 
observation.  We can match the detected occurence rate (33\%) if
\ion{He}{i} features are only visible within a $\ang{71}$ angle. This calculation assumes a spherically symmetric SN where circumstellar helium can only be detected on 33\% of the SN ejecta. 

Case in point, the ongoing discrepancy between SD interaction models and late-time observations warrants more robust radiative transport calculations through 3-D hydrodynamical stripping of hydrogen or helium from a companion star. The lack of visible hydrogen or helium emission at late-times highlights the need for SD models that can have both significant swept up or stripped mass and a lack of emission features at late-times. This could then explain the observed flux and mass limits that are well below those predicted by interaction models for SNe Ia and Iax. 

With true detections of helium in SN~Iax spectra, we have identified the need for luminosity-mass conversions in early-time spectra, not just in late-time phases. Such modeling would allow us to quantify the helium present in SN~2004cs, SN~2007J and potentially future SNe~Iax with helium detections. Calculating the helium mass present at early-times will further constrain the proposed C/O WD + He star progenitor system for SNe~Iax.

\subsection{SD Scenario and SNe~Ia}
\label{subsec:snIa_progenitors}

The pre-explosion detection of the progenitor system
for SN~2012Z \citep{mccully14} strongly indicates a SD progenitor
system for that SN, which by extension, may apply to other SNe~Iax.  Other observations
indicating a short delay time \citep[e.g.,][]{foley09, lyman13, lyman18},
also favor a SD progenitor channel for SNe~Iax.  The specific scenario
that best fits all observations is a near Chandrasekhar mass C/O WD
primary with a Roche-lobe filling He-star companion
\citep[e.g.,][]{foley13, jha17}.

A SN from such a system is expected to strip material from the
companion star, possibly sweep up additional circumstellar material,
and produce relatively narrow emission lines in late-time spectra
\citep{liu13}.  However, we have not yet seen any such emission lines
in our sample of 11 SNe~Iax with late-time spectra.  Moreover, we
have not detected the ``characteristic SD'' narrow emission lines in
SN~2012Z.  Therefore, the lack of narrow emission lines in the
late-time spectrum of a thermonuclear SN is poor evidence against
the SD scenario.  While it is currently unclear where the logical
argument breaks down, there are significant implications for SNe~Ia.

Several studies have searched for narrow emission lines in the
late-time spectra of SNe~Ia to see if there was evidence of
interaction with a non-degenerate companion star, and thus far, no
example has been found for a normal SN~Ia \citep[e.g.,][but see also
\citealt{graham18}]{mattila05, leonard07, shappee13, lundqvist13, maguire16, graham17, sand18, shappee18, dimitriadis18}.  While many of these studies concluded that the lack of
such lines is strong evidence against (at least a subset of) SD
scenarios for SNe~Ia, our experience with SNe~Iax suggests that such
strong conclusions are premature.

We suggest a full re-evaluation to determine the utility of these
observations for understanding the progenitor systems of SNe~Iax and
SNe~Ia.

\subsection{Future SN~Iax Observations}
\label{subsec:observations}

As more SNe~Iax are discovered, the number of objects with confirmed
helium emission in their spectra should increase. Consequently, it is
important that we collect high-cadence spectral observations of any
new SNe~Iax discovered with early-time \ion{He}{i} features. Since the
helium features in SN~2007J became stronger with time, it is clear
that very early spectra are insufficient to detect helium in at least
some SNe~Iax.  Therefore, continued monitoring of SNe~Iax may be
critical to detecting similar events.

SNe~2004cs and 2007J are relatively low-luminosity SNe~Iax.  Although
the number of events is still small, it behooves us to pay particular
attention to other low-luminosity SNe~Iax.

Of course when a SN~Iax is discovered with helium emission in its
spectrum, we should follow that object as long as possible.  Neither
SN~2004cs nor SN~2007J have late-time spectra.  It would be
particularly interesting to see if similar SNe~Iax have helium
emission features at late times.  Late-time spectra would also be
especially useful to determine if SNe~Iax with helium emission are
physically distinct from those without helium emission.

\section{Conclusions}
\label{sec:conclusions}

We model 110 spectra of 44 SNe~Iax using the spectral synthesis code
\texttt{SYNAPPS}.  We detect helium emission in two objects,
SNe~2004cs and 2007J, and do not detect any helium emission for any
other. We find the helium features in these objects to be poorly
described by a P-Cygni profile, but are well-fit by a Gaussian,
implying that the helium has a circumstellar origin.

For each spectrum without detected helium, we add helium features to
the spectrum until we make a detection, thus measuring a luminosity
limit for the helium features in each spectrum.  Only 16\% of the
SNe~Iax in our flux calibrated sample have sufficiently deep luminosity limits to rule
out helium emission at the level detected in SNe~2004cs and 2007J.
Using the subset of SNe~Iax where we could detect helium lines with a
similar luminosity as for SNe~2004cs and 2007J, we find
$33^{+11}_{-7}$\% of SNe~Iax have helium detections.

We examined 24 late-time spectra of 11 SNe~Iax for signs of stripped
hydrogen or helium from companion interaction. We find no evidence of
narrow hydrogen or helium emission lines in the late-time spectra.  We
measured luminosity limits for this emission.  Using the luminosity
limits, we provide swept-up mass limits using both a numerical
calculation \citep{botyanszki18} and an analytic formulation.  We find
that for both hydrogen and helium, the largest possible swept-up mass
for our sample ranges between $10^{-4}$ and $10^{-1} \ \Msun$, with
the typical value lower than theoretical predictions for SD SN~Iax
explosion models.  Considering the strong evidence for a SD progenitor
system for SNe~Iax, we suggest re-examining both these models and
similar conclusions for SN~Ia progenitor systems.

The helium emission in SNe~2004cs and 2007J is consistent with coming
from the ejecta of a relatively recent helium nova.  In particular,
the velocity of the material is similar to that of the Galactic helium
nova V445 Pup \citep{woudt09}.  As helium novae are only expected for
particular accretion rates \citep{piersanti14, brooks16}, this may be a simple
explanation for the fraction of SNe~Iax with detected helium emission.

More generally, SNe~Iax may all have similar progenitor systems, but
the exact progenitor system conditions at the time of explosion and/or
the details of the explosion may produce a variety of helium emission
strengths.  Factors such as explosion asymmetry, limited amount of
swept up He-rich material, or the high excitation energy of helium may
contribute to ``hidden'' helium in SNe~Iax.

On the other hand, the presence of detectable helium emission in only
a fraction of the SN~Iax sample may be suggestive of progenitor
diversity for this class.  As SNe~Iax are the only class of
thermonuclear SN with a detected progenitor system, it is particularly
ripe for detailed study.  Increasing the sample size of SNe~Iax with
spectral observations will help to constrain the current explosion
model.

\section*{Acknowledgements}

We thank the anonymous referee for their valuable
comments and suggestions. We thank P.\ Nugent and E.\ Ramirez-Ruiz for helpful
comments on this paper. We also thank Z.\ Liu, M.\ Magee, A.\ Miller,
Y.-C.\ Pan and L.\ Tomasella for providing spectral data used in this
research.

The UCSC group is supported in part by NSF grant AST-1518052, the
Gordon \& Betty Moore Foundation, and by fellowships from the David
and Lucile Packard Foundation to R.J.F.\ and from the UCSC Koret
Scholars program to W.V.J.-G.  Support for this work was provided by
NASA through Hubble Fellowship grant \# HST-HF2-51382.001-A awarded by
the Space Telescope Science Institute, which is operated by the
Association of Universities for Research in Astronomy, Inc., for NASA,
under contract NAS5-26555.  This research is supported at Rutgers
University through NSF award AST-1615455. 

This research used resources of the National Energy Research
Scientific Computing Center (NERSC), a U.S.\ Department of Energy
Office of Science User Facility operated under Contract No.\
DE-AC02-05CH11231.




\bibliographystyle{mnras}
\bibliography{references} 



\appendix

\section{Modeled SNe~Iax Sample}
\label{sec:all_spec}

Here we present all additional SNe~Iax spectra that were modeled with
\texttt{SYNAPPS}. Same as Figures~\ref{fig:combo_04cs} and \ref{fig:07J_combo}, we plot \texttt{SYNAPPS} mask fits in green, Gaussian profiles in red, and cross-correlation coefficients in blue. 

\begin{table}
	\caption{Spectral Data}
	\label{tab:spectra_table1}
	\begin{threeparttable}
	\begin{tabular}{cccc}
		\hline
		Object & Phase (days) & Phot. (Y or N) & Reference\\
		\hline
		SN~1991bj & $29-32$\tnote{a} & N & \cite{silverman12}\\
		SN~1991bj & $39-42$\tnote{a} & N & \cite{silverman12}\\
		SN~1999ax & $13-18$\tnote{a}  & N &  \cite{silverman12}\\
		SN~2002bp & $16-21$\tnote{a}  & N &  \cite{silverman12}\\
		SN~2002cx & 21 & Y &  \cite{silverman12}\\
		SN~2002cx & 26 & Y &  \cite{silverman12}\\
		SN~2002cx & 57 & Y &  \cite{silverman12}\\
		SN~2002cx & 232 & Y &  \cite{jha06}\\
		SN~2002cx & 284 & Y &  \cite{jha06}\\
		SN~2002cx & 317 & Y &  \cite{jha06}\\
		SN~2003gq & 62 & Y &  \cite{silverman12}\\
		SN~2004cs & 45 & Y &  \cite{foley09, foley13}\\
		SN~2004gw & $11-24$\tnote{a}  & N & \cite{foley09}\\
		SN~2004gw & $39-52$\tnote{a}  & N & \cite{foley09}\\
		SN~2005p & $62-84$\tnote{a}  & Y & \cite{silverman12}\\
		SN~2005p & 108 & Y & \cite{jha06}\\
		SN~2005cc & 5 & Y & \cite{matheson08}\\
		SN~2005cc & 10 & Y & \cite{matheson08}\\
		SN~2005cc & 15 & Y & \cite{matheson08}\\
		SN~2005cc & 20 & Y & \cite{matheson08}\\
		SN~2005cc & 35 & Y & \cite{matheson08}\\
		SN~2005hk & -8 & Y & \cite{phillips07}\\
		SN~2005hk & -4 & Y & \cite{phillips07}\\
		SN~2005hk & 0 & Y & \cite{phillips07}\\
		SN~2005hk & 4 & Y & \cite{phillips07}\\
		SN~2005hk & 13 & Y & \cite{phillips07}\\
		SN~2005hk & 15 & Y & \cite{phillips07}\\
		SN~2005hk & 22 & Y & \cite{phillips07}\\
		SN~2005hk & 24 & Y & \cite{phillips07}\\
		SN~2005hk & 28 & Y & \cite{phillips07}\\
		SN~2005hk & 38 & Y & \cite{phillips07}\\
		SN~2005hk & 43 & Y & \cite{phillips07}\\
		SN~2005hk & 52 & Y & \cite{phillips07}\\
		SN~2006hn & 21 & Y & \cite{silverman12}\\
		SN~2007J & $5-18$\tnote{a}  & Y & \cite{foley09, foley13, foley16}\\
		SN~2007J & $9-22$\tnote{a}  & Y & \cite{foley09, foley13, foley16}\\
		SN~2007J & $33-46$\tnote{a}  & Y & \cite{foley09, foley13, foley16}\\
		SN~2007J & $65-78$\tnote{a}  & Y & \cite{foley09, foley13, foley16}\\
		SN~2007ie & $21-26$\tnote{a}  & N & \cite{ostman11}\\ 
		SN~2007qd & 6 $\pm$ 3.36 \tnote{b} & Y & \cite{silverman12}\\
		SN~2008A & 20 & Y & \cite{matheson08}\\
		SN~2008A & 27 & Y & \cite{matheson08}\\ 
		SN~2008A & 29 & Y & \cite{silverman12}\\
		SN~2008A & 33 & Y & \cite{silverman12}\\
		SN~2008A & 42 & Y & \cite{matheson08}\\
		SN~2008A & 199 & Y & \cite{silverman12}\\
		SN~2008A & 224 & Y & \cite{silverman12}\\
		SN~2008A & 283 & Y & \cite{silverman12}\\
		SN~2008ae & 11 & Y & \cite{foley13}\\
		SN~2008ae & 15 & Y & \cite{foley13}\\
		SN~2008ae & 28 & Y & \cite{foley13}\\
		SN~2008ge & 37 $\pm$ 1.86 \tnote{b} & Y & \cite{silverman12}\\
		SN~2008ha & -1 & Y & \cite{valenti09}\\
		SN~2008ha & 8 & Y & \cite{foley09}\\
		SN~2008ha & 11 & Y & \cite{foley09}\\
		SN~2008ha & 16 & Y & \cite{valenti09}\\
		SN~2008ha & 23 & Y & \cite{foley09}\\
		SN~2008ha & 39 & Y & \cite{valenti09}\\
		SN~2008ha & 63 & Y & \cite{valenti09}\\
		SN~2009J & -5 & Y & \cite{foley13}\\
		SN~2009J & 0 & Y & \cite{foley13}\\
		SN~2009ku & 14 $\pm$ 1.86 \tnote{a}& Y & \cite{foley13}\\
		SN~2009ku & 16 $\pm$ 1.86 \tnote{a}& Y & \cite{foley13}\\
		SN~2009ku & 19 $\pm$ 1.86 \tnote{a}& Y & \cite{foley13}\\
		SN~2009ku & 38 $\pm$ 1.86 \tnote{a}& Y & \cite{foley13}\\	
	\end{tabular}
	\begin{tablenotes}
	\item[a] Phase calculated using SNID.
	\item[b] Phase converted using \cite{foley13}.
	\end{tablenotes}
	\end{threeparttable}
\end{table}

\begin{table}
	\caption{Spectral Data}
	\label{tab:spectra_table2}
	\begin{threeparttable}
	\begin{tabular}{cccc}
		\hline
		Object & Phase (days) & Phot. (Y or N) & Reference\\
		\hline
		SN~2010ae & -1 & Y & \cite{stritzinger14}\\
		SN~2010ae & 249 & Y & \cite{stritzinger14}\\
		SN~2011ay & 6 & Y & \cite{foley13}\\
		SN~2011ay & 15 & Y & \cite{foley13}\\
		SN~2011ay & 31 & Y & \cite{foley13}\\
		SN~2011ay & 55 & Y & \cite{foley13}\\
		SN~2011ay & 180 & Y & \cite{foley13}\\
		SN~2011ce & 14-20 & Y & \cite{foley13}\\
		SN~2011ce & 367 & Y & \cite{foley13}\\
		SN~2012Z & 6 & Y & \cite{stritzinger15}\\
		SN~2012Z & 9 & Y & \cite{yamanaka15}\\
		SN~2012Z & 11 & Y & \cite{stritzinger15}\\
		SN~2012Z & 15 & Y & \cite{stritzinger15}\\
		SN~2012Z & 17 & Y & \cite{yamanaka15}\\
		SN~2012Z & 21 & Y & \cite{stritzinger15}\\
		SN~2012Z & 30 & Y & \cite{yamanaka15}\\
		SN~2012Z & 32 & Y & \cite{yamanaka15}\\
		SN~2012Z & 34 & Y & \cite{stritzinger15}\\
		SN~2012Z & 194 & Y & \cite{stritzinger15}\\
		SN~2012Z & 215 & Y & \cite{stritzinger15}\\
		SN~2012Z & 248 & Y & \cite{stritzinger15}\\
		SN~2012Z & 255 & Y & \cite{stritzinger15}\\
		PS1-12bwh & -3 $\pm 2.82$ \tnote{a}  & Y & \cite{magee17}\\
		PS1-12bwh & 17 $\pm 2.82$ \tnote{a}  & Y & \cite{magee17}\\
		PS1-12bwh & 23 $\pm 2.82$ \tnote{a}  & Y & \cite{magee17}\\
		PS1-12bwh & 53 $\pm 2.82$ \tnote{a} & Y & \cite{magee17}\\
		LSQ12fhs & 23 & N & \cite{copin12}\\
		SN~2013dh & 14 & Y & \cite{childress16}\\
		SN~2013en & 4 & Y & \cite{liu15}\\
		SN~2013en & 10 & Y & \cite{liu15}\\
		SN~2013en & 14 & Y & \cite{liu15}\\
		SN~2013en & 45 & Y & \cite{liu15}\\
		SN~2013en & 60 & Y & \cite{liu15}\\
		SN~2013gr & $10-20$\tnote{b} & Y & \cite{childress16}\\
		SN~2013gr & $19-29$\tnote{b} & Y & \cite{childress16}\\
		SN~2013gr & $31-41$\tnote{b} & Y & \cite{childress16}\\
		SN~2014ck & 13 $\pm$ 1.86 \tnote{a}  & Y & \cite{tomasella16}\\
		SN~2014ck & 15 $\pm$ 1.86 \tnote{a}  & Y & \cite{tomasella16}\\
		SN~2014ck & 19 $\pm$ 1.86 \tnote{a}  & Y & \cite{tomasella16}\\
		SN~2014cr & $20-29$\tnote{b} & N & \cite{childress16}\\
		SN~2014dt & 20 $\pm$ 2.82 \tnote{a}  & Y & \cite{foley15}\\
		SN~2014dt & 140  & Y & \cite{foley16}\\
		SN~2014dt & 169  & Y & \cite{foley16}\\
		SN~2014dt & 200  & Y & \cite{foley16}\\
		SN~2014dt & 225  & Y & \cite{foley16}\\
		SN~2014dt & 268 & Y & \cite{foley16}\\
		SN~2014dt & 408 & Y & \cite{foley16}\\
		SN~2014ey & -$3-3$\tnote{b} & N & \cite{wiserep12}\\
		LSQ14dtt & $6-20$\tnote{b} & N & \cite{rosa14}\\
		SN~2015H & -1 $\pm$ 2.82 \tnote{a}  & Y & \cite{magee16}\\
		SN~2015H & 2 $\pm$ 2.82 \tnote{a} & Y & \cite{magee16}\\
		SN~2015H & 6 $\pm$ 2.82 \tnote{a}  & Y & \cite{magee16}\\
		SN~2015H & 26 $\pm$ 2.82 \tnote{a}  & Y & \cite{magee16}\\
		SN~2015H & 103 $\pm$ 2.82 \tnote{a}  & Y & \cite{magee16}\\
		PS15aic & $9-15$\tnote{b} & N & \cite{pan15}\\
		PS15csd & $4-13$\tnote{b} & N & \cite{harmanen15}\\
		SN~2015ce & -$5-1$\tnote{b} & N & \cite{balam17}\\
		SN~2016atw & $5-15$\tnote{b} & N & \cite{pan16}\\
		OGLE16erd & 0 $\pm$ 2.82 \tnote{a} & N & \cite{dimitriadis16}\\
		SN~2016ilf & $4-13$ & N & \cite{zhang16}\\
		iPTF16fnm & 1 $\pm$ 1.86 \tnote{a}  & Y & \cite{miller17}\\
		iPTF16fnm & 3 $\pm$ 1.86  \tnote{a}  & Y & \cite{miller17}\\
		iPTF16fnm & 7 $\pm$ 1.86  \tnote{a}  & Y & \cite{miller17}\\
		iPTF16fnm & 28 $\pm$ 1.86  \tnote{a}  & Y & \cite{miller17}\\
		SN~2017gbb & $6-19$\tnote{b} & N & \cite{lyman17}\\
		SN~2018atb & $10-29$\tnote{b} & N & \cite{heintz18}\\
	\end{tabular}
	\begin{tablenotes}
	\item[a] Phase converted using \cite{foley13}.
	\item[b] Phase calculated using SNID.
	\end{tablenotes}
	\end{threeparttable}
\end{table}

\newpage

\begin{figure*}
\subfigure[]{\includegraphics[width=.48\textwidth]{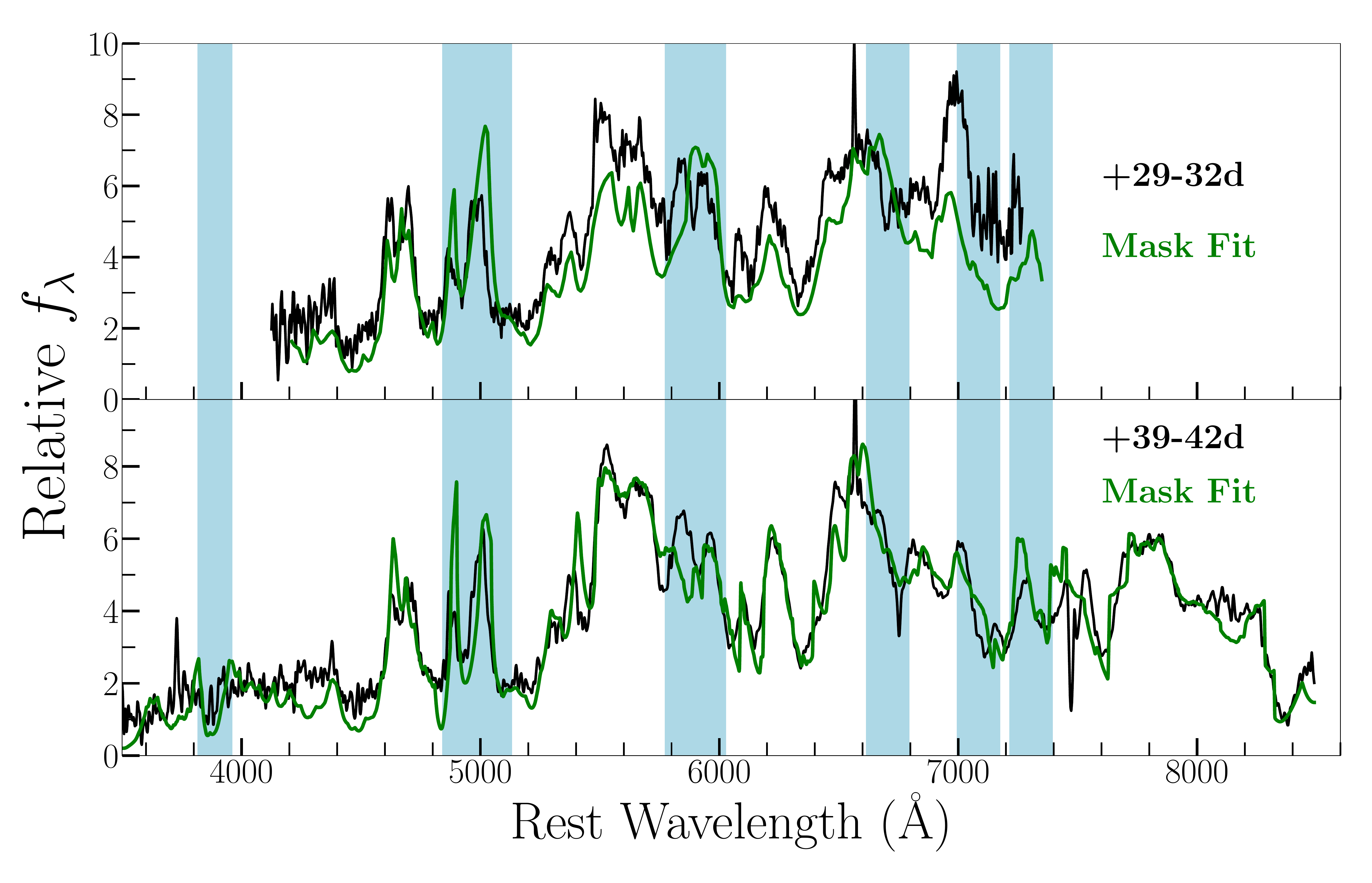}}
\subfigure[]{\includegraphics[width=.48\textwidth]{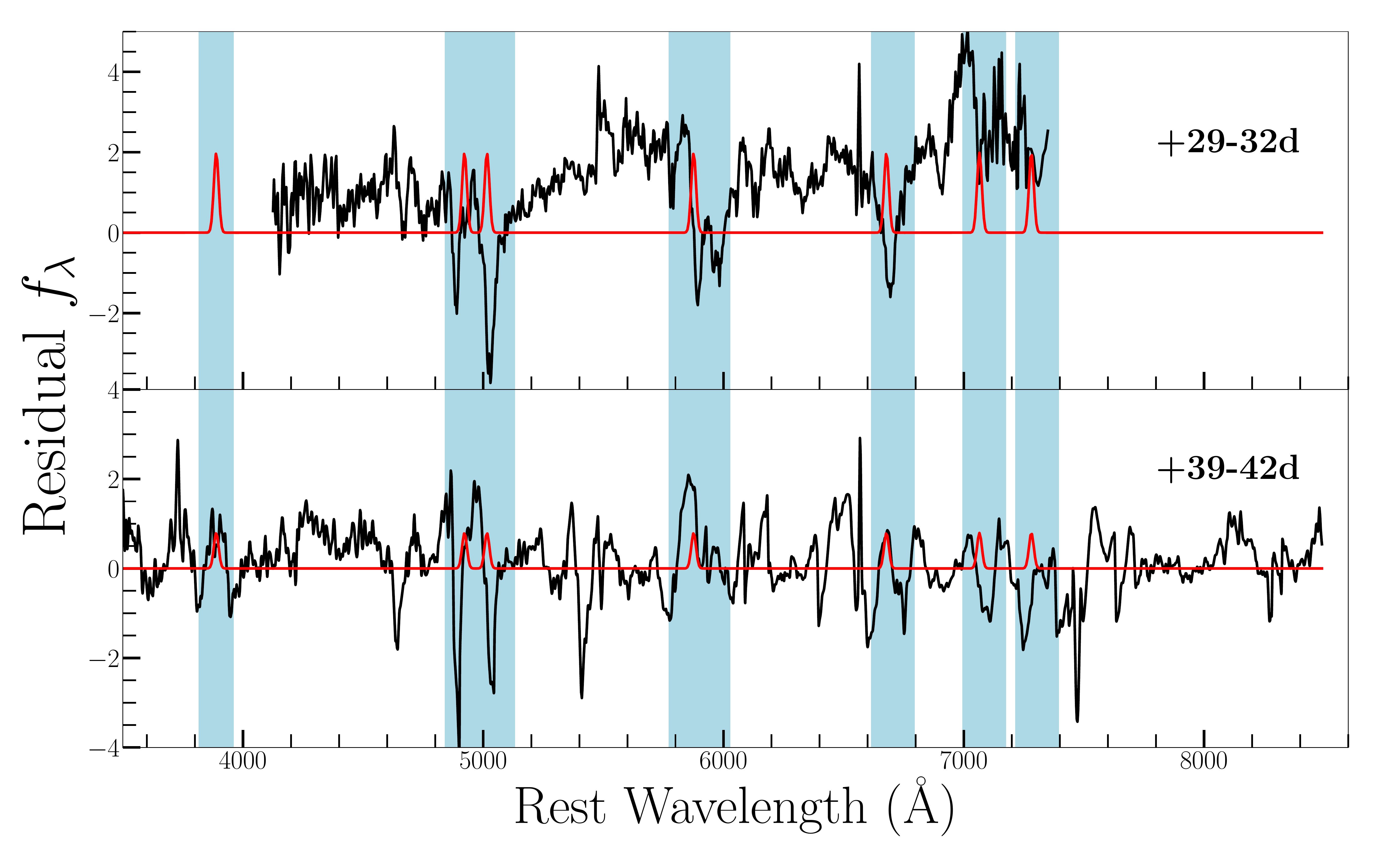}}\\[1ex]
\subfigure[]
{\includegraphics[width=0.8\textwidth]{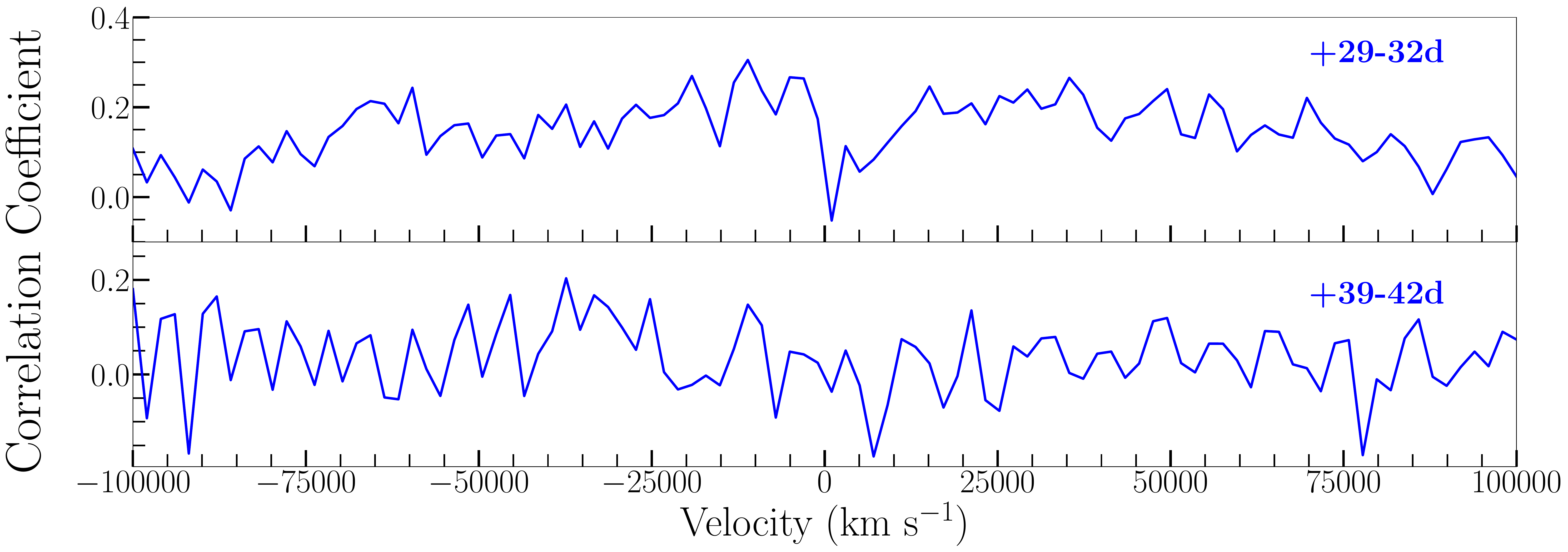}}
\caption{SN~1991bj. Phase relative to B band maximum and calculated using SNID. \label{fig:91bj_combo} }
\end{figure*}

\begin{figure}
\begin{center}
	\includegraphics[width=0.49\textwidth]{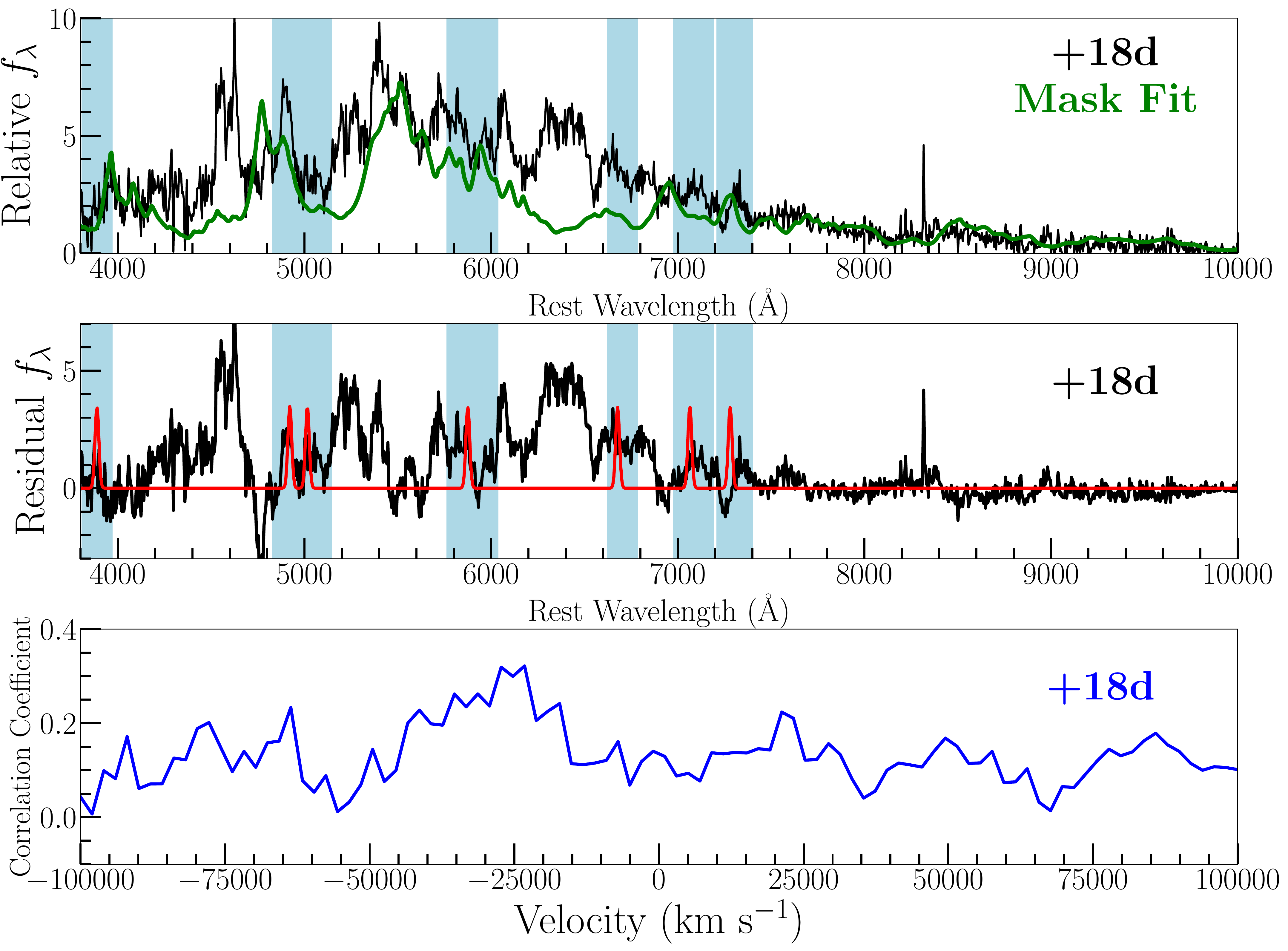}
    \vspace*{-5mm}
	\caption{SN~1999ax. Phase relative to B band maximum and calculated using SNID.} \label{fig:combo_99ax}
\end{center}
\end{figure}

\begin{figure}
\begin{center}
	\includegraphics[width=0.49\textwidth]{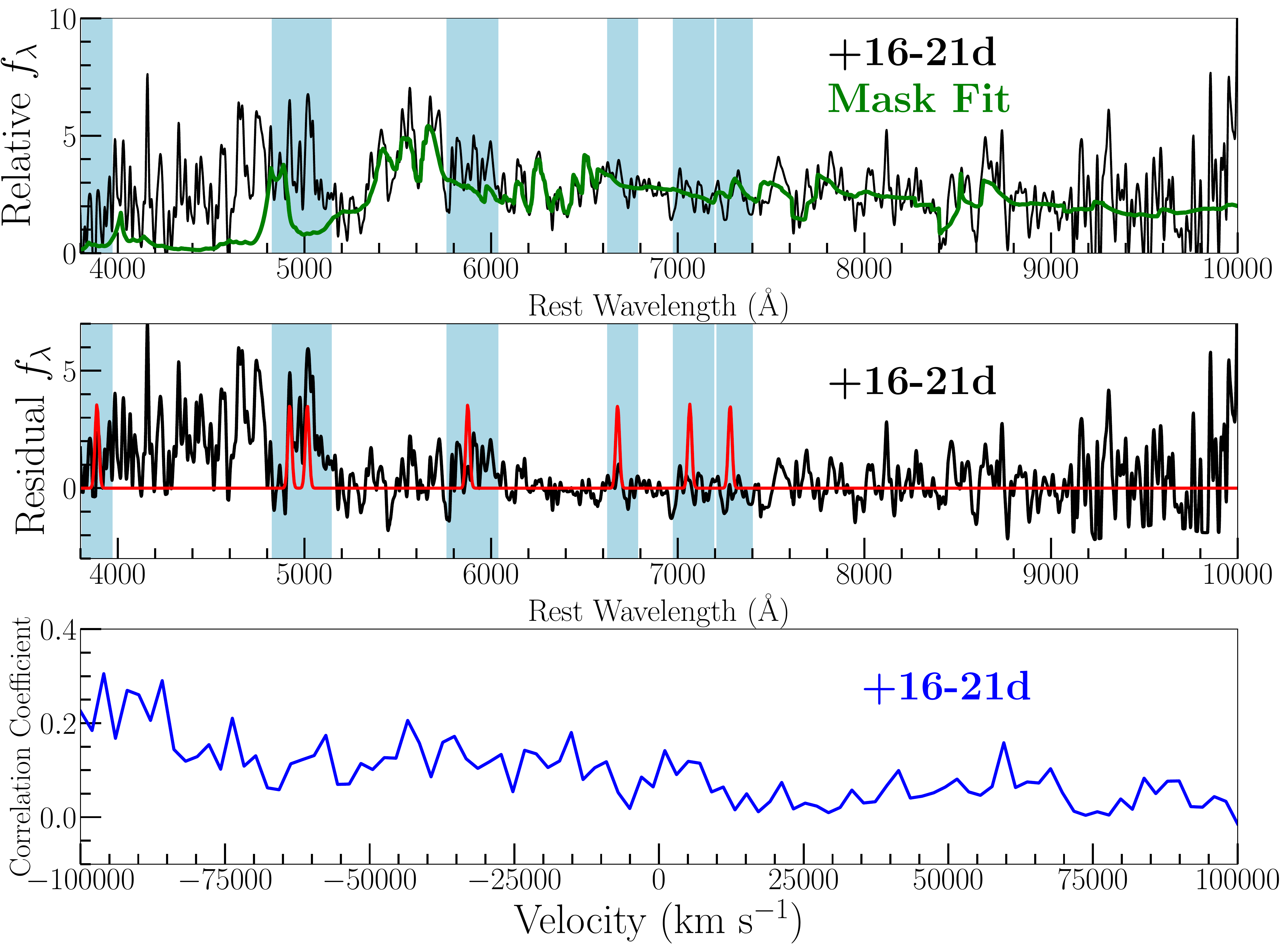}
    \vspace*{-5mm}
	\caption{SN~2002bp. Phase relative to B band maximum and calculated using SNID.} \label{fig:combo_02bp}
\end{center}
\end{figure}

\begin{figure*}
\subfigure[]{\includegraphics[width=.48\textwidth]{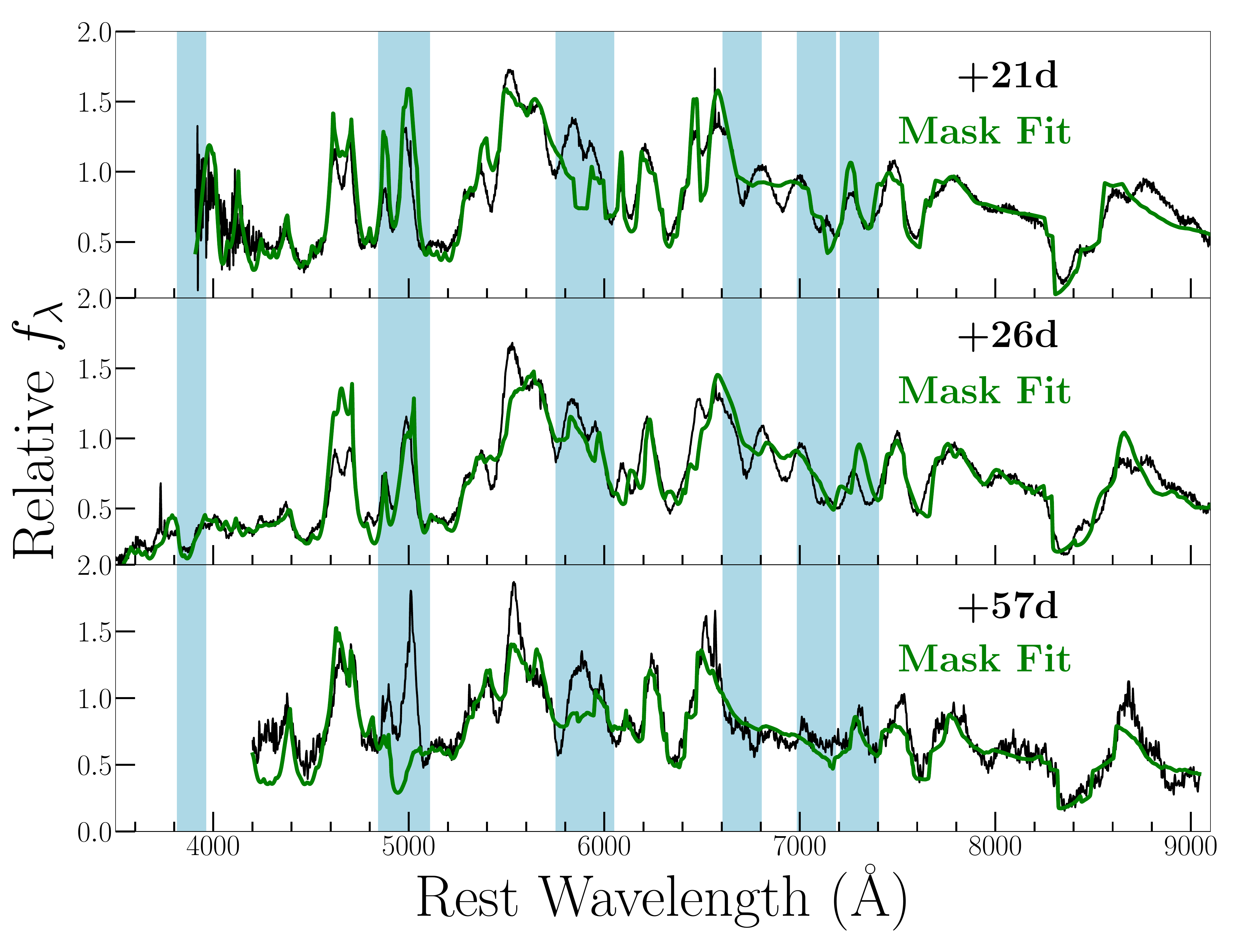}}
\subfigure[]{\includegraphics[width=.48\textwidth]{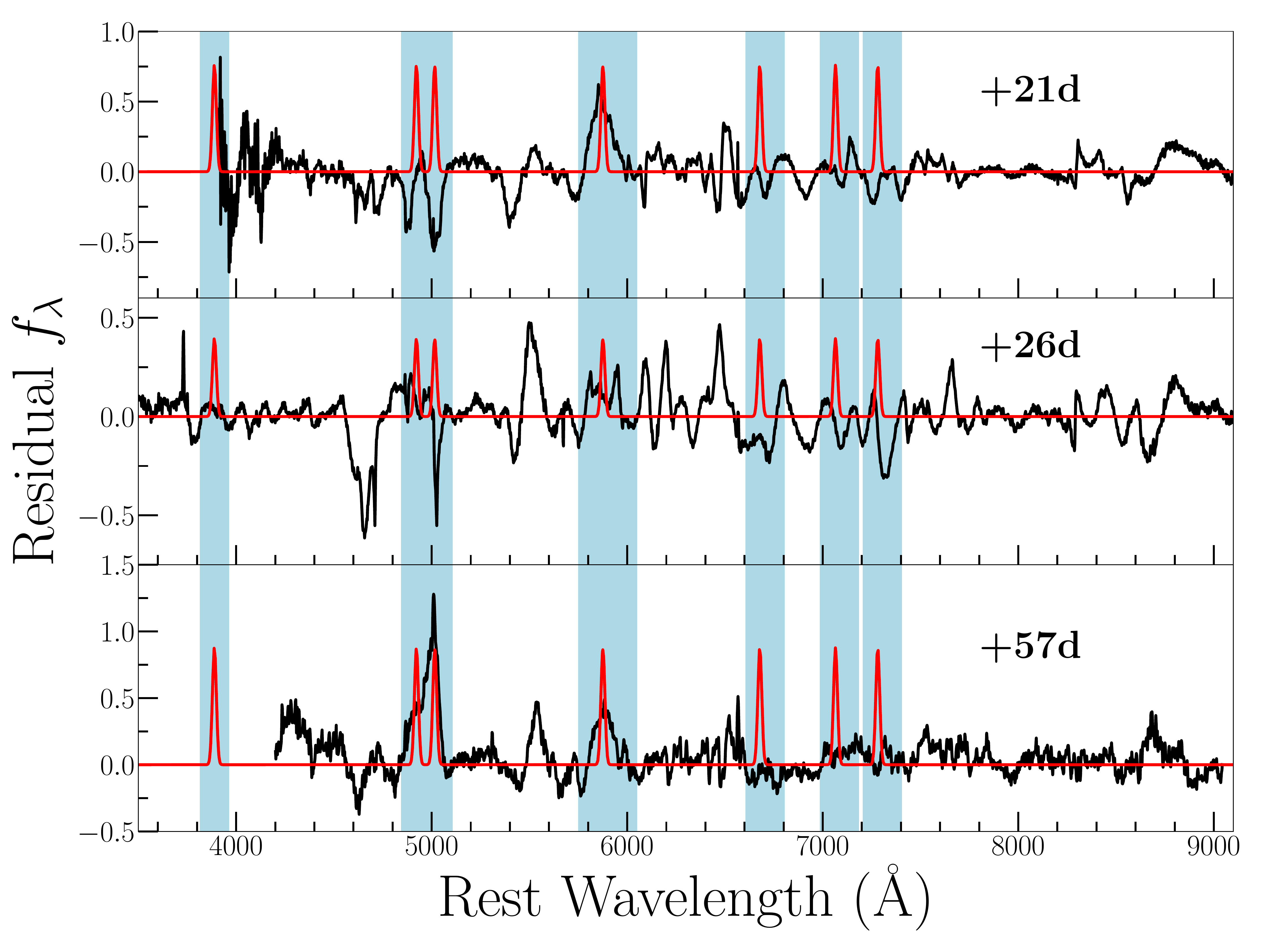}}\\[1ex]
\subfigure[]
{\includegraphics[width=0.8\textwidth]{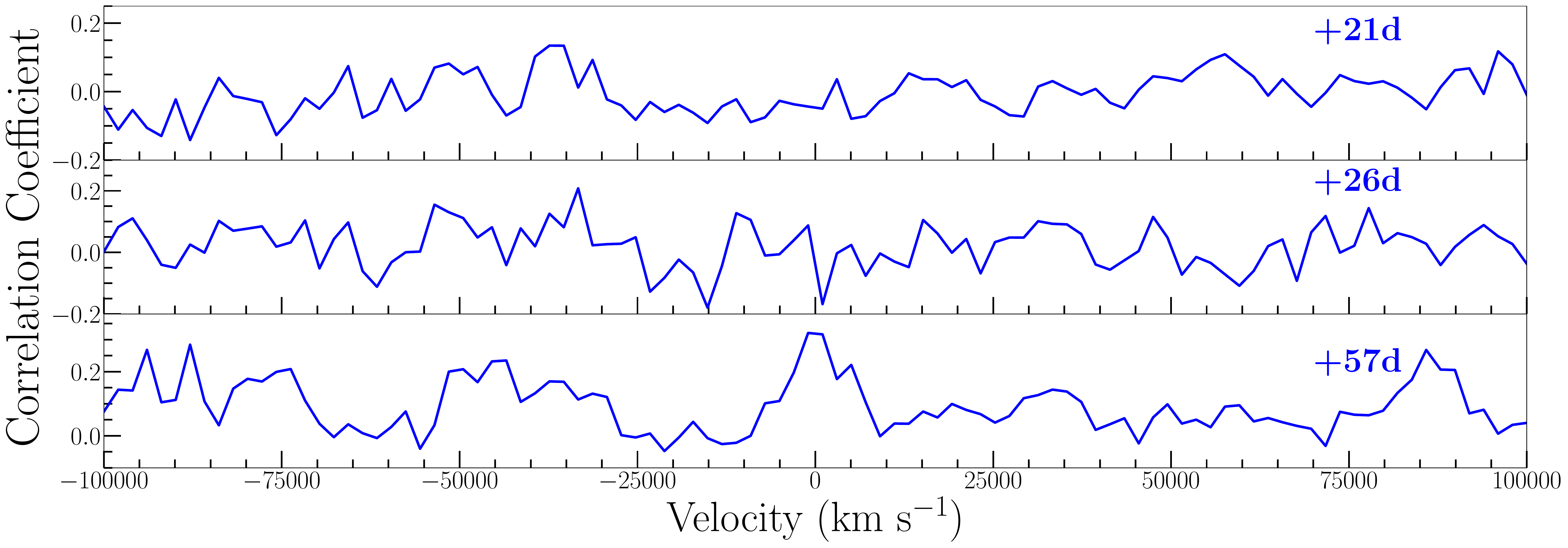}}
\caption{SN~2002cx. Phase relative to B band maximum. \label{fig:02cx_combo} }
\end{figure*}

\begin{figure}
\begin{center}
	\includegraphics[width=0.49\textwidth]{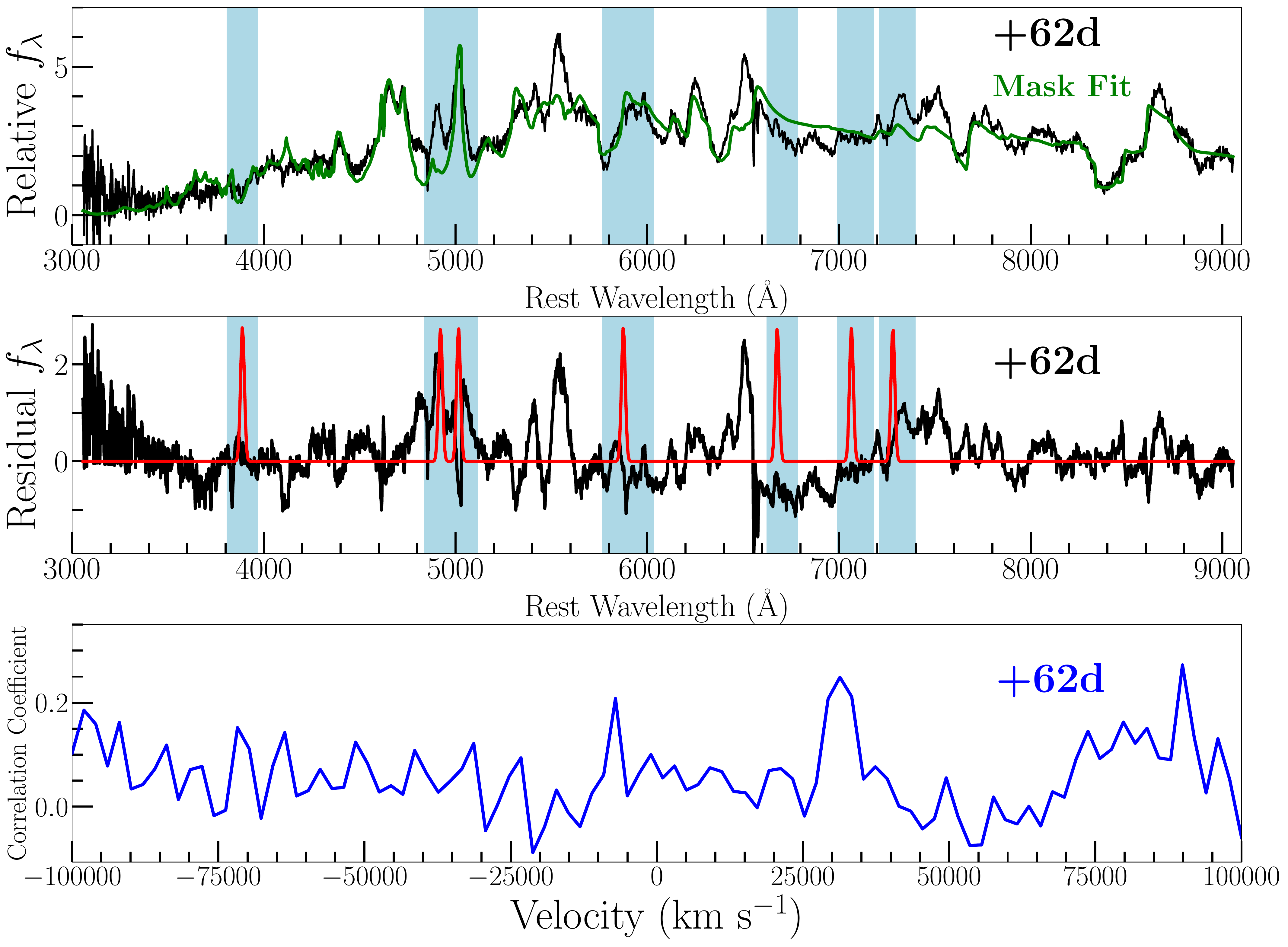}
    \vspace*{-5mm}
	\caption{SN~2003gq. Phase relative to B band maximum.} \label{fig:combo_03gq}
\end{center}
\end{figure}

\begin{figure*}
\subfigure[]{\includegraphics[width=.48\textwidth]{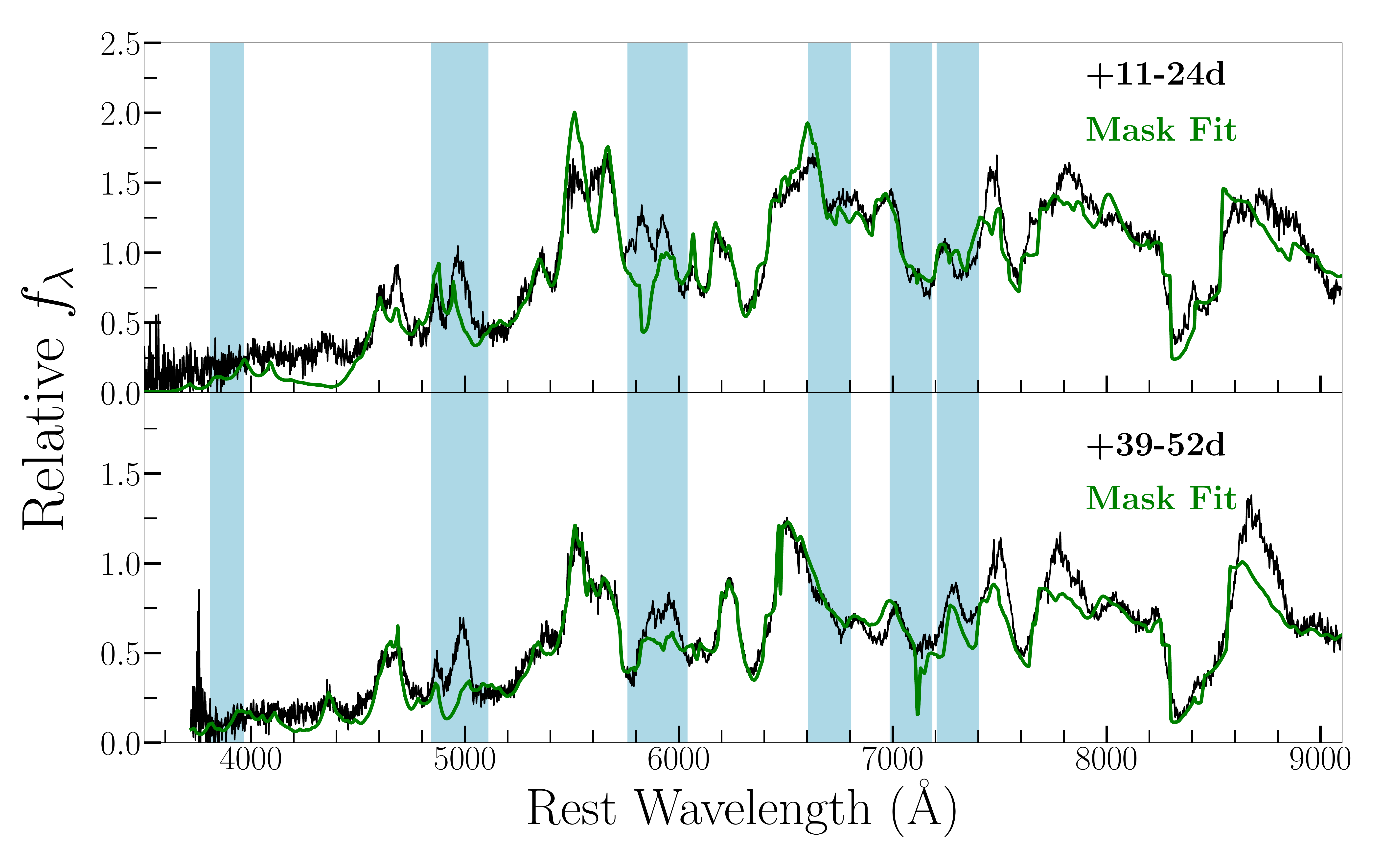}}
\subfigure[]{\includegraphics[width=.48\textwidth]{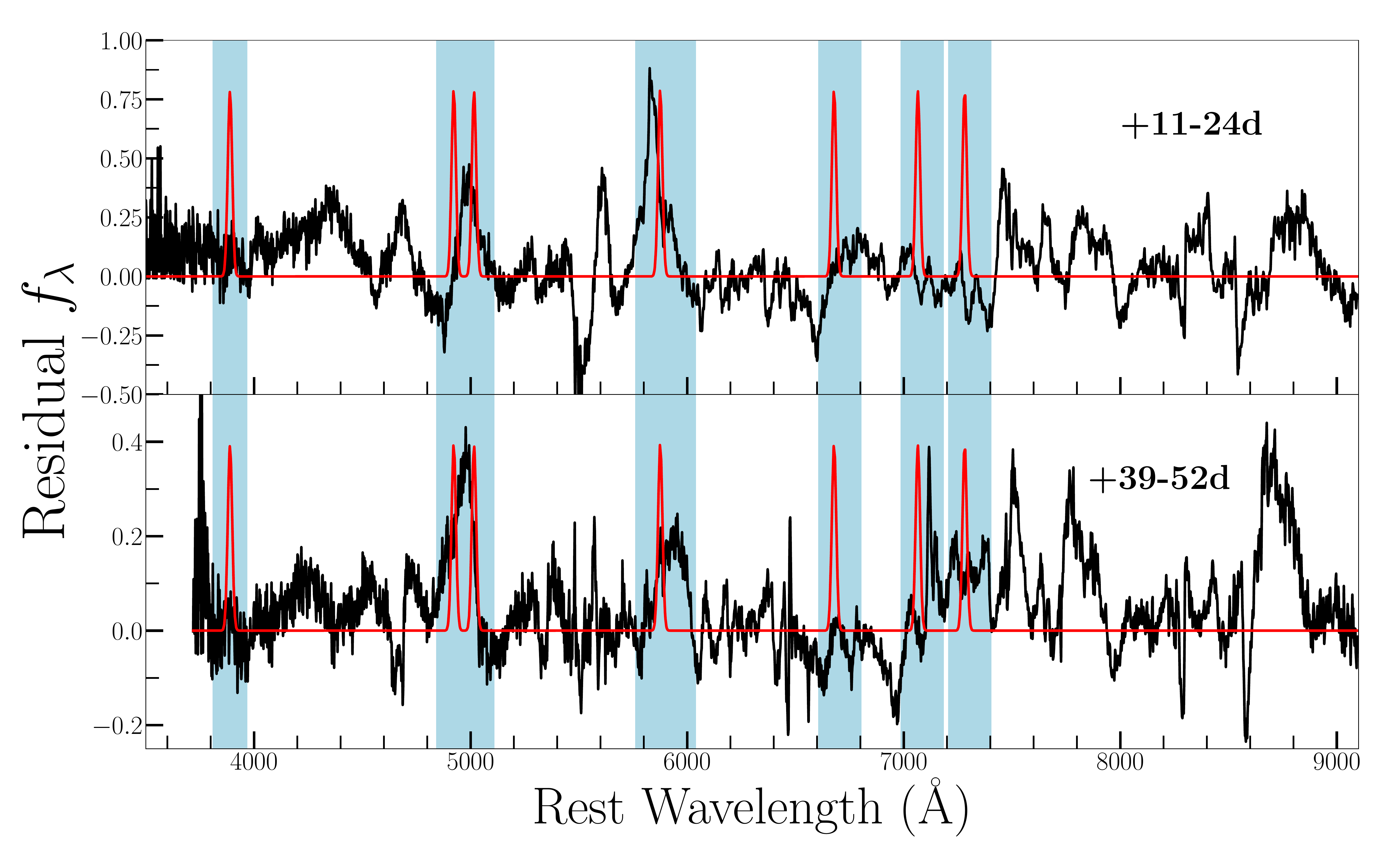}}\\[1ex]
\subfigure[]
{\includegraphics[width=0.8\textwidth]{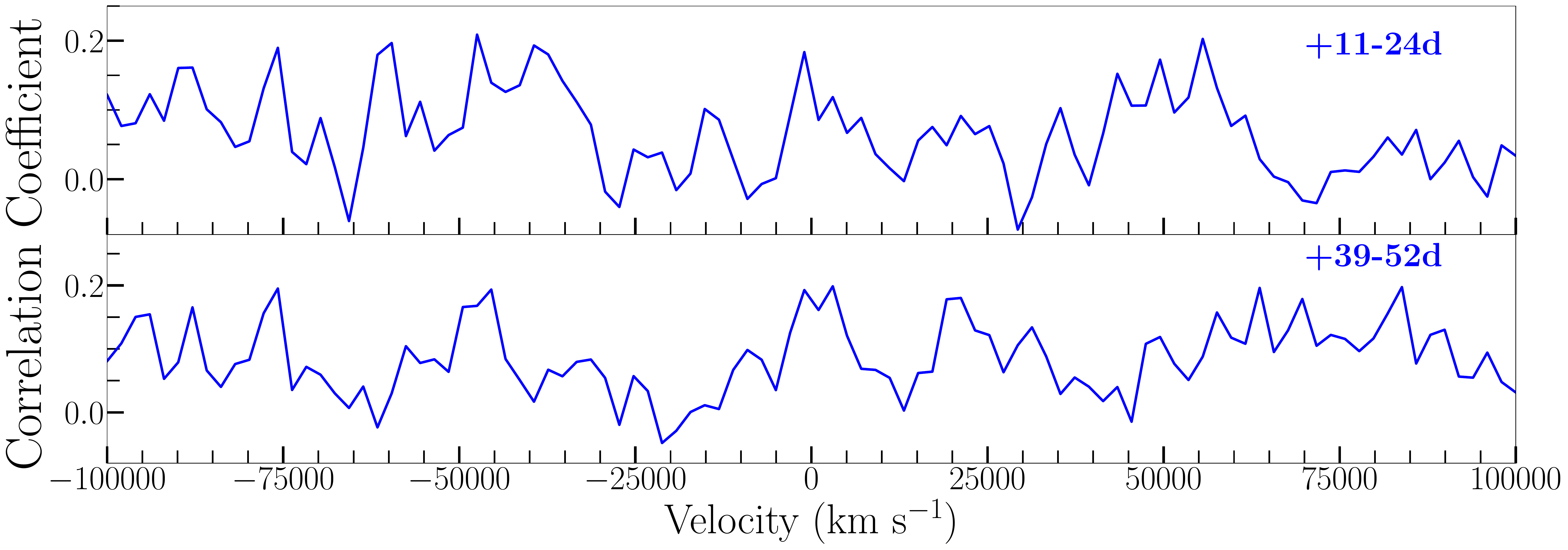}}
\caption{SN~2004gw. Phase relative to B band maximum and calculated using SNID. \label{fig:04gw_combo} }
\end{figure*}

\begin{figure}
\begin{center}
	\includegraphics[width=0.49\textwidth]{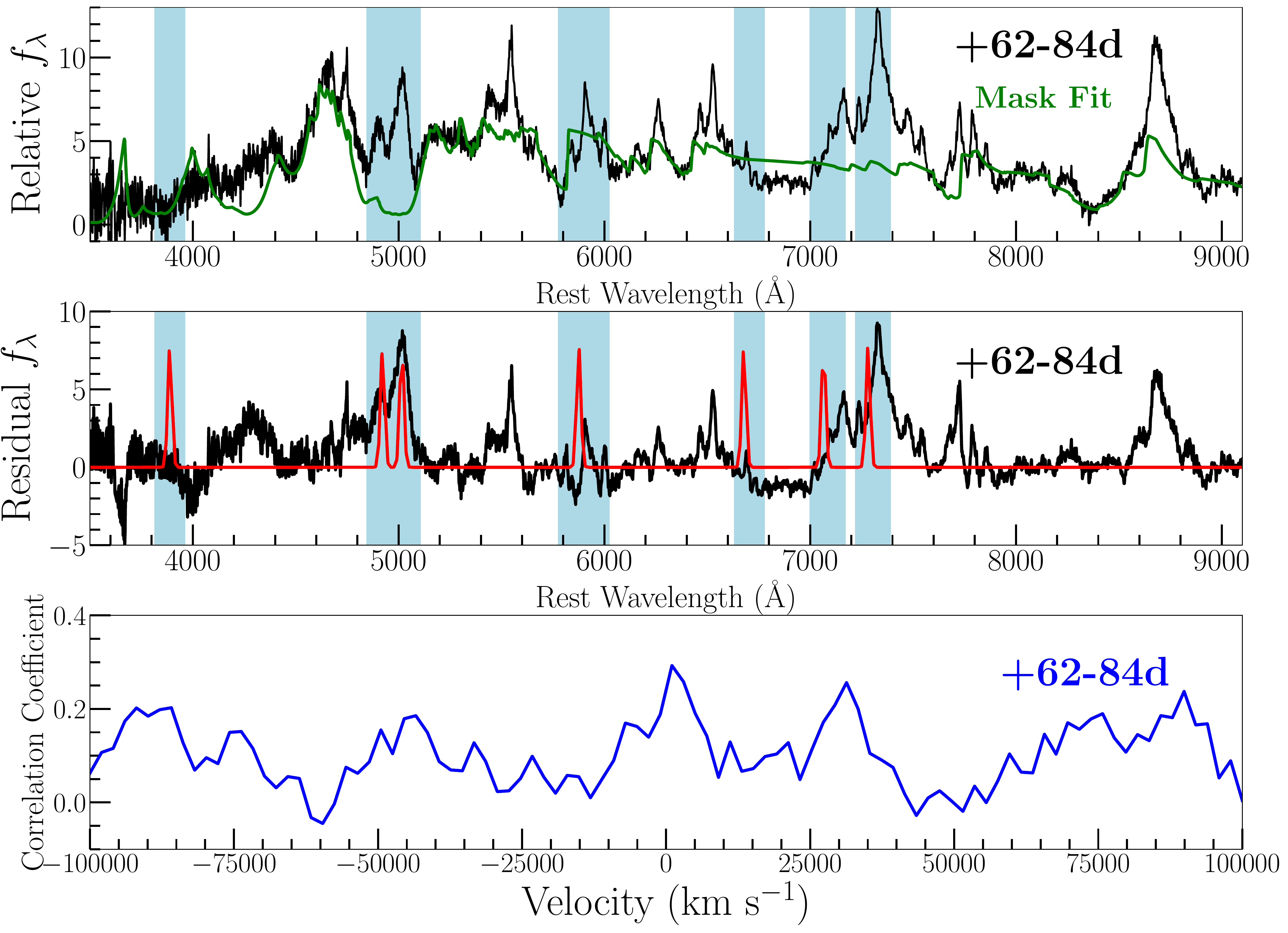}
	\caption{SN~2005p. Phase relative to B band maximum and calculated using SNID. Correlation peak not considered significant because it is generated by the [Ca II] feature.} \label{fig:combo_05p}
\end{center}
\vspace*{-5mm}
\end{figure}

\begin{figure*}
\subfigure[]{\includegraphics[width=.48\textwidth]{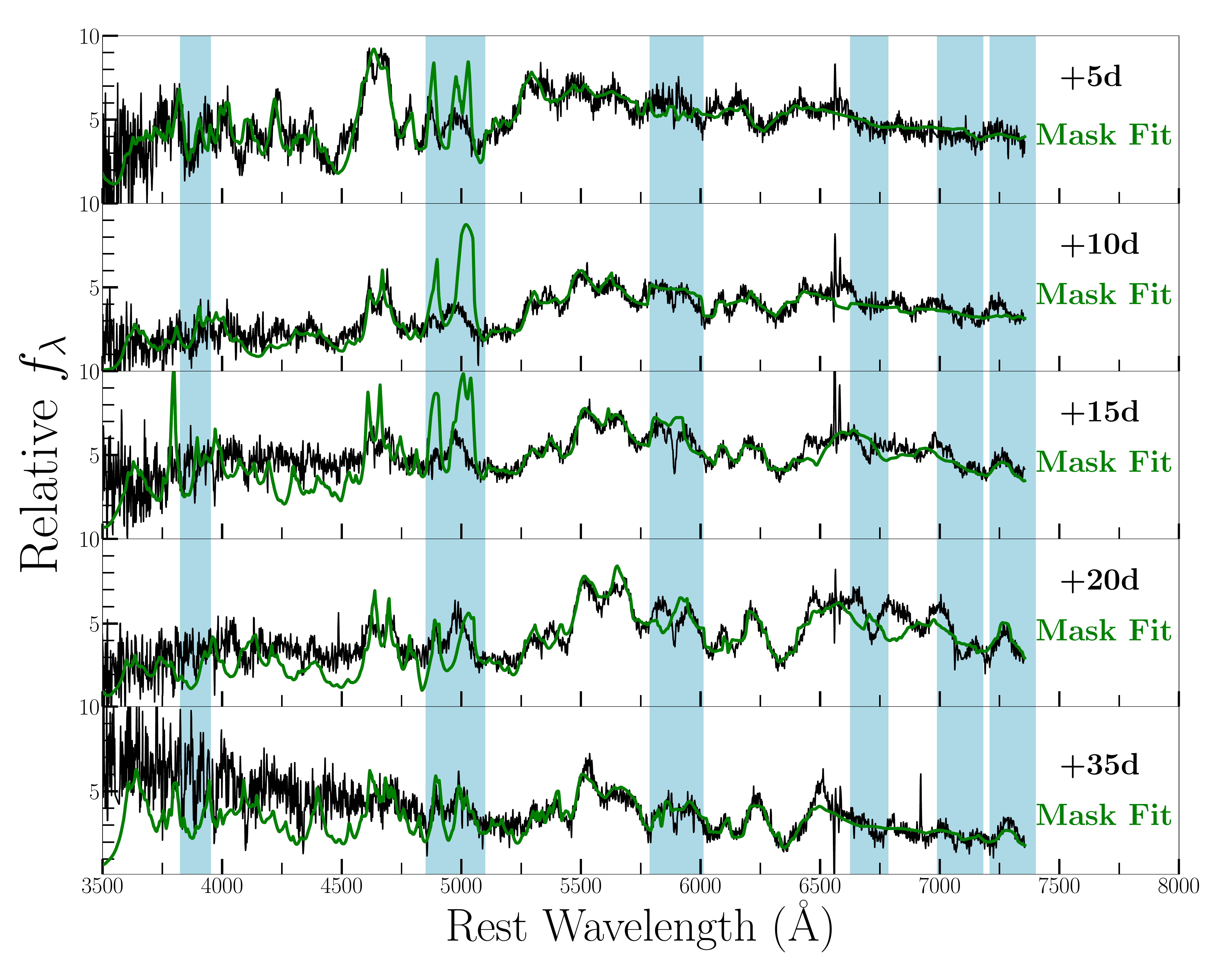}}
\subfigure[]{\includegraphics[width=.483\textwidth]{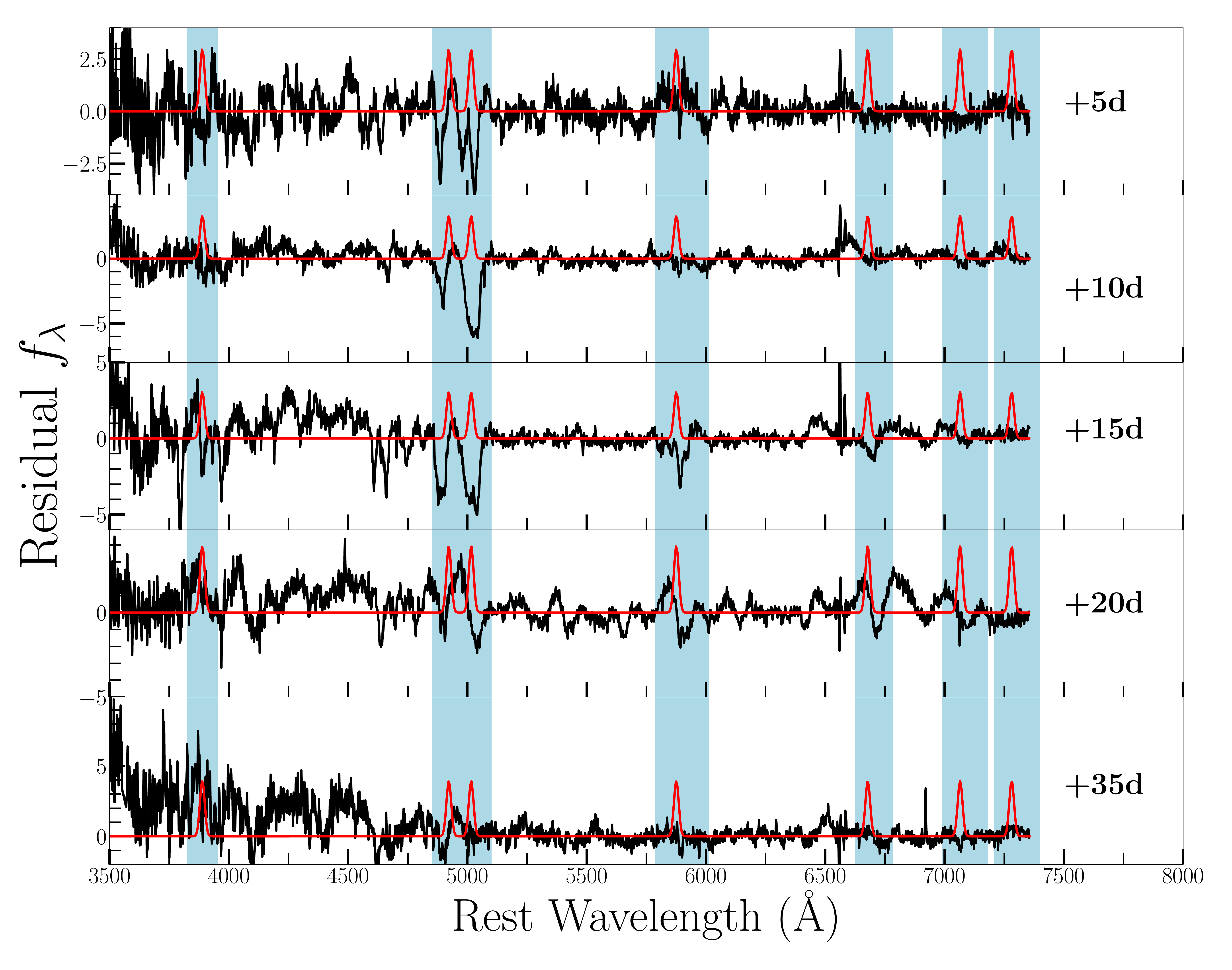}}\\[1ex]
\subfigure[]
{\includegraphics[width=0.8\textwidth]{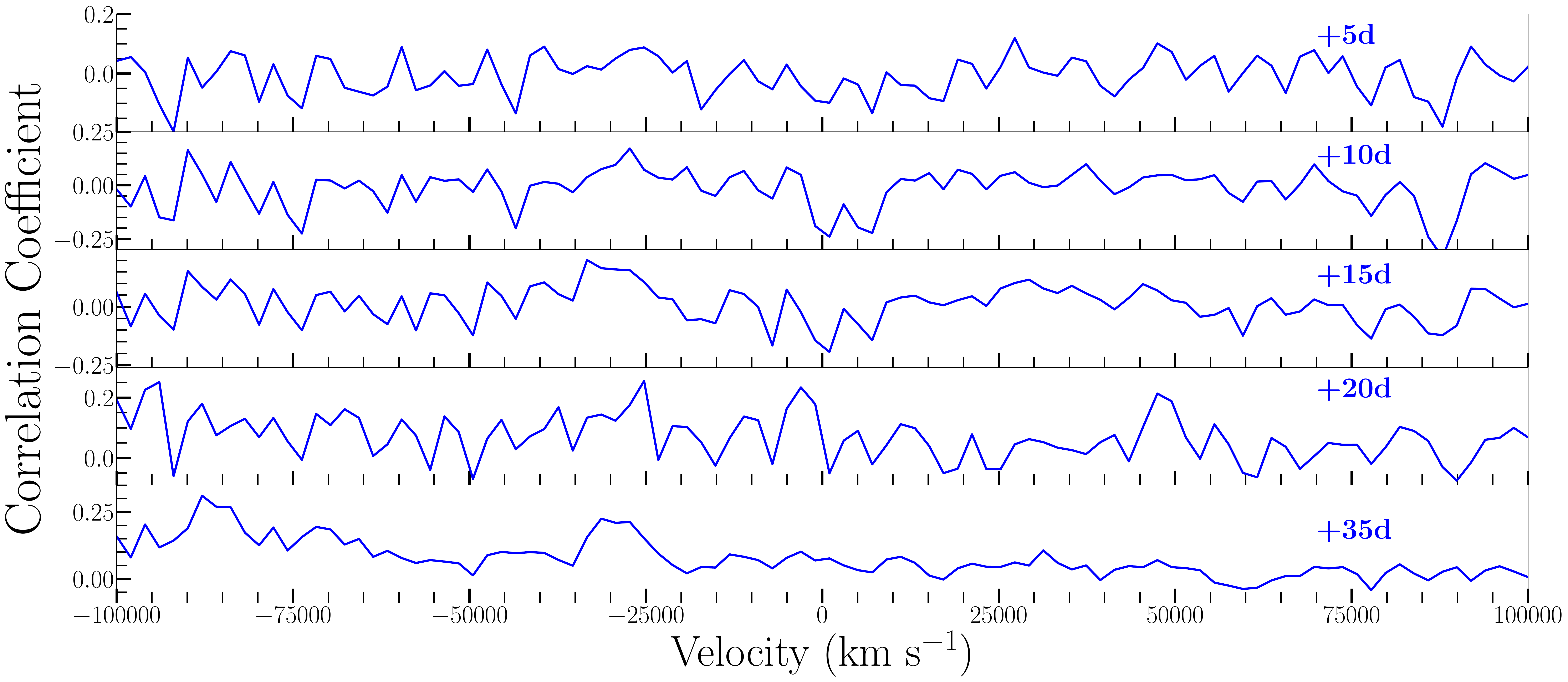}}
\caption{SN~2005cc. Phase relative to B band maximum. \label{fig:05cc_combo} }
\end{figure*}

\begin{figure*}
\subfigure[]{\includegraphics[width=.48\textwidth]{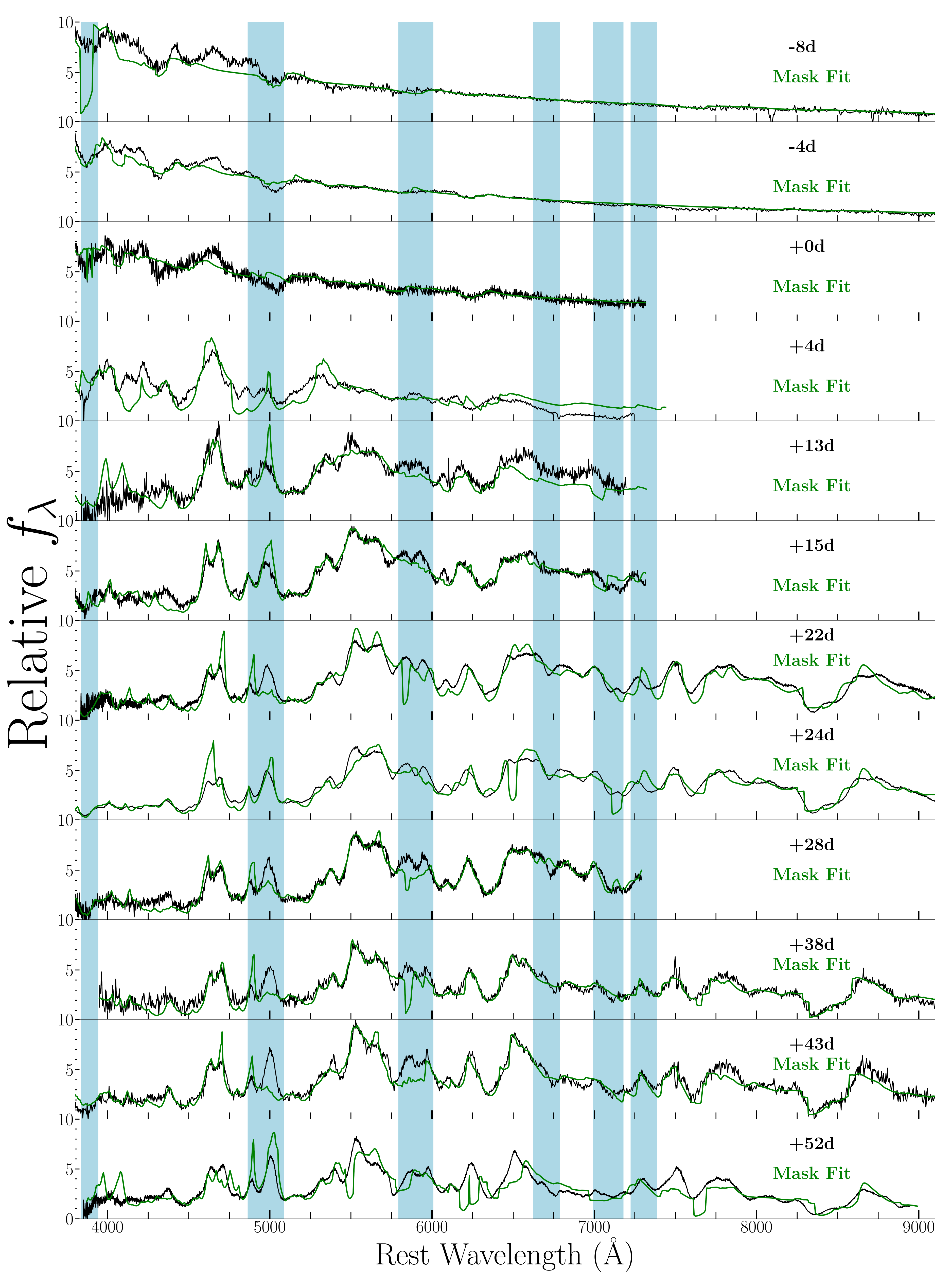}}
\subfigure[]{\includegraphics[width=.475\textwidth]{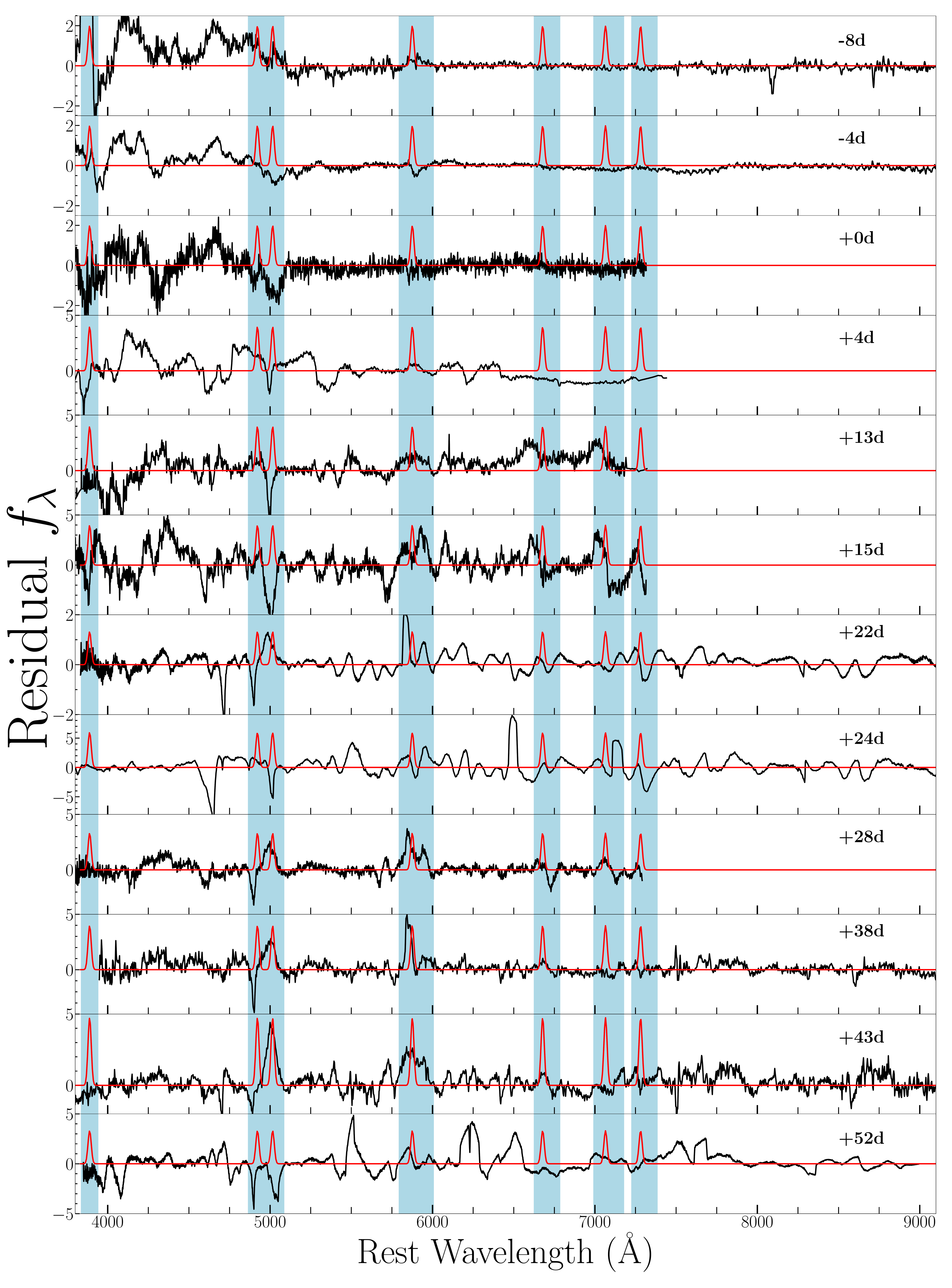}}\\[1ex]
\subfigure[]
{\includegraphics[width=0.8\textwidth]{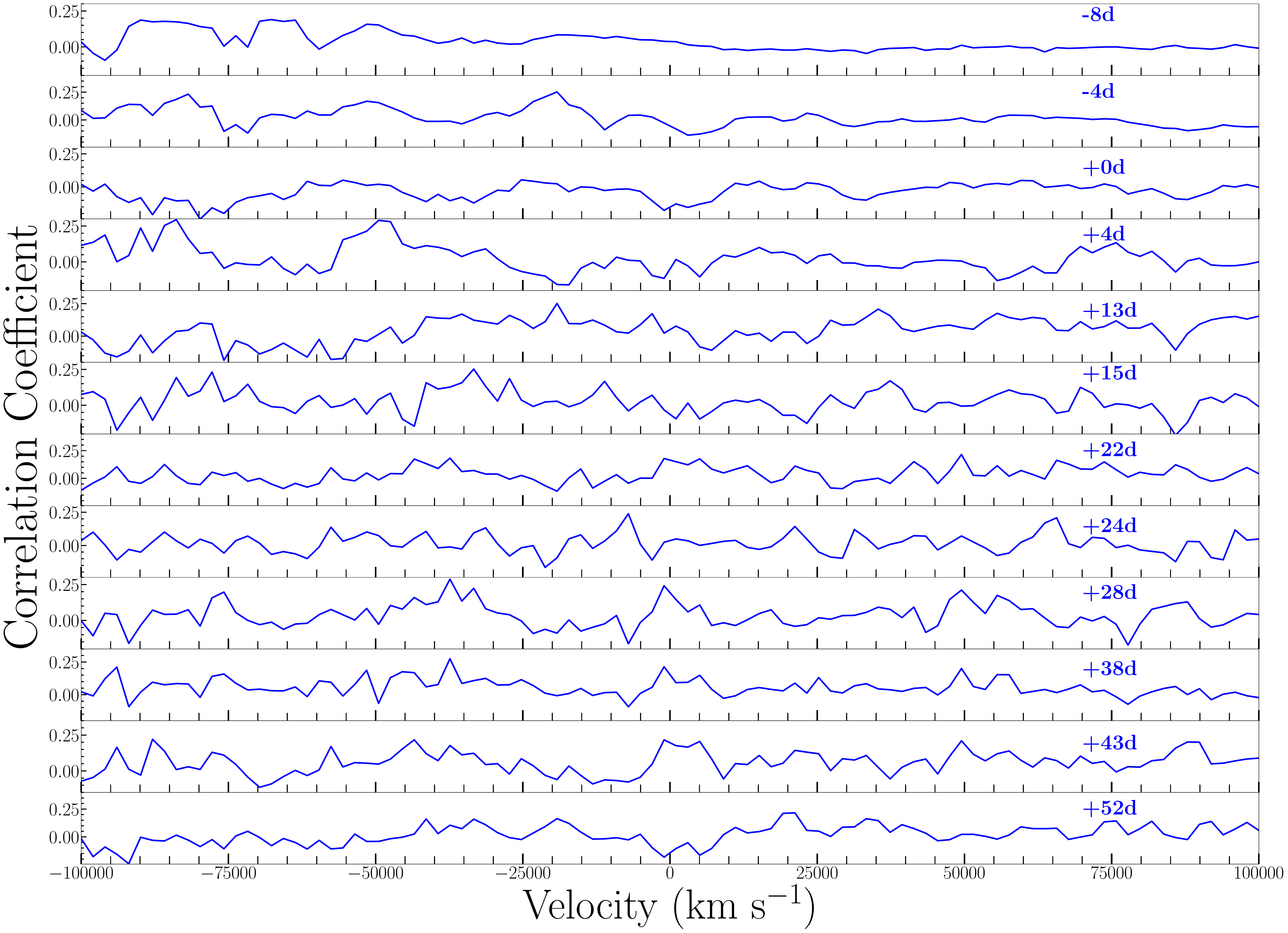}}
\caption{SN~2005hk. Phase relative to B band maximum. \label{fig:05hk_combo} }
\end{figure*}

\begin{figure}
\begin{center}
	\includegraphics[width=0.49\textwidth]{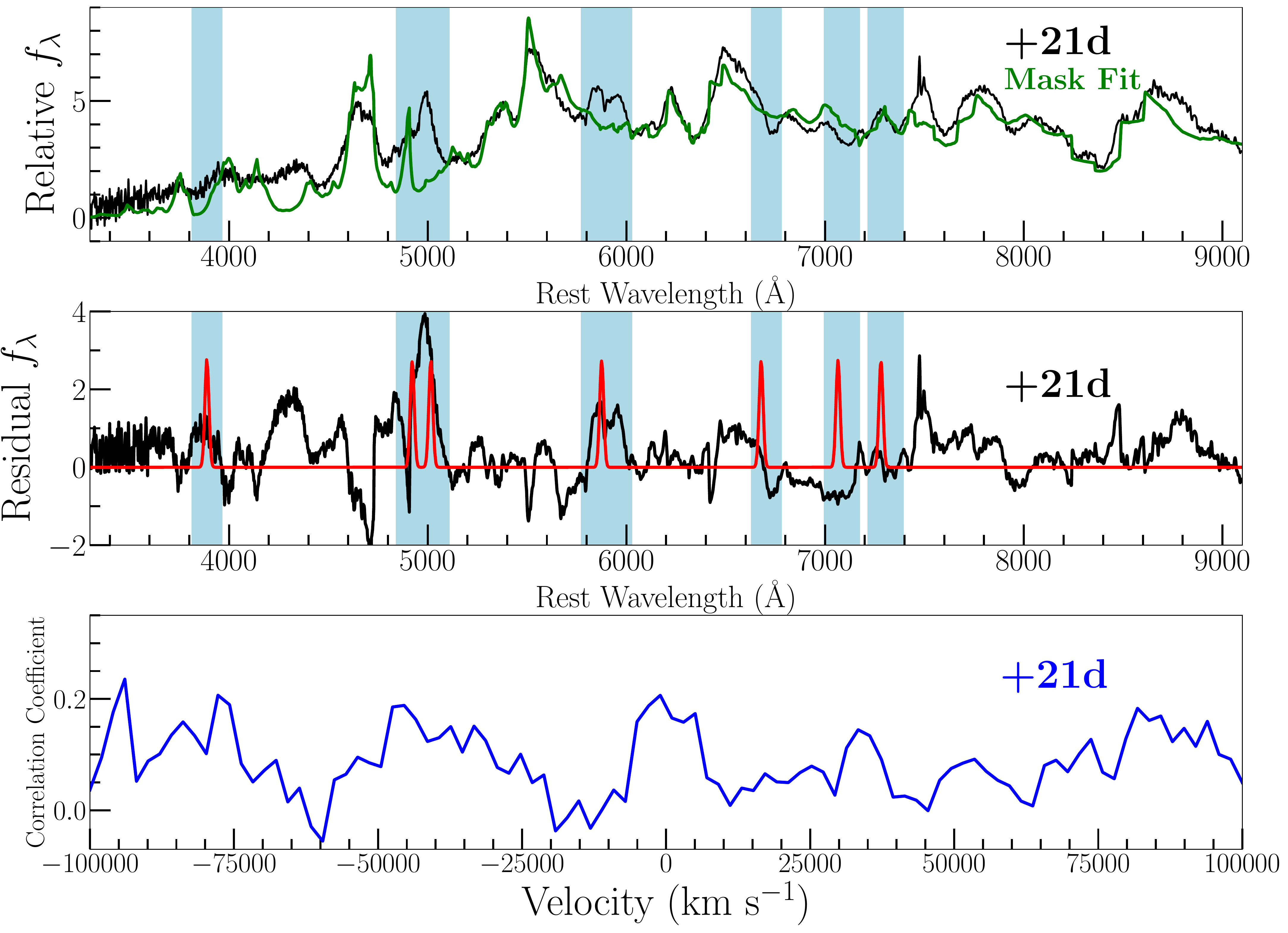}
	\caption{SN~2006hn. Phase relative to B band maximum. } \label{fig:combo_06hn}
\end{center}
\end{figure}

\begin{figure}
\begin{center}
	\includegraphics[width=0.49\textwidth]{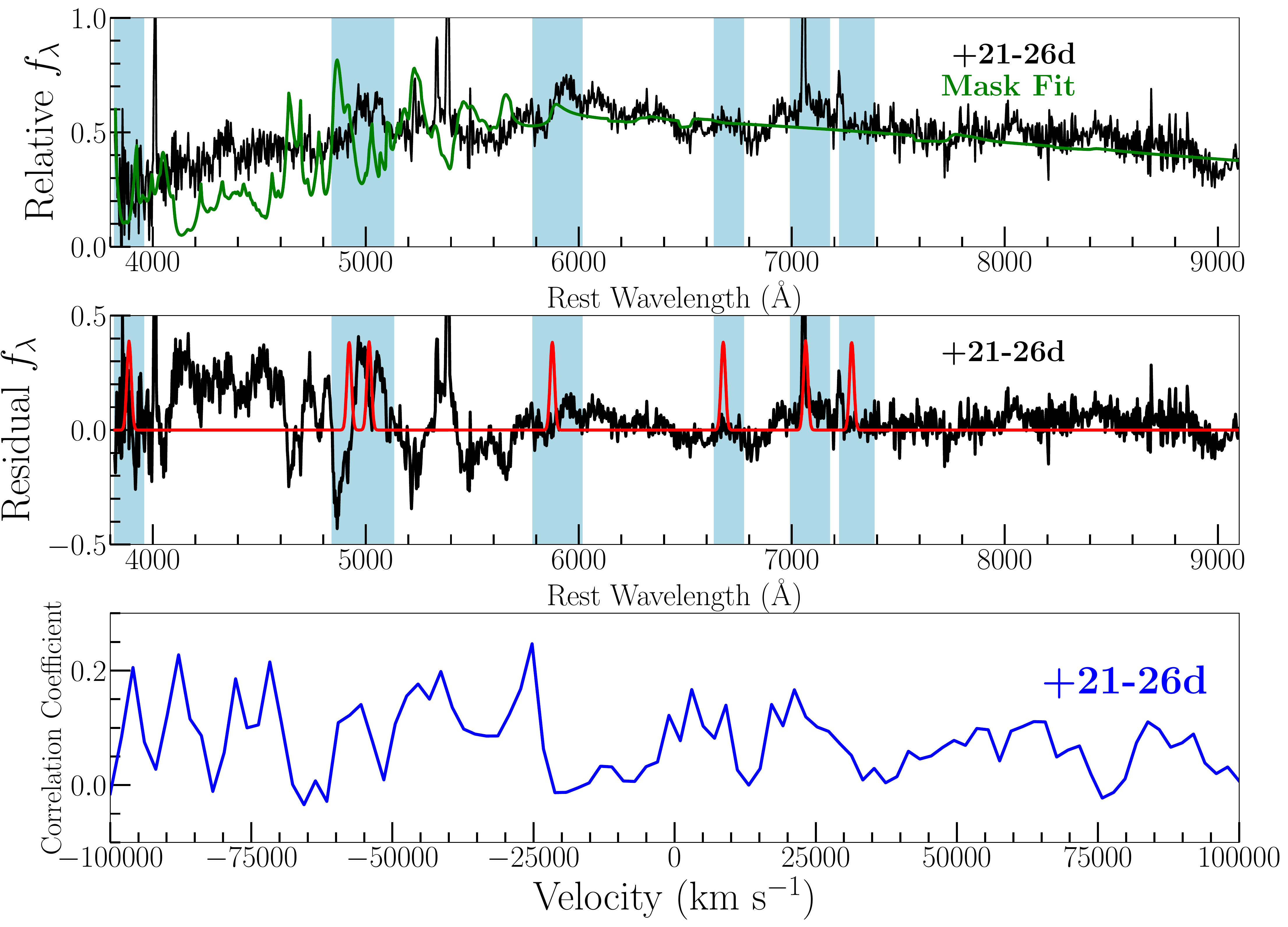}
	\caption{SN~2007ie. Phase relative to B band maximum and calculated using SNID.} \label{fig:combo_07ie}
\end{center}
\vspace*{-5mm}
\end{figure}

\begin{figure}
\begin{center}
	\includegraphics[width=0.49\textwidth]{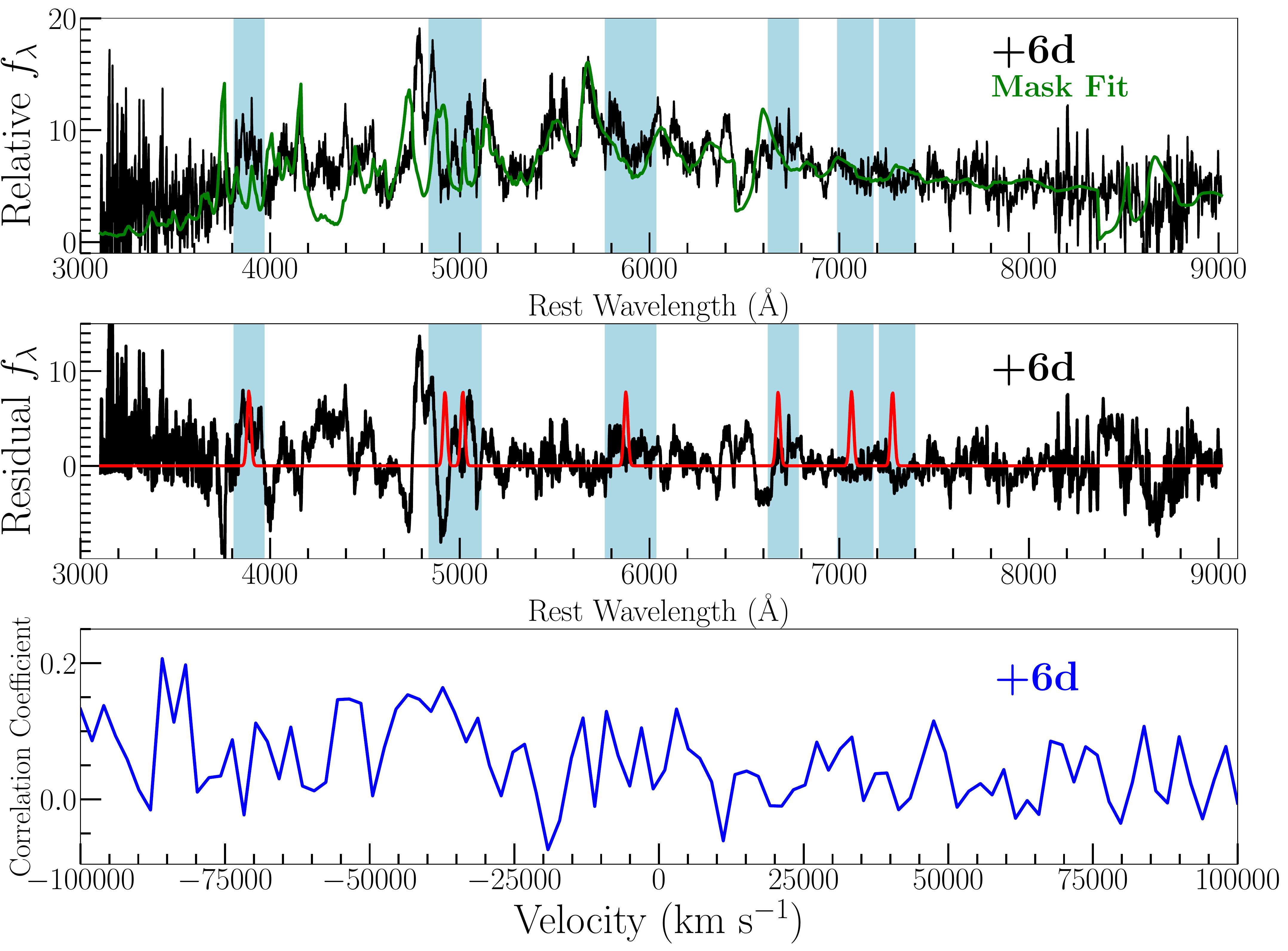}
	\caption{SN~2007qd. Phase relative to B band maximum.} \label{fig:combo_07qd}
\end{center}
\vspace*{-5mm}
\end{figure}

\begin{figure*}
\subfigure[]{\includegraphics[width=.47\textwidth]{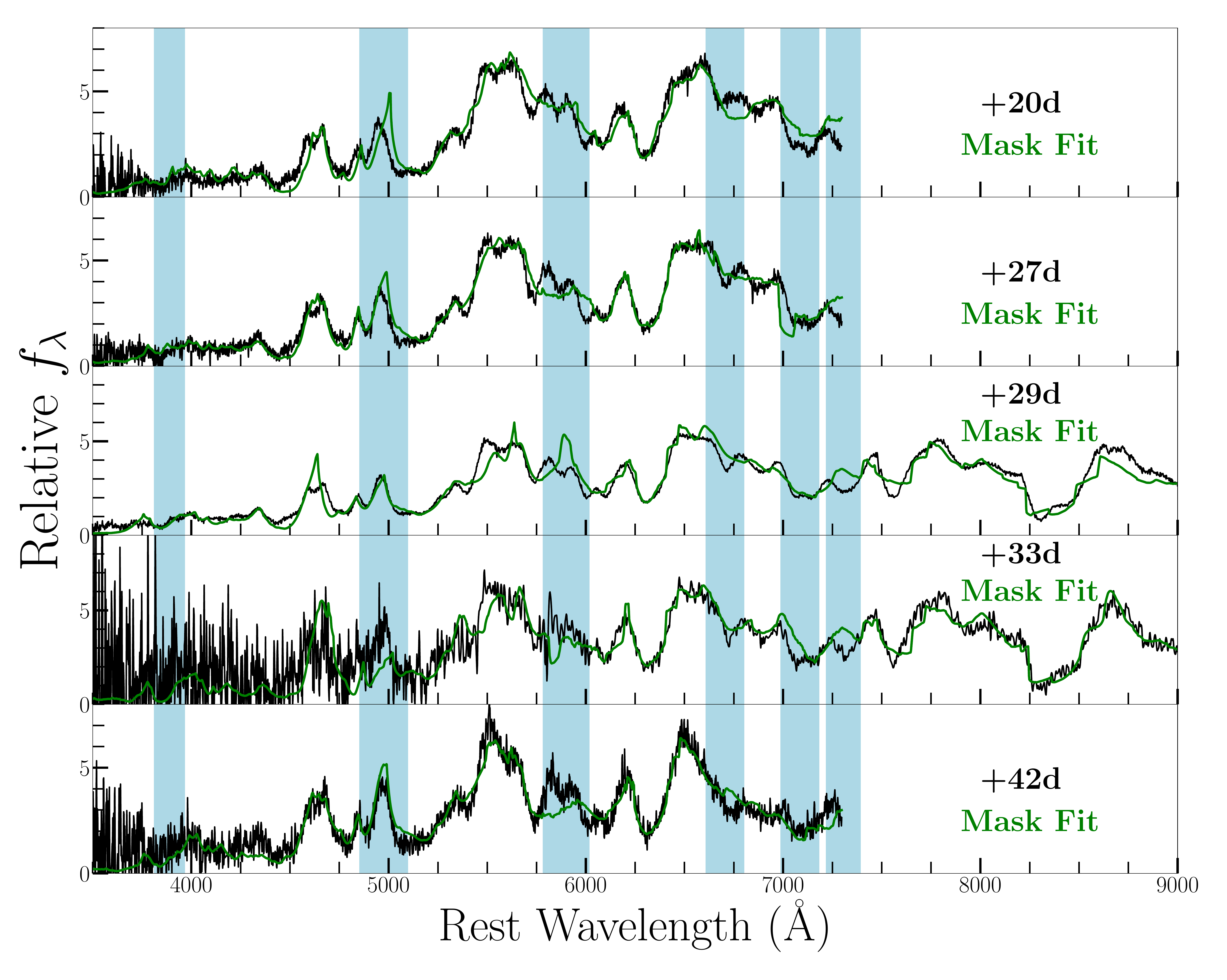}}
\subfigure[]{\includegraphics[width=.47\textwidth]{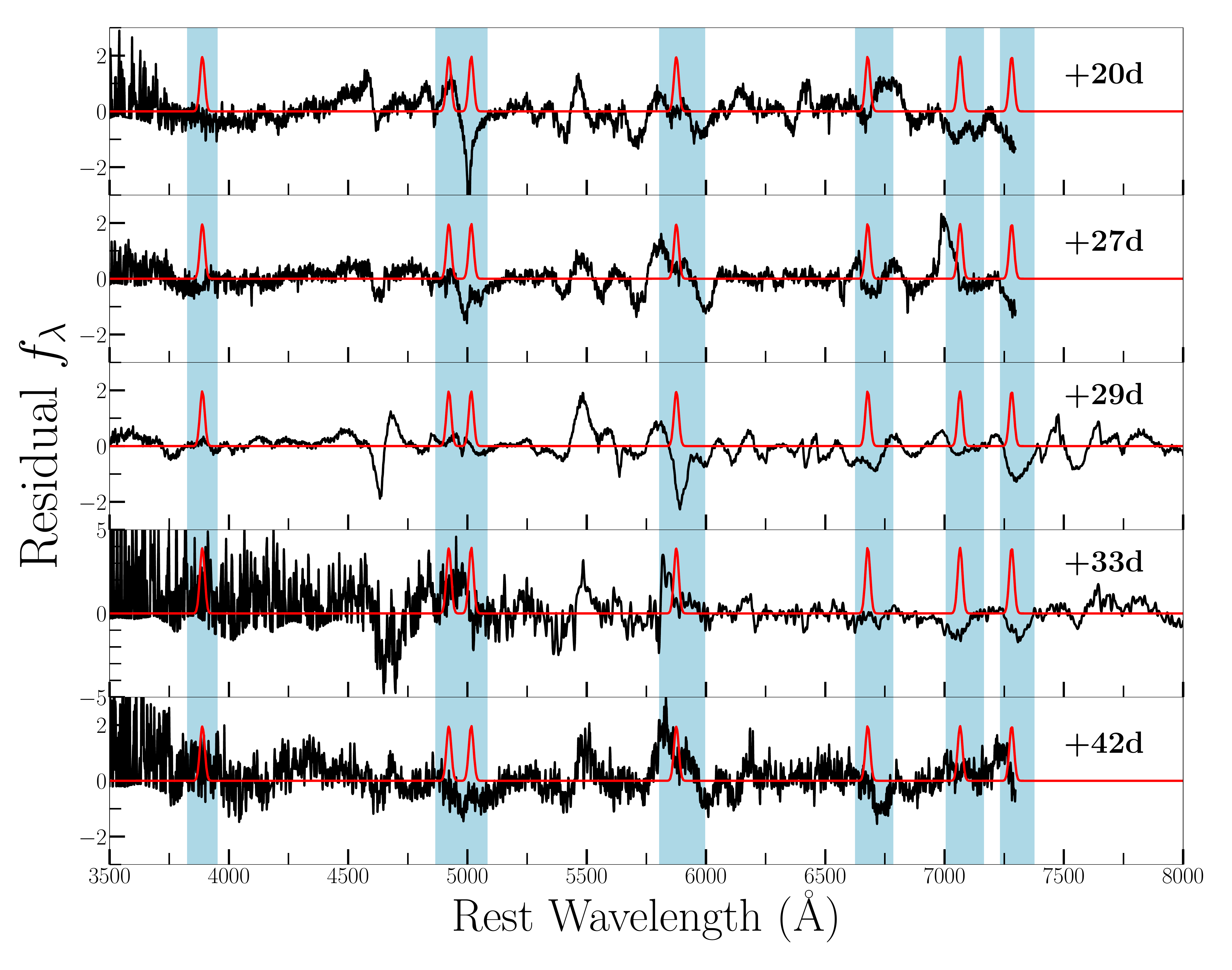}}\\[1ex]
\subfigure[]
{\includegraphics[width=0.8\textwidth]{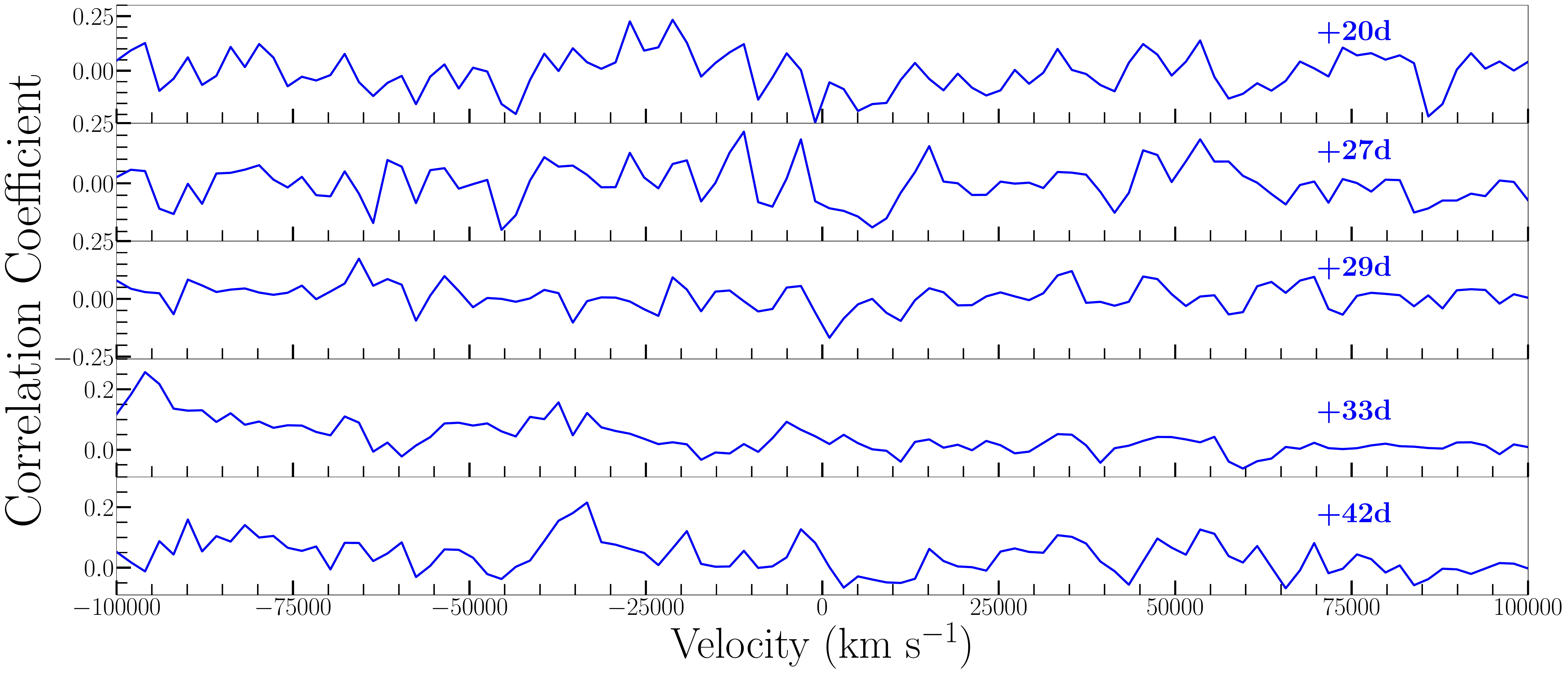}}
\caption{SN~2008A. Phase relative to B band maximum. \label{fig:08A_combo} }
\end{figure*}

\begin{figure*}
\subfigure[]{\includegraphics[width=.47\textwidth]{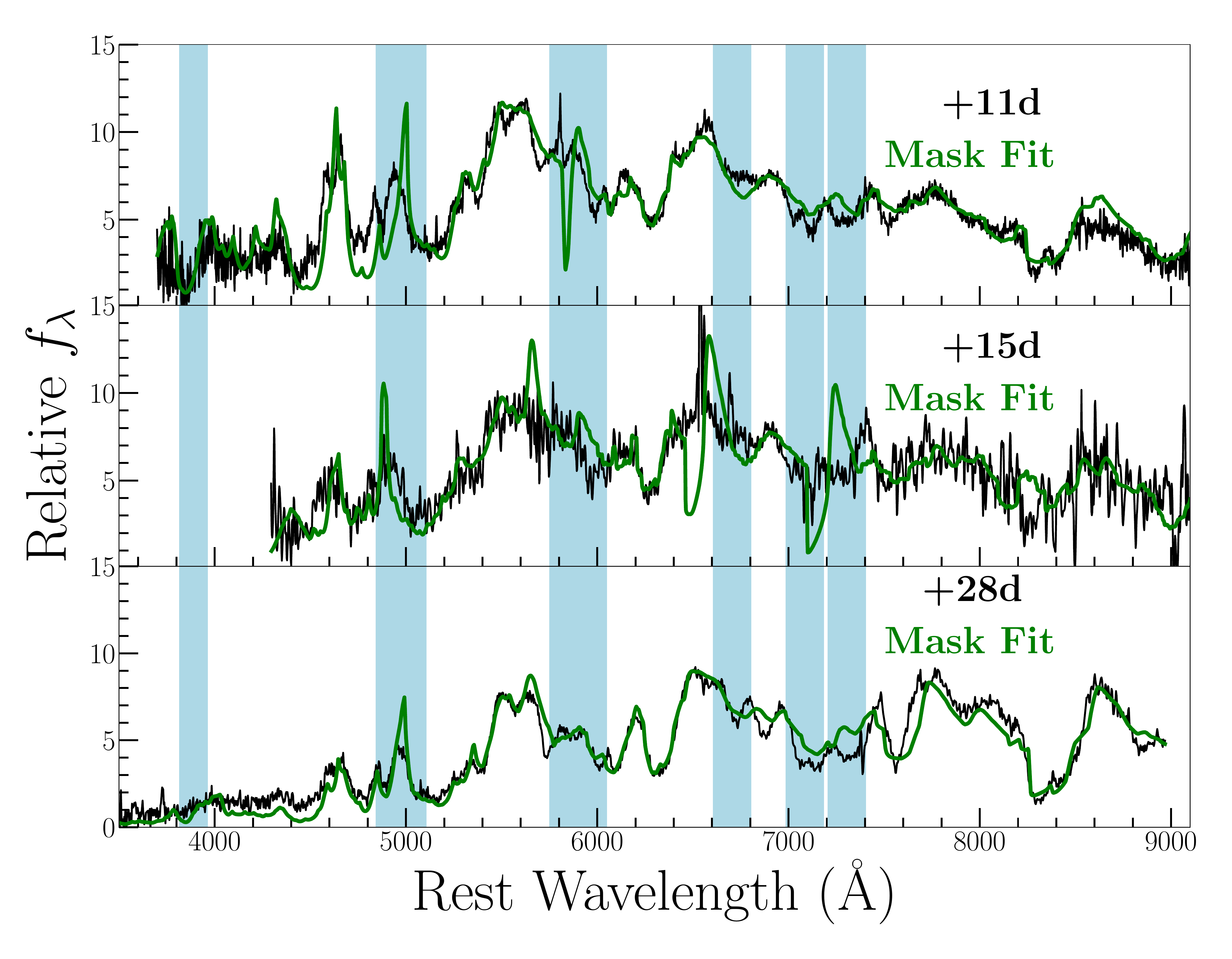}}
\subfigure[]{\includegraphics[width=.47\textwidth]{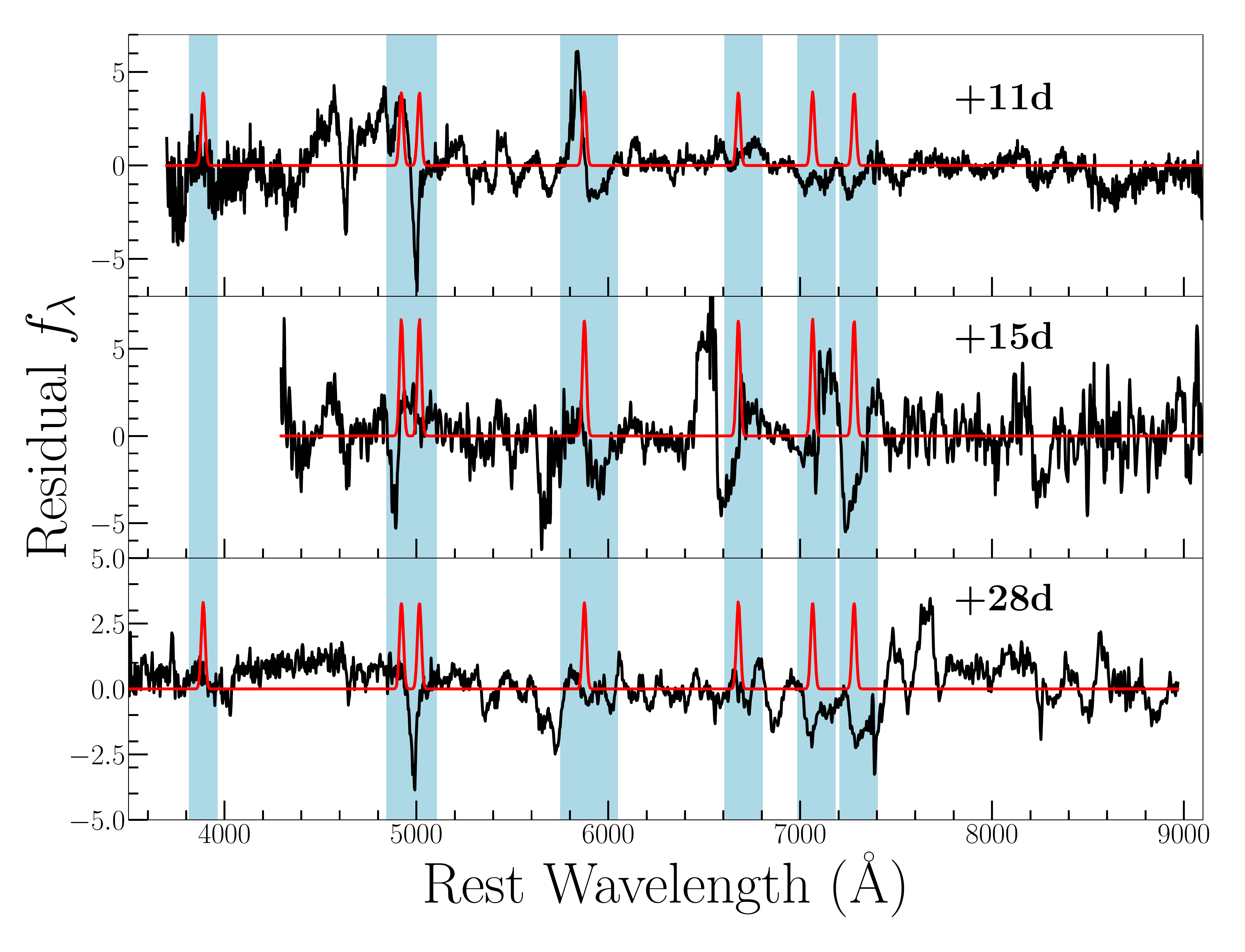}}\\[1ex]
\subfigure[]
{\includegraphics[width=0.8\textwidth]{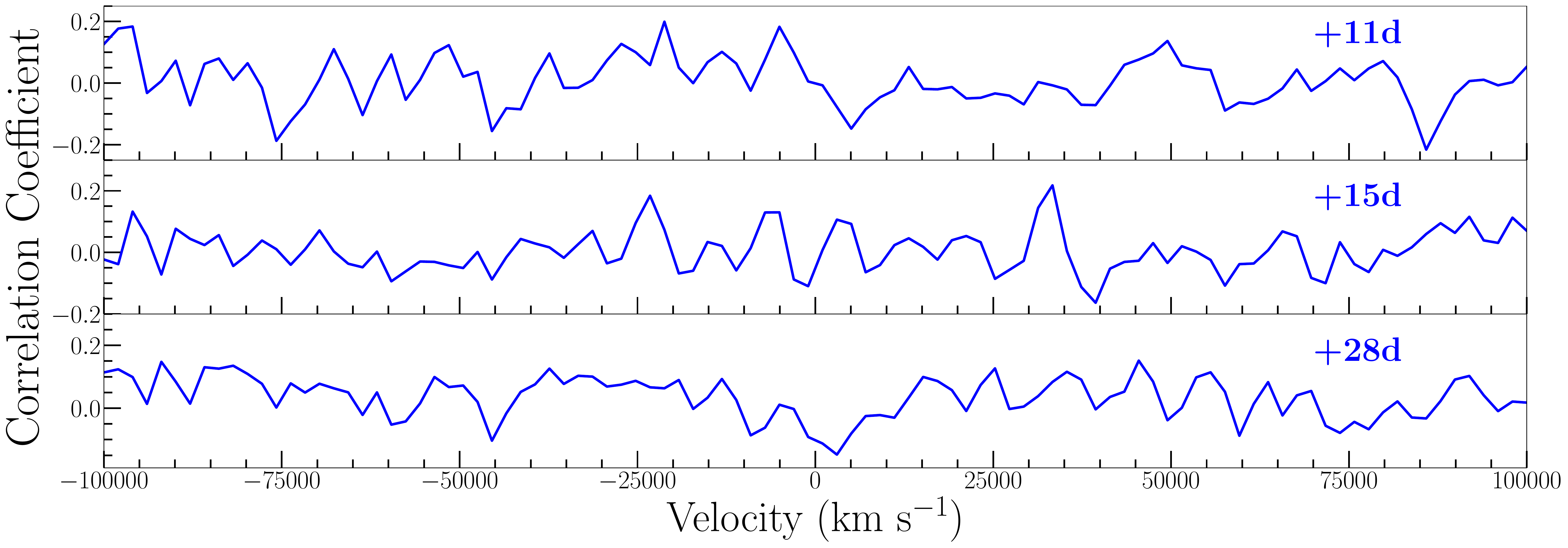}}
\caption{SN~2008ae. Phase relative to B band maximum. \label{fig:08ae_combo} }
\end{figure*}

\begin{figure}
\begin{center}
	\includegraphics[width=0.49\textwidth]{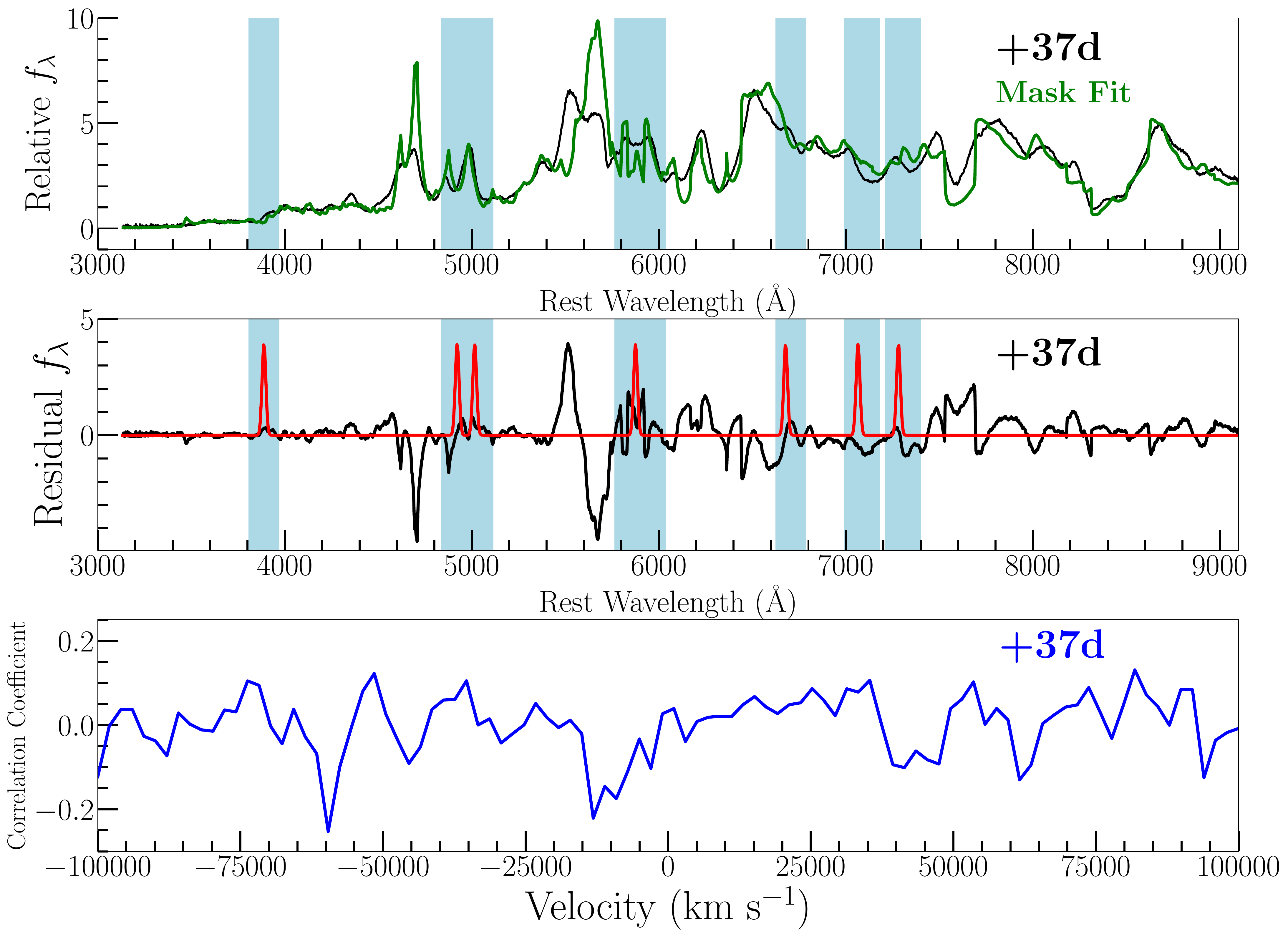}
	\caption{SN~2008ge. Phase relative to B band maximum.} \label{fig:combo_08ge}
\end{center}
\vspace*{-5mm}
\end{figure}

\begin{figure*}
\subfigure[]{\includegraphics[width=.47\textwidth]{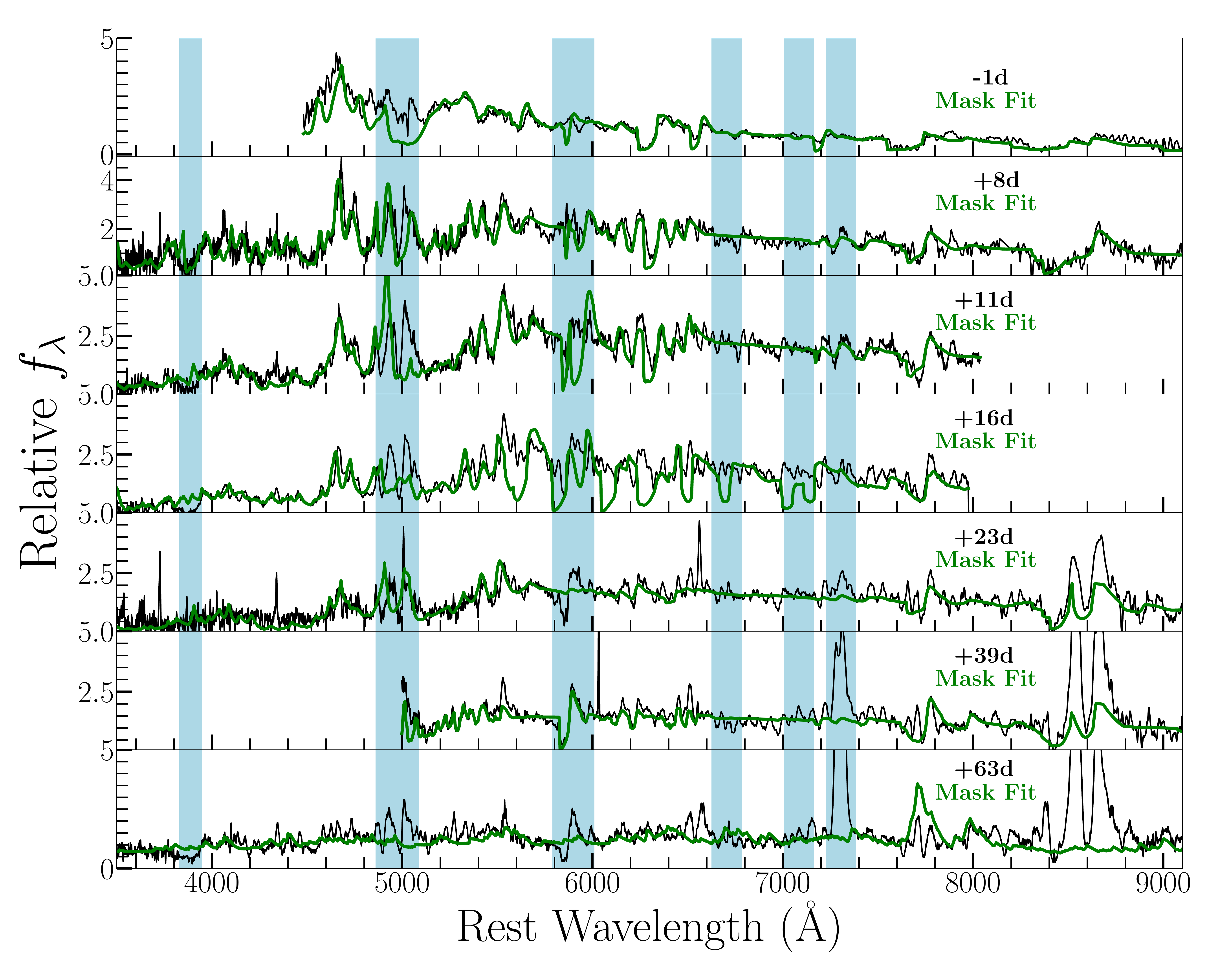}}
\subfigure[]{\includegraphics[width=.47\textwidth]{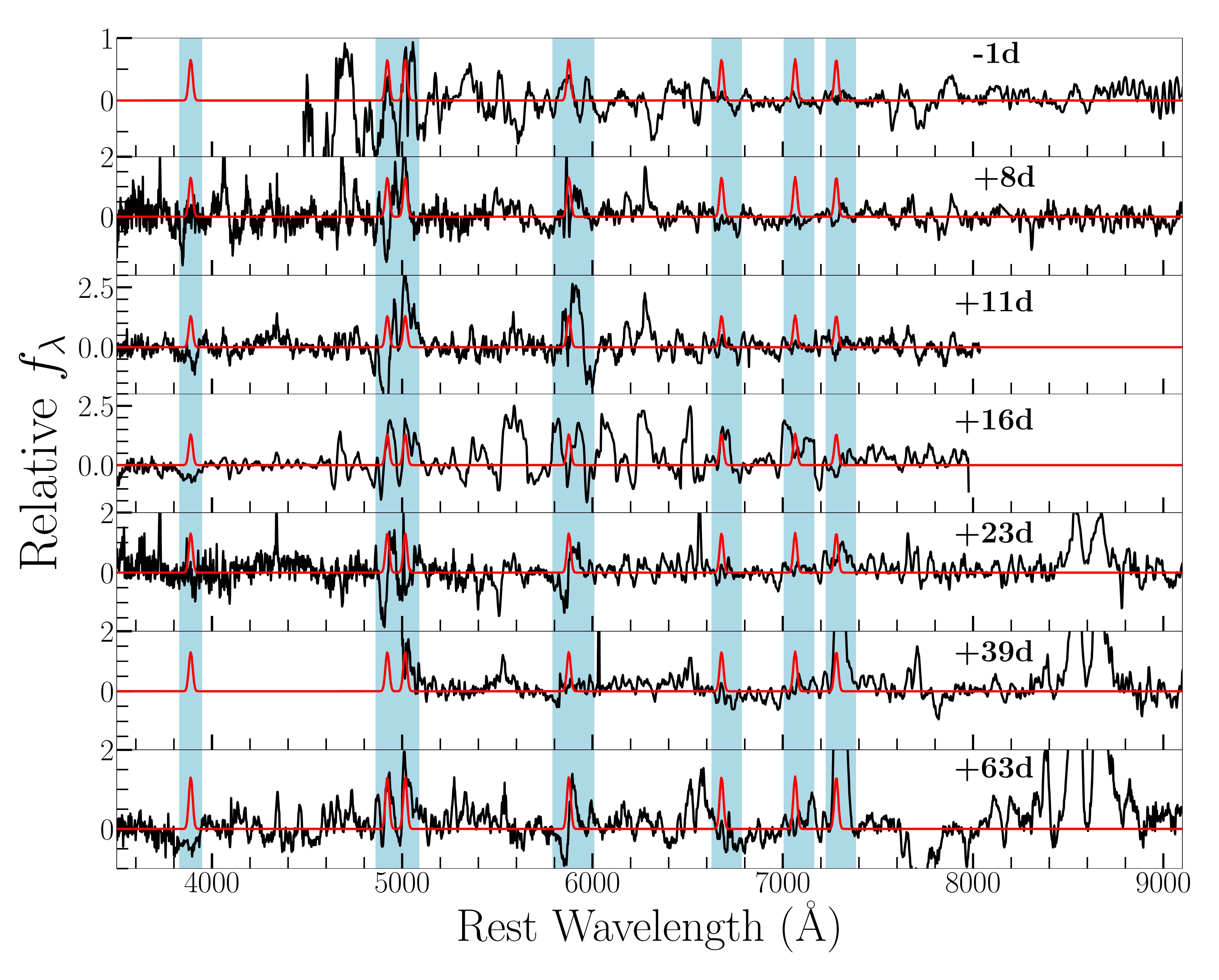}}\\[1ex]
\subfigure[]
{\includegraphics[width=0.8\textwidth]{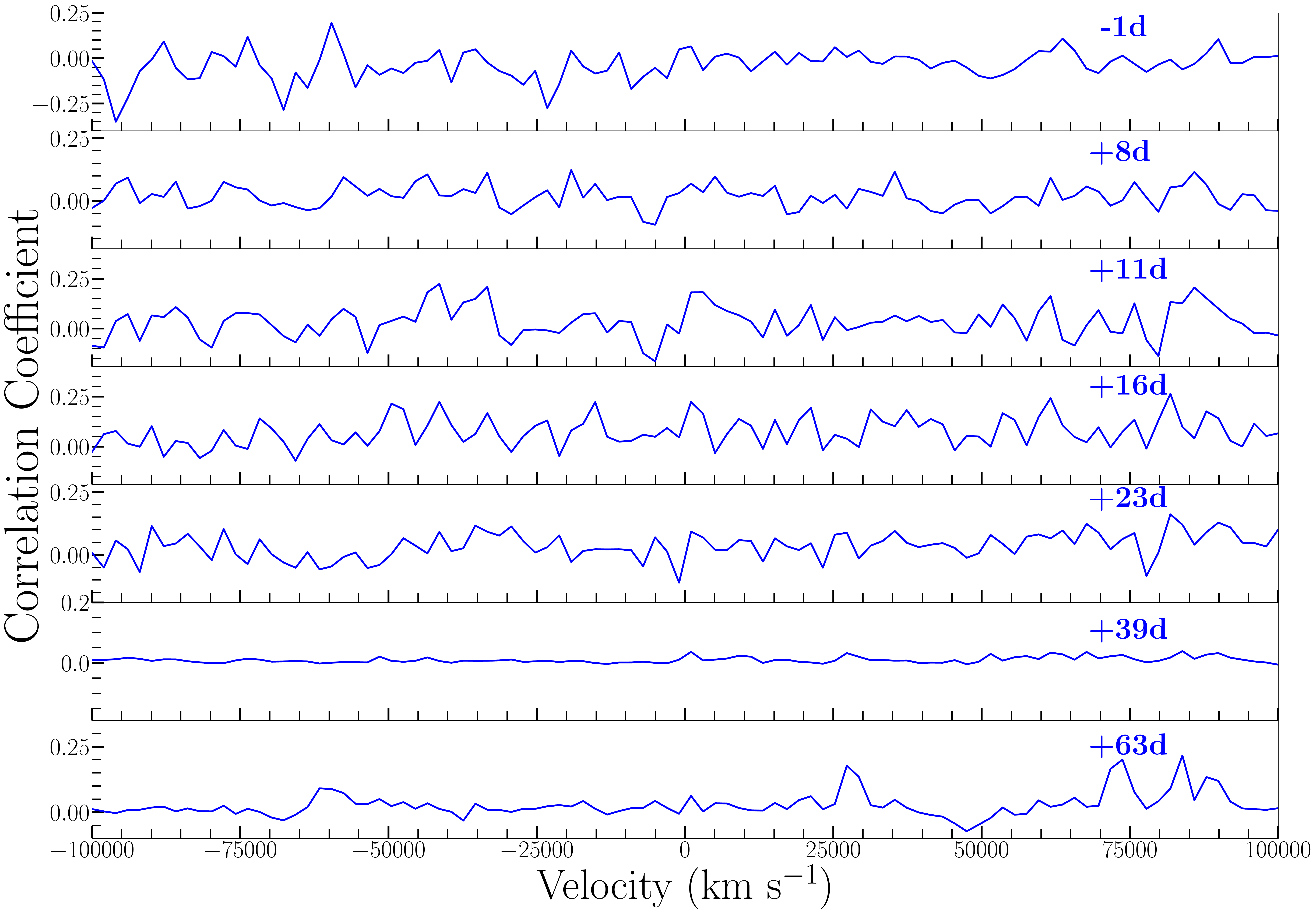}}
\caption{SN~2008ha. Phase relative to B band maximum. \label{fig:08ha_combo} }
\end{figure*}

\begin{figure*}
\subfigure[]{\includegraphics[width=.47\textwidth]{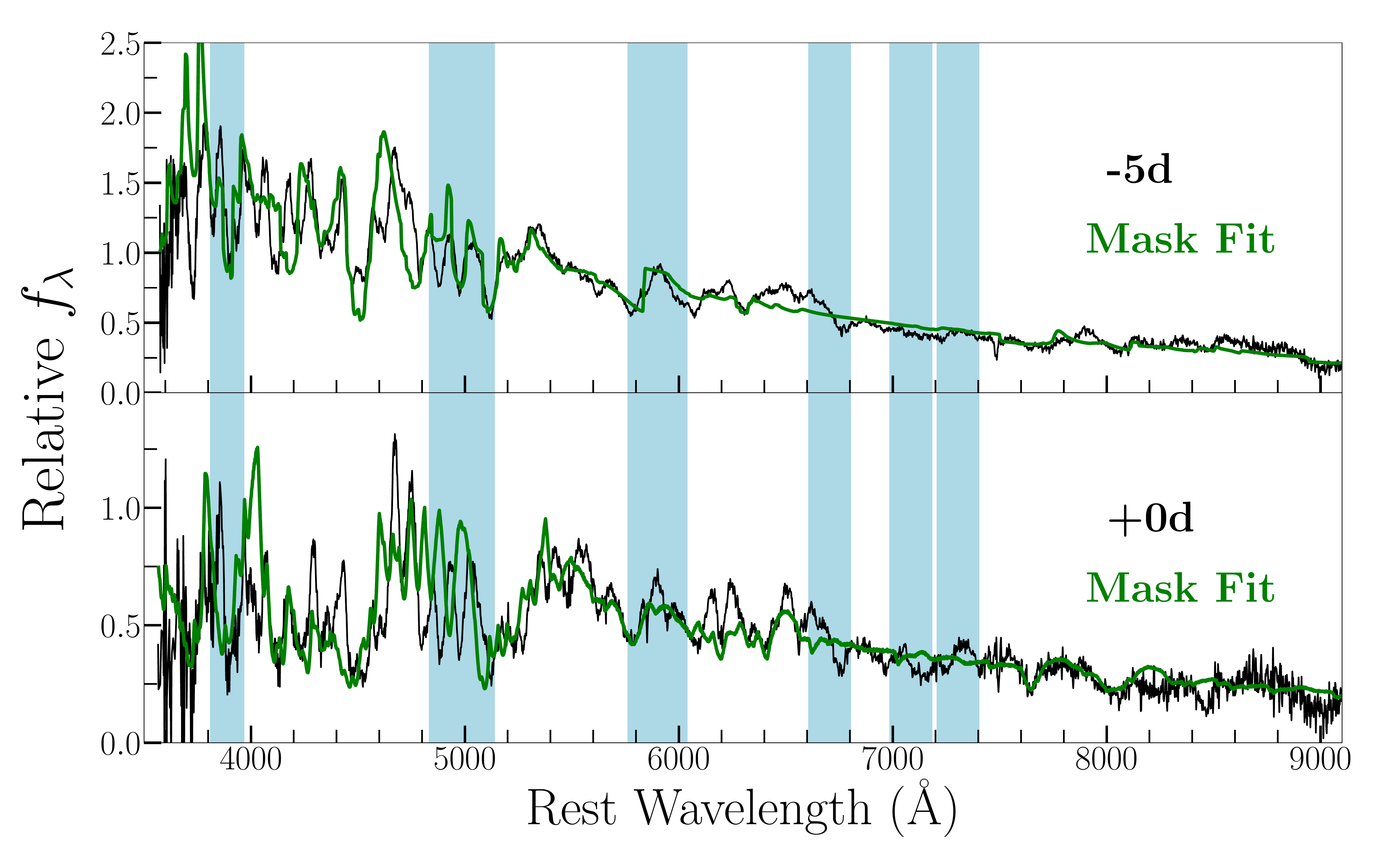}}
\subfigure[]{\includegraphics[width=.47\textwidth]{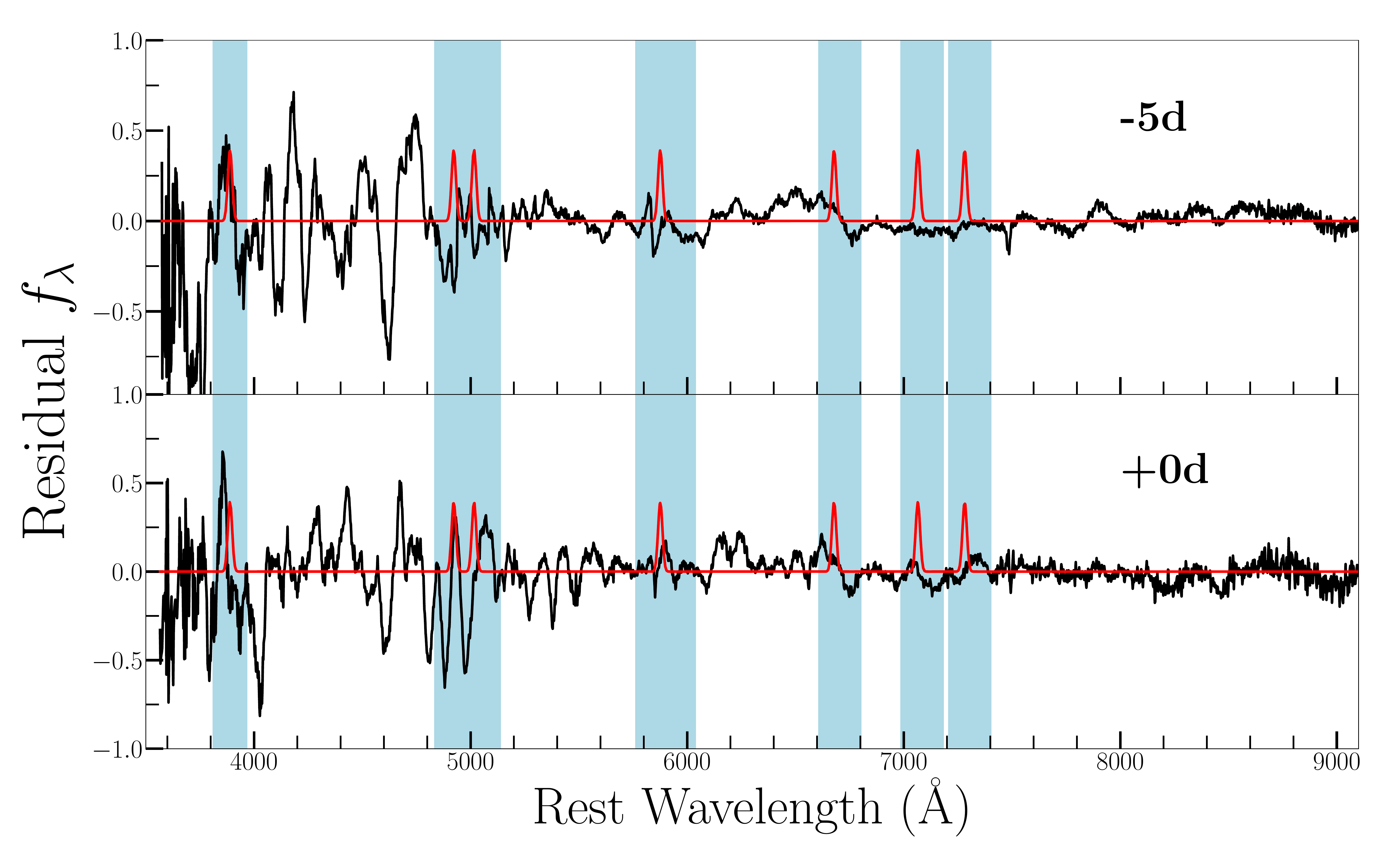}}\\[1ex]
\subfigure[]
{\includegraphics[width=0.8\textwidth]{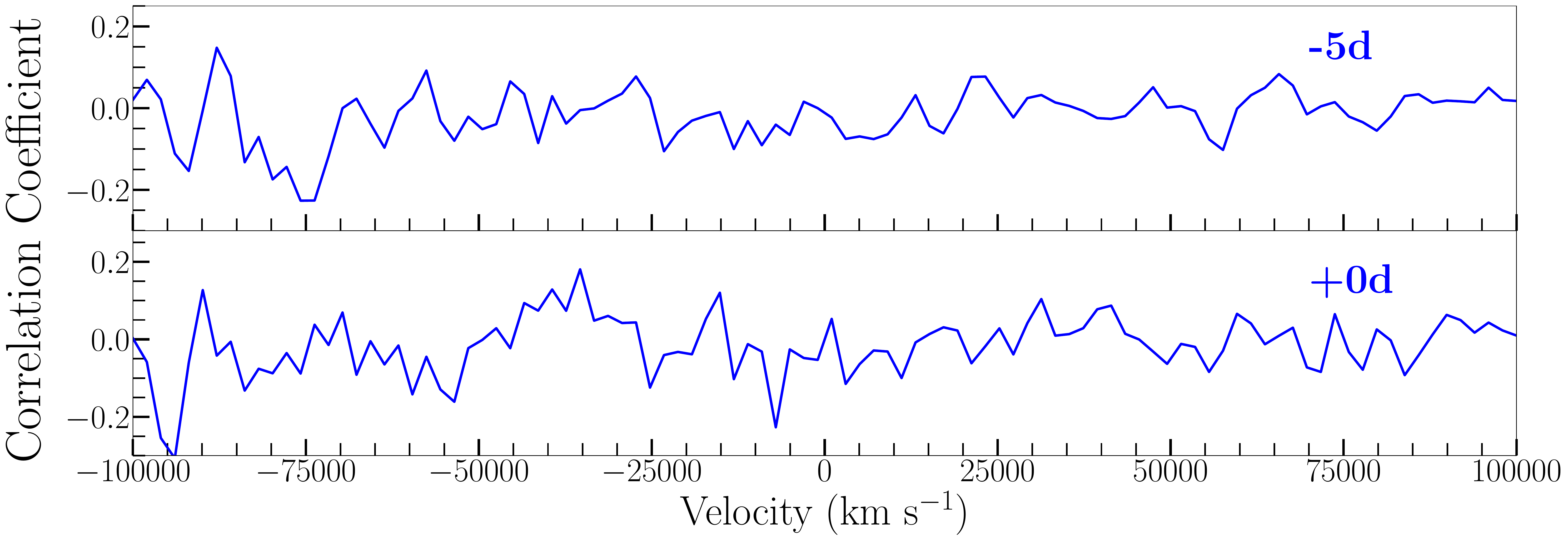}}
\caption{SN~2009J. Phase relative to B band maximum. \label{fig:09J_combo} }
\end{figure*}

\begin{figure*}
\subfigure[]{\includegraphics[width=.47\textwidth]{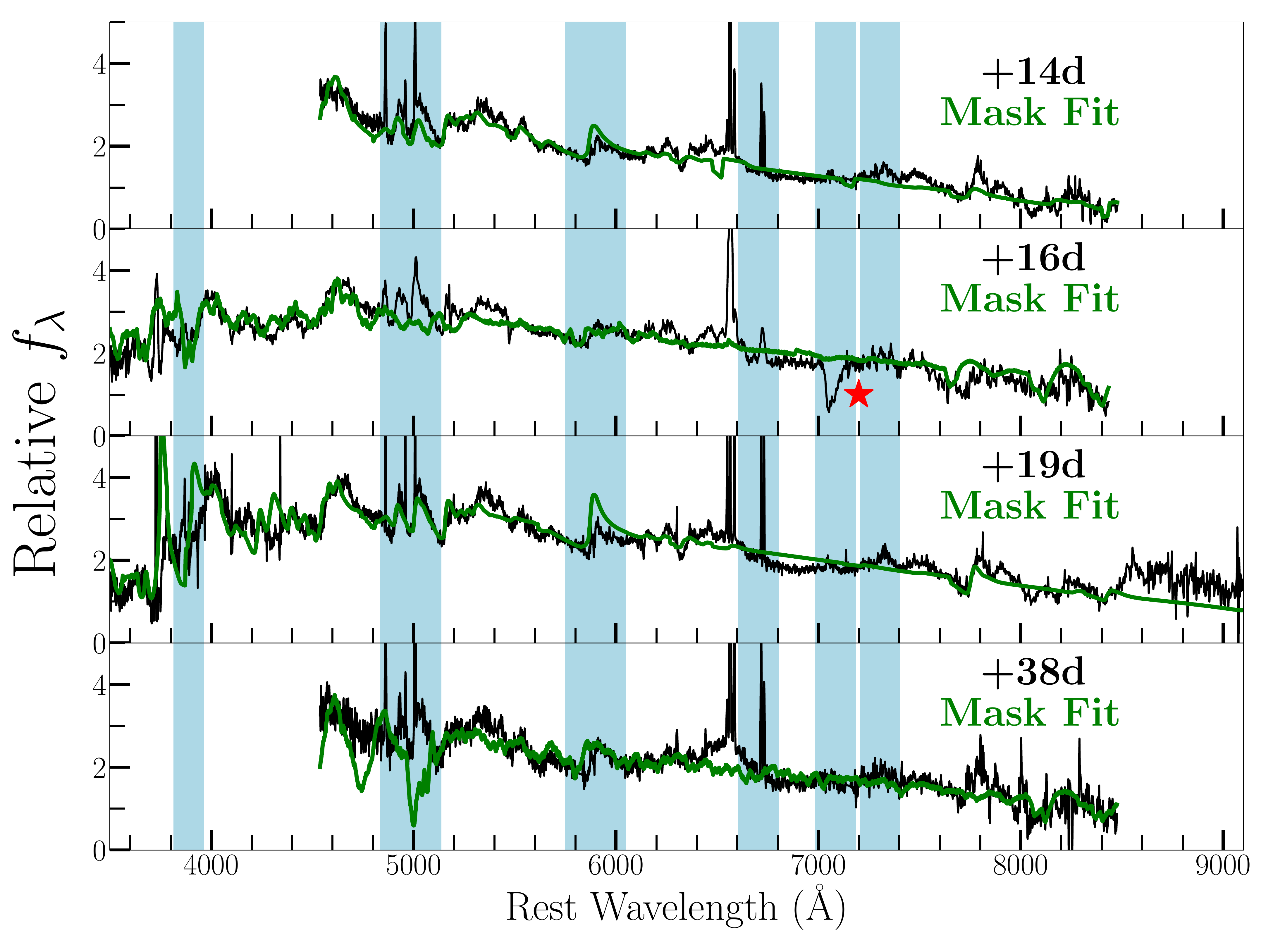}}
\subfigure[]{\includegraphics[width=.47\textwidth]{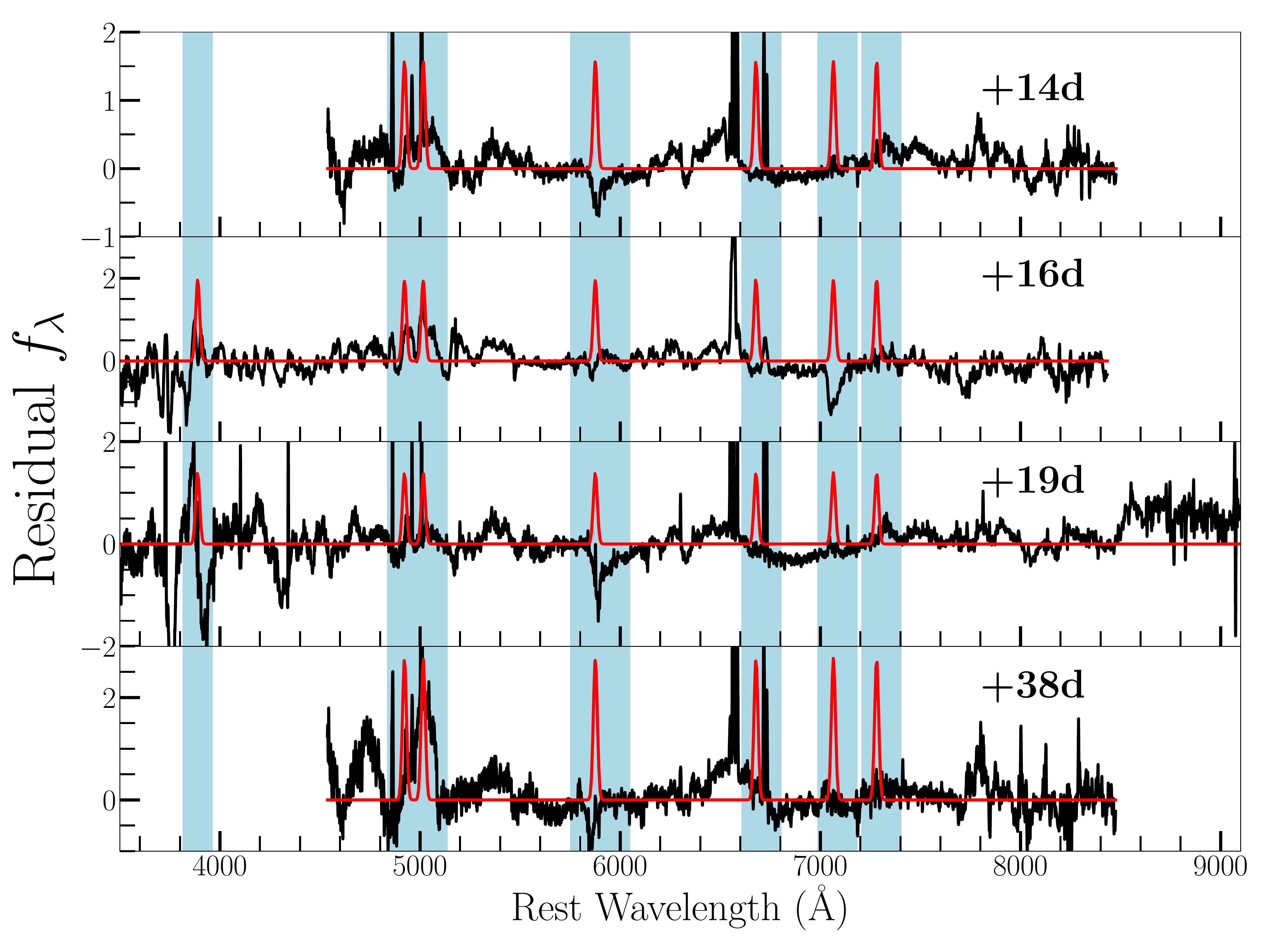}}\\[1ex]
\subfigure[]
{\includegraphics[width=0.8\textwidth]{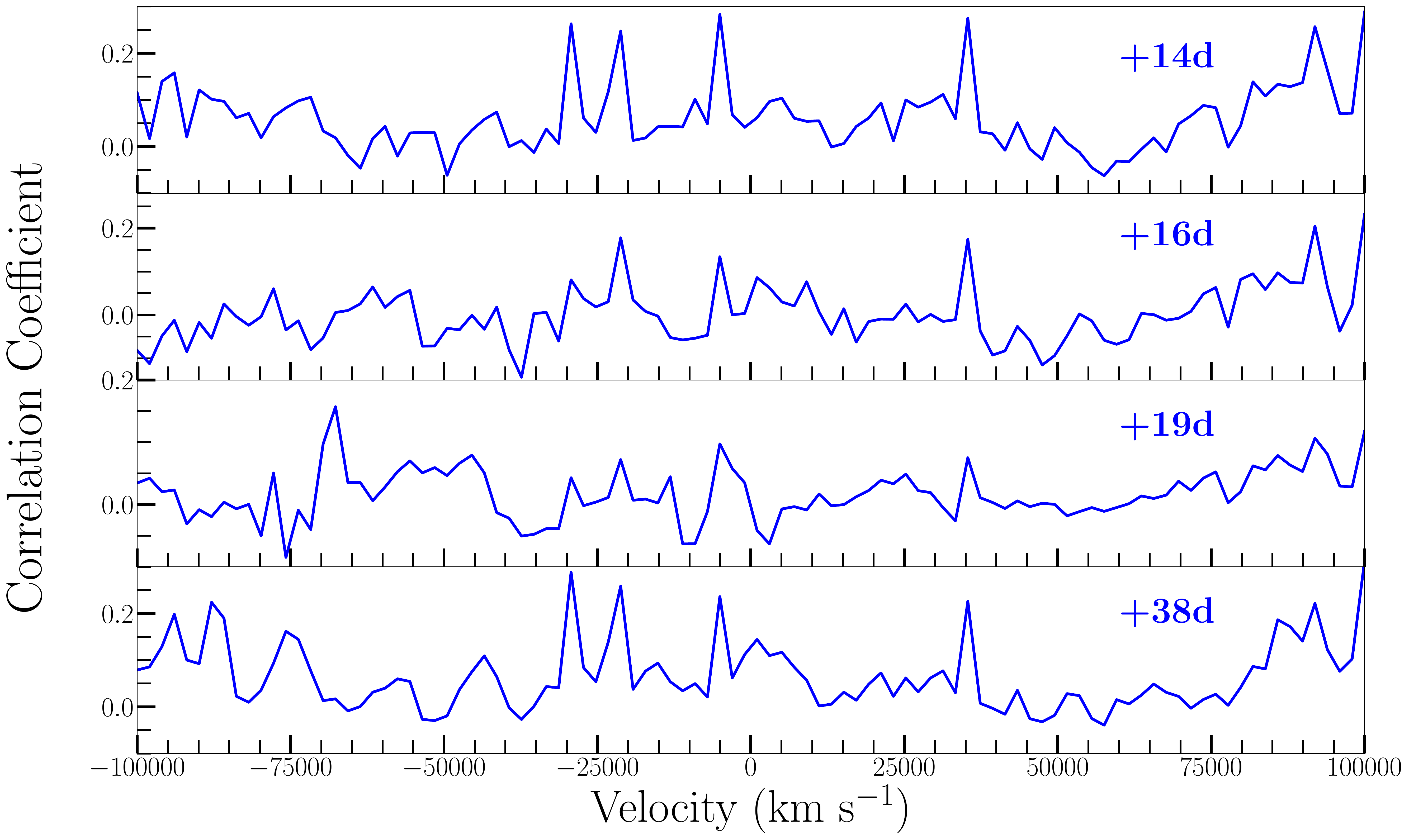}}
\caption{SN~2009ku. Phase relative to B band maximum. Red star indicates telluric absorption. \label{fig:09ku_combo} }
\end{figure*}

\begin{figure}
\begin{center}
	\includegraphics[width=0.49\textwidth]{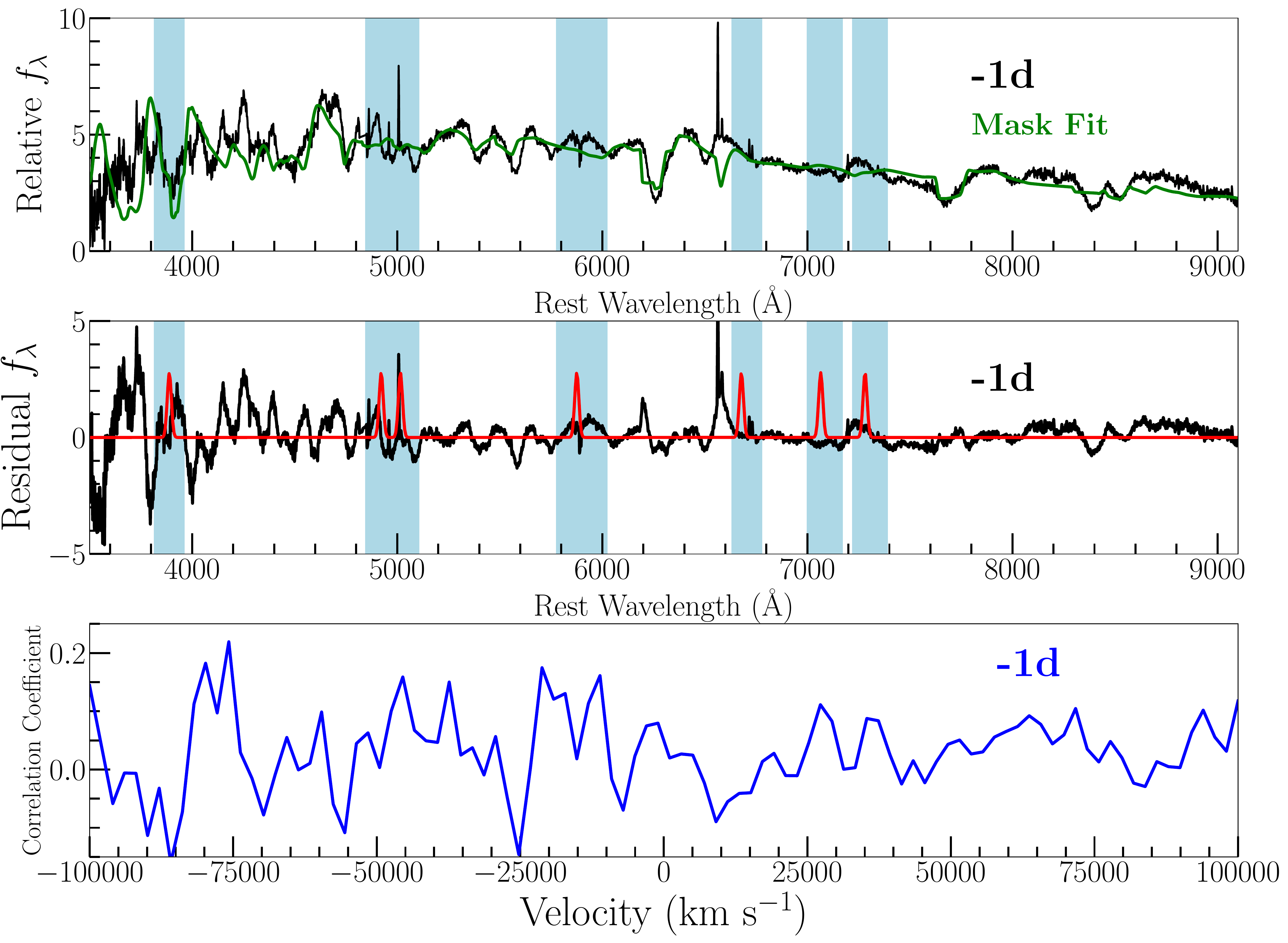}
	\caption{SN~2010ae. Phase relative to B band maximum.} \label{fig:combo_10ae}
\end{center}
\end{figure}

\begin{figure*}
\subfigure[]{\includegraphics[width=.47\textwidth]{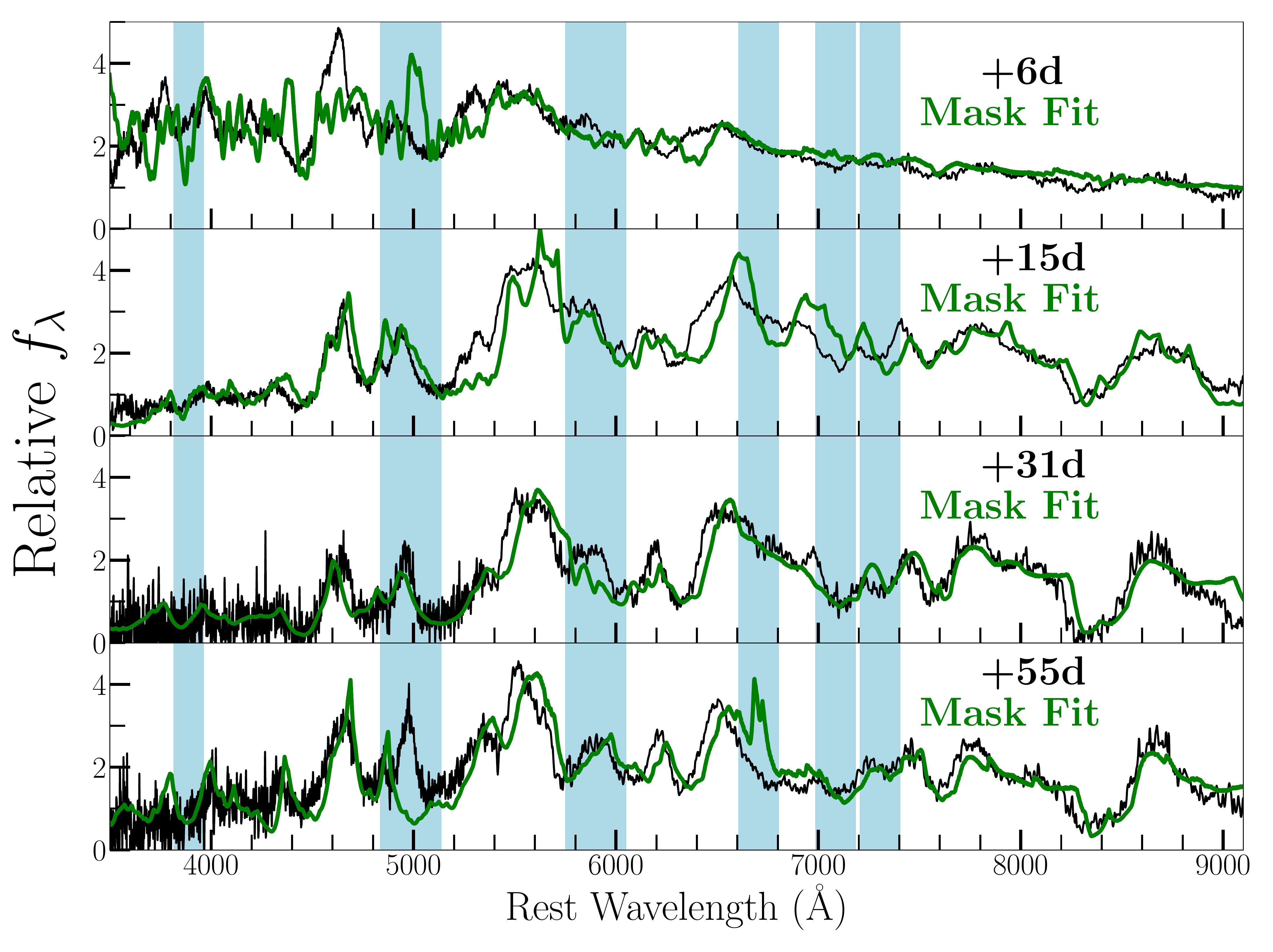}}
\subfigure[]{\includegraphics[width=.47\textwidth]{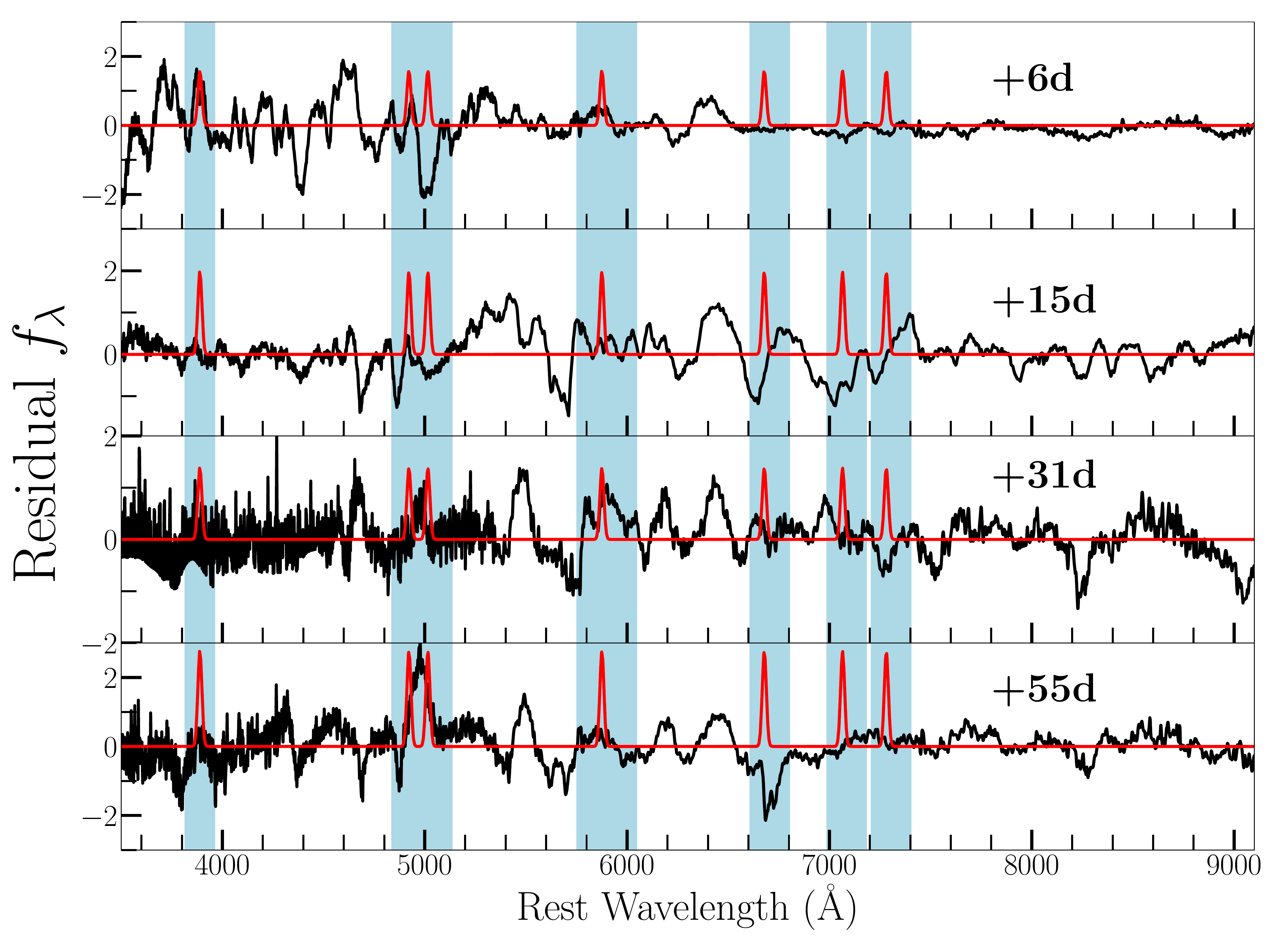}}\\[1ex]
\subfigure[]
{\includegraphics[width=0.8\textwidth]{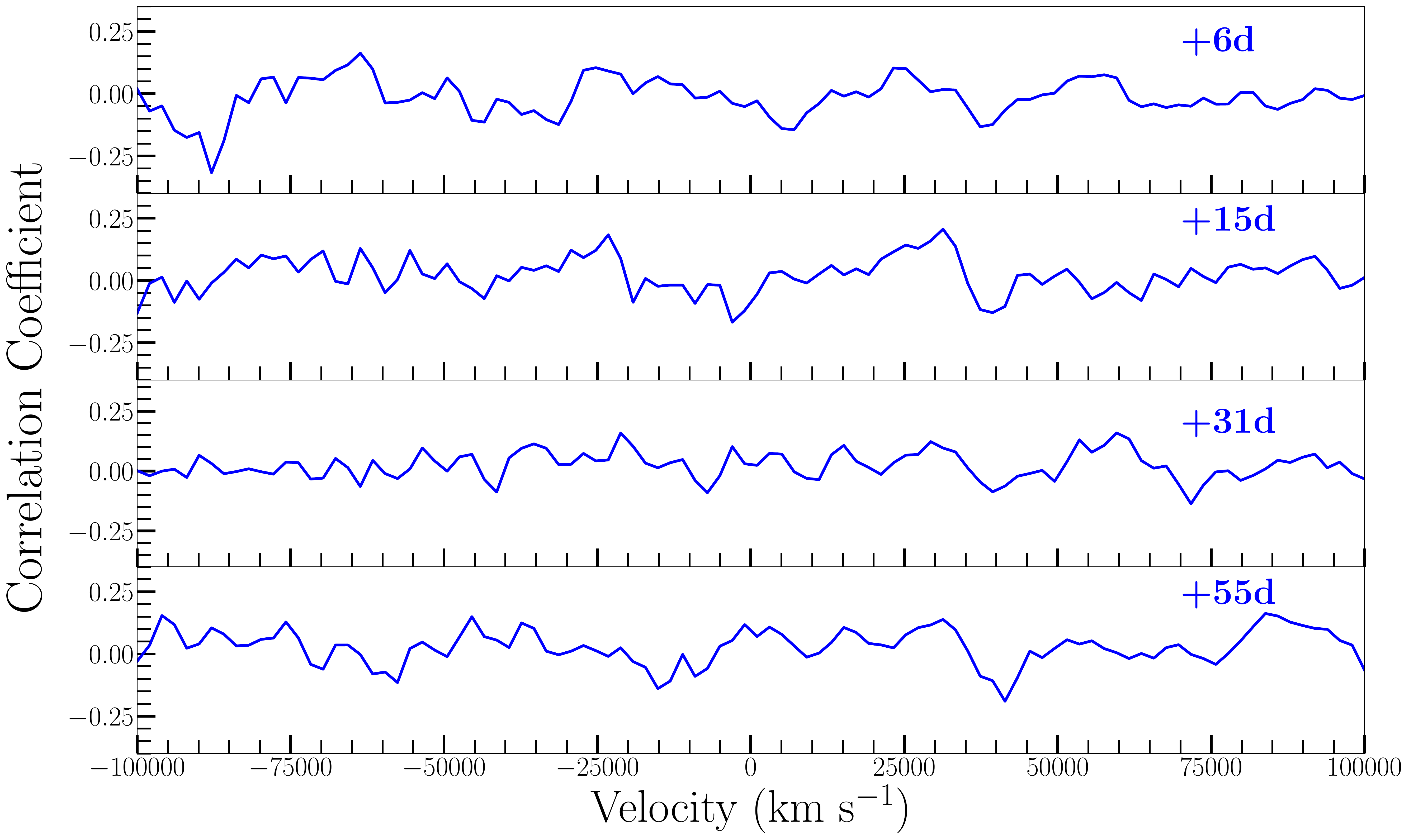}}
\caption{SN~2011ay. Phase relative to B band maximum. \label{fig:11ay_combo} }
\end{figure*}

\begin{figure}
\begin{center}
	\includegraphics[width=0.49\textwidth]{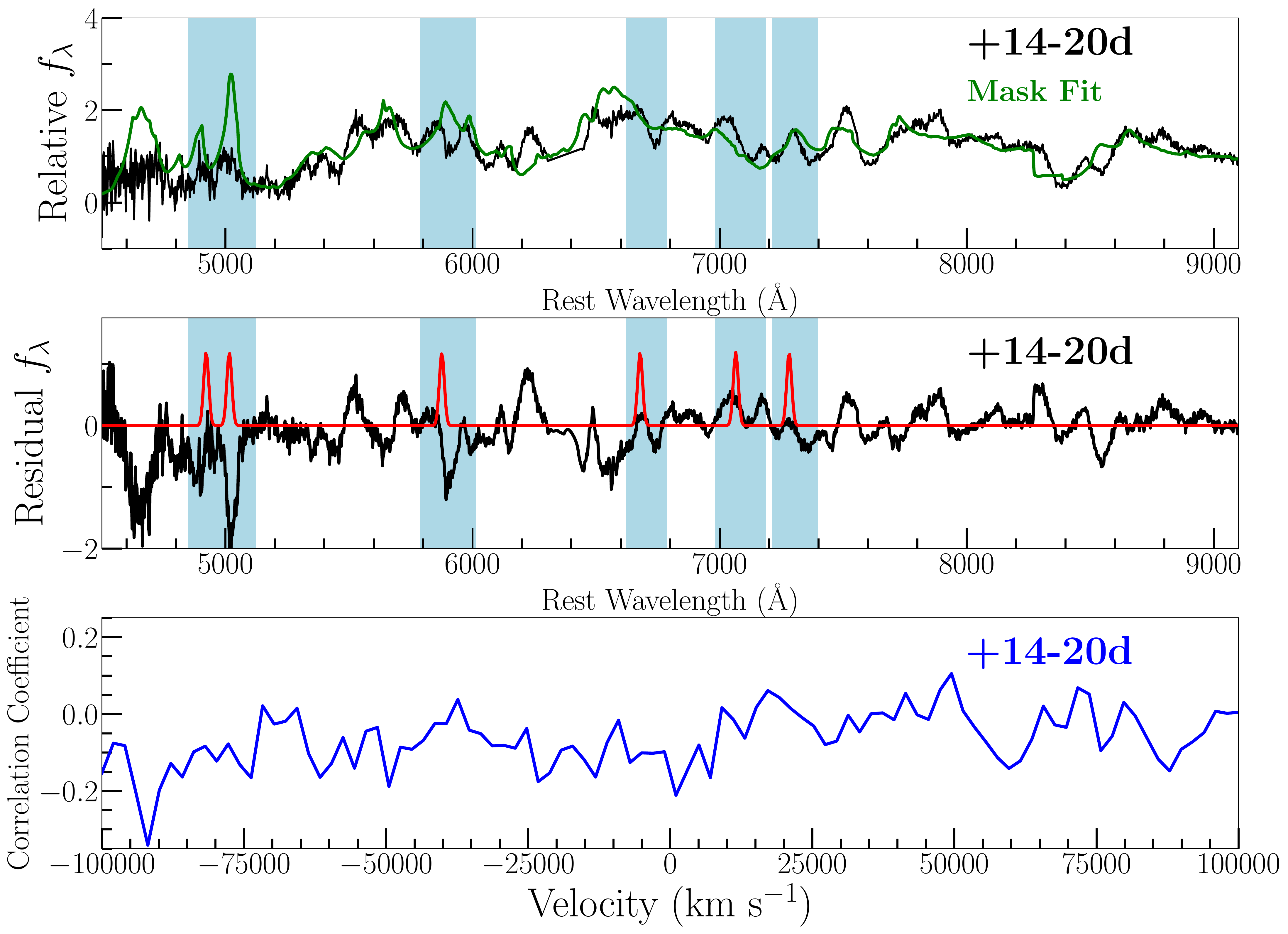}
	\caption{SN~2011ce. Phase relative to B band maximum and calculated using SNID.} \label{fig:combo_11ce}
\end{center}
\vspace*{-5mm}
\end{figure}

\begin{figure*}
\subfigure[]{\includegraphics[width=.47\textwidth]{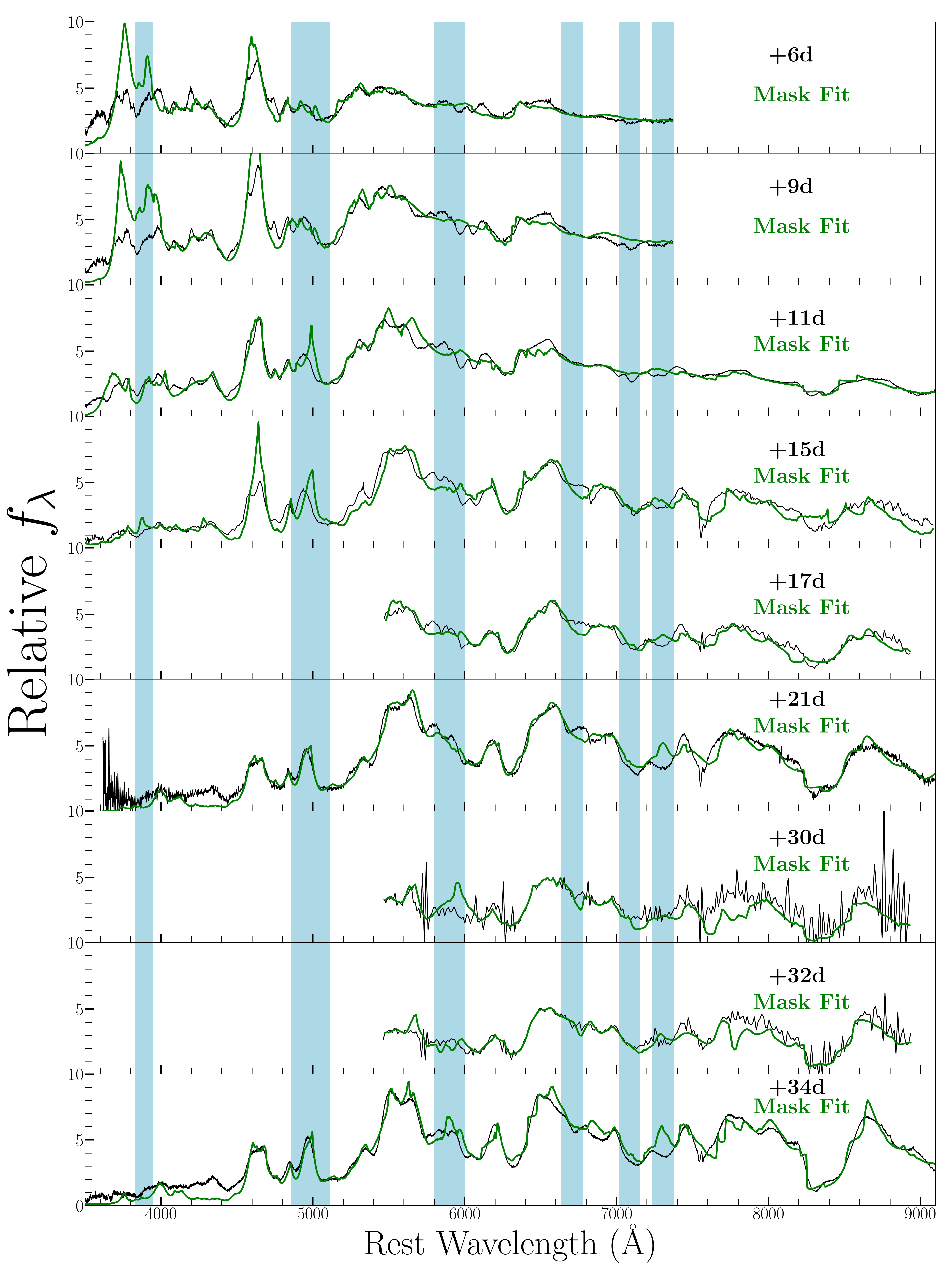}}
\subfigure[]{\includegraphics[width=.47\textwidth]{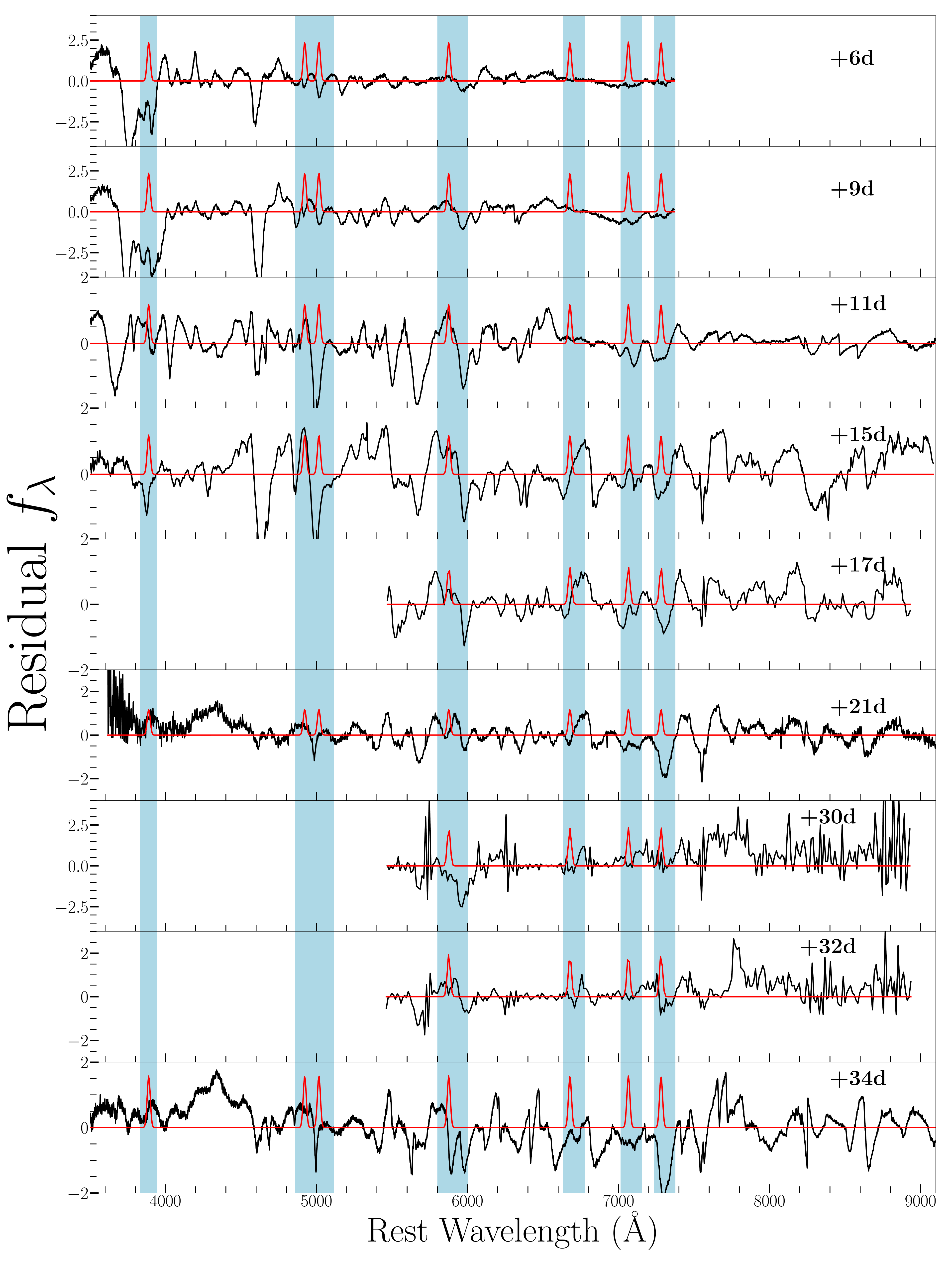}}\\[1ex]
\subfigure[]
{\includegraphics[width=0.8\textwidth]{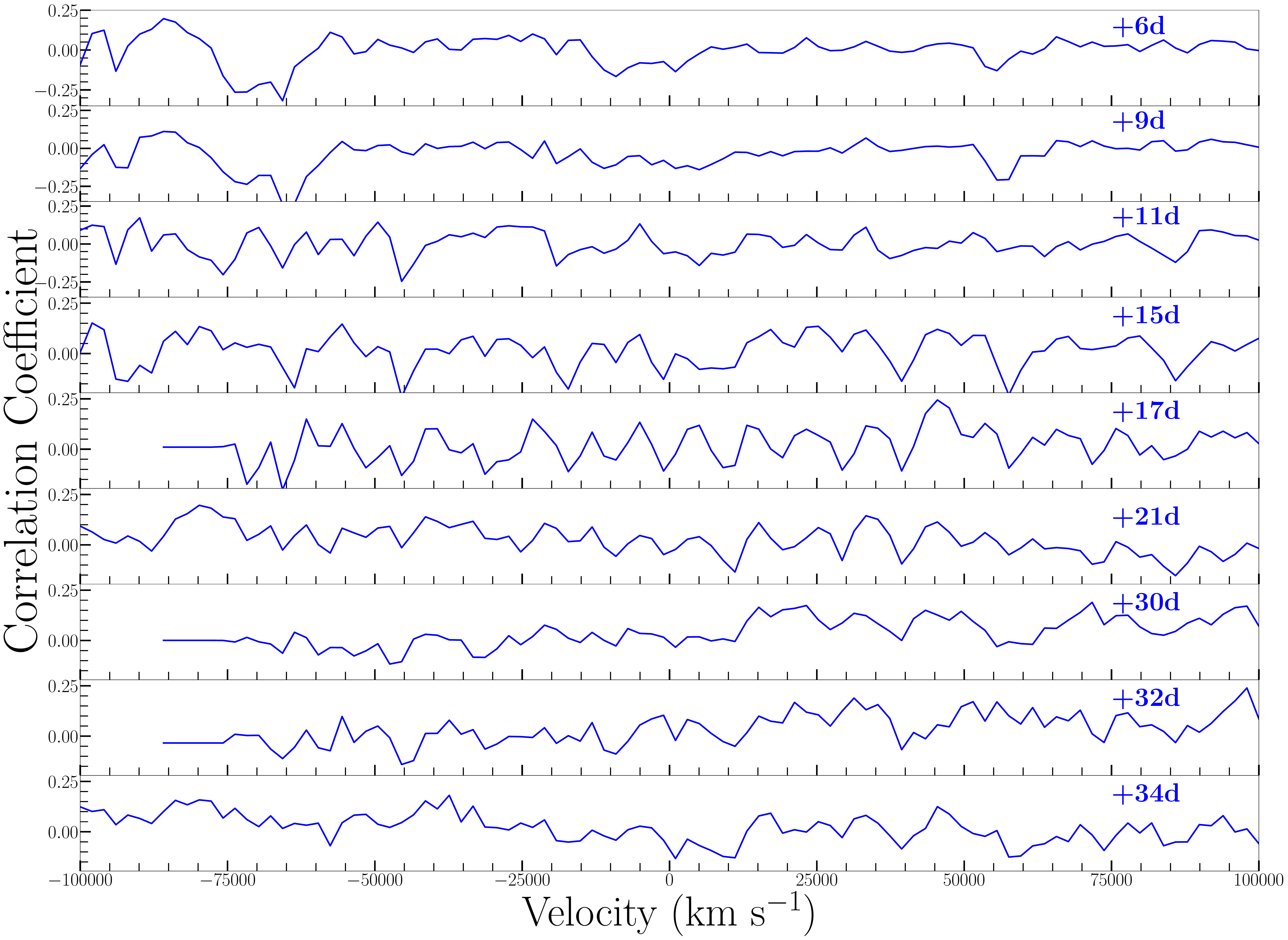}}
\caption{SN~2012Z. Phase relative to B band maximum. \label{fig:12Z_combo} }
\end{figure*}

\begin{figure*}
\subfigure[]{\includegraphics[width=.47\textwidth]{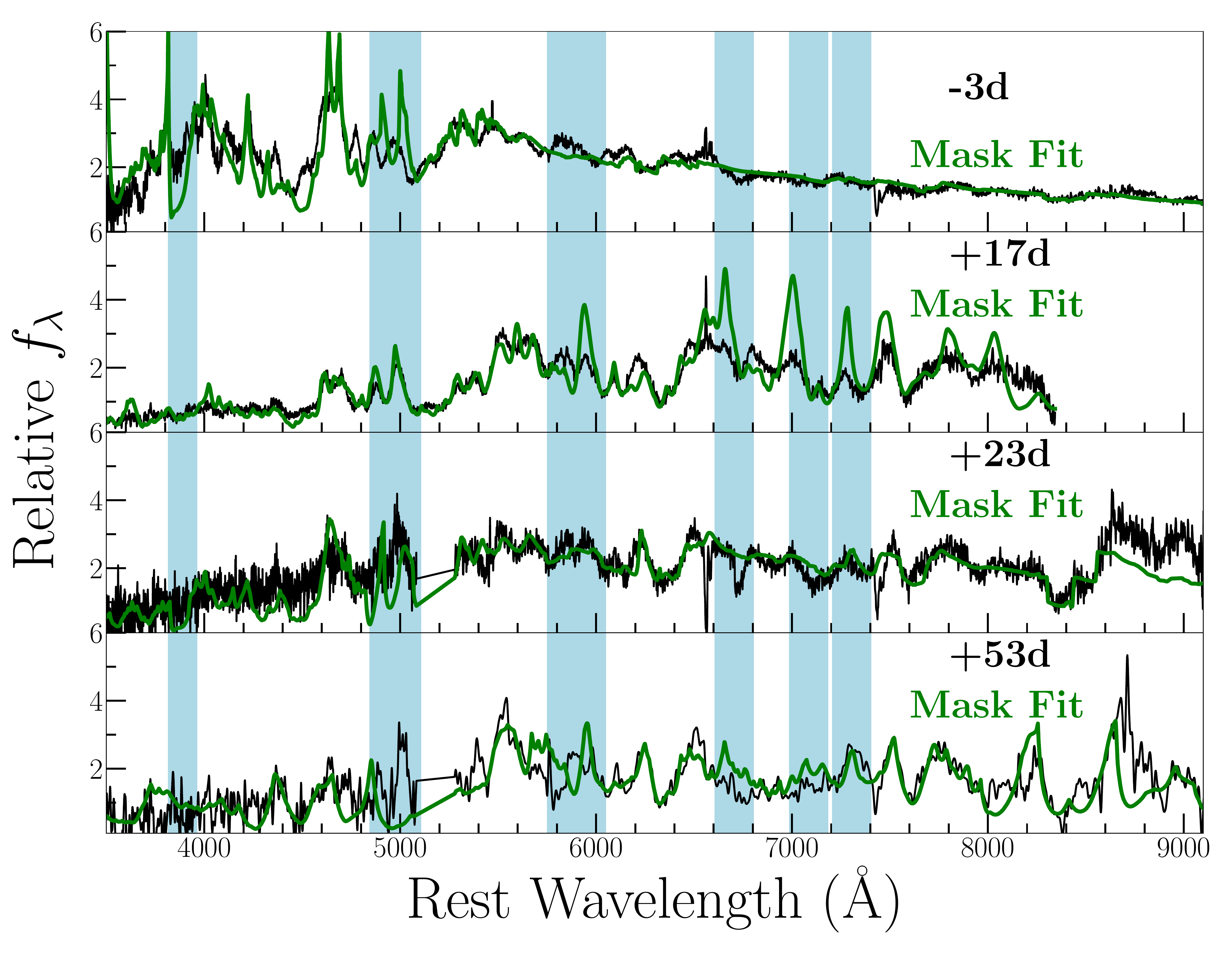}}
\subfigure[]{\includegraphics[width=.47\textwidth]{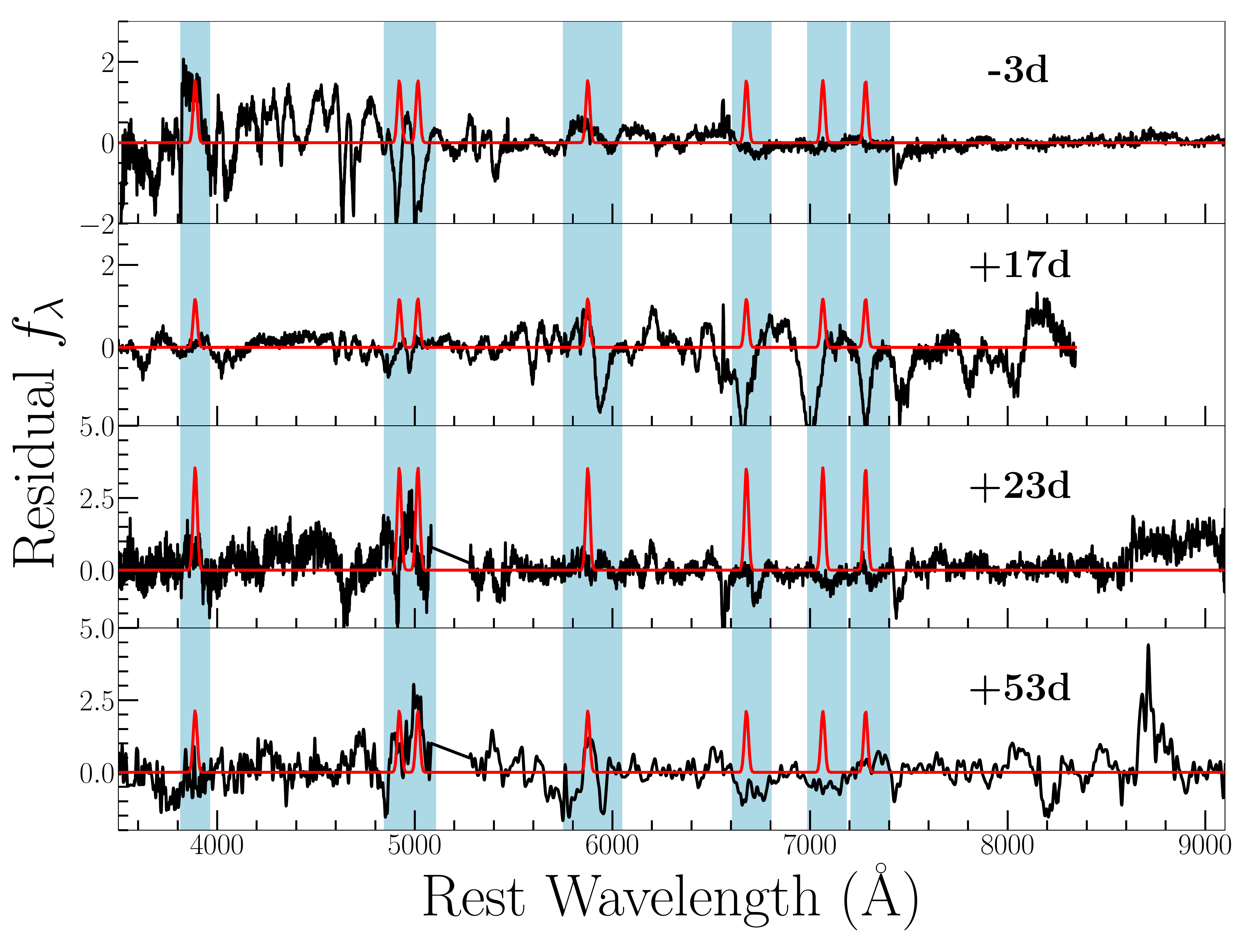}}\\[1ex]
\subfigure[]
{\includegraphics[width=0.8\textwidth]{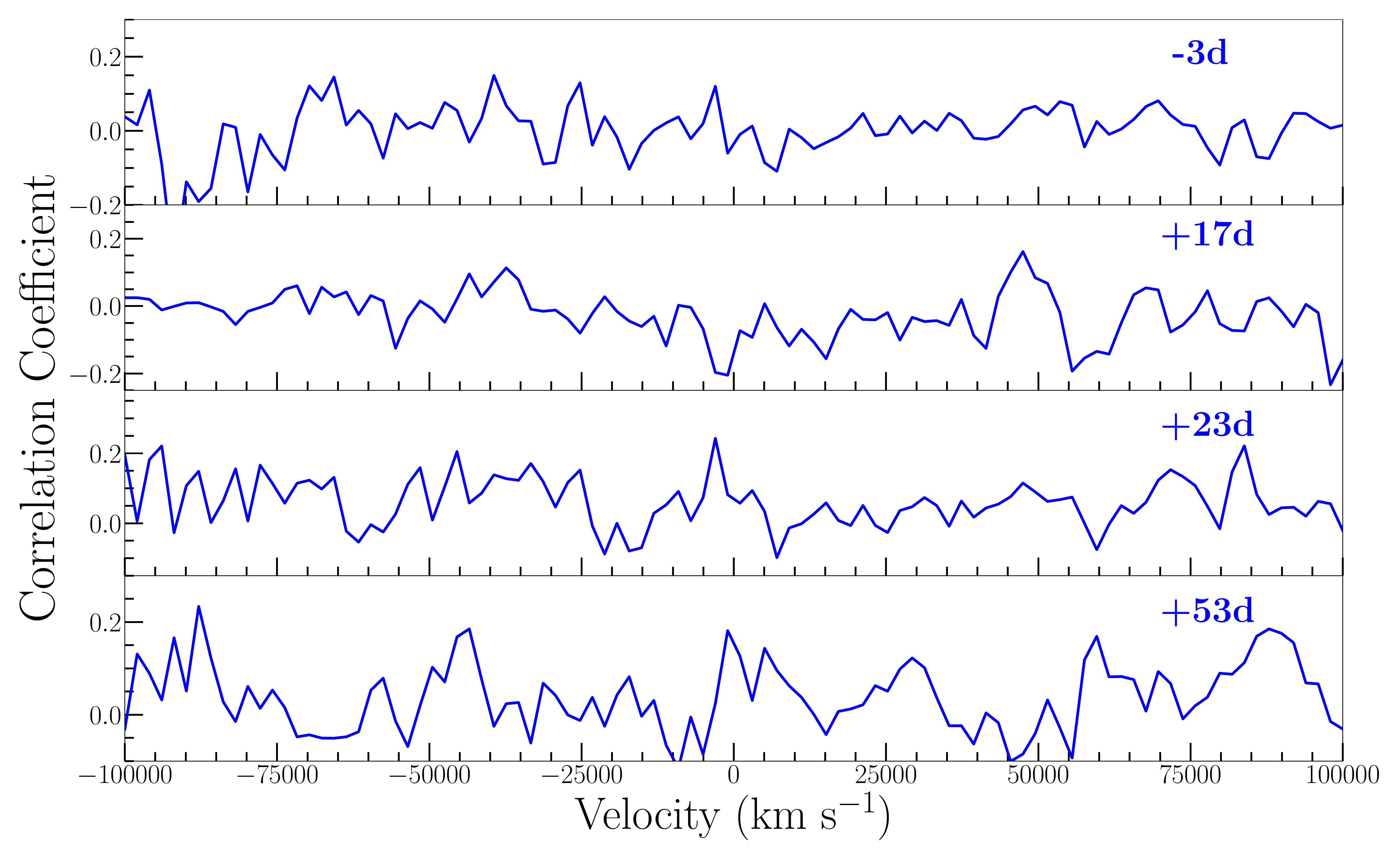}}
\caption{PS1-12bwh. Phase relative to B band maximum. \label{fig:PS1_12bwh_combo} }
\end{figure*}

\begin{figure}
\begin{center}
	\includegraphics[width=0.49\textwidth]{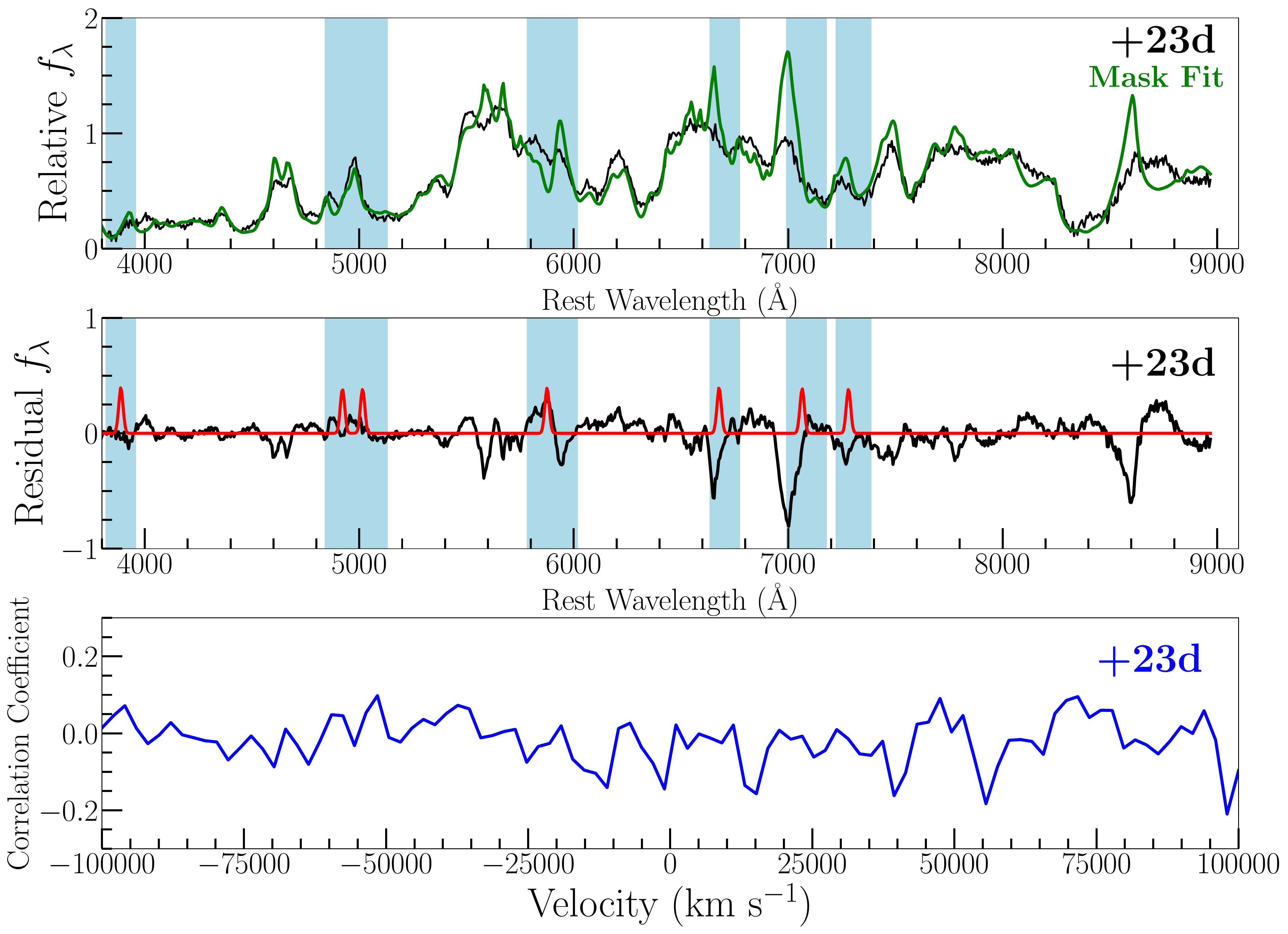}
	\caption{LSQ12fhs. Phase relative to B band maximum.} \label{fig:combo_12fhs}
\end{center}
\vspace*{-5mm}
\end{figure}

\begin{figure}
\begin{center}
	\includegraphics[width=0.49\textwidth]{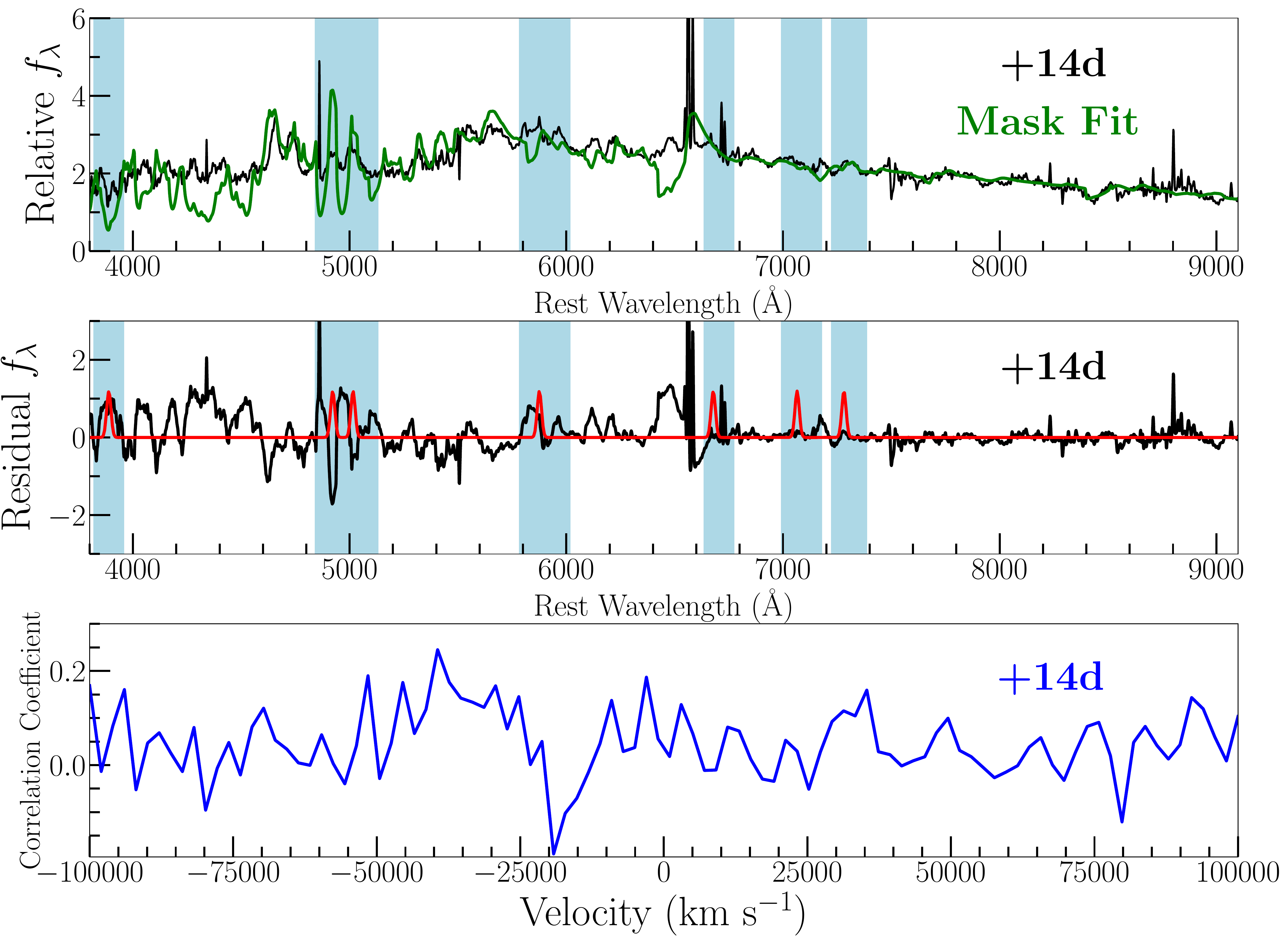}
	\caption{SN~2013dh. Phase relative to B band maximum.} \label{fig:combo_13dh}
\end{center}
\vspace*{-5mm}
\end{figure}

\begin{figure*}
\subfigure[]{\includegraphics[width=.47\textwidth]{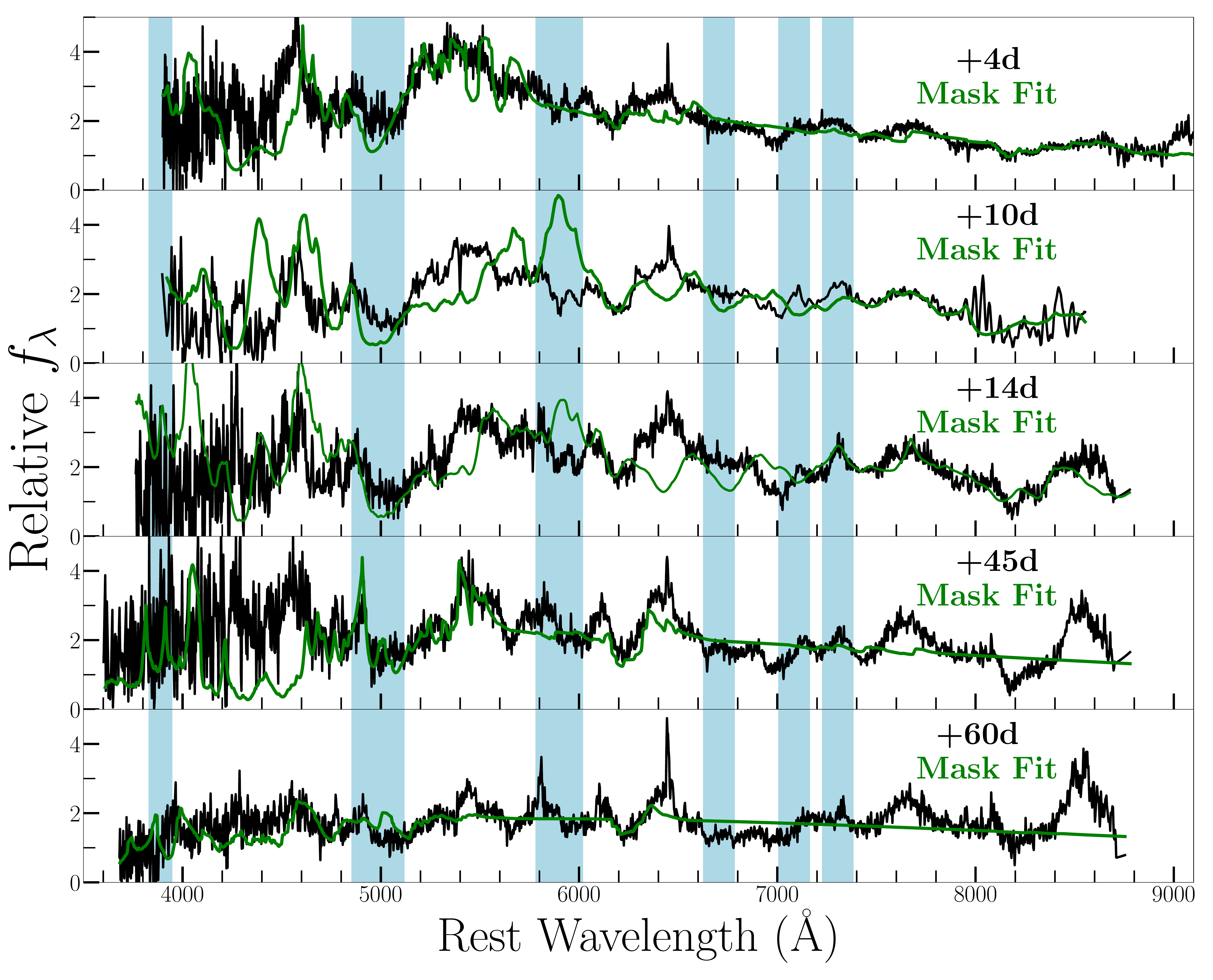}}
\subfigure[]{\includegraphics[width=.47\textwidth]{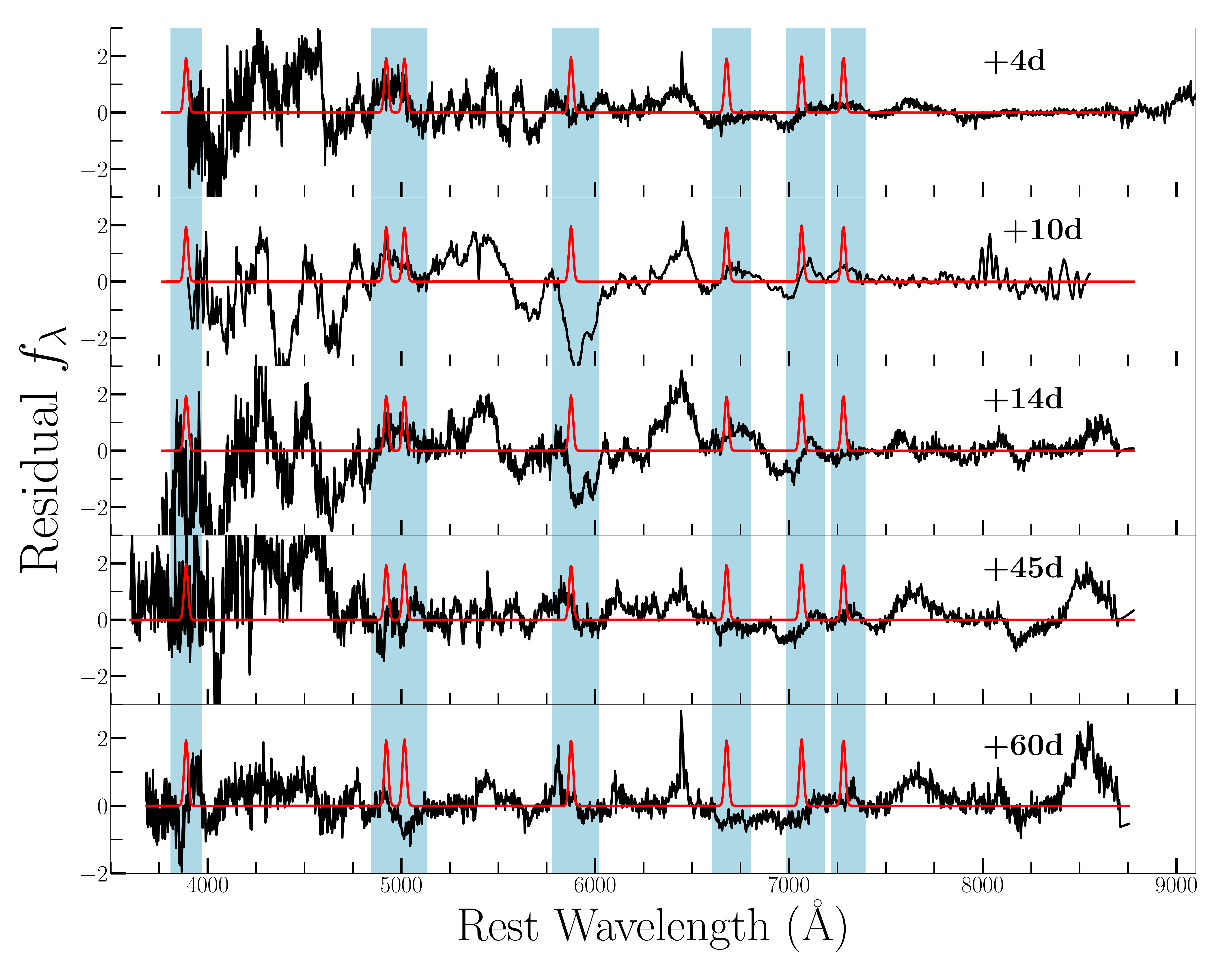}}\\[1ex]
\subfigure[]
{\includegraphics[width=0.8\textwidth]{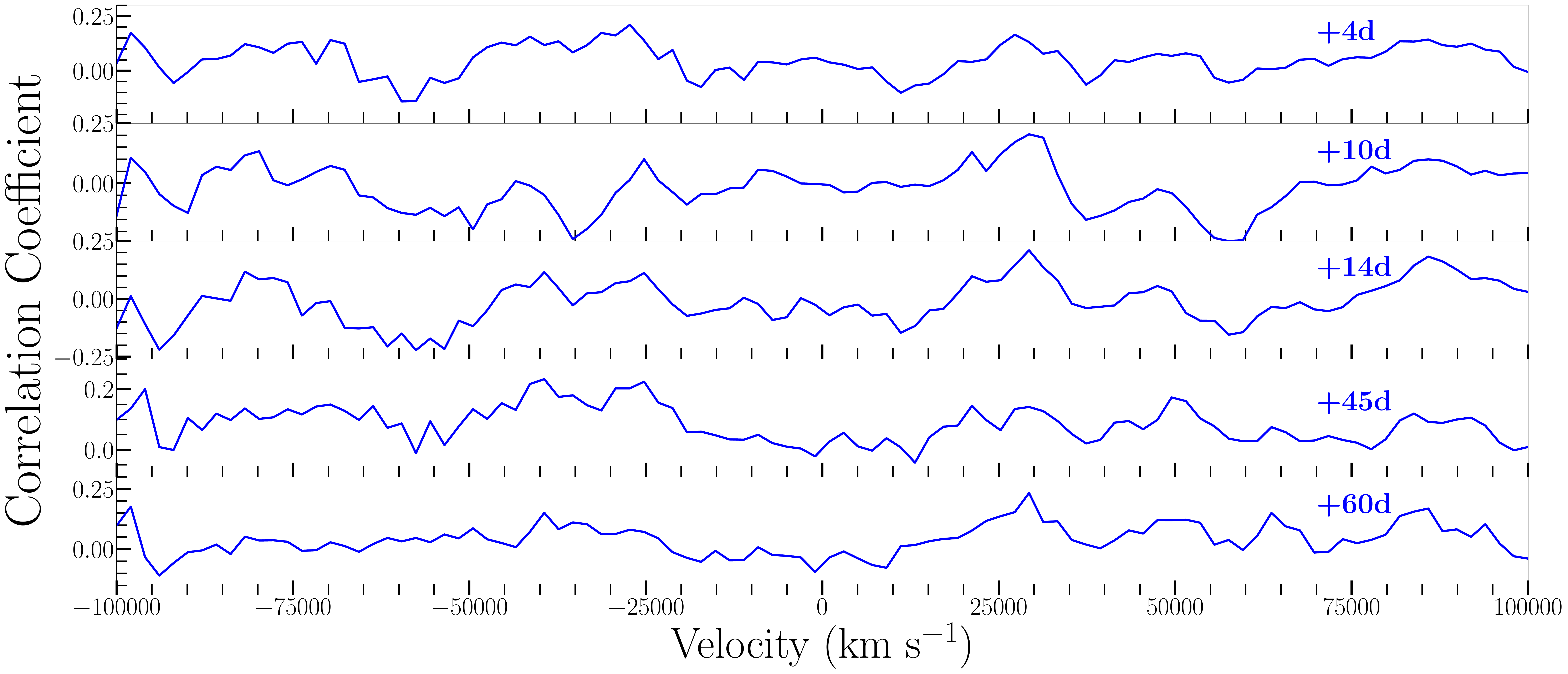}}
\caption{SN~2013en. Phase relative to B band maximum.  \label{fig:13en_combo} }
\end{figure*}

\begin{figure*}
\subfigure[]{\includegraphics[width=.47\textwidth]{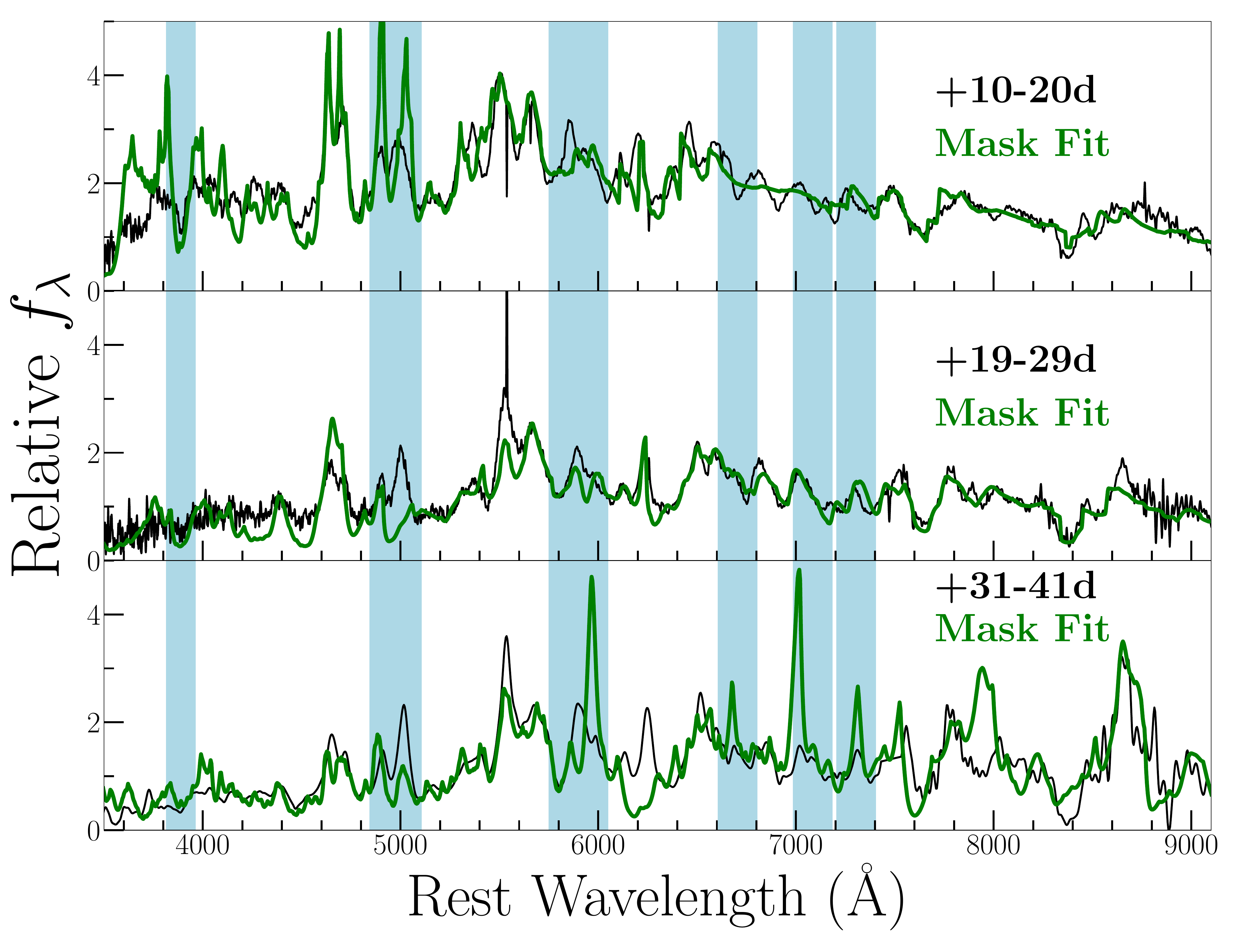}}
\subfigure[]{\includegraphics[width=.47\textwidth]{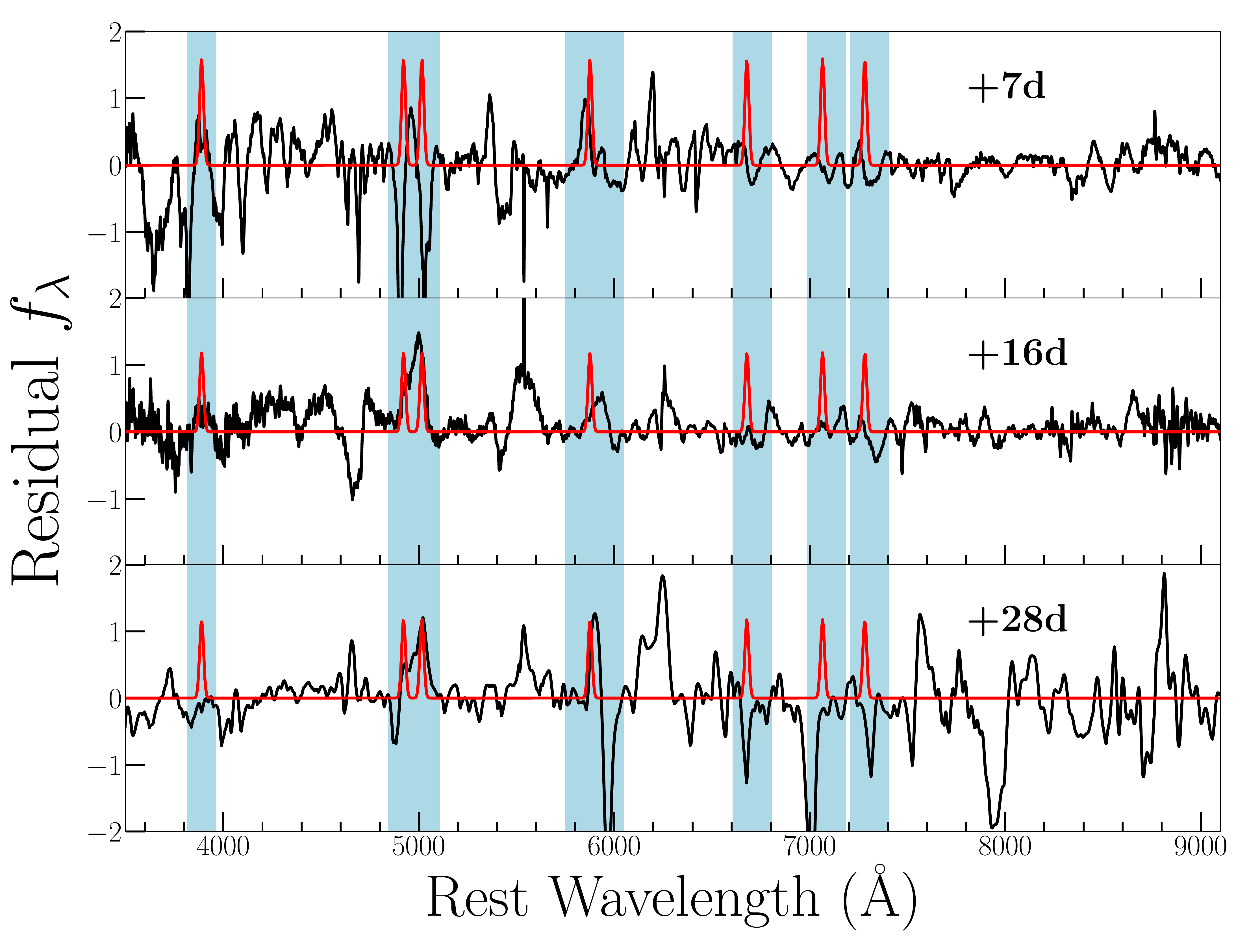}}\\[1ex]
\subfigure[]
{\includegraphics[width=0.8\textwidth]{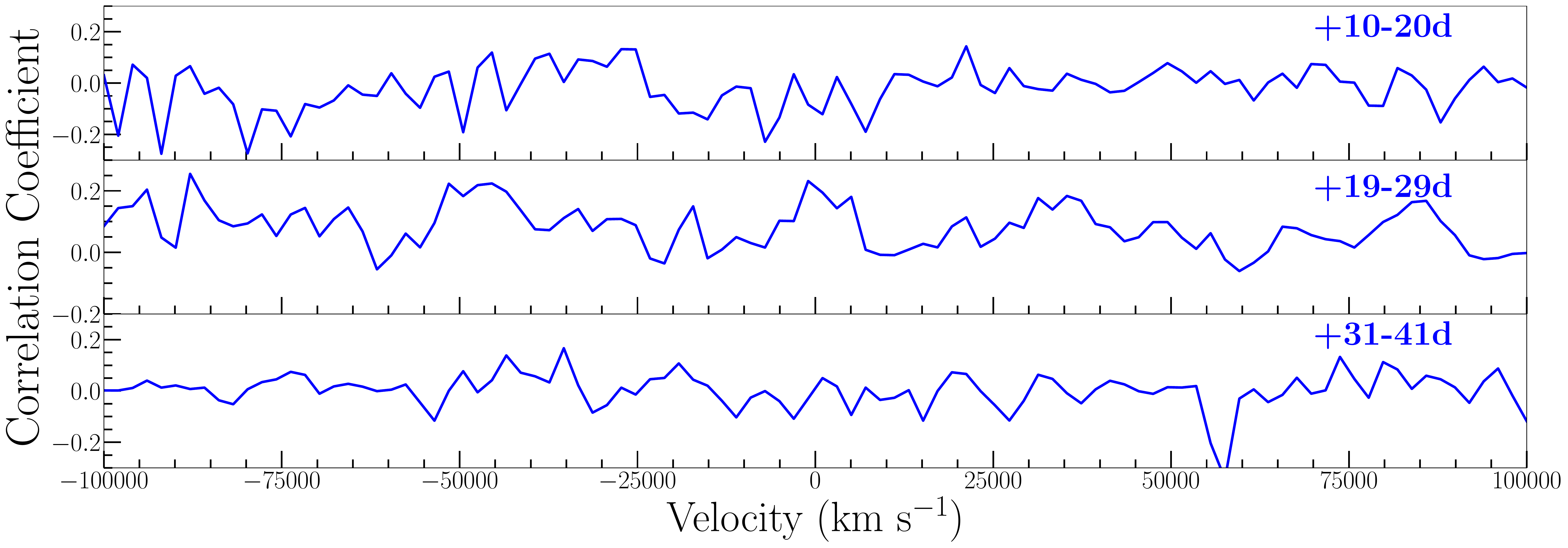}}
\caption{SN~2013gr. Phase relative to B band maximum and calculated using SNID.  \label{fig:13gr_combo} }
\end{figure*}

\begin{figure*}
\subfigure[]{\includegraphics[width=.47\textwidth]{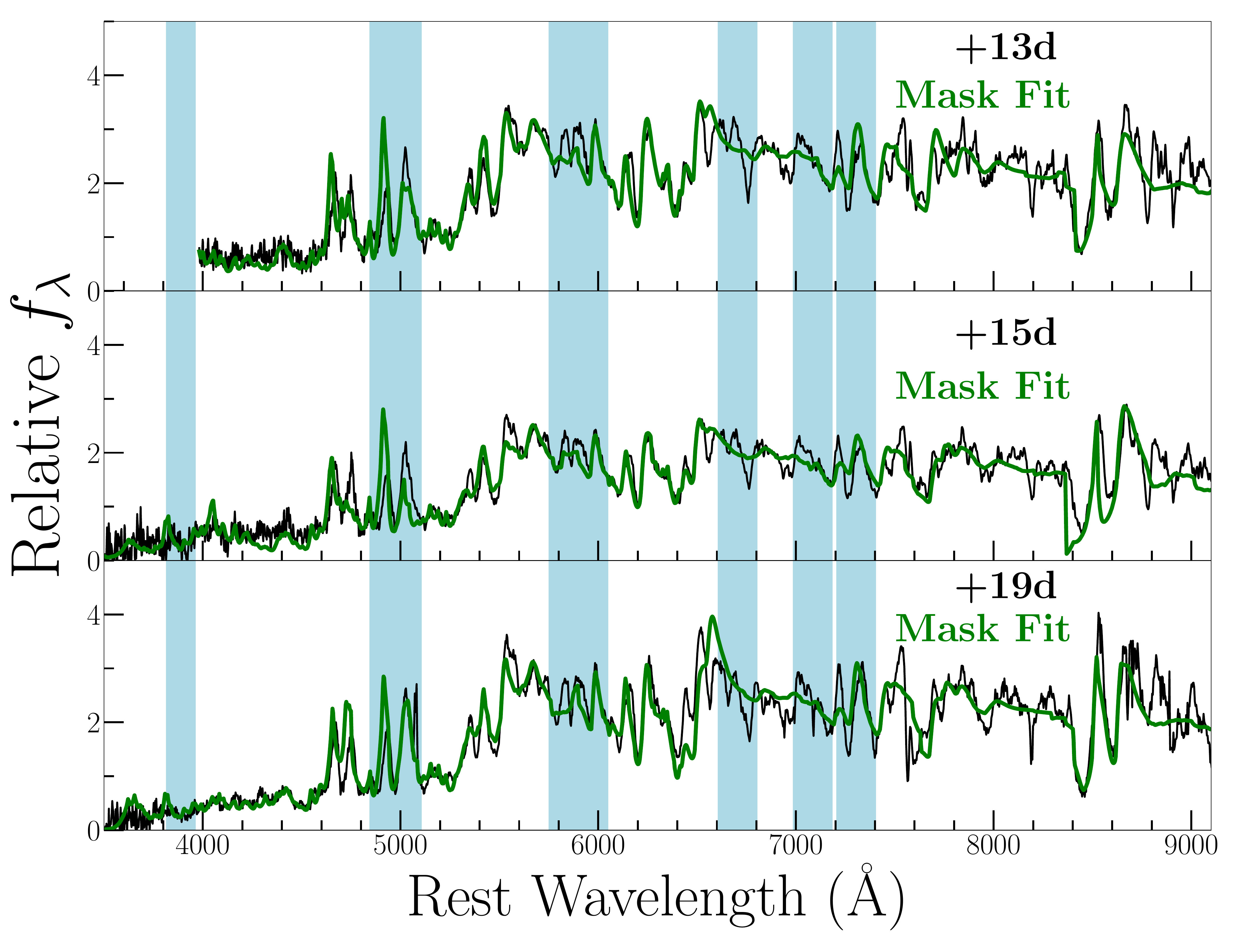}}
\subfigure[]{\includegraphics[width=.47\textwidth]{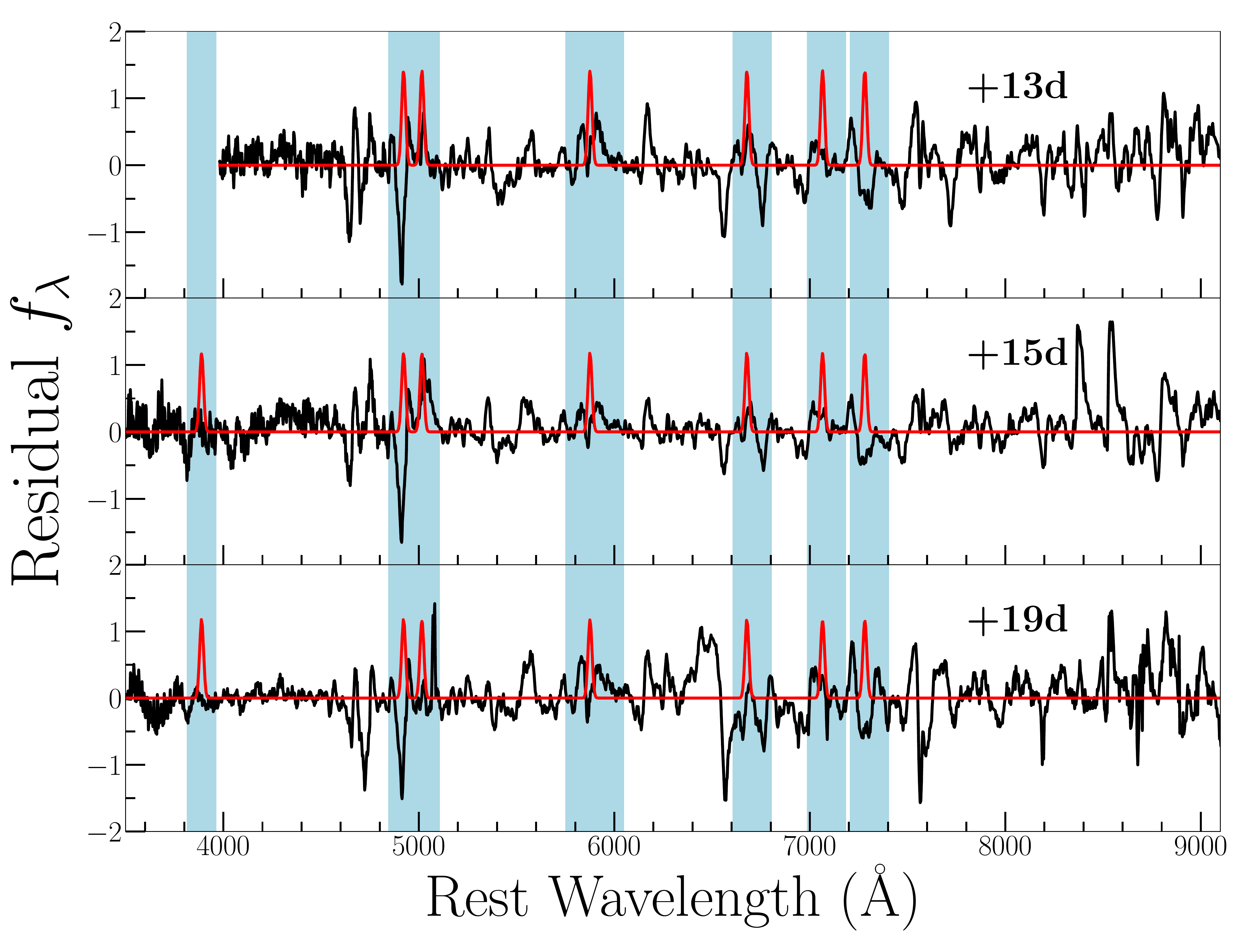}}\\[1ex]
\subfigure[]
{\includegraphics[width=0.8\textwidth]{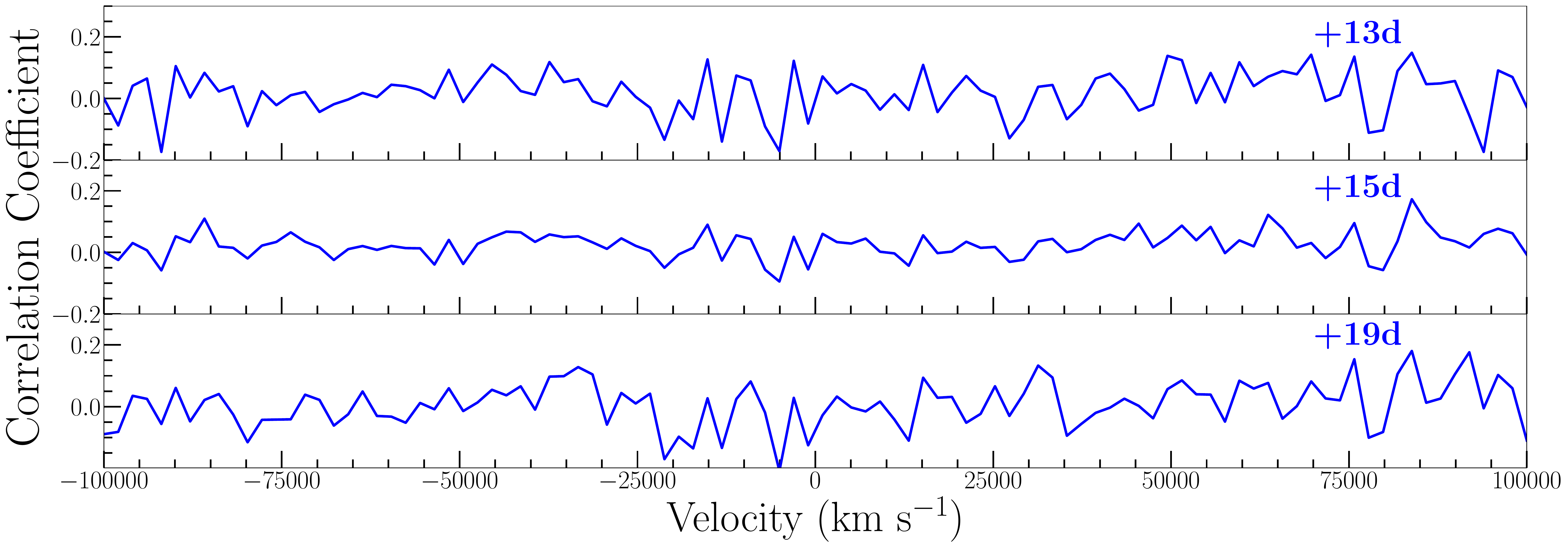}}
\caption{SN~2014ck. Phase relative to B band maximum.   \label{fig:14ck_combo} }
\end{figure*}

\begin{figure}
\begin{center}
	\includegraphics[width=0.49\textwidth]{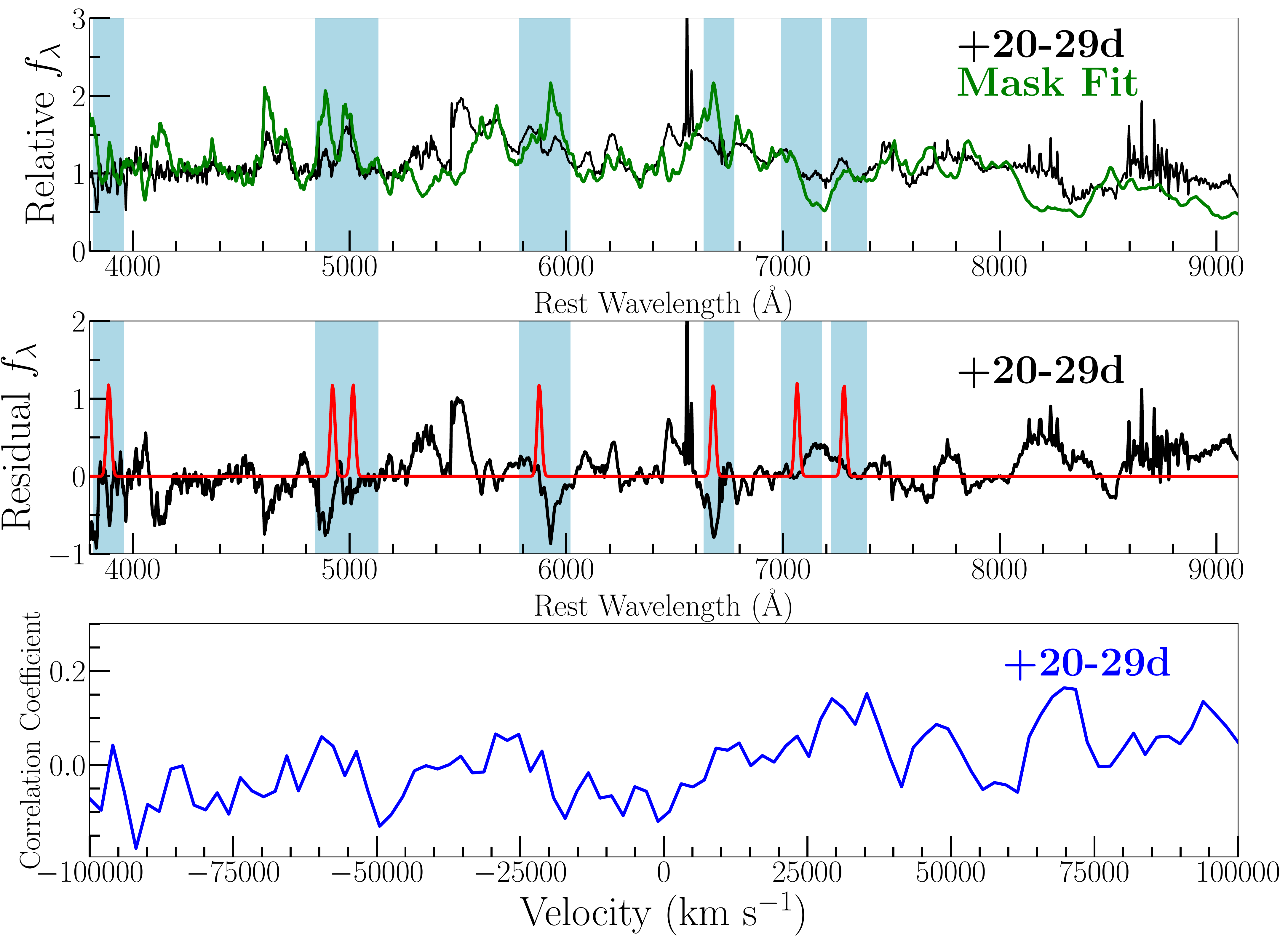}
	\caption{SN~2014cr. Phase relative to B band maximum and calculated using SNID.} \label{fig:combo_14cr}
\end{center}
\vspace*{-5mm}
\end{figure}

\begin{figure}
\begin{center}
	\includegraphics[width=0.49\textwidth]{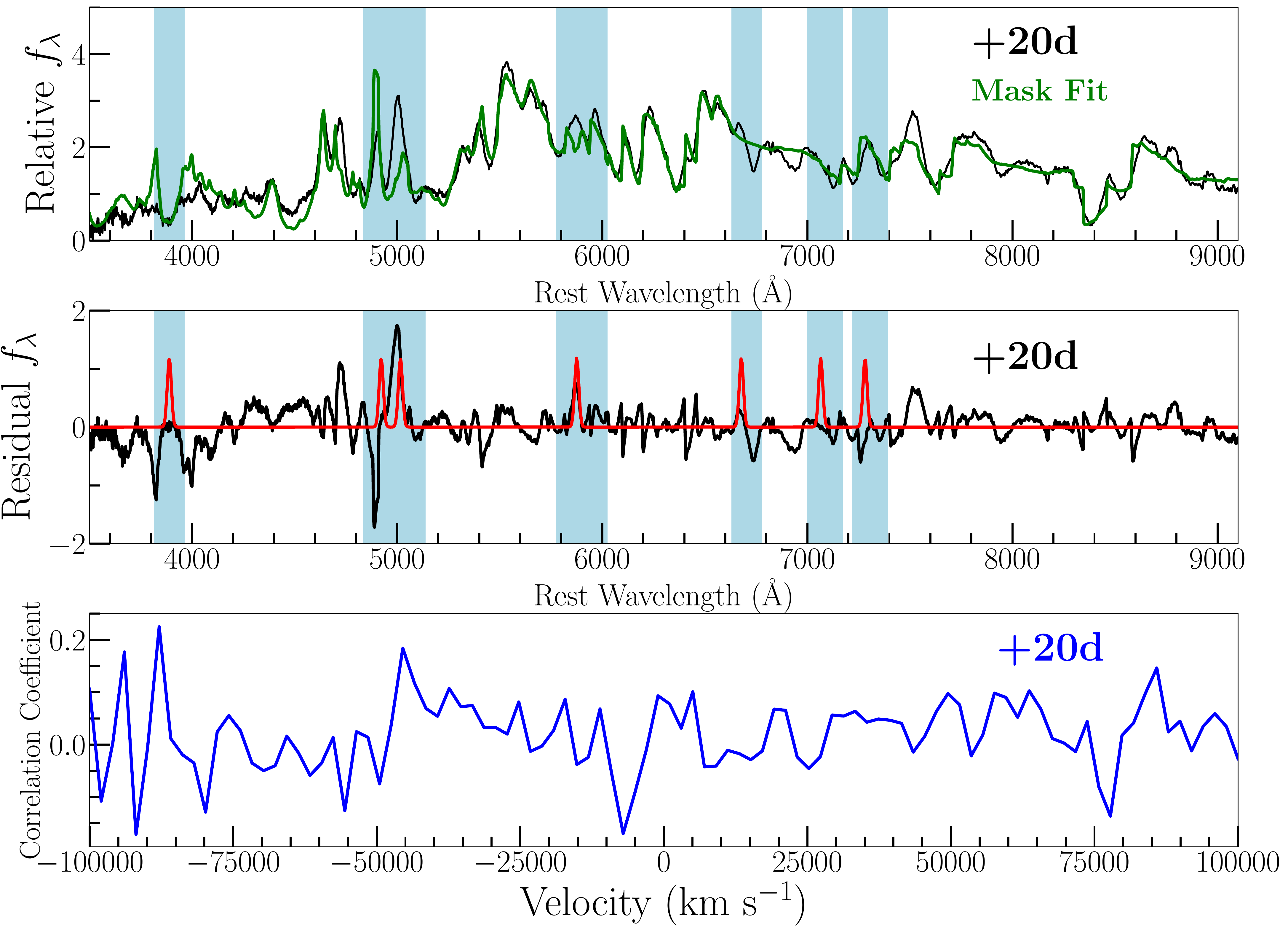}
	\caption{SN~2014dt. Phase relative to B band maximum.} \label{fig:combo_14dt}
\end{center}
\vspace*{-5mm}
\end{figure}

\begin{figure}
\begin{center}
	\includegraphics[width=0.49\textwidth]{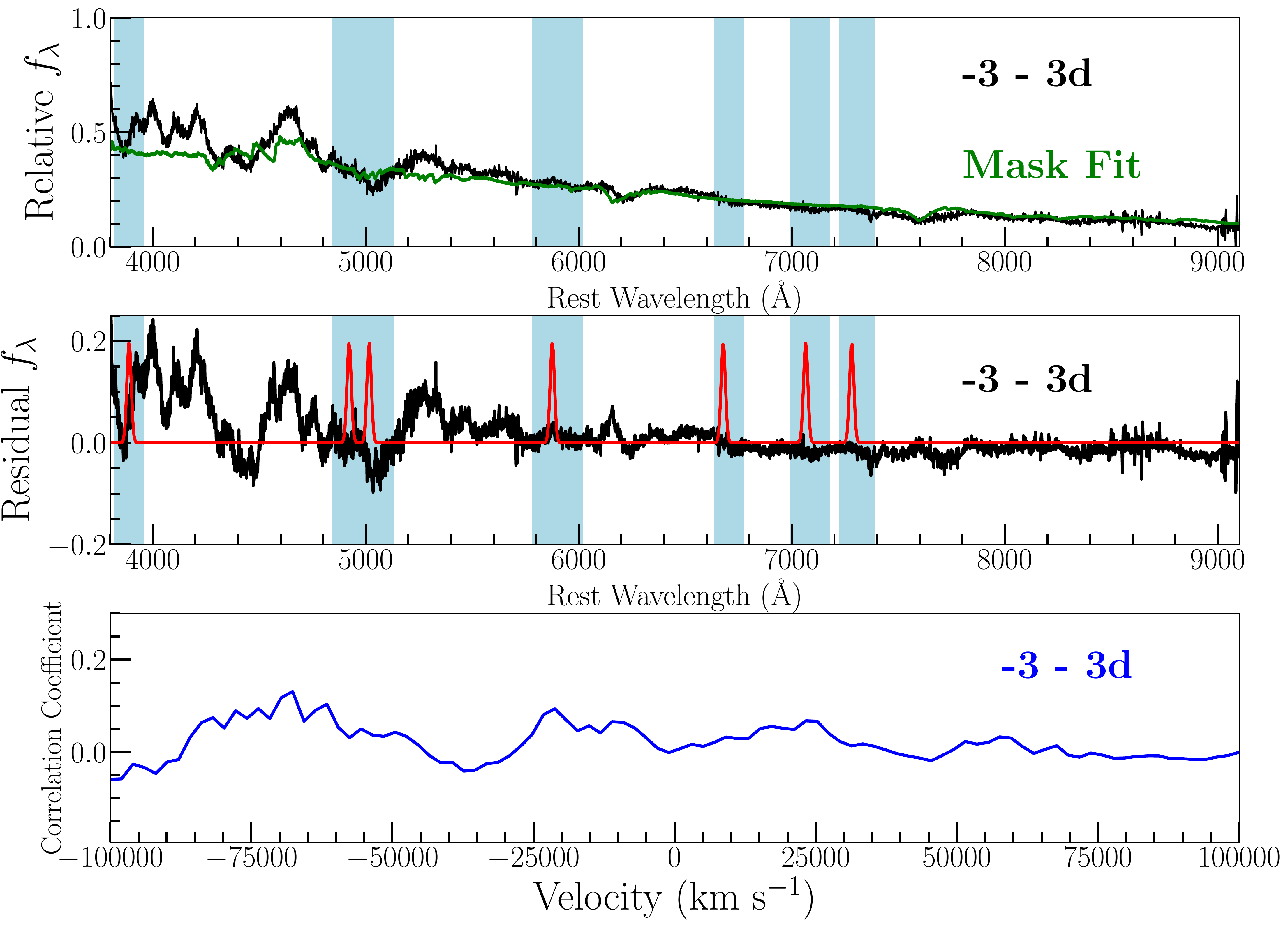}
	\caption{SN~2014ey. Phase relative to B band maximum and calculated using SNID.} \label{fig:combo_14ey}
\end{center}
\vspace*{-5mm}
\end{figure}

\begin{figure}
\begin{center}
	\includegraphics[width=0.49\textwidth]{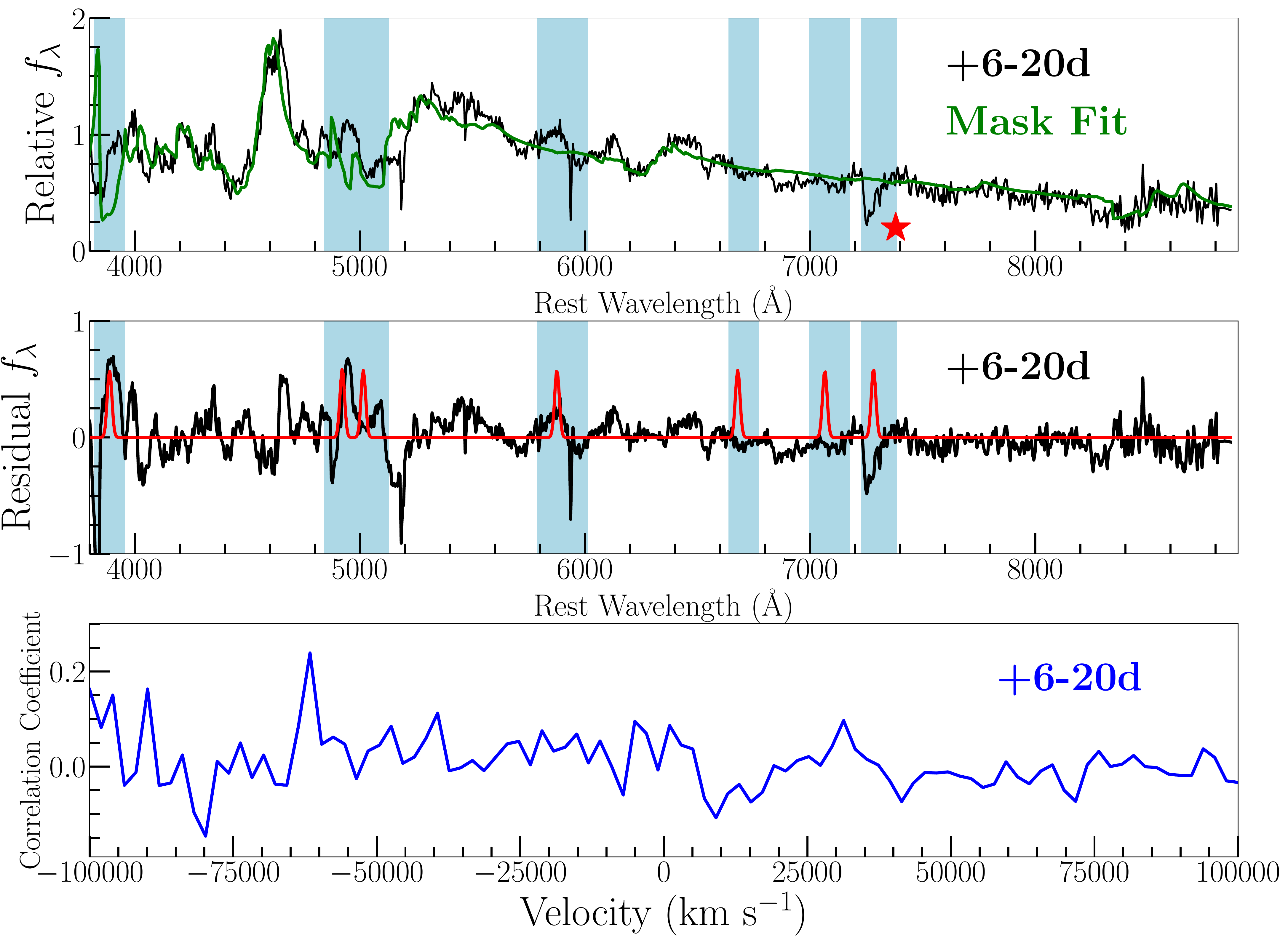}
	\caption{LSQ14dtt. Phase relative to B band maximum and calculated using SNID. Red star indicates telluric absorption.} \label{fig:combo_14dtt}
\end{center}
\vspace*{-5mm}
\end{figure}

\begin{figure*}
\subfigure[]{\includegraphics[width=.47\textwidth]{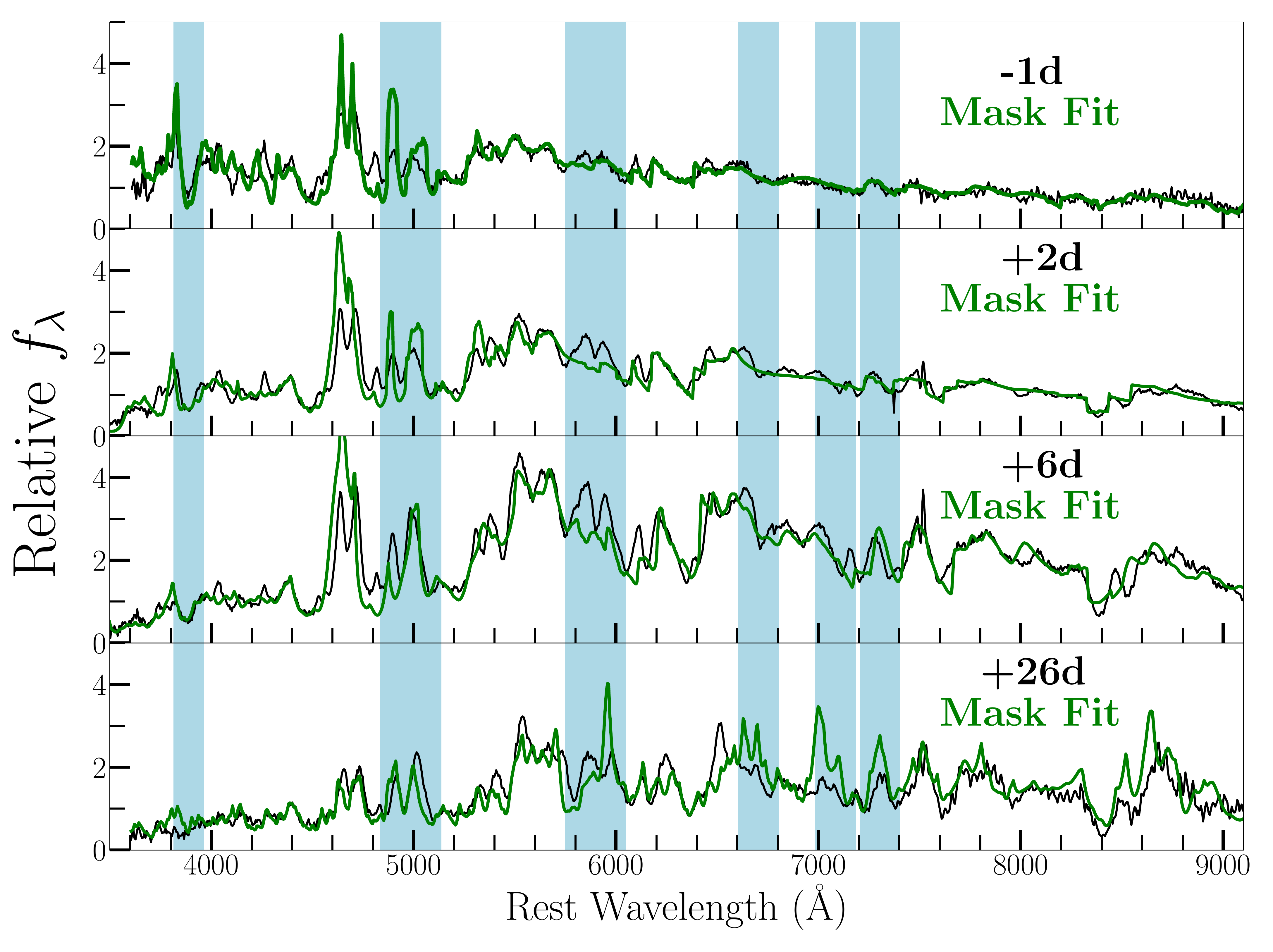}}
\subfigure[]{\includegraphics[width=.47\textwidth]{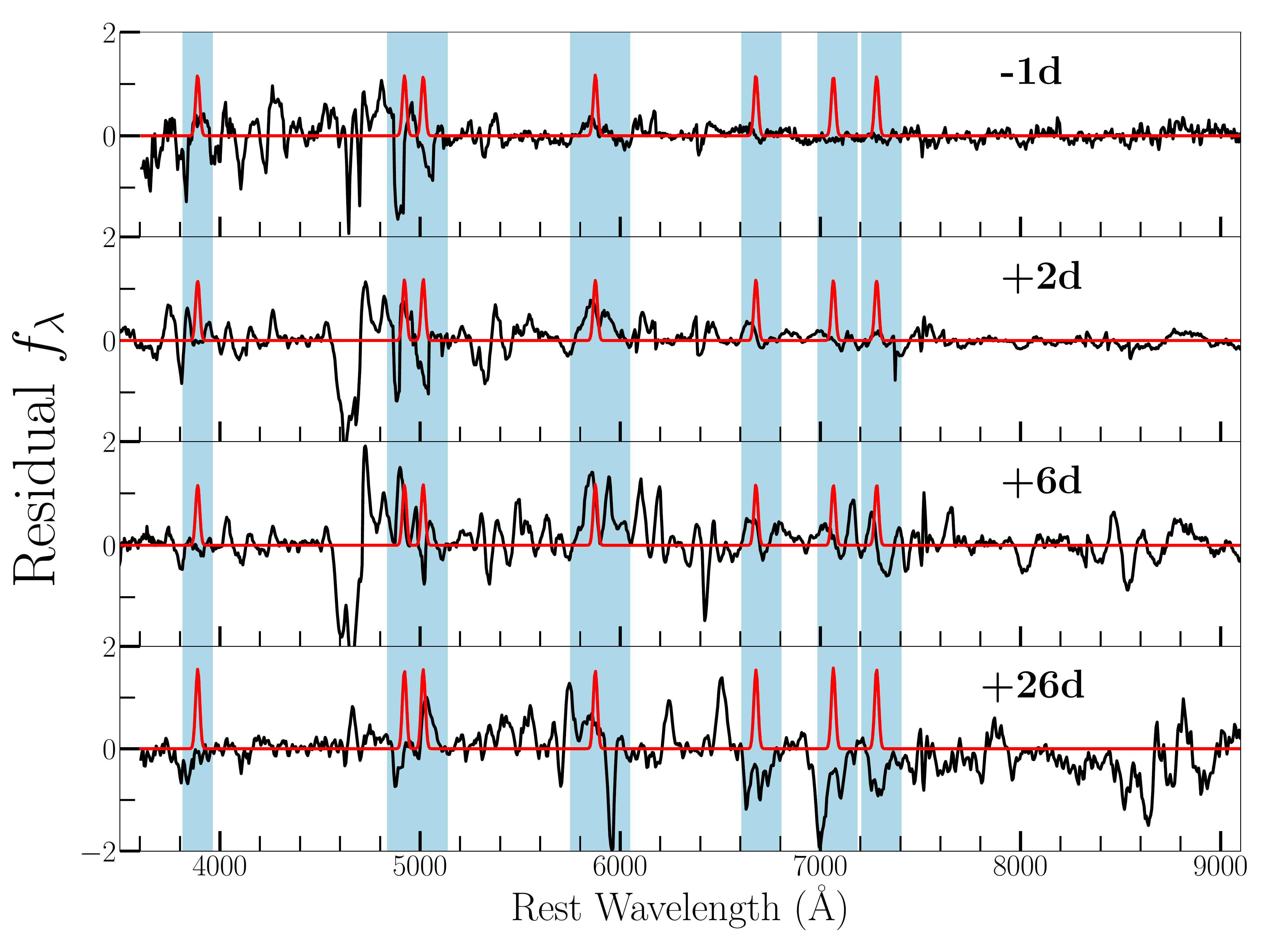}}\\[1ex]
\subfigure[]
{\includegraphics[width=0.8\textwidth]{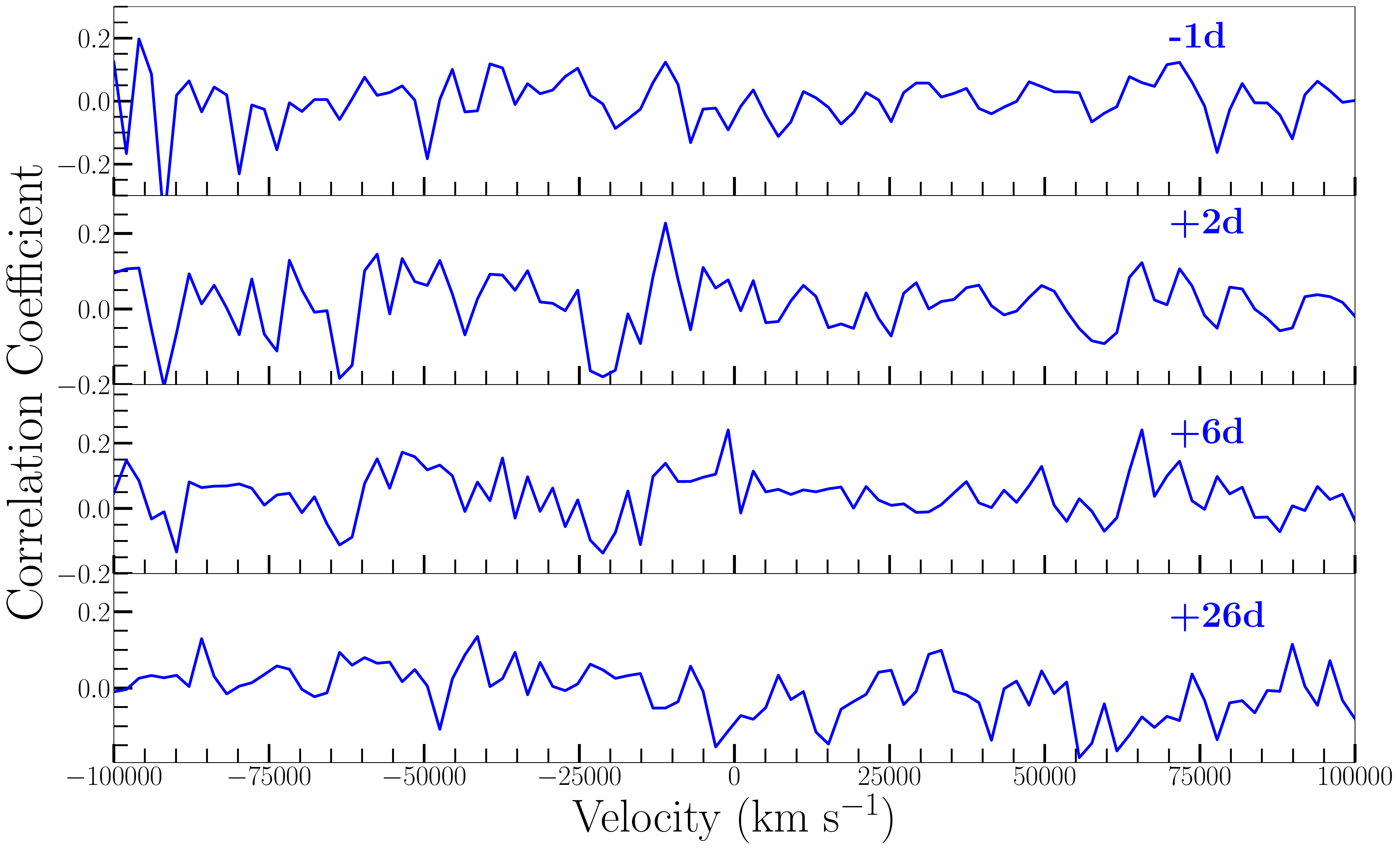}}
\caption{SN~2015H. Phase relative to B band maximum. \label{fig:15H_combo} }
\end{figure*}

\begin{figure}
\begin{center}
	\includegraphics[width=0.49\textwidth]{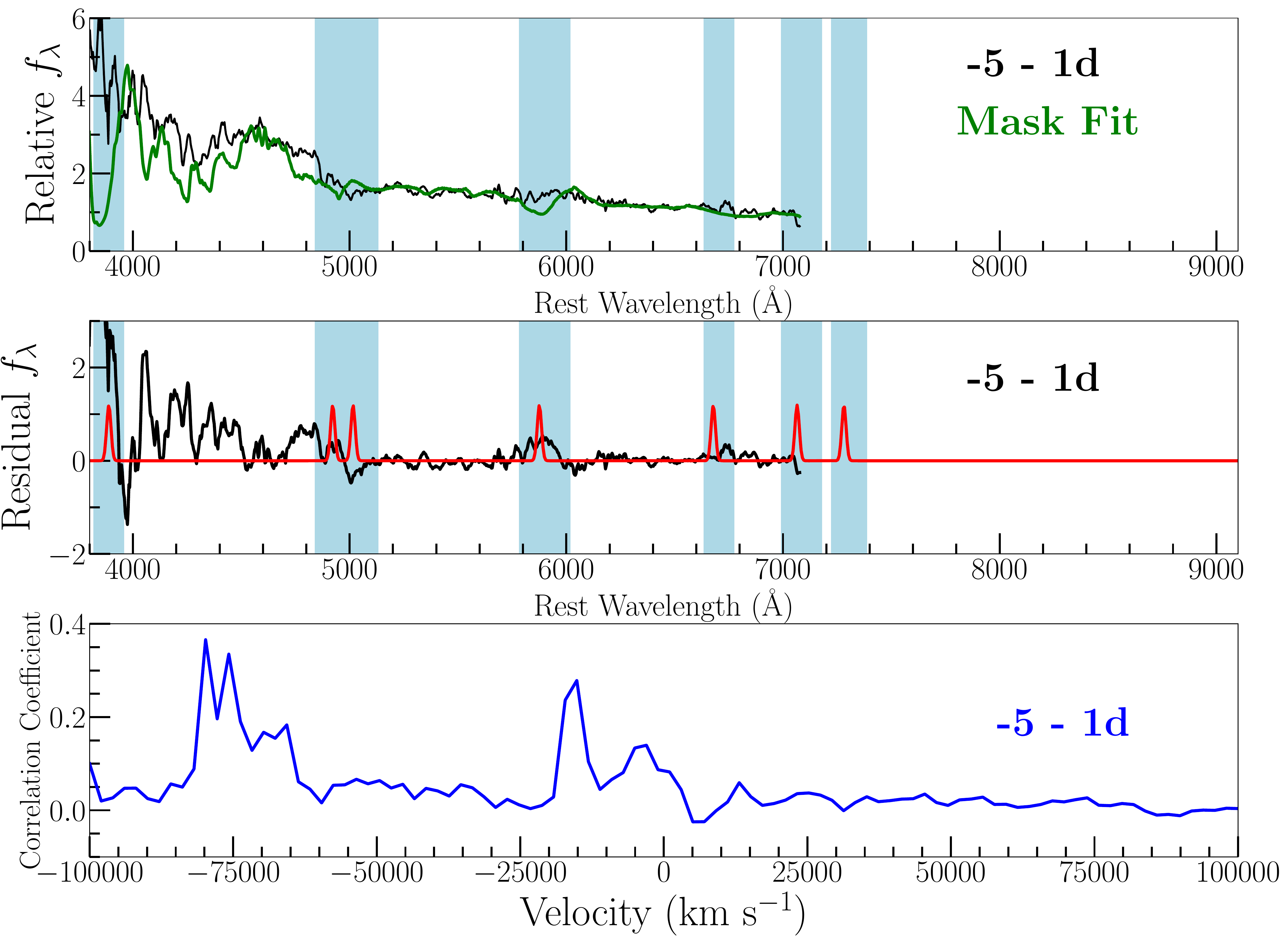}
	\caption{SN~2015ce. Phase relative to B band maximum and calculated using SNID.} \label{fig:combo_15ce}
\end{center}
\vspace*{-5mm}
\end{figure}

\begin{figure}
\begin{center}
	\includegraphics[width=0.49\textwidth]{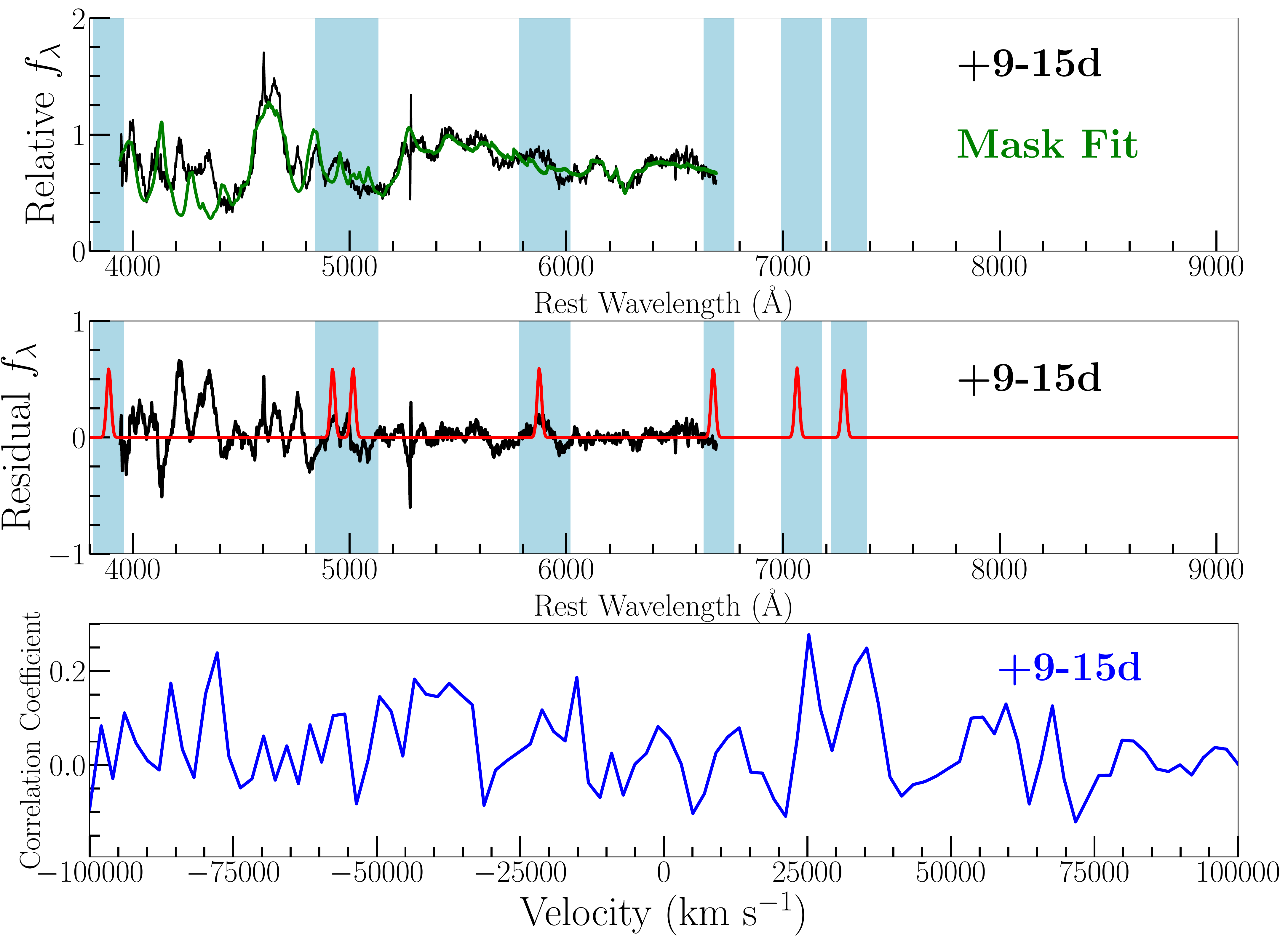}
	\caption{PS15aic. Phase relative to B band maximum and calculated using SNID.} \label{fig:combo_15aic}
\end{center}
\vspace*{-5mm}
\end{figure}

\begin{figure}
\begin{center}
	\includegraphics[width=0.49\textwidth]{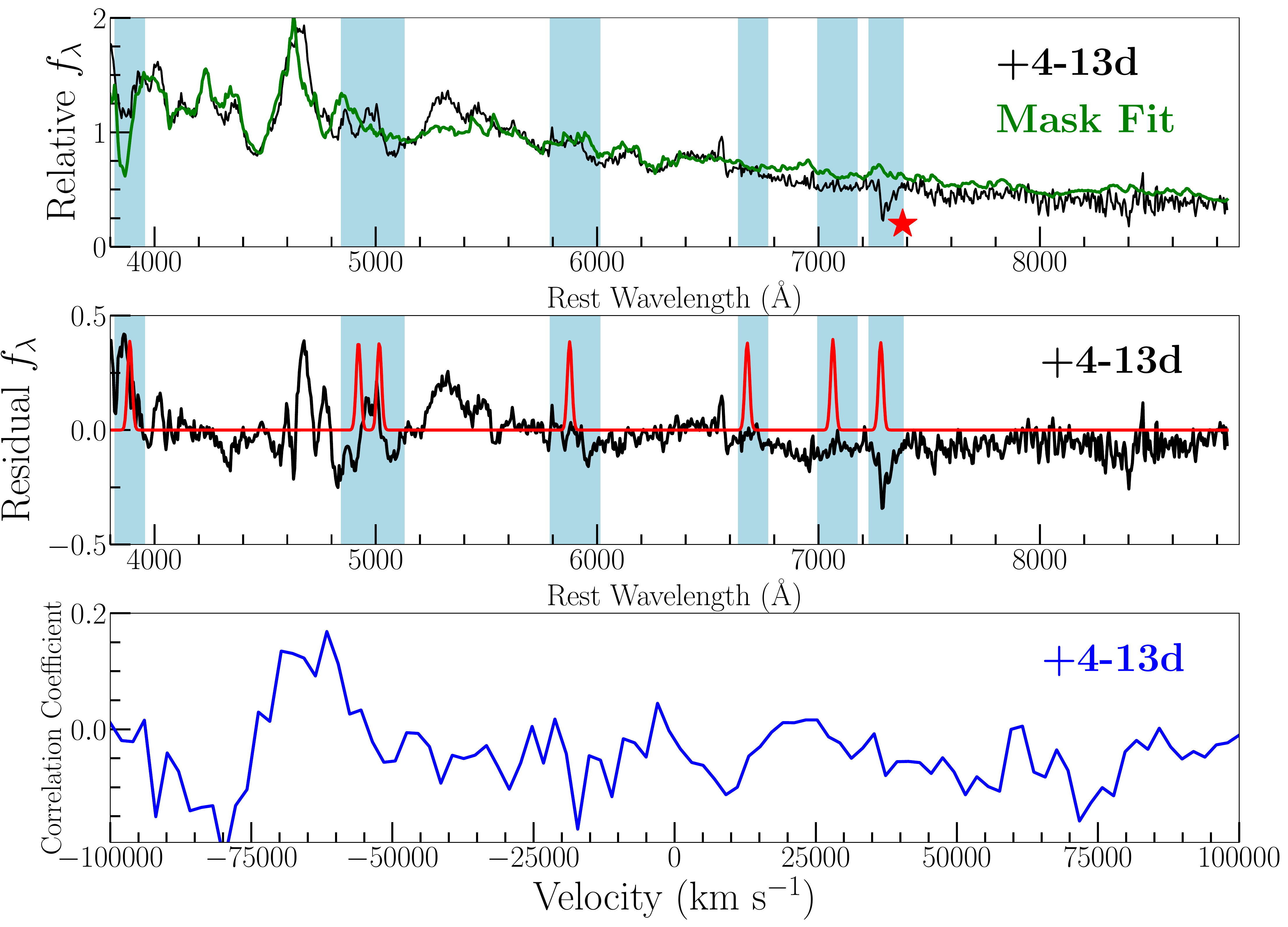}
	\caption{PS15csd. Phase relative to B band maximum and calculated using SNID. Red star indicates telluric absorption.} \label{fig:combo_15csd}
\end{center}
\vspace*{-5mm}
\end{figure}

\begin{figure}
\begin{center}
	\includegraphics[width=0.49\textwidth]{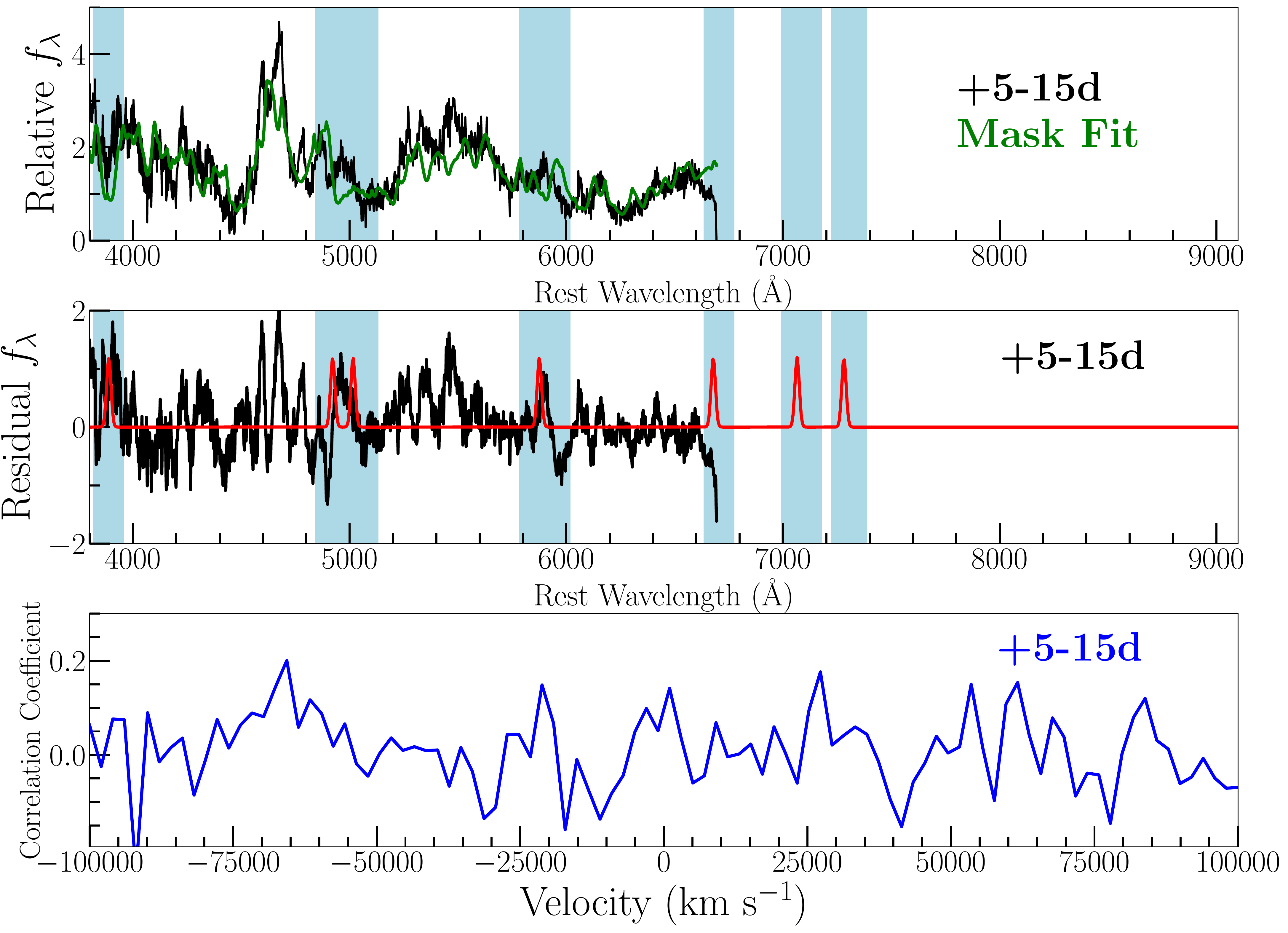}
	\caption{SN~2016atw. Phase relative to B band maximum and calculated using SNID.} \label{fig:combo_16atw}
\end{center}
\vspace*{-5mm}
\end{figure}

\begin{figure}
\begin{center}
	\includegraphics[width=0.49\textwidth]{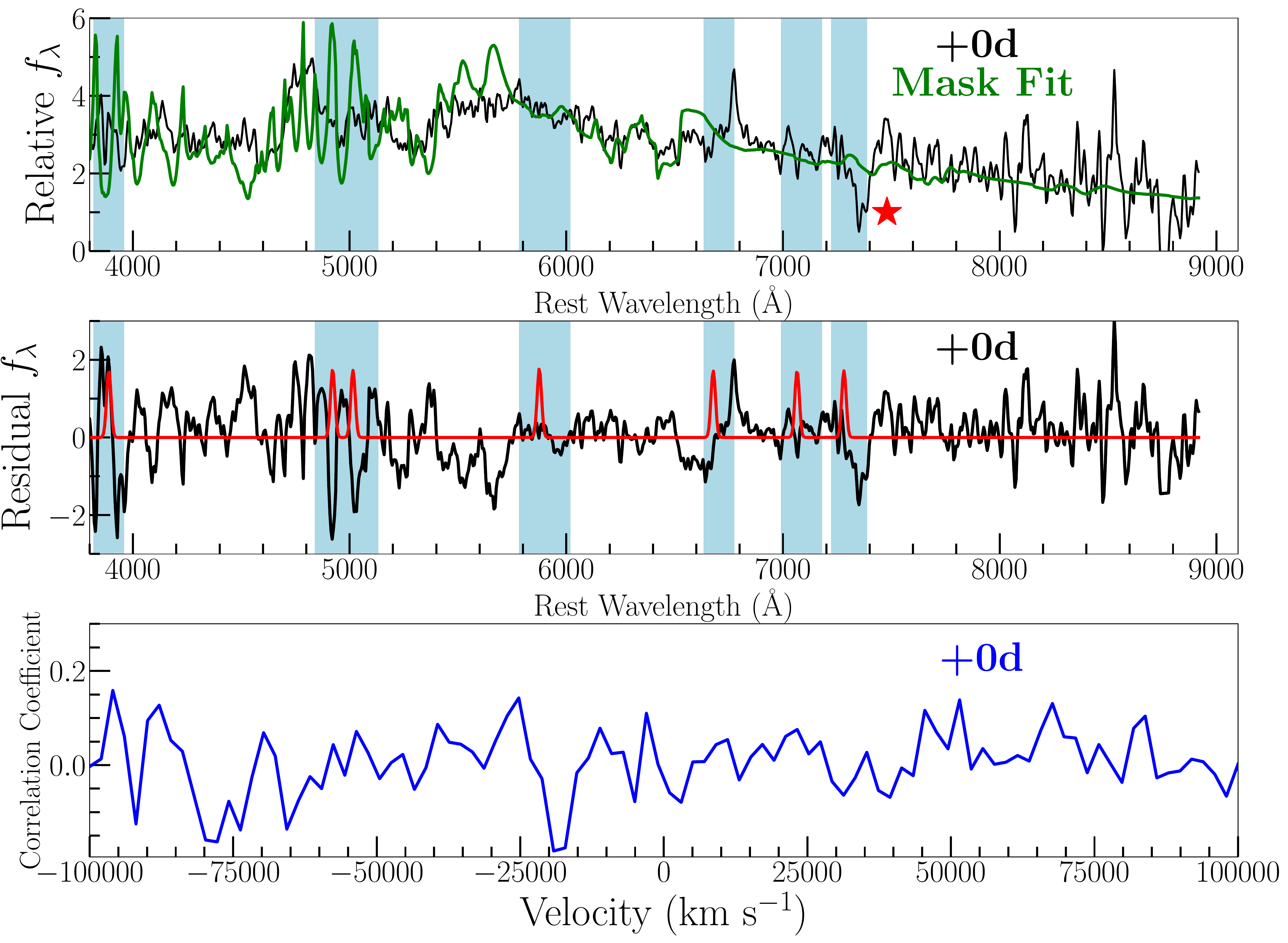}
	\caption{OGLE16erd. Phase relative to B band maximum. Red star indicates telluric absorption.} \label{fig:combo_16erd}
\end{center}
\vspace*{-5mm}
\end{figure}

\begin{figure}
\begin{center}
	\includegraphics[width=0.49\textwidth]{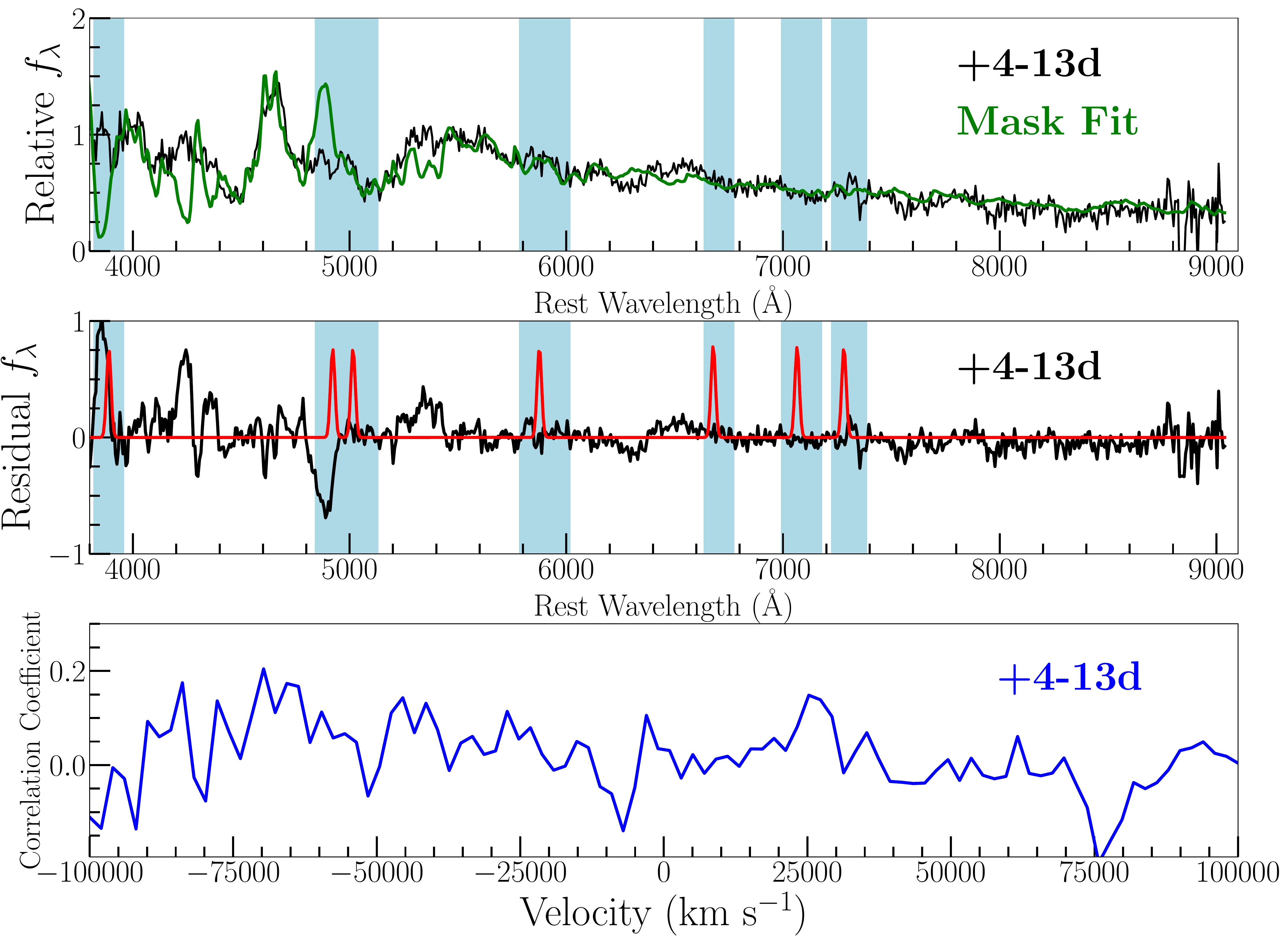}
	\caption{SN~2016ilf. Phase relative to B band maximum and calculated using SNID.} \label{fig:combo_16ilf}
\end{center}
\vspace*{-5mm}
\end{figure}

\begin{figure*}
\subfigure[]{\includegraphics[width=.47\textwidth]{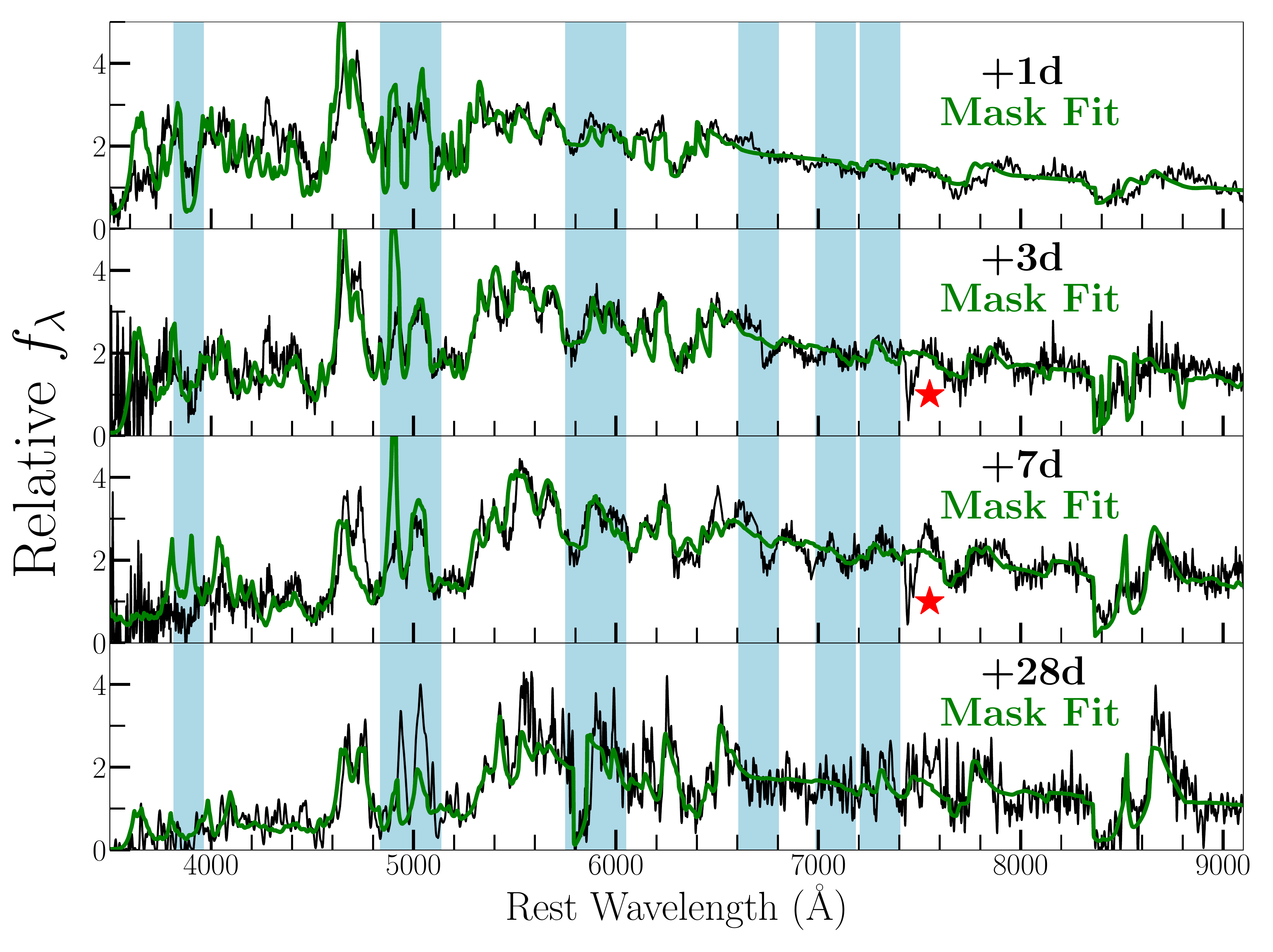}}
\subfigure[]{\includegraphics[width=.47\textwidth]{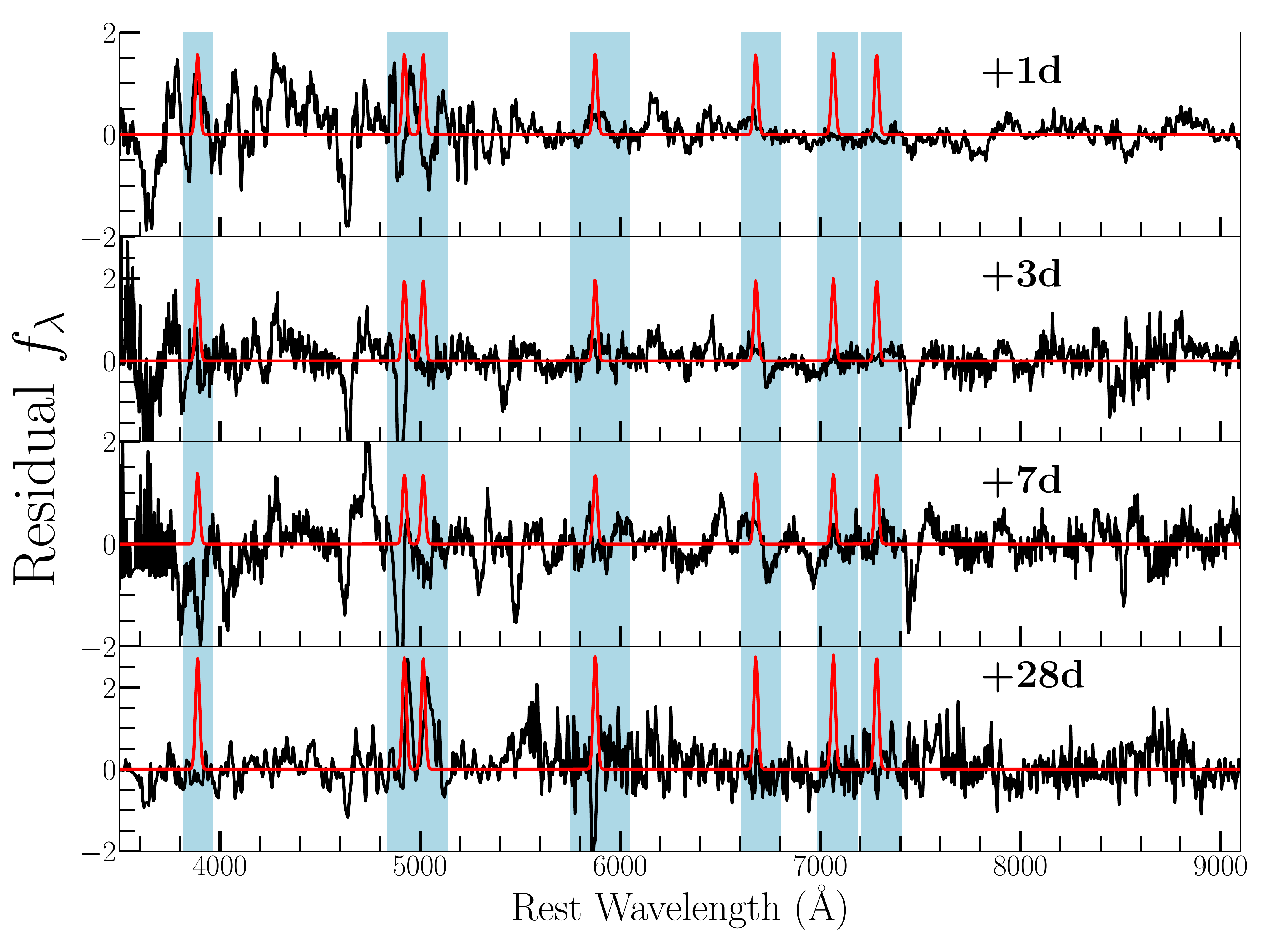}}\\[1ex]
\subfigure[]
{\includegraphics[width=0.8\textwidth]{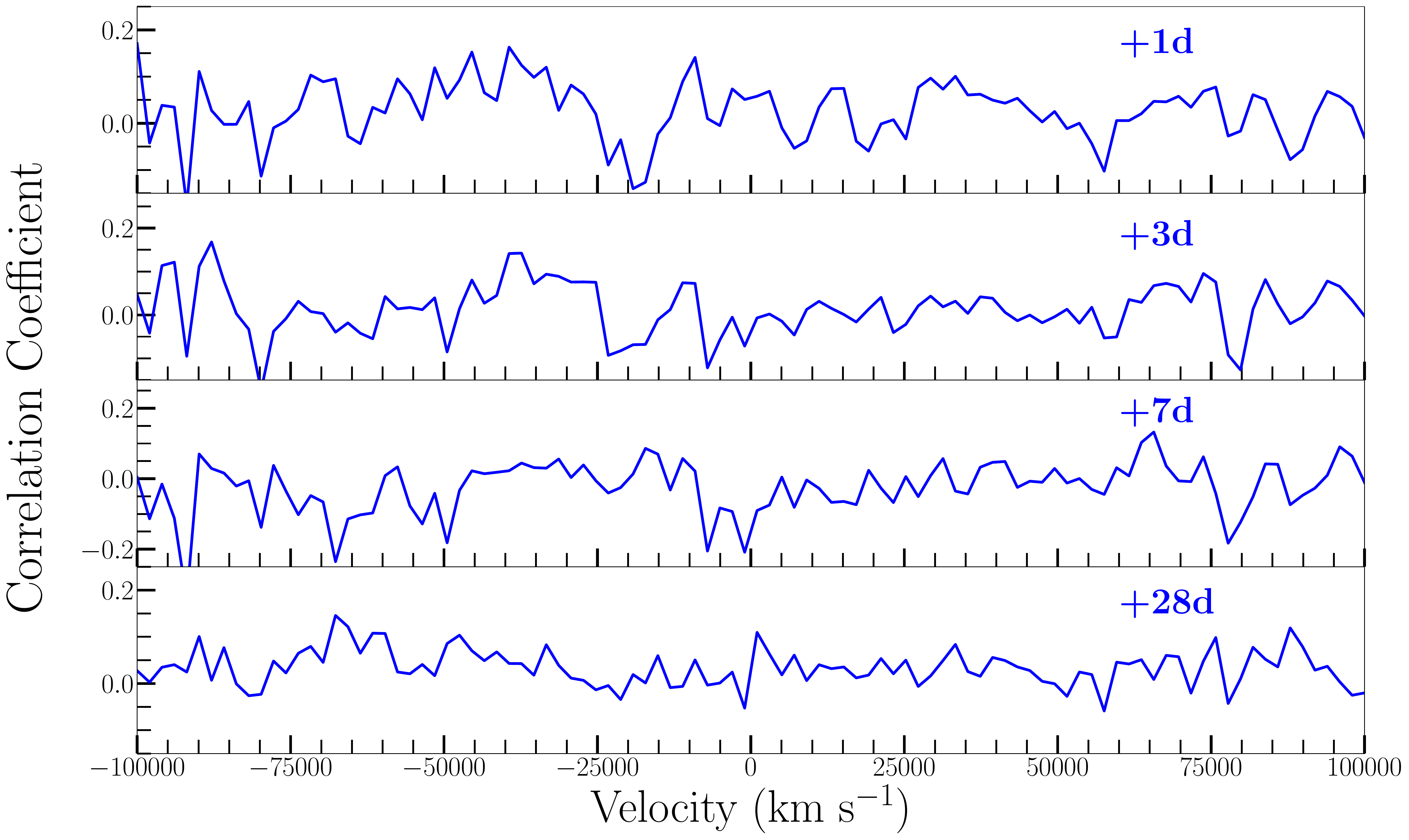}}
\caption{iPTF16fnm. Phase relative to B band maximum. Red star indicates telluric absorption.  \label{fig:16fnm_combo} }
\end{figure*}

\begin{figure}
\begin{center}
	\includegraphics[width=0.49\textwidth]{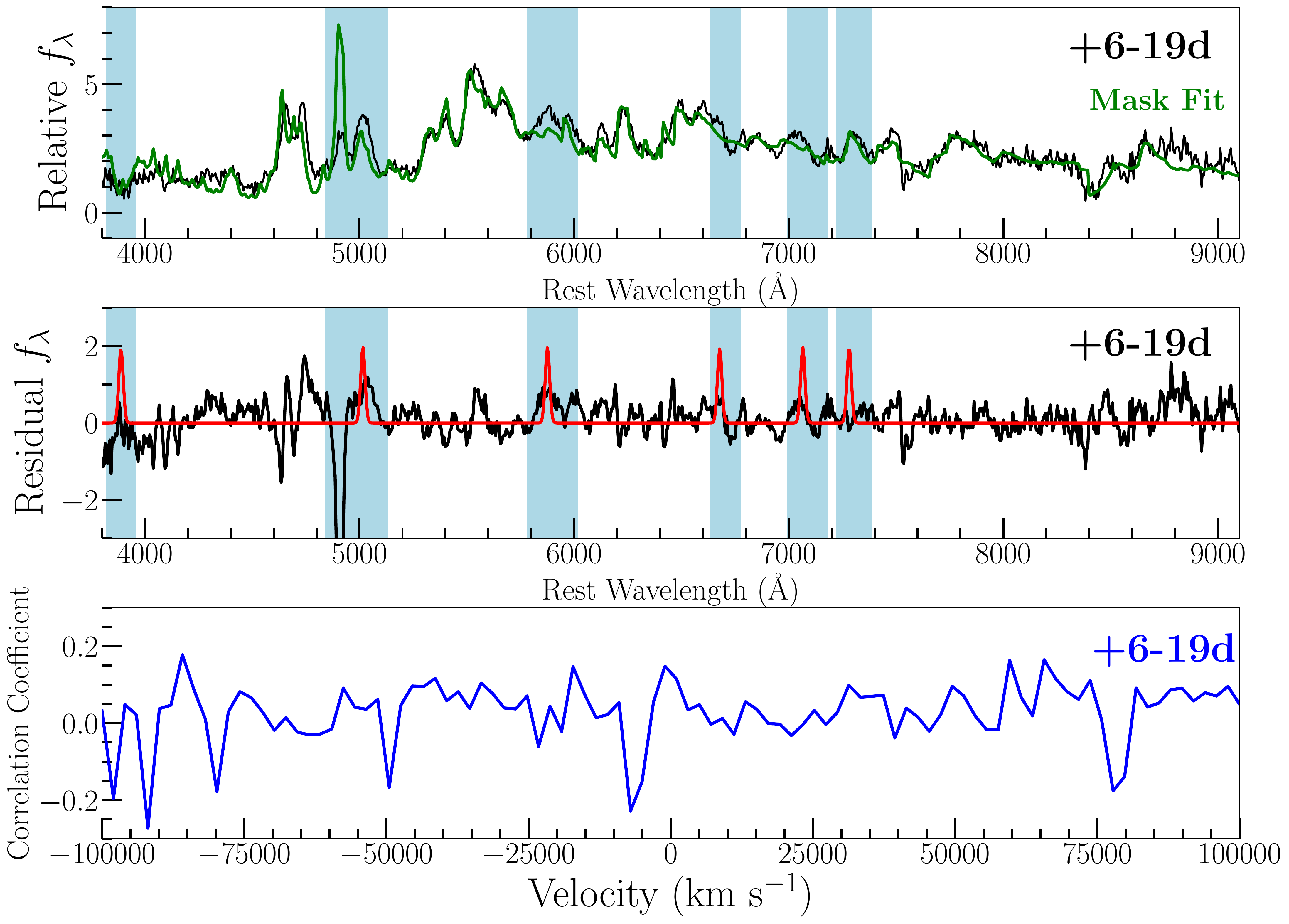}
	\caption{SN~2017gbb. Phase relative to B band maximum and calculated using SNID.} \label{fig:combo_17gbb}
\end{center}
\end{figure}

\begin{figure}
\begin{center}
	\includegraphics[width=0.49\textwidth]{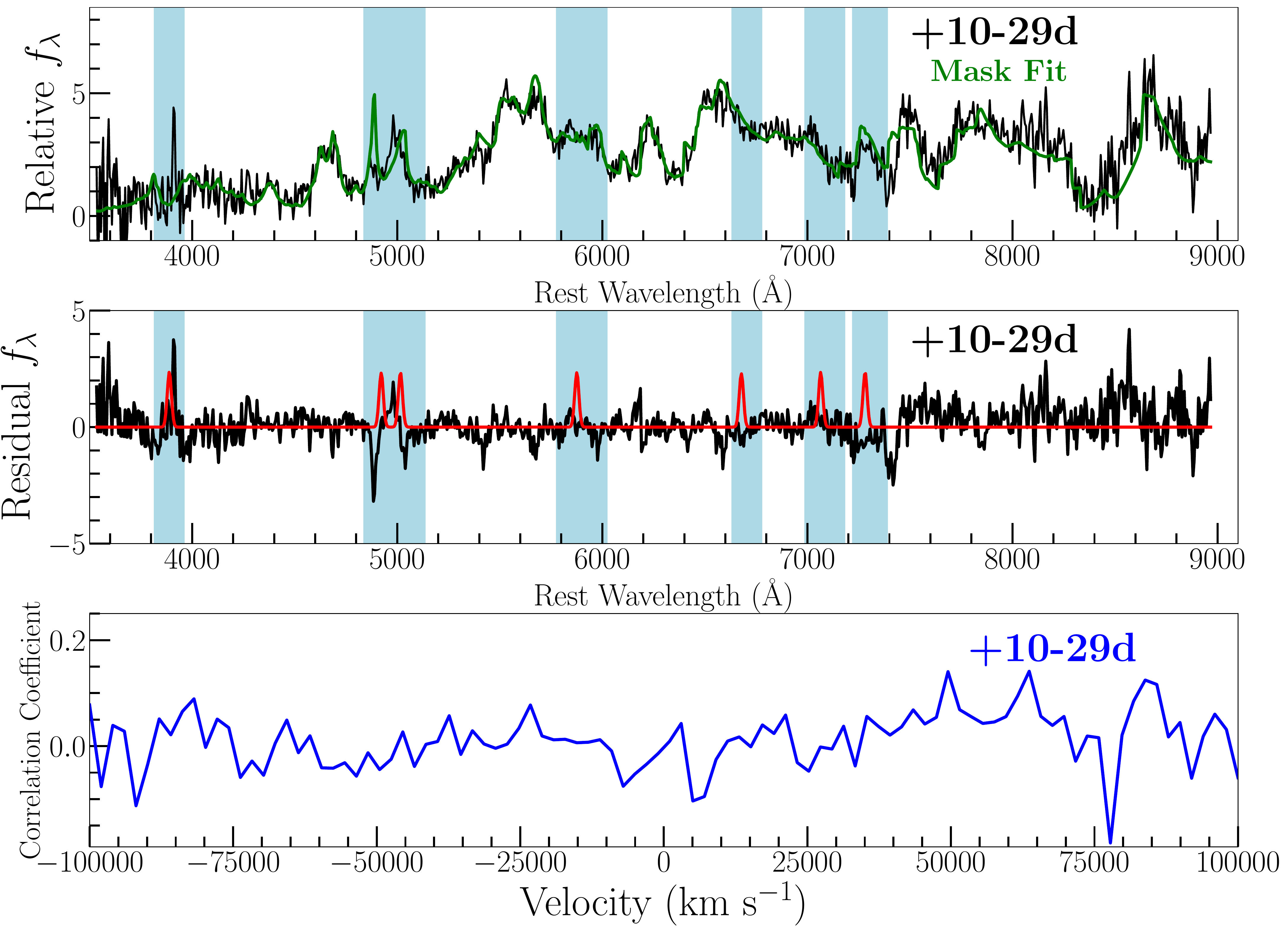}
	\caption{SN~2018atb. Phase relative to B band maximum and calculated using SNID.} \label{fig:combo_18atb}
\end{center}
\end{figure}


\bsp	
\label{lastpage}
\end{document}